\documentclass[a4paper,10pt,twoside]{article}

\usepackage{latexsym}
\usepackage{amssymb}
\usepackage{graphicx}
\usepackage{epsfig}

\newcommand{\Dm}[1]{\begin{equation} #1 \end{equation}}
\newcommand{\Ea}[1]{\begin{eqnarray} #1 \end{eqnarray}}

\newcommand{\Ds}{\displaystyle}

\newcommand{\Vec}[1]{\ensuremath{\mathbf{#1}}}

\newcommand{\Rea}{\ensuremath{\mathrm{Re}} }
\newcommand{\Imm}{\ensuremath{\mathrm{Im}} }
\newcommand{\To}{\ensuremath{\rightarrow}}

\newcommand{\kz}{\ensuremath{K^0} }
\newcommand{\kzb}{\ensuremath{\overline{K}^0} }
\newcommand{\ks}{\ensuremath{K_S} }
\newcommand{\kl}{\ensuremath{K_L} }
\newcommand{\dz}{\ensuremath{D^0} }
\newcommand{\dzb}{\ensuremath{\overline{D}^0} }
\newcommand{\bz}{\ensuremath{B^0} }
\newcommand{\bzb}{\ensuremath{\overline{B}^0} }

\newcommand{\beps}{\ensuremath{\overline{\epsilon}} }
\newcommand{\epseps}{\ensuremath{\epsilon'/\epsilon}}

\newcommand{\etal}{\emph{et al.}}

\title{Measurements of direct $CP$ violation}
\author{M. S. Sozzi\thanks{CERN, EP div., CH-1211 Geneva 23, 
Switzerland. E-mail: \texttt{marco.sozzi\@cern.ch}. On leave from Scuola 
Normale Superiore, Pisa, Italy.} and I. Mannelli\thanks{Scuola Normale 
Superiore, Piazza dei Cavalieri 7, 56126, Pisa, Italy. E-mail: 
\texttt{mannelli@sns.it}}}
\date{June 2003}

\begin{document}

\maketitle


\tableofcontents


\section{Introduction}

The experimental fact that not all natural phenomena are insensitive to the
mirror-reversal of their spatial arrangement and the simultaneous replacement
of all the particles with the corresponding antiparticles, \emph{i.e.} the 
violation of $CP$ symmetry (or $CP$ violation in short), is a feature of the 
fundamental physical laws which is not yet satisfactorily explained.

Since its discovery \cite{FitchCronin} 39 years ago, $CP$ violation has been a 
very active field of research at the heart of particle physics. The reason is
that $CP$ violation is deeply related to such fundamental and diverse issues 
\cite{FitchCronin_nobel} as the microscopic time-reversibility of physical 
laws and the origin of the baryonic asymmetry of the universe; it is the only 
known phenomenon which allows an absolute distinction between particles and 
anti-particles \cite{Sakurai}.
The very successful Standard Model of particle physics (often indicated as SM 
in what follows) can accommodate $CP$ violation within its framework, but it 
does not shade any real light on its origin.

Since $CP$ violation was discovered in the decays of flavoured neutral mesons,
which remain to date the only systems in which it has been observed, the term 
``\emph{direct} $CP$ violation'' was coined to indicate effects appearing in 
the amplitudes describing the physical decays of such particles, as opposed to 
effects in the peculiar meson-antimeson oscillations which such systems
exhibit, dubbed as manifestations of ``\emph{indirect} $CP$ violation''.
For quite some time, all measured $CP$ violation effects could be related to a 
single phenomenological parameter describing an asymmetry in the mixing of the 
neutral kaon and its antiparticle (indirect $CP$ violation), despite many 
experiments devoted to searches for effects in other systems, and in 
particular differences in the decay properties of particles and anti-particles.

The 1990's marked important progress for the understanding of $CP$ violation, 
with the first experimental evidence of such phenomenon outside the kaon 
system coming just a few years after the proof of the existence of direct $CP$ 
violation in neutral kaon decays was obtained.

The unequivocal evidence of the occurrence of \emph{direct} $CP$ violation in 
Nature, recently obtained after a search several decades long, clears the way 
to the interpretation of the lack of symmetry under $CP$ as a truly universal 
property of weak interactions. As such it is expected to be present, albeit to 
different degree, in several aspects of Nature governed by these interactions.

To date, the original observation of direct $CP$ violation in neutral kaon 
decays remains the only detected manifestation of this kind of effect.
A review of the current situation, and of the long road which led to it, 
seems therefore appropriate at this time, when important milestones have been 
achieved, and several new experimental studies are underway or in preparation.

The literature on $CP$ violation is extremely vast, and even a compilation of 
the most significant papers would be a daunting task; introductory discussions 
can be found in any particle physics textbook, and many review articles exist. 
The literature can be traced starting from recent \cite{Branco}, 
\cite{BigiSanda} or by now classic \cite{Sachs} books on the subject. 
First-hand accounts of the early years of $CP$ violation can be read in 
\cite{FitchCronin_nobel}, \cite{CP_reviews_blois}, \cite{CP_reviews_chicago}.
Previous review articles specifically devoted to the search for direct $CP$ 
violation include \cite{Gollin} and particularly \cite{Winstein}, while a 
comprehensive review of $CP$ violation effects in kaon decays can be found in 
\cite{Isidori}.

The emphasis of this article is on the description of the experimental 
progress in the field, from its origins to the present experiments. 
The approach is a phenomenological one, and any attempt to make justice to the 
huge theoretical literature would be outside the scope of this work.

The plan of the paper is as follows: in section 2 we review the basic concept 
of direct $CP$ violation, focusing on light and heavy meson decays.
Section 3 - the longest one - is devoted to the search and experimental 
evidence for direct $CP$ violation in the neutral kaon system. 
Section 4 describes searches in charged kaon decays, while section 5 accounts 
for the relevant experimental activities in heavier flavoured meson systems 
($D$ and $B$). 
In section 6 some searches for direct $CP$ violation effects in other 
processes are described, while section 7 briefly touches on the related issues 
of $CPT$ and $T$ violation.
Some conclusions and perspectives are presented in section 8.

\section{Direct $CP$ violation}

$CP$ violation is a lack of symmetry of a physical process when all the 
spatial coordinates are inverted (parity operation, P) and particles are 
replaced by their anti-particles (charge conjugation, C). 
As such it has connections with the odd number of (ordinary) spatial 
dimensions, and with the roots of the current quantum field theoretical 
paradigm for explaining Nature at the microscopic level.
Both the $P$ and $C$ symmetries have been found experimentally to be valid 
with good accuracy in all processes driven by the strong and electromagnetic 
interactions, and badly broken by the weak interactions, which however 
approximately respect their combination $CP$.

A general property of any Lorentz invariant quantum field theory based on a
hermitian, local Lagrangian, is that the combined operation $CPT$, in which T
represents the time reversal operation, is a valid symmetry 
\cite{LudersPauli}. No violation of this law has ever been detected.
Throughout this paper we will assume the validity of $CPT$ symmetry, except 
where explicitly noted; experimental tests of its validity will be briefly 
discussed in section 7. 

The first manifestation of $CP$ violation in Nature, to be detected 
experimentally, was the decay of the long-lived neutral $K$ meson ($\kl$) in 
both two- and three- pion states \cite{FitchCronin}, in 1964. The number of 
proposed ``exotic'' explanations for this so-called ``Princeton effect'' (see 
\emph{e.g.} \cite{Kabir} for an excellent review written few years after the 
discovery) clearly indicates the physicists' uneasiness\footnote{``What shook 
all concerned now was that with $CP$ gone there was nothing elegant to replace 
it with.'', A. Pais in \cite{CP_reviews_blois}.} 
in definitely dropping the concept of Nature's left-right symmetry, shaken at 
the roots but not yet entirely invalidated after the discovery of parity 
violation \cite{Parity} in 1956, since it could be apparently restored by 
complementing space inversion with charge conjugation.

No evidence for $CP$ violation has ever been found in processes induced by the 
strong or electromagnetic interactions, and so it is usually assumed
that such an effect arises uniquely in weak interactions.

Within the approach of quantum field theory, the appearance of $CP$ violation 
is linked to the presence of ineliminable complex terms in the Lagrangian 
density. 
Given the real nature of observables, $CP$ violation can be experimentally
detected only when measurable quantities are affected by the 
quantum-mechanical interference of two or more terms which have relative 
phases different from $0$ and $\pi$; this somewhat indirect link between the
effect and its mathematical description is one of the main reasons for the
complexity of the formalism, which sometimes tends to obscure the relevant
physical facts.

The simplest manifestation of $CP$ violation is the spontaneous transformation 
of a system which is in a well-defined $CP$ eigenstate into another state 
which is either a $CP$ eigenstate with a different eigenvalue or not a $CP$ 
eigenstate at all. 
Although the term has received slightly different meaning during the years, 
this is indeed what is called \emph{direct} $CP$ violation. 
Since its discovery, $CP$ violation has been experimentally accessible only in 
the weak decays of neutral mesons with flavour, in which case it can manifest 
itself also in a different and subtler, but sometimes dominant way, 
\emph{i.e.} through virtual meson-antimeson transitions \cite{Gell-Mann}, the 
so-called \emph{indirect} $CP$ violation.
For flavoured neutral mesons $M$, distinguished from their anti-particles 
$\overline{M}$ by the flavour eigenvalue $F = \pm 1$, indirect $CP$ violation 
appears in $\Delta F = 2$ virtual processes, as an asymmetry between the 
\mbox{$M \To \overline{M}$} and \mbox{$\overline{M} \To M$} transition rates, 
while direct $CP$ violation can be defined \cite{Nir} as that occurring in 
$\Delta F = 1$ physical decays\footnote{$CP$ violation in $\Delta F=0$ 
transitions, such as $K^{*} \To K \pi$ decays, would also be referred to as 
direct $CP$ violation.}.

It should be stressed that indirect $CP$ violation is a feature of neutral 
mesons only, since conserved quantum numbers make particle-antiparticle mixing 
impossible for charged particles, baryons and leptons\footnote{As long as the 
relevant conservation laws are valid, which actually appears not to be 
exactly the case for baryons (and possibly for neutrinos as well), even within 
the Standard Model.},
while most of the possible $CP$ violation effects which can be considered in 
microscopic phenomena are of the direct type\footnote{Intrinsic electric 
dipole moments of elementary particles would be flavour-conserving 
$CP$-violating static properties.}.
It happens however that the most straightforward manifestation of $CP$ 
violation (namely direct $CP$ violation) is somewhat hidden and masked by a 
larger effect (indirect $CP$ violation) in the system in which the phenomenon 
was actually discovered.

The importance of distinguishing among $CP$ violation in the $\Delta F=2$ and 
$\Delta F=1$ transitions lies in the fact that the former only appears in the 
mixing of flavoured mesons and anti-mesons, which is observable as an intrinsic 
property of the physical meson states, induced by their mutual virtual 
interactions (which in the Standard Model are among the few measurable 
manifestation of electroweak interactions at the second order), while the 
latter is a general property of the flavour-changing weak interactions at
first order.


The Standard Model provides us with some justification for the smallness 
of observed $CP$ violation effects, and it also leads us to expect direct $CP$ 
violation to be an ubiquitous feature in all phenomena controlled by weak 
interactions\footnote{It is interesting to note that the existence of direct 
$CP$ violation can also test the validity of the fundamental concepts of 
quantum mechanics against some ``hidden variable'' alternatives 
\cite{Benatti}.}; the measurable effects are expected to be larger in heavy 
meson systems, which are unfortunately more difficult to study from an 
experimental point of view.

The description of $CP$ violation in the Standard Model is that it originates 
in both forms from a single irreducible complex term in the mixing matrix 
relating ``physical'' quarks (mass eigenstates) to the states which 
participate in the charged-current weak interactions (flavour eigenstates). 
It is well known that such a complex term can have physical significance only 
when at least three (non-degenerate) quark generations are present \cite{KM}; 
this implies in turn that the magnitude of $CP$ violation effects in hadrons 
is necessarily linked to the amount of mixing among the quark generations 
themselves \cite{Jarlskog}. Such mixing happens to be small, 
therefore suppressing most $CP$ violation effects within the Standard Model. 
The above picture also implies that, for such effects to occur at all in a 
given process, their amplitudes must receive contributions from all three 
quark generations.

The subtle and elusive nature of $CP$ violation is such that its dominant 
manifestation for light quark systems requires a quantum-mechanical interplay 
of particle-antiparticle mixing.
This is the reason why $CP$ violation was actually discovered in neutral 
kaons, being the lightest system with enough complexity to allow for such 
phenomena.
Remarkably, the neutral kaon system actually exhibits a very rich 
phenomenology of $CP$ violation, with all possible kinds of $CP$ violation 
contributing, depending on the circumstances under investigation.
There is no \emph{a priori} reason for direct $CP$ violation to be smaller than
its indirect counterpart, but this turns out to be the case for neutral kaons.

In heavier flavoured meson systems ($D,B$ mesons) some kinds of $CP$ violation 
are believed to be irrelevant in the framework of the Standard Model, while 
the searches for asymmetries in exclusive decay channels, with tiny branching 
ratios, and the difficulties in relating the measurements to the fundamental 
parameters of the underlying theory, make the study of $CP$ violation in these 
systems a very challenging, although promising, enterprise.

The smallness of $CP$ violation effects contributes in making theoretical 
predictions difficult and in many cases quite uncertain, the main reason being 
the present limited capability to deal with hadronic states in the 
non-perturbative QCD regime. 
To overcome as much as possible these problems, a massive development of 
theoretical and numerical techniques took place; the interested reader is 
referred to recent reviews such as \cite{Buras} or \cite{Bertolini}.

\subsection{Meson decays}

The general features and experimental signatures of $CP$ violation are 
discussed in any review paper on the subject; we just recall the main points 
in the following.

Due to the almost complete matter-antimatter asymmetry of the known universe,
it is clear that detection of $CP$-violating effects can only be achieved 
by studying microscopic phenomena, such as the static properties of 
``elementary'' particles and their simplest transformations, namely 
spontaneous decays (the study of asymmetries in multi-particle interactions is
also possible in some cases, but is experimentally more challenging). 

The most straightforward indication of $CP$ violation would be a rate 
asymmetry between two $CP$-conjugate decays; it is well known that $CPT$ 
symmetry by itself forces the equality of the fundamental basic properties of 
particles and anti-particles (charge conjugation partners, or $CP$-partners 
since parity does not play a role for space-integrated observables), such as 
the total decay rate. This $CPT$-enforced equality also extends to partial 
decay rates for classes of final states which are not distinguished by the 
$CP$-conserving interactions (assumed to be the strong and electromagnetic 
ones). 
This means that, assuming $CPT$ symmetry as we do in most of the following, 
any $CP$-violating partial rate asymmetry must show up in at least two related 
decay channels, in such a way that they compensate each other when summed 
together: if $f_1,f_2,\ldots,f_n$ make up a complete set of allowed final 
states for the decay of a particle $A$, which are not distinguished by the 
strong and electromagnetic interactions, one must have for the partial rates 
$\Gamma$:
\Dm{
  \Gamma(A \To f_1) - \Gamma(\overline{A} \To \overline{f}_1) = 
  - \sum_{i=2}^{n} \left[ \Gamma(A \To f_i) - \Gamma(\overline{A} \To 
  \overline{f}_i) \right]
}

In several cases, conservation laws other than the one related to $CP$ symmetry
(\emph{i.e.} baryon number, lepton number) hinder the possibility of having 
$CP$-violating decays, and therefore flavoured (not self-$CP$-conjugate) 
mesons are the natural candidates for searches of $CP$ violation.

Neutral mesons $M$ can be coincident with their anti-particles 
(self-conjugate), as is the case for the $\pi^0$ or, when they are 
characterised by some non-zero additive quantum number (such as flavour), they 
can have distinct anti-particles $\overline{M}$, and in this case they exhibit 
a richer phenomenology.
If one defines
\Ea{
  CP | M \rangle = | \overline{M} \rangle \quad && \quad
  CP | \overline{M} \rangle = | M \rangle
}
the $CP$ eigenstates are 
\Ea{
  & | M_1 \rangle \equiv \frac{1}{\sqrt{2}} 
    \left( | M \rangle + | \overline{M} \rangle \right) \quad (CP=+1) \\
  & | M_2 \rangle \equiv \frac{1}{\sqrt{2}} 
    \left( | M \rangle - | \overline{M} \rangle \right) \quad (CP=-1) 
}
If $CP$ symmetry is valid, such states coincide with the physical eigenstates.
The mutual coupling of $M$ and $\overline{M}$ through weak interactions gives 
origin to a mass shift and a mass splitting $\Delta m$, so that in the time 
evolution of free mesons the phenomenon of flavour oscillations takes place, 
with a characteristic period $\hbar/\Delta m c^2$.

Considering a single decay channel $f$ for a non-self-conjugate (\emph{e.g.}
charged) meson $M$ (and its antiparticle $\overline{M}$ decay to 
$\overline{f}$), when the decay can proceed through two different elementary 
amplitudes $A_1,A_2$ one can write, for the total decay amplitude:
\Ea{
  && A(M \To f) = |A_1| e^{i\phi_1} e^{i\delta_1} +   
     |A_2| e^{i\phi_2} e^{i\delta_2} \\
  && A(\overline{M} \To \overline{f}) = |A_1| e^{-i\phi_1} e^{i\delta_1} +   
     |A_2| e^{-i\phi_2} e^{i\delta_2} 
}
in which a part of the amplitude phase ($\phi_i$) changes sign for the 
$CP$-conjugate decay (the actual $CP$-violating phase, often called ``weak'' 
phase) while another part ($\delta_i$) does not (often called the ``final 
state interaction'', ``strong'' or ``re-scattering'' phase).
One can easily show that the asymmetry of the partial decay rates is given by
\Ea{
  & A_{CP}^{(f)} \equiv 
    \frac{\Gamma(\overline{M} \To \overline{f})-\Gamma(M \To f)}
      {\Gamma(\overline{M} \To f)+\Gamma(M \To \overline{f})} = \nonumber \\
  & \frac{2 |A_1| |A_2| \sin (\delta_1 -\delta_2) \sin(\phi_1 - \phi_2)}
    {|A_1|^2 + |A_2|^2 + 2|A_1||A_2| \cos (\delta_1 -\delta_2) 
    \cos(\phi_1 - \phi_2)}
\label{eq:direct_asym}
}
It should be clear that any single phase in an amplitude is irrelevant, since
it can be always redefined away, and only phase \emph{differences} are 
significant, so that the presence of at least two interfering amplitudes is 
required to have an effect.
Moreover, expression (\ref{eq:direct_asym}) for the asymmetry clearly displays 
that in order to have an observable effect in the rates (squared modulus 
amplitudes in quantum mechanics) induced by the difference of the 
$CP$-violating phases $\phi_i$, these two amplitudes must also have different 
final-state interaction phases.
The latter actually provide the ``reference'' phases against which the sign
change of the weak phases can be detected, and are usually induced by the 
($CP$-symmetric) strong or electromagnetic interactions of the final-state 
particles, as indicated above. 
These strong phases are often related to the experimentally measurable phases, 
which describe the scattering of the final state particles, via the 
Fermi-Watson theorem \cite{FermiWatson}; they are difficult to compute from 
first principles, and this leads to the difficulty of predicting the partial 
rate asymmetries or extracting the weak phases from the measured values of
such asymmetries. 

From the above expression it is also clear that, in order to have a 
large asymmetry, the two interfering amplitudes must be of comparable 
magnitude: the best decay modes in which to look for direct $CP$ violation are 
therefore the ones in which the dominant, lowest-order (\emph{tree} level)  
graphs are suppressed because of the small inter-generation mixing (``Cabibbo 
suppressed''), or by some other selection rule, so that one is left with 
higher-order graphs (\emph{penguin}) of comparable magnitude.

\begin{figure}[hbt!]
\begin{center}
\begin{minipage}{0.47\textwidth}{
  \epsfig{file=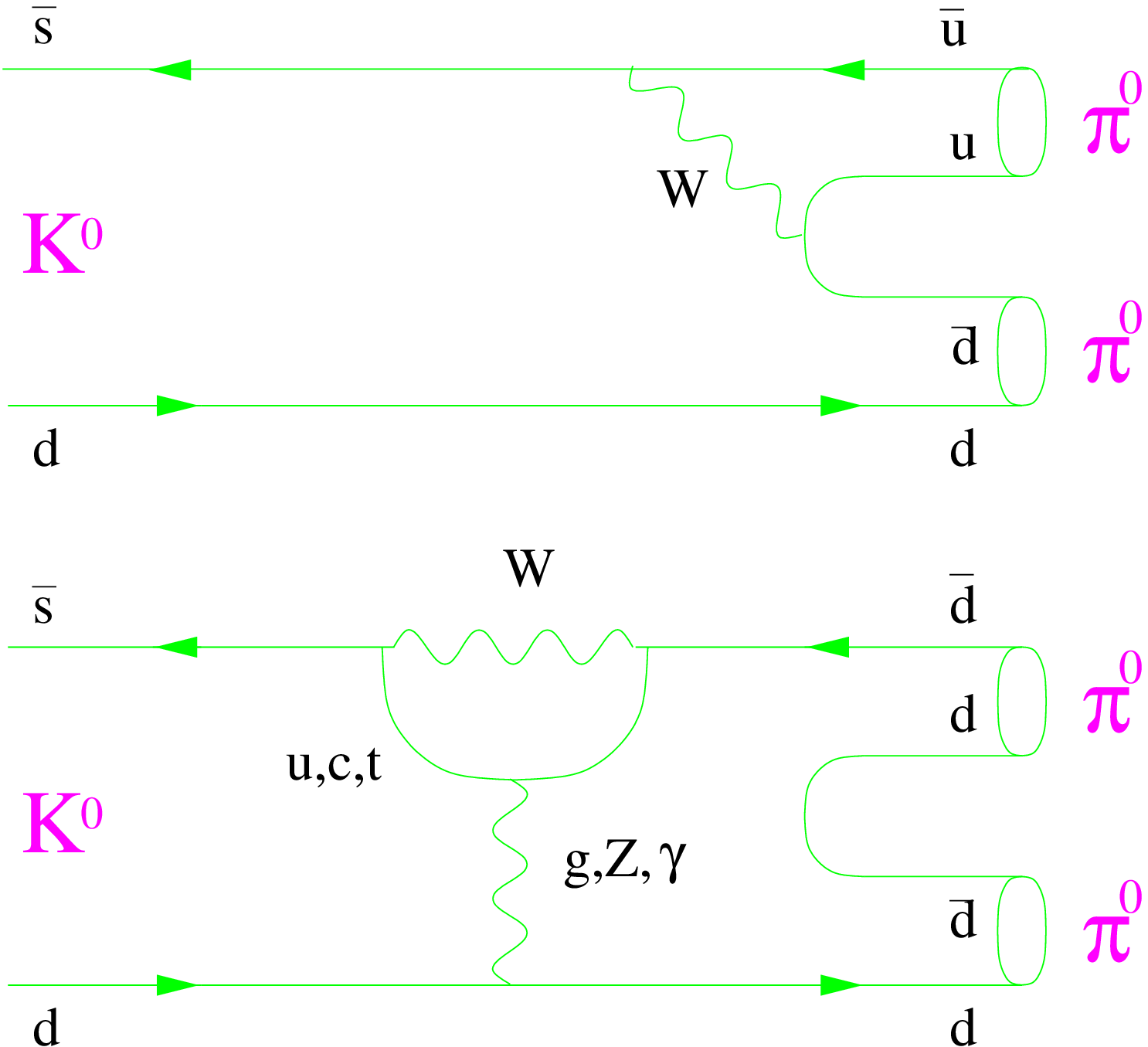,width=1.0\textwidth}
}
\end{minipage}
\begin{minipage}{0.47\textwidth}{
  \epsfig{file=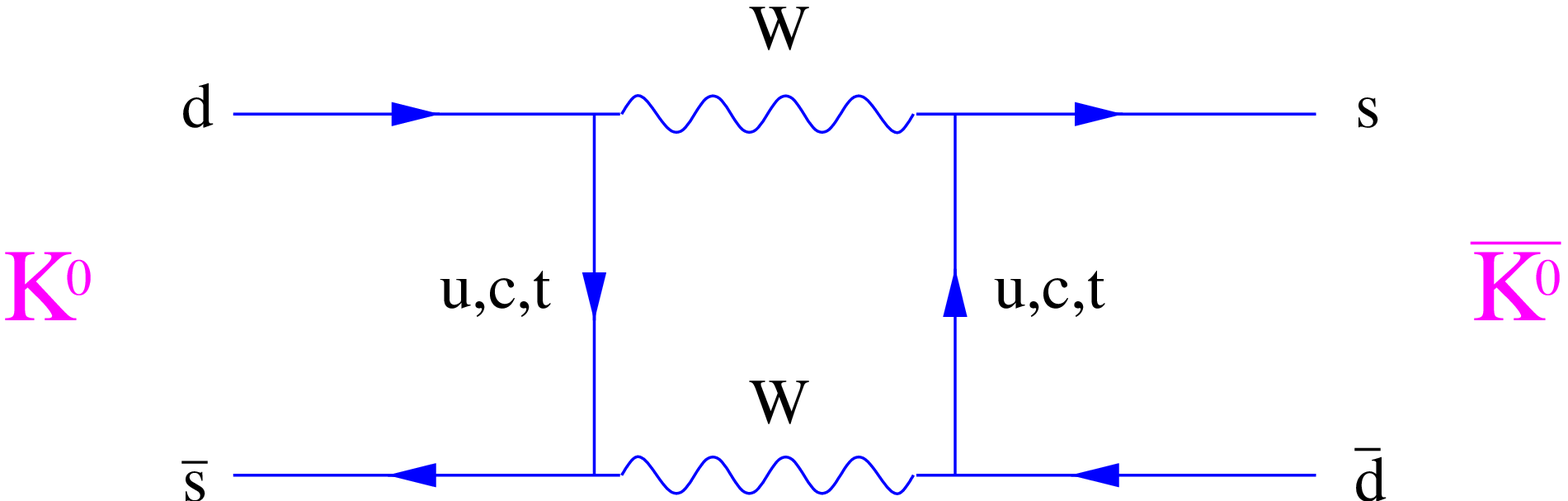,width=1.0\textwidth}
}
\end{minipage}
\caption{Left: the ``tree'' and ``penguin'' diagrams originating $\kz \To 
\pi^+ \pi^-$ decays in the Standard Model. Right: The ``box'' diagram 
originating $\kz--\kzb$ transitions in the Standard Model.}
\label{fig:tree_box}
\end{center}
\end{figure}

As an example, $CP$ violation in $\kz \To \pi\pi$ ($\Delta S=1$, where $S$ is
strangeness) decays can arise in the Standard Model because of different weak 
phases in the tree amplitude and the (electro-weak or gluonic) penguin one, 
shown diagrammatically in fig. \ref{fig:tree_box}.
Since the first one contributes to the isospin $I=0,2$ $\pi\pi$ final states, 
while the second one only to the $I=0$ state, different strong phases are 
expected.

The above requirements for the observability of a $CP$-violating effect are 
not specific to partial decay rate asymmetries, but apply to several other 
particle-antiparticle asymmetries.
The two interfering amplitudes can be due to competing elementary processes in
the same decay, possibly inducing so-called $CP$ violation ``in the decay''.
Decays of self-conjugate neutral mesons into $CP$ eigenstates with different
$CP$ eigenvalues, such as \mbox{$\eta \To \pi \pi$}, would also be indications 
of this kind of $CP$ violation.

In neutral flavoured meson systems one can have also $CP$ violation purely 
``in the mixing'', actually the dominant effect in the neutral kaon system.
This is an asymmetry in the virtual transitions \mbox{$M \To \overline{M}$}  
with respect to \mbox{$\overline{M} \To M$}, originating physical states which 
do not contain the same amount of $M$ and $\overline{M}$ components, 
\emph{i.e.} states which are not $CP$ eigenstates. 
This kind of $CP$ asymmetry is therefore an intrinsic property of the meson 
system itself, driven by its (virtual) interactions with other states.
If $CP$ symmetry is approximately valid, the physical states will be largely 
composed of a single $CP$ eigenstate, with a small ``impurity'' component 
belonging to the opposite $CP$ eigenvalue. 
In the Standard Model, $\Delta F=2$ transitions occur at second order in the 
weak interactions, and are described by the so-called \emph{box} diagrams 
(fig. \ref{fig:tree_box}, for the neutral kaon system).

$CP$ violation in the mixing can be singled out in the decays of the physical 
$M--\overline{M}$ coherent mixtures to \emph{flavour-specific} final states 
(\emph{i.e.} states to which either $M$ or $\overline{M}$ can decay but not 
both, due to some selection rule) for which only a single elementary 
amplitude can contribute.
In the Standard Model, semi-leptonic decays of neutral mesons ($M \To 
X^\mp l^\pm \nu(\overline{\nu})$, where $X$ is an hadronic state and $l^\pm$ a 
charged lepton) are of this kind, and as such they can be exploited for the 
study of $CP$ violation in the mixing. As long as the $\Delta F = \Delta Q$ 
rule, linking the change in flavour eigenvalue and in electric charge for the 
hadrons involved in the semi-leptonic decay (this rule being justified by the 
tree level quark diagrams in the Standard Model) is valid, these decays are 
\emph{flavour-specific}. As an example only $\kz \rightarrow \pi^- l^+ \nu$ 
and $\kzb \rightarrow \pi^+ l^- \overline{\nu}$ decays are allowed, but not 
the ones in which the roles of $\kz$ and $\kzb$ are exchanged.
Lacking in each case a second amplitude which can interfere, $CPT$ symmetry 
forces the two amplitudes to be equal; by observing the decays of the physical 
flavour mixtures, any rate difference among the decay to states with positive 
or negative leptons reflects the asymmetry in the composition of the mesons, 
\emph{i.e.} measures $CP$ violation in the mixing.

In the decays of neutral flavoured mesons, the amplitudes for reaching a final 
state $f$ (common to $M$ and $\overline{M}$, \emph{i.e.} not flavour-specific) 
without and with mixing (\mbox{$M \To f$} and \mbox{$M \To \overline{M} \To 
f$}) can interfere; in this case the ``reference'' phase, which allows the 
effect of a non-zero phase in the decay amplitude to be observed, is the one 
of the mixing amplitude, and no final-state interactions are required to have 
an asymmetry.
This is the so-called $CP$ violation ``in the interference of decay with and
without mixing'' (or ``in the interference of mixing and decay'', or simply 
``mixing-induced''). 

\begin{figure}[hbt!]
\begin{center}
    \epsfig{file=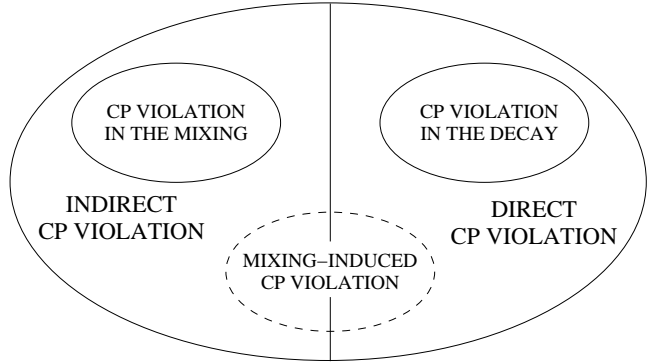,width=0.7\textwidth}
  \caption{Graphical schematization of different kinds of $CP$ violation in 
  meson processes.}
  \label{fig:cp_types}
\end{center}
\end{figure}

While $CP$ violation in the decay is clearly a manifestation of direct $CP$ 
violation, and $CP$ violation in the mixing is clearly a manifestation of 
indirect $CP$ violation, in the mixing-induced case one cannot ascribe in 
a physically meaningful way the $CP$ violation to either the $M--\overline{M}$ 
mixing or the $M,\overline{M} \rightarrow f$ decay, so that one is dealing 
with an undefined mixture of direct and indirect $CP$ violation. 
However, it should be clear that, being the mixing an intrinsic property of 
the decaying meson system, any difference between the amount of mixing-induced 
$CP$ violation in decays to different final states must be due to direct $CP$ 
violation; this allows to identify this component without ambiguities by 
measuring several channels to which mixing-induced $CP$ violation can 
contribute.
Figure \ref{fig:cp_types} summarises in a graphic way the different kinds of
$CP$ violation effects in mesons, which will be discussed in more detail in
section 5.

The quantitative understanding of meson decays within a theoretical model is 
clearly easier when approximate selection rules or dynamical effects are such 
that these processes are driven by a single elementary amplitude, so that the 
measurements of the decay parameters (including the branching ratio itself) 
can be related in a straightforward way to the fundamental parameters of the 
model.
For this reason, the theoretical and experimental studies of $CP$ violation in 
heavy meson decays has been focused initially on the mixing-induced one 
($CP$ violation in the mixing being small in these systems).
The recent measurements of $CP$ violation in neutral $B_d$ meson decays
\cite{CDF_ind} \cite{BABAR_ind} \cite{BELLE_ind}, providing the first 
evidence of $CP$ violation outside the neutral kaon system, are actually an 
example of a class of mixing-induced $CP$ violation phenomena which can be 
related to the underlying parameters of the Standard Model in a reliable way. 

From an experimental point of view, the mixing properties of neutral kaons can 
be studied rather easily, since such mesons appear in Nature as two flavour 
superpositions with very different lifetimes; this (accidental) property is 
not at all shared by heavier neutral meson systems, for which the physical 
states have very similar lifetimes. 
$CP$ violation in heavy flavoured neutral mesons is therefore better studied 
in the ``orthogonal reference frame'' of \emph{tagged} flavour eigenstates, 
\emph{i.e.} states whose flavour eigenvalue at production time is known, by 
exploiting either associate production or the production of coherent 
meson-antimeson pairs; such approach has been used also for kaons.

While charged mesons are the most straightforward testing ground for direct 
$CP$ violation, a possibility of extracting direct $CP$ violation signals in 
neutral meson decays lies in the accurate measurement of the time-dependent 
decay asymmetries for non flavour-specific final states, which can be fed by 
both $M$ and $\overline{M}$, as will be discussed in detail when addressing 
heavy meson decays.
It should be clear that, if the flavour oscillations are not too fast, in the
limit of short proper decay times any $CP$ violation effect due to mixing 
would be relatively less important than the ones due to decay, which manifest 
themselves in a time independent way.

For channels with only two (pseudo-)scalar mesons or a (pseudo-)scalar and a 
vector meson in the final state, a $CP$ asymmetry can only become apparent as 
a difference of the partial decay rates, while for states with more complex 
spin content or larger number of particles in the final state, asymmetries in 
the distributions of kinematic variables can also be used to search for $CP$ 
violation.

Summarising, evidence for direct $CP$ violation in meson decays could come
from either: 
\begin{itemize}
\item 
Any difference between the properties of distinct $CP$-conjugate decays.
\item
Any difference between the decay properties of $CP$-conjugate states to a 
common final state, in cases in which particle-antiparticle mixing cannot 
occur (\emph{i.e.} any $CP$ asymmetry for charged particles).
\item 
Any difference between the amount of $CP$ violation in decays of a given
physical meson to different final states, \emph{e.g.} differences in the 
mixing-induced neutral meson decay asymmetries to different final states.
\item 
The study of the time-dependent decay asymmetries in neutral meson decays.
\item 
Any non-zero $CP$-violating decay asymmetries of \emph{flavour-tagged} mesons 
to ``right-sign'' \emph{flavour-specific} final states.
\end{itemize}
The above signatures will be discussed in the following.

\subsection{$T$-odd correlations}

Experiments on transverse muon polarization in $K_{\mu3}$ decays, to be
discussed in the following, are an example of a class of measurements 
searching for $CP$ violation as non-zero values of $T$-odd observables 
(assuming $CPT$ symmetry); such searches were performed on several systems 
(see \emph{e.g.} references in \cite{Golowich}), so far with null results. 

When event (decay) rates are measured for an isolated system, Lorentz 
invariance only allows them to depend on scalar (or pseudoscalar) quantities; 
just as the rate dependence of beta decay on the pseudoscalar quantity 
$\Vec{S} \cdot \Vec{p}$ ($\Vec{S}$ being a polarization vector and $\Vec{p}$ a 
three-momentum) gave evidence for parity violation, any rate dependence on a 
$T$-odd quantity could be expected to indicate $T$ violation. 
Lorentz symmetry requires that the simplest of such $T$-odd quantities involve 
three vectors as a triple product; at least three particles are required to 
get a non-trivial value for such a quantity, \emph{e.g.} $\Vec{S} \cdot 
\Vec{p}_1 \times \Vec{p}_2$ in a three-body decay, in which a polarization 
vector and two three-momenta are involved.
In order to avoid using spin information, at least four particles are required 
to form a non-null $T$-odd correlation, such as $\Vec{p}_1 \cdot \Vec{p}_2 
\times \Vec{p}_3$, because the vectors have to be independent; moreover, no 
identical particles should appear in the final state, otherwise the quantity 
vanishes trivially.

An important feature of $T$-odd correlations is that final-state interactions
can themselves generate a non-null value for such quantities, even in absence
of $T$ violation, so that the measurement of a non-zero expectation value is 
not, by itself, evidence for violation of $T$ symmetry\footnote{This happens 
because the inversion of spins and momenta (sometimes called ``naive'' time 
reversal) is not the only effect of time reversal, and therefore such a 
non-zero $T$-odd correlation can actually be invariant under $T$, as indeed 
the part due to strong or electromagnetic final-state interactions is.}.
In order to detect a true signal of $T$ violation, one should therefore 
either subtract the effect due to final-state interactions, if known, or study 
processes in which such interactions are known to have negligible effects.
Conversely, it should be noted that $T$-odd correlations provide an example of 
how direct $CP$ violation (assuming $CPT$) can appear even in absence of any 
final-state interactions, contrary to the case of partial rate or angular 
distribution asymmetries, in which ``strong'' phases are an essential 
requirement to make $CP$ violation observable. 
When such phases are indeed present but small in magnitude, triple product 
correlations may happen to be the observables most sensitive to $CP$ violation.

This class of measurements (assuming $CPT$ symmetry) is actually complementary 
to that of partial decay rate asymmetries, for the decay modes in which no 
final-state interactions appear, when any measurement of a non-zero value for 
a $T$-odd correlation directly indicates $CP$ violation, are actually the ones 
in which the comparison of the decay rates for particles and anti-particles 
give no information on the validity of $CP$ symmetry, their equality being 
enforced by $CPT$.

It is important to note that even when, possibly uncontrollable, final-state 
interactions are present, the study of $T$-odd correlations is relevant to $CP$
symmetry because their value for the $CP$-conjugate process is just the 
opposite of that for the original one, if $CP$ symmetry is valid.
One can therefore detect an unambiguous signal of $CP$ violation by comparing 
non-zero values of $T$-odd correlations for $CP$-conjugate processes, even in
presence of unknown final-state interaction effects.
Clearly, such comparisons have to be performed in absence of instrumental 
$CP$-asymmetric effects: the study of states produced in particle-antiparticle 
collisions (the initial state being a $CP$ eigenstate) is therefore a good 
environment for such kind of experiments.

\section{Neutral kaon decays}

The system of neutral kaons, $\kz$ and $\kzb$, is an extraordinary physics 
laboratory, being the lightest system of two coupled particles (as 
``elementary'' as isolated observation allows) which are distinguished only by 
a quantum number (flavour, strangeness $S$ in this case) which is conserved
by the strong and electromagnetic interactions but \emph{not} by the weak 
interactions responsible for their decay. As such, neutral kaons allow the 
direct observation of characteristic quantum-mechanical interference effects
of a small magnitude.

The central role of kaons in the development of particle physics cannot be 
overstated: as the lightest particles with a flavour quantum number not to
be found in ordinary matter but readily produced at GeV energies, they 
stimulated the study of such diverse issues as flavour mixing, parity 
non-conservation, the existence of the \emph{charm} quark, before providing 
the first (and, for a long time, unique) evidence of $CP$ non-conservation.
For the neutral kaon system the peculiar features of particle-antiparticle 
state mixing can be easily observed, and a large variety of experimental 
measurements can be performed, since both the flavour eigenstates $\kz,\kzb$ 
and the physical mass eigenstates $\ks,\kl$ are separately accessible for 
experimentation: the system offers the unique possibility of observing in
principle arbitrary superpositions of flavour eigenstates.
This occurs due to the relatively small ratio of the $\kz,\kzb$ mass to the 
pion masses, which together with the approximate validity of $CP$ symmetry 
leads to a very large lifetime difference for the physical states: $\tau_S = 
8.9 \cdot 10^{-11}$ s for $\ks$, for which the dominant hadronic decay channel 
$\pi \pi$ is allowed, and $\tau_L = 5.2 \cdot 10^{-8}$ s for $\kl$, for which 
the $\pi \pi$ decay channel is largely suppressed by the (approximate) $CP$ 
symmetry.
This lifetime difference makes easy the experimentation with pure $\kl$,
obtained by considering the neutral kaons surviving undecayed after a long 
($\gg \beta \gamma c \tau_S$) decay region.
$\ks$ decays can be statistically studied in a neutral kaon beam close to the
production point, which initially contains the same amount of $\ks$ and $\kl$ 
components (as long as the $CPT$ symmetry is valid). To experiment with pure 
$\ks$, the coherent production of neutral kaon pairs (\emph{e.g.} from $\phi$
decays) can be exploited.

However, since within the Standard Model $CP$ violation requires the 
involvement of all three quark families, which in the kaon system only occurs 
through virtual processes, and not in first order, the magnitude of the 
observable effects is generally small.

\subsection{Phenomenology}

The discussion of the phenomenology of the neutral kaon system and $CP$ 
violation can be found in standard particle physics textbooks and in several 
reviews (see \emph{e.g.} \cite{Isidori} \cite{Maiani} \cite{Belusevic}).
Unfortunately, while the discussion of the formalism has its common roots in 
the Weisskopf-Wigner approach to coupled unstable systems 
\cite{WeisskopfWigner} and in the classic paper of Wu and Yang \cite{WuYang}, 
the variety of different conventions and approximations can be the source of 
some confusion: as an example, at least 5 different definitions of the direct 
$CP$ violation parameter $\epsilon'$ in \mbox{$K \To \pi \pi$} decays can be 
found in the literature, and they coincide only when some approximations are 
made, which are often justified only in some class of phase conventions for 
the meson states.
$CP$ violation is expressed in the formalism by the presence of complex 
quantities, but since the mapping of quantum-mechanical states to physical 
states is to within a phase, it is crucial to separate the irrelevant 
(convention-dependent) phases from the ``physical'' ones which characterise 
$CP$ violation\footnote{``Anyone who has played with these invariances knows 
that it is an orgy of relative phases'', A. Pais in \cite{InwardBound}.} 
(see \emph{e.g.} \cite{Tsai} and references therein).

We now summarise the consistent definitions that we use in the following 
discussion. Instead of using more specific but uncommon formalism (see 
\emph{e.g.} \cite{Kostelecky} \cite{Paschos}) we adopt the usual one, trying
to state clearly the approximations and the phase conventions relevant to 
deduce a given relation.

The relative phase of $\kz$ and $\kzb$ states is not measurable and can be 
fixed by convention (even independently from the quark field phases). 
Given the phase convention
\Ea{
  & CP | \kz \rangle = e^{i\xi_{CP}} | \kzb \rangle \\
  & CP | \kzb \rangle = e^{-i\xi_{CP}} | \kz \rangle
}
(with real $\xi_{CP}$) the $CP$ eigenstates are defined to be 
\Ea{
  && | K_1 \rangle \equiv \frac{1}{\sqrt{2}} 
    ( e^{-i\xi_{CP}/2} | \kz \rangle + 
      e^{i\xi_{CP}/2}  | \kzb \rangle ) \quad (CP=+1) \\
  && | K_2 \rangle \equiv \frac{1}{\sqrt{2}} 
    ( e^{-i\xi_{CP}/2} | \kz \rangle - 
      e^{i\xi_{CP}/2}  | \kzb \rangle ) \quad (CP=-1)
}
The $\xi_{CP}$ phase can be seen as the parameter of a transformation whose 
generator is the strangeness operator: since $\kz,\kzb$ states are defined by 
the strangeness-conserving strong interactions, no observable can depend on 
the value of such quantity.
The simplest phase convention, adopted in the following (except where 
explicitly stated) is $\xi_{CP}=0$, \emph{i.e.} 
\Ea{
  CP | \kz \rangle = | \kzb \rangle \quad && \quad 
  CP | \kzb \rangle = | \kz \rangle 
} 
In this case the ``physical''\footnote{Although still non-observable in 
principle.} states $\ks,\kl$ (short- and long-lived, respectively) can be 
written in terms of the strangeness eigenstates $\kz,\kzb$ as
\Ea{
  & | \ks \rangle = \frac{1}{\sqrt{1+|\epsilon_S|^2}} 
     ( | K_1 \rangle + \epsilon_S | K_2 \rangle ) = \quad \quad \nonumber \\
  & \quad \quad \frac{1}{\sqrt{2(1+|\epsilon_S|^2)}}
     \left[ (1+\epsilon_S) | \kz \rangle + 
     (1-\epsilon_S) | \kzb \rangle \right] \\
  & | \kl \rangle = \frac{1}{\sqrt{1+|\epsilon_L|^2}} 
    ( | K_2 \rangle + \epsilon_L | K_1 \rangle ) = \quad \quad \nonumber \\
  & \quad \quad \frac{1}{\sqrt{2(1+|\epsilon_L|^2)}}
     \left[ (1+\epsilon_L) | \kz \rangle - 
     (1-\epsilon_L) | \kzb \rangle \right] 
}
$CPT$ symmetry requires $\epsilon_S = \epsilon_L$, leaving a single (complex) 
$CP$ impurity parameter to parameterise the composition of physical states:
\Dm{
  \beps \equiv (\epsilon_S+\epsilon_L)/2
}
This parameter would be zero if $CP$ symmetry were valid in $\kz--\kzb$ state 
mixing.
The impurity parameter $\beps$ is not physically measurable and depends on the 
choice of phase convention; a phase convention independent term which 
parameterises $CP$ violation in the mixing of states is the measure of the 
non-orthogonality of the physical states:
\Dm{
  \langle \ks | \kl \rangle = \frac{2 \Rea(\beps)}{1+|\beps|^2}
}
The statement that the real part of $\beps$ is a measurable quantity 
parameterising the amount of indirect $CP$ violation is seen to be 
approximately valid in every so called ``physical'' phase convention in which 
$|\beps| \ll 1$; that such a class of phase conventions is allowed at all 
clearly follows from the fact that $CP$ violation is experimentally known to 
be small with respect to the dominant weak interaction effects, so that a 
formalism in which all the parameters related to $CP$ violation are small must 
exist, although a phase convention in which $|\beps| \gg 1$ could also be 
adopted at will (see \emph{e.g.} \cite{Wu}).
In a different formalism, widely used for heavier meson systems, one defines
\Dm{
  \frac{q}{p} \equiv \frac{1-\beps}{1+\beps}
}  

Due to their coupling through the strangeness-violating weak interactions, 
the time evolution of $\kz$ and $\kzb$ has to be described together, and the
appropriate formalism to treat such coupled systems is that developed by 
Weisskopf and Wigner (see \emph{e.g.} \cite{Kabir}). Considering the 
projection onto the two-dimensional subspace of $\kz$ and $\kzb$ components, 
the time evolution of a generic state $\Psi$ is described by a non-hermitian 
quasi-Hamiltonian $\mathbf{\mathcal{H}}$:
\Dm{
  i \hbar \frac{\partial}{\partial t} \Psi = \mathbf{\mathcal{H}} \Psi
}
The $2 \times 2$ matrix $\mathcal{H}$ can be written in full generality as
\Dm{
  \mathbf{\mathcal{H}} \equiv \mathbf{M} - \frac{i}{2} \mathbf{\Gamma}
}
where $\mathbf{M}$ and $\mathbf{\Gamma}$ (the ``mass matrix'' and the ``decay 
matrix'') are both hermitian matrices, whose components, at lowest order in 
the small (weak) Hamiltonian $H_I$ responsible for $\kz,\kzb$ not being mass 
eigenstates, are\footnote{In the following index 1 refers to \kz and 2 to 
\kzb and $\hbar = c = 1$ unless explicitely noted.}
\Ea{
  & M_{11} = m_K + \langle \kz | H_I | \kz \rangle + O(H_I^2) \\
  & M_{22} = m_K + \langle \kzb | H_I | \kzb \rangle + O(H_I^2) \\
  & M_{12} = M_{21}^* = \langle \kz | H_I | \kzb \rangle + O(H_I^2) \\
  & \Gamma_{11} = 2\pi \sum_f | \langle f | H_I | \kz \rangle |^2 
    \delta(m_K - E_f) + O(H_I^4) \\
  & \Gamma_{22} = 2\pi \sum_f | \langle f | H_I | \kzb \rangle |^2 
    \delta(m_K - E_f) + O(H_I^4) \\
  & \Gamma_{12} = \Gamma_{21}^* = 2\pi \sum_f \langle \kz | H_I | f \rangle 
    \langle f | H_I | \kzb \rangle \delta(m_K - E_f) + O(H_I^4)
}
where $m_K$ is the (common) mass of $\kz$ and $\kzb$, as defined by the 
$CP$-conserving strong interactions, and $f$ is any final state (of energy 
$E_f$) accessible to $\kz$ or $\kzb$ in physical decays.

$CPT$ symmetry requires the diagonal elements of the quasi-Hamiltonian to be 
equal: $M_{11} = M_{22}$, $\Gamma_{11} = \Gamma_{22}$.
The off-diagonal elements $M_{12}$ and $\Gamma_{12}$ are responsible for 
strangeness oscillations, and $T$ symmetry requires them to be real.
$CP$ symmetry imposes the same constraints on these quantities, namely the 
equality of the diagonal matrix elements and the reality of the off-diagonal 
ones.
It is interesting to note that a Hamiltonian $H_I$ which changes strangeness 
by one unit $(|\Delta S| = 1)$, such as the one describing weak interactions 
in the Standard Model, can only contribute to $\mathbf{\Gamma}$ and not to 
$\mathbf{M}$ in lowest order.

The states $\ks,\kl$ of definite mass and lifetime are the eigenstates of
$\mathbf{\mathcal{H}}$:
\Dm{
  \mathbf{\mathcal{H}} | K_{S,L} \rangle = \lambda_{S,L} | K_{S,L} \rangle
}
where the eigenvalues are $\lambda_{S,L} \equiv m_{S,L} - i\Gamma_{S,L}/2$
($m_{S,L}, \Gamma_{S,L}$ real), which (assuming $CPT$ symmetry) are given by 
\Dm{
  \lambda_{S,L} = \mathcal{H}_{11} \pm \sqrt{\mathcal{H}_{12} \mathcal{H}_{21}}
}
As mentioned, the decay widths are very different, with $\Gamma_L/\Gamma_S 
\approx 600$.
The physical states $\kl,\ks$ have also a tiny mass difference, $m_L - m_S 
\simeq 3.5 \cdot 10^{-12}$ MeV/$c^2$, induced in the Standard Model by 
second-order weak interactions, according to the Glashow-Iliopoulos-Maiani 
mechanism, and a correspondingly long flavour oscillation length in the 
laboratory.

Of the eight real parameters defining the matrix $\mathbf{\mathcal{H}}$, one 
is an irrelevant global phase, four can be chosen as the masses and lifetimes 
of the physical states, two vanish if $CPT$ symmetry is valid, and the last 
one expresses $CP$ violation: it can be chosen to be the phase convention 
independent part of $\beps$. One has
\Dm{
  \frac{1-\beps}{1+\beps} \equiv \frac{q}{p} = 
  \sqrt{\frac{\mathcal{H}_{21}}{\mathcal{H}_{12}}} =
  \frac{\Delta \lambda}{2 \mathcal{H}_{12}}
}
(where $\Delta \lambda \equiv \lambda_S - \lambda_L$), and in the limit 
$|\beps| \ll 1$ (possible for small $CP$ violation) one has
\Dm{
  \beps \simeq \frac{\mathcal{H}_{12}-\mathcal{H}_{21}}{2 \Delta \lambda} = 
  \frac{\Imm(M_{12})-i\Imm(\Gamma_{12})/2}{i \Delta m -\Delta \Gamma/2}
}
where, as is conventional in the case of the $\kz--\kzb$ system, $\Delta m$ 
and $\Delta \Gamma$ are defined to be positive and, in the limit of small
$CP$ violation: 
\Ea{
  && \Delta m \equiv m_L - m_S \simeq -2 \Rea(M_{12}) \\
  && \Delta \Gamma \equiv \Gamma_S - \Gamma_L \simeq 2 \Rea(\Gamma_{12})
}
In the neutral kaon system the following empirical relation holds 
\Dm{
  \Delta m \simeq \Gamma_S/2 \simeq \Delta \Gamma/2
}
and therefore one obtains in the end the expression
\Dm{
  \beps \simeq -\frac{e^{i\pi/4}}{\Delta m \sqrt{2}} 
  \left[\Imm(M_{12})-i\Imm(\Gamma_{12})/2 \right] \simeq
  \frac{e^{i\pi/4}}{2 \sqrt{2}} 
  \left[\frac{\Imm(M_{12})}{\Rea(M_{12})}+
  i \frac{\Imm(\Gamma_{12})}{\Rea(\Gamma_{12})} \right]
}
so that indeed the $CP$ impurity parameter $\beps$ is linked to the 
off-diagonal elements of $\mathbf{M}$ and $\mathbf{\Gamma}$ being complex. 
It can be easily seen that what actually determines the presence of $CP$ 
violation in the quasi-Hamiltonian $\mathcal{H}$ describing meson-antimeson 
mixing is a non-zero \emph{relative} phase between $M_{12}$ and $\Gamma_{12}$: 
$CP$ violation is present in the quasi-Hamiltonian if and only if 
$|\mathcal{H}_{12}| \neq |\mathcal{H}_{21}|$, and 
\Dm{
  \langle \ks | \kl \rangle \propto \Imm(M_{12}^* \Gamma_{12})
}
$CP$ symmetry would imply that this relative phase is zero, and the 
coefficients of the two flavour eigenstates in the physical states differ at 
most by a pure phase:
\Dm{
  \left| \frac{1-\beps}{1+\beps} \right| \equiv \left| \frac{q}{p} \right| = 1
}
In presence of small $CP$ violation one has instead 
\Dm{
  \left| \frac{1-\beps}{1+\beps} \right| \equiv 
  \left| \frac{q}{p} \right| \simeq 1 - \frac{1}{2} 
  \sin \left[ \arg \left( \frac{\Gamma_{12}}{M_{12}} \right) \right]
}

The charge asymmetry for semi-leptonic $\kl$ or $\ks$ decays is a pure 
measurement of $CP$ violation in the mixing, as long as $CPT$ symmetry and the 
$\Delta S = \Delta Q$ rule\footnote{Experimental information on the validity 
of this rule come from the limits on \mbox{$BR(\Sigma^+ \To n e^+ \nu)$} and 
the detailed study of semi-leptonic $\kz$ decays by CPLEAR which give 
\cite{PDG_2002} \mbox{$A(\Delta S = -\Delta Q)/A(\Delta S = \Delta Q)$} = 
\mbox{$(-0.002 \pm 0.006)$} \mbox{$+i(0.0012 \pm 0.0021)$}.} are valid:
\Dm{
  \delta_{L,S}^{(l)} \equiv 
  \frac{\Gamma(\kl,\ks \To \pi^- l^+ \nu) - 
        \Gamma(\kl,\ks \To \pi^+ l^- \overline{\nu})}
       {\Gamma(\kl,\ks \To \pi^- l^+ \nu) + 
        \Gamma(\kl,\ks \To \pi^+ l^- \overline{\nu})} = 
  \langle \ks | \kl \rangle
}
Semi-leptonic decays of $\kl$ are easily accessible from an experimental point
of view, the branching ratios being $\simeq 39\%$ for $\pi^\pm e^\mp \nu$ 
($K_{e3}$) and $\simeq 27\%$ for $\pi^\pm \mu^\mp \nu$ ($K_{\mu3}$); the
measured charge asymmetries are \cite{PDG_2002} \cite{KTeV_deltal} 
\cite{NA48_deltal}:
\Ea{
  & \delta_L^{(e)} = (0.3322 \pm 0.0055)\% \\
  & \delta_L^{(\mu)} = (0.304 \pm 0.025)\% 
}

So far the only measurement of the charge asymmetry in semi-leptonic $\ks$ 
decays has been carried out by the KLOE experiment. A preliminary result based 
on 170 pb$^{-1}$ (out of the $\sim$ 500 collected until 2002) gave 
\cite{Gatti}:
\Dm{
  \delta_S^{(e)} = (1.9 \pm 1.8)\%
}
If $CPT$ symmetry and the $\Delta S = \Delta Q$ rule are valid, one expects 
$\delta_L^{(l)} = \delta_S^{(l)} = 2 \Rea(\beps)/(1+|\beps|^2)$, from 
which $\Rea(\beps) = (1.655 \pm 0.003) \cdot 10^{-3}$ is obtained.

Within a few years of the discovery of $CP$ violation, the crucial question
became that of whether the single parameter $\beps$ describing the asymmetry
of $\kz--\kzb$ mixing in the effective quasi-Hamiltonian 
$\mathbf{\mathcal{H}}$ could account for all the $CP$ non-conservation effects 
in Nature.
Already in 1964, L. Wolfenstein pointed out \cite{WolfensteinSW} that a 
hypothetical ``super-weak'' interaction, capable of driving $\kz--\kzb$ 
transitions in first order (\emph{i.e.} satisfying a $\Delta S = 2$ selection 
rule) would induce $\kl \rightarrow \pi \pi$ decays through the $CP$ impurity 
of the physical states, and that the smallness of the coupling required to 
give the measured amount of $CP$ violation through this mechanism would 
effectively confine all the measurable effects of such a new interaction to 
the very sensitive neutral kaon system.
The introduction of a small complex, non-diagonal term in the effective 
quasi-Hamiltonian, induced by a new Hamiltonian $H_{SW}$
\Dm{
  M_{12}^{(SW)} = \langle \kz | H_{SW} | \kzb \rangle = i \Delta_{SW}
}
would give by diagonalization
\Dm{
  \beps \simeq \frac{-\Delta_{SW}}{\Delta m} \frac{1+i}{2}
}
and inserting the value of the $\kl-\ks$ mass difference one obtains (in an 
appropriate phase convention) $|\Delta_{SW}| \simeq \sqrt{2} |\beps| \Delta m 
= 1.1 \cdot 10^{-8}$ eV. Writing $|\Delta_{SW}| \sim G_{SW} \, m_K^3$, so to 
have a super-weak coupling constant $G_{SW}$ dimensionally comparable to 
$G_F$, one has $G_{SW} \sim 10^{-11} G_F$. 
$\Delta_{SW}$ is suppressed by a factor $|\beps|$ with respect to $\Delta m$, 
which is a second-order weak interaction effect: the tiny mass difference 
between $\kl$ and $\ks$ effectively boosts the effects of an extremely weak 
interaction.

Although the super-weak scenario followed the familiar pattern of having weaker
fundamental interactions respecting all but a few of the symmetries which are
valid for the stronger ones (a fact which has to do with the logical structure 
of our physical theories and tells little about Nature), it actually stood 
more as a paradigm for classifying theories of $CP$ violation rather than a 
realistic model.
In recent times the concept of ``super-weak $CP$ violation'' acquired a more 
blurred meaning: its extension beyond the neutral kaon sector can be done in 
different ways, \emph{i.e.} implying either no $CP$ violation in other meson 
systems, or $CP$ violation being limited to $\Delta F=2$ flavour-changing 
interactions in every system. 
Various models actually implement the super-weak idea in different ways, 
leading to different estimates of $CP$ violation outside the kaon system.
Moreover, when considering $CP$ violation effects in neutral meson systems 
induced by the interference of decays with and without mixing, it is a matter
of convention whether the $CP$ violation phases are assumed to be in the 
$\Delta F=1$ (direct) or in the $\Delta F=2$ (indirect) sector for a  
\emph{single}, given decay mode: only by comparing different decays an 
unambiguous signature for direct $CP$ violation can be found, and super-weak 
$CP$ violation indicates a framework in which $CP$ violation in \emph{any} 
decay mode can be ascribed to the $\Delta F=2$ mixing.

The Standard Model does not belong to the super-weak class, $CP$ violation 
arising in an ubiquitous phase appearing in most weak decays driven by the 
charged current of quarks.

Always assuming $CPT$ symmetry, considering a $CP$ eigenstate $f$ with 
eigenvalue $CP=+1$ (such as $\pi\pi$), so that the $K_2$ state would not 
decay to it in absence of $CP$ violation, the measure of $CP$ violation is 
usually expressed in terms of 
\Dm{
  \eta_f = |\eta_f| e^{i\phi_f} \equiv \frac{A(\kl \To f)}{A(\ks \To f)} 
  \quad (CP|f\rangle = +|f\rangle)
}
Both the modulus and the phase of $\eta_f$ can be measured\footnote{The phase
convention independent expression for the physical parameter $\eta_f$ should
be properly written as $\eta_f = \frac{A(\kl \To f)}{A(\ks \To f)} 
\frac{\langle \kz | \ks \rangle}{\langle \kz | \kl \rangle}$, where the second 
factor is 1 with the standard phase convention.}: the first from a 
branching ratio measurement, and the second by studying the $\kl-\ks$ 
interference term in the decay rate as a function of proper time $t$, in the 
region where the contribution by both physical states are comparable (around 
$t \approx 12 \tau_S$): this requires a good knowledge of the admixture of \kz 
and \kzb in the beam used to perform the measurement, such as is obtained by 
using pure \kl or \ks beams.

To first order in the (small) parameter $|\beps|$ (\emph{i.e.} in an 
appropriate phase convention) one has
\Dm{ 
  \eta_f \simeq \beps + \epsilon_f
}
where 
\Dm{
  \epsilon_f \equiv \frac{A(K_2 \To f)}{A(K_1 \To f)}  \quad
  (CP|f\rangle = +|f\rangle)
}
The above decomposition of $\eta_f$ depends on the choice of phase convention;
$\epsilon_f$ represents the direct $CP$ violation part of the amplitude ratio, 
while $\beps$ represents the part proceeding through the $CP$-conserving decay
of the small component with opposite $CP$ eigenvalue ($K_1$ in this case). 

For the experimentally less accessible case of final states which are $CP$ 
eigenstates with $CP=-1$, the measure of $CP$ violation due to $\ks$ decays is
similar:
\Dm{
  \eta_f = |\eta_f| e^{i\phi_f} \equiv \frac{A(\ks \To f)}{A(\kl \To f)} 
    \quad (CP|f\rangle = -|f\rangle) 
}
\Dm{
  \epsilon_f \equiv \frac{A(K_1 \To f)}{A(K_2 \To f)}  \quad
  ( CP|f\rangle = -|f\rangle)
}

While in principle the measurement of a single $\eta_f$ could give a 
measurement of $\epsilon_f$ by comparison with the value of $\beps$ obtained 
from the $\kl$ semi-leptonic charge asymmetry, $\epsilon_f$ turns to be so 
small that such an approach cannot reach the required precision, being 
affected by systematic errors which cannot be controlled at the level 
required. 
Moreover, as already stated, for a single final state $f$ the amount of $CP$ 
violation ascribed to $\epsilon_f$ is phase-convention dependent, and only a
measurement of two different decay channels allows a definite answer to be
given on the existence of $CP$ violation.

The $\kl$ decay channels with largest branching ratio which can support direct 
$CP$ violation are $\pi^+ \pi^-$ and $\pi^0 \pi^0$, for which respectively 
\cite{PDG_2002} \cite{KTeV_eprime2}:
\Ea{ 
  |\eta_{+-}| = (2.287 \pm 0.017) \cdot 10^{-3} \quad && \quad
  \phi_{+-} = (43.4 \pm 0.7)^\circ \\
  |\eta_{00}| = (2.23 \pm 0.011) \cdot 10^{-3} \quad && \quad 
  \phi_{00} = (43.2 \pm 1.0)^\circ
}

When considering neutral kaon decays to $\pi \pi$, which are the dominant 
final states common to both $\ks$ and $\kl$ (through $CP$ violation), one has 
to take into account the fact that the decay channels which are independent 
from the point of view of the $CP$-conserving strong interactions are actually 
the states of definite isospin $I=0,2$ ($I=1$ being forbidden by Bose 
symmetry, since kaons have zero spin). 
One can define the (unmeasurable) amplitude ratios to a $\pi \pi$ final state 
of definite isospin:
\Ea{
  \eta_I \equiv 
  \frac{A(\kl \To (\pi \pi)_I)}{A(\ks \To (\pi \pi)_I)} 
}
which for the dominant amplitude, the one for the isospin 0 state indicated 
as $(\pi\pi)_{I=0}$, is usually named $\epsilon$:
\Dm{
  \epsilon \equiv \eta_0
}

In the limit of isospin symmetry, the Fermi-Watson theorem (based on $CPT$ and 
unitarity) allows to factor out from the decay amplitudes the strong phases 
$\delta_I$, corresponding to $\pi \pi$ scattering in the isospin $I$ 
eigenstate at energy equal to the $\kz$ 
mass:
\Ea{
  & A_I \equiv A(\kz \rightarrow (\pi \pi)_I) = a_I e^{i\delta_I} \\
  & \overline{A}_I \equiv A(\kzb \rightarrow (\pi \pi)_I) = 
    \overline{a}_I e^{i\delta_I} 
}
which is useful because, due to $CPT$ symmetry one has $\overline{a}_I = 
a_I^*$, implying that $CP$ violation is just described by the imaginary part 
of $a_I$.

The experimental fact that the $I=0$ isospin state of $\pi\pi$ dominates over 
the $I=2$ one in $K$ decays (as can readily be seen from the ratio of $\pi^+
\pi^-$ and $\pi^0 \pi^0$ branching ratios) constitutes the so-called 
$\Delta I =1/2$ ``rule'', for which no convincing theoretical explanation has 
been found so far. 
The violation of this ``rule'' is parameterised by the quantity
\Dm{
  \omega \equiv \frac{A(\ks \To (\pi\pi)_{I=2})}{A(\ks \To (\pi\pi)_{I=0})}
}
and experimentally (neglecting possible amplitudes with $|\Delta I| > 3/2$, 
as allowed by the data) $|\omega| \simeq 1/22.2$ (although including isospin 
breaking effects one could get $|\omega| \simeq 1/29.5$ \cite{Gardner}).

Given this strong dominance, a practical choice of phase convention is 
the one in which the corresponding decay amplitude is real (Wu-Yang phase 
convention); only in this case the $\kz--\kzb$ mixing parameter $\beps$ 
coincides with $\epsilon$, while in general (dropping terms of order $\beps \, 
\Rea(a_2)/Rea(a_0), \beps \, \Imm(a_2)/\Rea(a_0)$ and in the limit $|\beps| 
\ll 1$) one has:
\Dm{
  \epsilon = \beps +i\frac{\Imm(a_0)}{\Rea(a_0)}
}
The parameter $\epsilon$ is independent from the choice of phase convention;
its real part $\Rea(\epsilon) \simeq \Rea(\beps)$ describes $CP$ violation in 
the mixing, while its imaginary part expresses $CP$ violation in the 
interference of mixing and decay for the dominant isospin 0 $\pi\pi$ channel, 
which as such is not a definite signal of direct $CP$ violation.
The phase of $\epsilon$ is determined by $CPT$ symmetry to be close to the 
so-called ``super-weak phase'' \cite{PDG_2002} \cite{KTeV_eprime2}:
\Dm{
  \phi_{SW} \equiv 
  \arctan \left[ \frac{2 \Delta m}{\Delta \Gamma} \right] \simeq
  (43.46 \pm 0.05)^\circ
}
the two phases being equal in the limit in which no direct $CP$ violation 
occurs in $\Delta S=1$ decay amplitudes of neutral kaons, except for the 
dominant one to the $(\pi\pi)_{I=0}$ state (and small overall $CP$ violation), 
since 
\Dm{
  \epsilon \simeq 
    \frac{\Imm(\Gamma_{12}^* M_{12})}{(\Delta m)^2 \sqrt{2}} e^{i\phi_{SW}} +
    i \left[ \frac{\Imm(\Gamma_{12})}{\Delta \Gamma} + 
      \frac{\Imm(a_0)}{\Rea(a_0)} \right]
}
and the second term cancels in the above limit.

Direct $CP$ violation in $\pi \pi$ decays of neutral kaons can arise as a 
difference between the amount of $CP$ violation in different isospin channels, 
and is parameterised by
\Dm{
  \epsilon' \equiv \frac{\omega}{\sqrt{2}} (\eta_2 - \eta_0) =
  \frac{\eta_0}{\sqrt{2}} \left[ \frac{A(\kl \To (\pi \pi)_{I=2})}
  {A(\kl \To (\pi \pi)_{I=0})} -
  \frac{A(\ks \To (\pi \pi)_{I=2})}{A(\ks \To (\pi \pi)_{I=0})} \right]
}
Both isospin states contribute (differently) to the two measurable channels
$\pi^+ \pi^-$ and $\pi^0 \pi^0$: from the isospin decomposition (neglecting 
possible $\Delta I > 3/2$ amplitudes, see \cite{Gardner})
\Ea{
  & A(\kz \To \pi^+ \pi^-) = 
    \frac{1}{\sqrt{3}} ( \sqrt{2} A_0 + A_2) \\
  & A(\kz \To \pi^0 \pi^0) = 
    \frac{-1}{\sqrt{3}} ( A_0 - \sqrt{2} A_2) 
}
and one obtains
\Ea{ 
  \Ds
  & \eta_{+-} = \epsilon + \frac{\epsilon'}{1+\omega/\sqrt{2}} \\
  & \eta_{00} = \epsilon - \frac{2\epsilon'}{1-\omega \sqrt{2}} 
}
without any approximation.
All the above parameters are independent from the choice of phase convention.

The experimental similarity of $\eta_{+-}$ and $\eta_{00}$, both in modulus 
and phase, indicate that $|\epsilon'| \ll |\epsilon|$, \emph{i.e.} direct $CP$ 
violation is relatively small in kaon decays. 
 
The above formul\ae\, can be simplified by introducing some approximations.
Since $|\omega| \ll 1$ one has 
\Ea{ 
  \eta_{+-} \simeq \epsilon + \epsilon' \quad && 
  \quad \eta_{00} \simeq \epsilon - 2\epsilon'
}
which in the literature are sometimes promoted (inconsistently) to exact 
relations defining (suitably different) $\epsilon$ and $\epsilon'$ parameters.

The smallness of $CP$ violation ($|\epsilon|,|\epsilon'| \ll 1$) allows 
(together with $CPT$ symmetry) to approximate
\Ea{ \Ds
  & \omega \simeq e^{i(\delta_2-\delta_0)} 
    \Rea \left( \frac{a_2}{a_0} \right) \\
  & \epsilon' \simeq \frac{i}{\sqrt{2}} e^{i(\delta_2 - \delta_0)}
    \frac{\Rea(a_2)}{\Rea(a_0)} \left[ \frac{\Imm(a_2)}{\Rea(a_2)} - 
    \frac{\Imm(a_0)}{\Rea(a_0)} \right] \simeq 
    \frac{i}{\sqrt{2}} e^{i(\delta_2 - \delta_0)} 
    \Imm \left( \frac{a_2}{a_0} \right)
}
which are valid in the limit in which all the quantities $|\beps|$, $|\beps \, 
\Rea(a_2)/\Rea(a_0)|$, $|\beps \, \Imm(a_2)/\Rea(a_0)|$, $|\beps \, 
\Imm(a_0)/\Rea(a_0)|$ are much lower than 1.

In this approximation the phase of $\epsilon'$ is determined by the strong
phases $\delta_I$; from the analysis of $\pi \pi$ interaction data, one 
obtains \cite{Colangelo}: $\delta_2 - \delta_0 = (-47.7^\circ \pm 1.5^\circ)$ 
as the best estimate of this phase difference\footnote{The long-standing 
problem of the discrepancy of such measurements with the values obtained by 
the isospin analysis of the neutral and charged kaons decay amplitudes to two 
pions, which are affected by larger theoretical uncertainties (see \emph{e.g.} 
\cite{Cirigliano} \cite{Gardner} and references therein), might be clarified 
by better measurements of the neutral kaon $\pi\pi$ branching ratios, 
\emph{i.e.} $\delta_2-\delta_0 \simeq (-47.8 \pm 2.8)^\circ$ from KLOE 
\cite{KLOE_delta_pipi}, \cite{KLOE_R}.}.

The reader could notice that the parameter $\epsilon'$ does not vanish even if 
the strong scattering phases $\delta_I$ for $I=0,2$ are equal to each other. 
This is so because $\epsilon'$ is not a measure of $CP$ violation in the decay 
into \emph{a single} channel, but actually refers to decay modes receiving 
contributions from two distinct amplitudes, corresponding to definite isospin 
states $I=0,2$. $CP$ violation in the decay is singled out in the real part of 
$\epsilon'$ which, in order to be non-zero, requires the strong phases for the 
two channels to be different ($\delta_2-\delta_0 \neq 0,\pi$). 
The imaginary part of $\epsilon'$ instead, does not disentangle $CP$ violation 
in the decay from the mixing-induced one, but since it involves the difference 
of $CP$ violation in two decay modes, it is an unambiguous indicator of direct 
$CP$ violation nevertheless.

For a final state $f$ which is a $CP$ eigenstate one has in general
\Dm{
  \frac{\Gamma(\kz \To f) - \Gamma(\kzb \To f)}
  {\Gamma(\kz \To f) + \Gamma(\kzb \To f)} \simeq 2 \Rea(\epsilon_f)
}
(valid in the limit $|\eta_f| \ll 1$ if $|\beps| \ll 1$) and for $\pi^+ \pi^-$ 
or $\pi^0 \pi^0$ states the $\epsilon_f$ parameters are $\epsilon_{+-} \simeq 
\epsilon'$, $\epsilon_{00} \simeq -2 \epsilon'$, respectively.

Since $\epsilon'$ arises from the interference of two amplitudes, one of which
is empirically found to be suppressed by the rather small factor $\omega$, one
can see that this measure of direct $CP$ violation is suppressed with respect 
to the ``natural'' magnitude which could be expected from the size of the 
complex phases in the elementary amplitudes.

It could also be noticed that the (conventional) pseudoscalar nature of kaons 
and pions implies that the state which decays dominantly into $\pi\pi$ is the
$C$-odd one, so that indirect $CP$ violation described by $\epsilon$ is 
$C$-odd and $P$-even, while direct $CP$ violation described by $\epsilon'$ is 
$C$-even and $P$-odd.

\begin{figure}[hbt!]
\begin{center}
    \epsfig{file=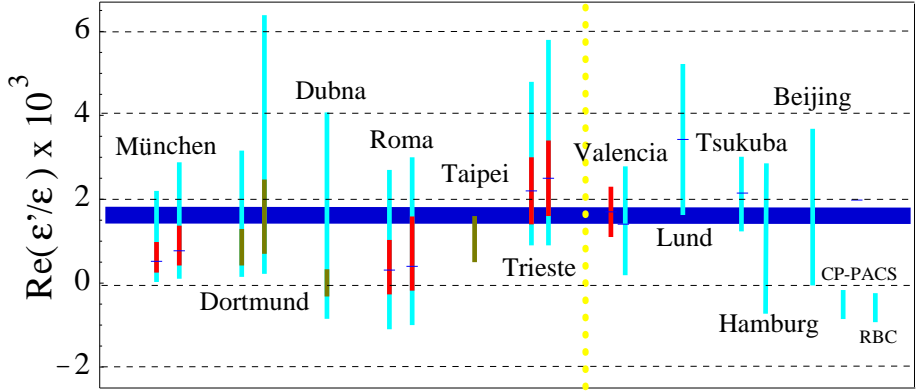,width=1.0\textwidth}
  \caption{Recent Standard Model predictions for $\Rea(\epseps)$ by different 
groups along the years. The vertical dotted line indicates the time at which 
the results from the latest generation of experiments become available; the 
horizontal band is the experimental world average. Where darker error bars are 
shown, they refer to independent ``Gaussian'' variation of the input 
parameters, while the larger ones correspond to the full ``scanning'' of the 
parameter space.}
  \label{fig:theory}
\end{center}
\end{figure}

A great deal of theoretical effort was devoted to the computation of 
$\epsilon'$ in the Standard Model (see fig. \ref{fig:theory}), which turns out 
to be a daunting task (see \cite{epseps_theory} for a recent review): no 
reliable approach exists yet to compute all required matrix elements, in which 
strong interaction effects play a relevant role. The two dominant terms which 
contribute to $\epsilon'$ turn out to have similar phases, so that an 
additional cancellation makes the final result both smaller and more sensitive 
to theoretical uncertainties.

It should be clear, therefore, that the smallness of $\epsilon'/\epsilon$ is
to a large extent accidental, and as such it does not preclude the fact that 
direct $CP$ violating effects can be larger, and even dominant, in other kinds
of processes. 

It should be noticed that the phase of $\epsilon'$ is quite close to that of 
$\epsilon$ $(\simeq \phi_{SW})$.
This accidental similarity of phases (which would not be valid if the 
short-lived neutral kaon were heavier than the long-lived one, instead of the 
opposite) implies that for small direct $CP$ violation ($|\epsilon'| \ll 
|\epsilon|$), the $CPT$-enforced equality of the phases $\phi_{00}$ and 
$\phi_{+-}$ with $\phi_{SW}$ is already well constrained.
From the above discussion it should be clear that, in first approximation, the 
measurement of the single real number $\Rea(\epseps)$ is sufficient to get
information on direct $CP$ violation, while a significant non-zero value of 
$\Imm(\epseps)$ would be a signal of $CPT$ violation.

Small violations of Bose statistics, not experimentally excluded otherwise, 
have been discussed \cite{Greenberg} as possible contributions to the 
difference among $\eta_{+-}$ and $\eta_{00}$: by allowing a small component  
of the isospin state $I=1$ in $\pi^+ \pi^-$ with zero relative angular 
momentum, one can have a $CP$-conserving contribution to \mbox{$\kl \To \pi^+ 
\pi^-$}. Therefore, the measured difference of $CP$ violation in the two 
$\pi \pi$ decay channels, usually ascribed to direct $CP$ violation 
($\epsilon' \neq 0$) can be used to give an upper bound to the probability 
$\beta_{B}^2$ of wrong-statistics admixture, in the limit in which no direct 
$CP$ violation is present, namely
\Dm{
        \beta_{B}^2 \le 2.7 \cdot 10^{-6}
}

\subsection{Experiments on $K \To \pi \pi$ decays}

If the only mechanism for $CP$ violation would be the super-weak one, inducing 
an asymmetry in $\kz--\kzb$ mixing, all $\pi \pi$ decays of neutral kaons 
would be $CP$-conserving decays of the $K_1$ component, both for $\ks$ and 
$\kl$: in this case all their properties, such as the ratio of $\pi^+ \pi^-$ 
(``charged'') to $\pi^0 \pi^0$ (``neutral'') decays, should be the same for 
both physical states.
In other words, if $CP$ violation only manifests itself as a ``constituent'' 
property of the decaying meson itself, it should appear to be the same through
any kind of decay process; considering the $\pi \pi$ decays in which 
$CP$ violation was originally found, the $CP$-violating amplitude ratios 
$\eta_{+-}$ and $\eta_{00}$ should be exactly equal in this case.
The ratio of these two $CP$-violating amplitude ratios is thus a quantity 
sensitive to the presence of (direct) $CP$ violation in the decay amplitudes.

The squared modulus of the ratio of the above mentioned $CP$-violating 
parameters is an experimentally accessible quantity, being the double ratio 
$R$ of partial decay amplitudes
\Dm{
  R \equiv \frac{\Gamma(\kl \rightarrow \pi^0 \pi^0)}
                {\Gamma(\ks \rightarrow \pi^0 \pi^0)}
           \frac{\Gamma(\ks \rightarrow \pi^+ \pi^-)}
                {\Gamma(\kl \rightarrow \pi^+ \pi^-)} = 
           \left| \frac{\eta_{00}}{\eta_{+-}} \right|^2
}
Experiments measuring $R$ are basically counting experiments, which must 
accurately measure the number of $\pi \pi$ events from $\kl$ and $\ks$, within 
regions of same volume in phase space, which for such 2-body decays means kaon 
momentum, decay position, and centre of mass angular coordinates.

Given the approximate validity of the $\Delta I = 1/2$ ``rule'' ($|\omega| \ll
1$), and the smallness of direct $CP$ violation ($|\epsilon'| \ll 1$), the 
following approximate relation holds
\Dm{
  \Rea(\epseps) \simeq \frac{1}{6}(1-R)
\label{eq:R}
}

The extraction of $\epseps$ from the comparison of $\eta_{+-}$ or $\eta_{00}$ 
with the mixing parameter $\epsilon$, as obtained from the charge asymmetry 
$\delta_L$ in semi-leptonic $\kl$ decays, is not precise enough to draw any 
conclusion on direct $CP$ violation: neglecting $|\omega|$ one has 
\Ea{
  \Ds
  & \Rea(\epsilon'/\epsilon) \simeq 
  \left| \frac{\epsilon'}{\epsilon} \right| \simeq 
  \frac{2|\eta_{+-}|\cos(\phi_{+-})}{\delta_L}-1 = 
  (17 \pm 39) \cdot 10^{-3} \\
  & \Rea(\epsilon'/\epsilon) \simeq 
  \left| \frac{\epsilon'}{\epsilon} \right| \simeq 
  \frac{1}{2}-\frac{|\eta_{00}|\cos(\phi_{00})}{\delta_L} = 
  (11 \pm 36) \cdot 10^{-3} 
}
while assuming $CPT$ symmetry (as well as $\phi(\epsilon)=\phi_{SW}$) and 
$|\epsilon'|^2 \ll 1$ one has also 
\Ea{
  \Ds
  & \Rea(\epsilon'/\epsilon) \simeq \frac{1}{2} 
  \left[ \frac{4|\eta_{+-}|^2}{\delta_L^2 (1+\tan^2 \phi_{SW})} -1 \right] =
  (15 \pm 39) \cdot 10^{-3} \\
  & \Rea(\epsilon'/\epsilon) \simeq \frac{1}{4} 
  \left[ 1 - \frac{4|\eta_{00}|^2}{\delta_L^2 (1+\tan^2 \phi_{SW})} \right] =
  (5 \pm 30) \cdot 10^{-3}
}

From an experimental point of view, initially the main difficulty in a 
measurement of the double ratio $R$ was the determination of 
\mbox{$\Gamma(\kl \To \pi^0 \pi^0)$} with sufficient precision. 
The difficulties of measuring accurately the energy and direction of the 
final-state photons, and the presence of the severe background due to the 
$CP$-conserving \mbox{$\kl \To 3\pi^0$} decays, 200 times more frequent, made 
such measurements quite challenging. 

In experiments in which the $\ks$ flux is not a limiting factor, the 
statistical error on the double ratio is usually limited by the number 
$N^L_{00}$ of collected \mbox{$\kl \To \pi^0 \pi^0$} decays, and is roughly 
given by $1.2/\sqrt{N^L_{00}}$ if the experimental acceptances for $\pi^+
\pi^-$ are comparable to the the ones for $\pi^0 \pi^0$.
Intense kaon beams are therefore a primary requirement for any experiment 
which aims at measuring $R$ with high precision.

$\kl$ decays are readily obtained from a neutral beam produced at a 
sufficiently large distance from the region of observation: such beams usually
contain also large amounts of neutrons and photons from $\pi^0$ decays.

One way of producing $\ks$ decays is that of exploiting the phenomenon of kaon 
regeneration in matter (see \emph{e.g.} \cite{Belusevic}).
Early experiments used regenerators placed on neutral ($\kl$) beams to 
produce $\ks$. The proper decay time $t$ distribution into a final state $f$ 
with $CP=+1$ can be described in terms of the magnitude and phase of the 
$CP$-violating amplitude ratio $\eta_f$ and of the regeneration amplitude 
$\rho$:
\Dm{
  I_f(t) \propto 
  |\rho|^2 e^{-\Gamma_S t} + 2 |\rho| |\eta_f| \cos (\Delta m \, t 
  -\phi_f + \phi_\rho) e^{-(\Gamma_S+\Gamma_L)t/2} + |\eta_f|^2 e^{-\Gamma_L t}
}
where $f=\pi^+\pi^-$ or $\pi^0\pi^0$, and $\phi_\rho$ is the phase of the 
regeneration amplitude $\rho$ (which can be measured in dedicated 
experiments); the expression exhibits the oscillating term due to the 
interference of $\ks$ and $\kl$ decay amplitudes. 

It is evident from the above expression that even with regeneration amplitudes 
typically below 10\%, most of the \mbox{$K \To \pi \pi$} decay rate downstream 
of the regenerator is due to the \ks component.
A major concern with this technique is due to the incoherently regenerated 
$\ks$ events which, being produced at a finite angle, can be measured in the 
detector with different acceptance relative to $\kl$; the fraction of such 
events has to be carefully estimated, or measured, and then subtracted.

Without making use of a regenerator, a neutral beam containing both $\ks$ and
$\kl$ is readily available close to the production target; thanks to $CPT$ 
symmetry, $\ks$ are produced in the same amount as $\kl$, and the proper-time 
distribution of decays to a given $\pi\pi$ state $f$ is 
\Dm{
  I_f(t) \propto e^{-\Gamma_S t} + 2D(p_K) |\eta_f| \cos (\Delta m t -\phi_f)
  e^{-(\Gamma_S+\Gamma_L)t/2} + |\eta_f|^2 e^{-\Gamma_L t}
}
In the above formula the ``dilution factor'' $D(p_K)$ is defined as the ratio 
of (incoherent) $\kz$ and $\kzb$ production at the target, for a given value 
of the kaon momentum $p_K$
\Dm{
  D(p_K) \equiv \frac{N_{\kz}(p_K) - N_{\kzb}(p_K)}
                     {N_{\kz}(p_K) + N_{\kzb}(p_K)}
\label{eq:dilution}
}
which depends on the details of the interaction (targeting angle, etc.) and 
can also be fitted from the $I_f(t)$ distribution for either decay mode.

Due to the large $\kl/\ks$ lifetime ratio and $|\eta_f| \ll 1$, the $\pi \pi$ 
decays occurring in a region close to the target ($t \lesssim$ a few $\tau_S$) 
are largely dominated by the $\ks$ component.
A practical concern in this case, in which one has to collect decays close to
the production target, is the required shielding of the detector from the high 
rate of particles generated in the collision.
On the other hand, not using a regenerator frees from the need of knowing (or 
fitting) the regeneration amplitude in order to perform a measurement of the
decay parameters.

Pure $\ks$ beams can only be obtained by exploiting the correlated production 
of $\ks \kl$ pairs, as is done in $e^+ e^-$ collider experiments working at 
the energy of the $\phi$ resonance, using a detected $\kl$ decay in the 
opposite hemisphere to ``tag'' a $\ks$, as discussed in section 3.5.

Instead of working with kaon mass eigenstates, associate production of neutral 
kaons such as the study of the proton-antiproton annihilation reactions 
\mbox{$p\overline{p} \To \pi^+ K^- \kz$} and \mbox{$p\overline{p} \To \pi^- 
K^+ \kzb$} allows to compare asymmetries as a function of time in decays of 
neutral kaons tagged to be initially in a strangeness eigenstate, as discussed 
in section 3.6.

The early results suggested that $CP$ violation in $\pi^0 \pi^0$ decays was 
about twice as large than in $\pi^+ \pi^-$ decays, and therefore $\epsilon'$ 
was a large number. Improved experiments however showed that $|\eta_{00}|$ was
close to $|\eta_{+-}|$, and when $\phi_{00}$ was measured in interference
experiments and shown to be comparable to $\phi_{+-}$, the smallness of
$\epsilon'$ was proved.

The results of the first two experiments specifically designed to detect direct
$CP$ violation by comparing $\pi \pi$ decays in a pure $\kl$ beam with those 
in a beam where the great majority of such decays are due to the $\ks$ 
component were reported in 1972. Despite exploiting the best experimental 
techniques then available, both could collect only a few dozen 
\mbox{$\kl \To \pi^0 \pi^0$} decays. Given this basic statistical limitation, 
only an upper limit for $\epsilon'/\epsilon$ of historical interest could be 
inferred, but the methods employed were relevant for the subsequent 
developments. 

With no hint to the size of the effect, no measurements of direct $CP$ 
violation were performed for some time. With the advent of the 
Cabibbo-Kobayashi-Maskawa model for weak interactions, definite predictions 
became available ($\epseps \sim 1/450$ \cite{Ellis} or even $10^{-1} \div 
10^{-2}$ \cite{Gilman}) and an intense experimental activity started again in 
the '80s.

\subsection{\bf First experiments on direct $CP$ violation in $K \To \pi\pi$
decays}

\subsubsection*{\emph{BNL-AGS experiment (1972)}}

The measurement of the Princeton group at the Brookhaven AGS \cite{Banner72} 
used a 20 cm thick movable uranium regenerator to measure the \mbox{$\ks \To 
\pi \pi$} decay rates; the regeneration amplitude cancels in forming the 
double ratio $R$.
The experimental setup was switched between detection of charged and neutral 
decay modes, so that all four decay rates entering the double ratio 
were eventually measured, for kaon momenta in the \mbox{3-10 GeV/$c$} range.
For neutral events, at least one of the four photons from $\pi^0 \pi^0$ decays
was required to convert to an $e^+ e^-$ pair and be measured in a 
spark chamber magnetic spectrometer; the other 3 photons were detected, and 
their impact positions were measured, in a large lead plate chamber. 
The (subtracted) background fraction in the neutral mode, due to \mbox{$\kl 
\To 3\pi^0$} decays was at the level of 2.5\%. 
For charged decays a Cerenkov counter and a muon veto were also used, in order 
to reject semi-leptonic $\kl$ decays.

The optimal thickness of the photon converter ($\sim$ 0.1 radiation lengths)
necessary for triggering and decay vertex determination, coupled with the 
limited solid angle through the spectrometer magnet, limited in an important
way the acceptance for $\pi^0 \pi^0$ decays.
Regenerator ($\ks$) events were weighted as a function of the regenerator 
position along the beam, in order to get a decay-point distribution similar to 
the one due to $\kl$. 
Incoherently regenerated $\ks$ were statistically subtracted; their fraction 
was directly measured by the transverse momentum distribution for charged 
decays, and used for neutral events after estimating the neutral/charged 
acceptance ratio by Monte Carlo simulation. 
$\kl$ flux normalisation was performed using beam target activity monitors.

With a sample of 124 \mbox{$\kl \To \pi^0 \pi^0$} decays, the result was 
$\Rea(\epseps)=$ \mbox{$(-10 \pm 24)$} $\cdot 10^{-3}$.

\subsubsection*{\emph{CERN-PS experiment (1972)}}

The experiment of the Aachen-CERN-Torino group at the CERN PS \cite{Holder72}
also alternated movable-regenerator runs and ``vacuum'' runs (no regenerator on
the $\kl$ beam) to measure the most challenging \mbox{$K_{S,L} \To \pi^0 
\pi^0$} decays with a non-magnetic detector based on a lead-glass calorimeter, 
with kaons in the 2-6 GeV/$c$ momentum range. 
As for the BNL experiment, data was collected with the regenerator placed at 
several different longitudinal positions, in order to match the decay vertex 
distribution of $\kl$ along the 4 m decay volume.
By requiring at least two photons to convert in thin lead foils before 
entering the detector and be measured in spark chambers, the directions of all 
four photons hitting the calorimeter could be known. In this way the resolution
on the reconstructed $\gamma \gamma$ invariant mass was about 10 MeV/$c^2$, 
and the subtracted \mbox{$\kl \To 3\pi^0$} background was kept below 2\%.

With this method, the diffractively regenerated $\ks$ component, limited by
the destructive interference with the coherently regenerated one due to the 
choice of regenerator thickness, could be directly measured from the angular 
distribution in the data, and was found to be around 11\%. 
Flux normalisation was performed by using \mbox{$\kl \To 3\pi^0$} decays, 
mostly unaffected by regeneration, from which however a large ($\simeq$ 33\%) 
diffracted component had to be subtracted.

With 167 \mbox{$\kl \To \pi^0 \pi^0$} decays, and using independent 
measurements of $|\eta_{+-}/\rho|^2$ obtained with a different detector placed 
on the same beam, the result was $\Rea(\epseps)=$ \mbox{$(0 \pm 20)$} $\cdot 
10^{-3}$. 
The largest systematic error was actually due to the error on the knowledge of 
the regeneration amplitude, other important contributions being due to the
knowledge of the regenerator position and the $\kl$ scattering subtraction.

\subsubsection*{\emph{Second BNL-AGS experiment (1979)}}

A second experiment at the AGS by a New York group \cite{Christenson79} 
exploited the detection of both charged and neutral $\pi \pi$ decays from a 
beam of neutral kaons of 8-18 GeV/$c$ momentum, using a detector arrangement 
which could be moved along the beam direction, placed at short distance 
($\approx$ 7 m) from the production target. 

The good proper decay time resolution of $\simeq 0.4 \tau_S$ achieved also for 
neutral decays, allowed the measurement of the $\ks--\kl$ interference in both 
the $\pi^0 \pi^0$ and $\pi^+ \pi^-$ modes, in the proper decay time intervals 
4-11 $\tau_S$ (for neutral) and 4-18 $\tau_S$ (for charged). 
From the fit of the $\pi \pi$ decay distributions, using the known value of 
$\Delta m$ as input, both the moduli and phases of $\eta_{+-}$ and 
$\eta_{00}$ were measured (as well as the dilution factor).

Charged and neutral decays were collected with alternate detector settings:
a magnetic spectrometer based on proportional wire chambers providing a 9 
MeV/$c^2$ kaon mass resolution for the former, and a lead-glass calorimeter 
providing a 21 MeV/$c^2$ $\pi^0$ mass resolution for the latter. 
The detection of $\pi^0 \pi^0$ decays required at least two of the decay 
photons to be converted to $e^+e^-$ in a set of three lead sheets, for 
triggering and measurement purposes. 

Backgrounds from 3-body $\kl$ decays were important, of the order of 10\% for 
the neutral mode and even larger for the charged one; backgrounds from 
coherent regeneration and diffractive scattering in the photon converter and
collimators also had to be taken into account.

The comparison of the amplitude ratios for the two $\pi \pi$ decay modes from 
the same experiment was free from most systematic errors at the level of 
accuracy allowed by the statistical sample. 
Measurements performed at two longitudinal positions of the detector, shifted
by 1 m, were combined after correcting for the acceptance dependence; their 
consistency was used as a systematic check.

The result on direct $CP$ violation was $\Rea(\epseps)=$ \mbox{$(0 \pm 30)$} 
$\cdot 10^{-3}$, based on 85000 \mbox{$K_{L,S} \To \pi^0 \pi^0$} decays.

\subsubsection*{\emph{BNL E749 (1985)}}

In 1979 an interesting proposal for the measurement of direct $CP$ violation 
at the AGS with a different experimental approach was submitted to BNL by a
Yale-Brookhaven group \cite{Adair}. 
Although the experiment was not actually performed with the approach initially
discussed, it is interesting to consider its proposed scheme.

The idea is to search for a difference in the charge ratio ($\pi^0 \pi^0$ to 
$\pi^+ \pi^-$) for $\pi \pi$ decays of pure $\kl$ and $\ks$, induced by the 
direct $CP$ violating \mbox{$K_2 \To \pi \pi$} decay.
The dominance of $CP$ violation induced by the $\kz--\kzb$ mixing implies that 
most of the $\pi \pi$ decays, both from $\kl$ and $\ks$ are due to their $K_1$ 
components, while a possible small fraction of decays of the $K_2$ component 
could have different properties (such as the charge ratio).

Considering decays after a regenerator placed on a $\kl$ beam, the $\pi \pi$ 
decay amplitude from coherently regenerated $\ks$ at any given point is 
coherent with the ones due to $\kl$ (through $K_1$ or $K_2$), their phase 
difference varying linearly with the distance of the decay point from the 
regeneration point due to the $\kl-\ks$ mass difference (see \emph{e.g.} 
\cite{Kabir}); the ratio of regenerated $\ks$ to transmitted $\kl$ amplitudes
at the exit face of the regenerator of thickness $L$ is:
\Dm{
  \frac{\rho}{A_L} = \frac{2\pi n}{m_K (\lambda_S-\lambda_L)} 
  \frac{f(0)-\overline{f}(0)}{2} [1-e^{-i(\lambda_S-\lambda_L)L/\beta \gamma}]
}
where $n$ is the density of scattering centres, $f(0)$ and $\overline{f}(0)$
are the forward scattering amplitudes for $\kz$ and $\kzb$ respectively, and 
$\beta,\gamma$ the relativistic factors for the kaon.
The corresponding $\pi^+ \pi^-$ decay intensity in vacuum as a function of 
proper time $t$ (with $t=0$ at the exit face of the regenerator) is therefore
\Ea{
  & R(\pi^+ \pi^-;t) \propto 
    \left[ \left| \frac{\rho}{A_L} \right|^2 e^{-\Gamma_S t} + 
    |\eta_{+-}|^2 e^{-\Gamma_L t} + \right. \nonumber \\
  & \left. 2 \left| \frac{\rho}{A_L} \right| |\eta_{+-}| 
    \cos(\Delta m t - \phi_{+-} + \phi_\rho) 
    e^{-(\Gamma_S+\Gamma_L)t/2} \right]
}
where the regeneration phase, containing a term due to the regenerator
thickness 
\Dm{ 
  \phi_\rho = \arg[i(f(0)-\overline{f}(0))] + 
  \arg[(1-e^{-i \Delta \lambda L/\beta\gamma})/i \Delta \lambda]
} 
can be experimentally determined.

In presence of a direct $CP$ violating amplitude (\mbox{$K_2 \To \pi \pi$}), 
the charge ratio of $\pi \pi$ decays could be expected to vary along the beam.
With an appropriate choice of regenerator, in some proper decay time interval 
there could be significant destructive interference between the regenerated 
$\ks$ ($\approx K_1$) and (indirect $CP$-violating) $K_1$ amplitudes from 
$\kl$; in this region any effect on the charge ratio due to a direct 
$CP$-violating \mbox{$K_2 \To \pi\pi$} decay would be strongly put in evidence.
Given the fact that for $K_1$ decays the $\pi^0 \pi^0/\pi^+ \pi^-$ ratio is 
$\simeq 1/2$ (due to the dominance of the isospin $I=0$ component in the final 
$\pi\pi$ state), while for $K_2$ it would be expected to be $\simeq 2$ 
($\pi \pi$ in the $I=2$ isospin state being required), and that any change in 
the charge ratio could only be due to direct $CP$ violation, this proposal 
represented an appealing strategy for its detection.
Unfortunately, knowing that \mbox{$\kl \To \pi\pi$} decays proceed 
dominantly through indirect $CP$ violation, the same decay time interval under
consideration is also the one in which the number of expected $\pi \pi$ events 
will be smallest.

The plan was to make $\pi\pi$ yield measurements with different settings of a 
(thick) fixed and a (thin) movable regenerator at a distance $\sim$ 6 m from 
the production target, in order to measure the charge ratio for pure $\kl$, 
almost pure $\ks$, and a coherent mixture in which the indirect $CP$-violating 
component would be made small by the above discussed interference.
The goal was to measure the charge ratios with an accuracy better than 1\%, 
potentially reaching a sensitivity of $\simeq 1.6 \cdot 10^{-3}$ on 
$|\epseps|$.

The experiment was performed at BNL \cite{Black85}, with a magnetic detector 
based on proportional wire chambers and a lead-glass calorimeter which could 
measure simultaneously both charged and neutral decays, requiring a converted 
photon for the latter.
Anti-coincidence counters and muon veto counters helped in reducing the 3-body
background from $\kl$ decays.
The detector was placed at $\sim$ 10 m from the target.
An 80 cm thick graphite regenerator was placed on the neutral beam at short 
regular intervals, to alternate the collection of $\kl$ and $\ks$ decays in
the 1.2 m long fiducial region, for a 7-14 GeV/$c$ kaon momentum range.

Residual backgrounds for the neutral mode were at the level of 1.5\% from 
incoherent $\ks$ regeneration, and 17.5\% from $\kl$ 3-body background,
and much smaller for charged decays.
The $\kl-\ks$ acceptance difference cancellation was achieved by analyzing
the data in bins of kaon energy and longitudinal decay position, which was 
determined with good accuracy also for neutral decays by extrapolating the 
converted photon trajectory to its intersection with the narrow neutral beam.

With 1122 \mbox{$\kl \To \pi^0 \pi^0$} decays, a result of $\Rea(\epseps)=$ 
\mbox{$(1.7 \pm 7.2 \pm 4.3)$} $\cdot 10^{-3}$ was obtained with the double 
ratio method, the first error quoted being statistical and the second 
systematic, dominated by the uncertainty in the $\kl$ background subtraction.

\subsubsection*{\emph{FNAL E617 experiment (1985)}}

About at the same time of the BNL E749 experiment, the Chicago-Saclay 
experiment E617 at FNAL was performed \cite{Bernstein85}, using a double-beam 
technique to reduce the systematic uncertainties linked to the separate 
data-collection of $\kl$ and $\ks$ decays. 
In this approach, two identical neutral beams, produced by 800 GeV/$c$ protons 
impinging on a single target, entered the 13 m long evacuated decay volume 
side-by-side. 
A thick regenerator at fixed longitudinal position was set on one of them, 
alternating from one beam to the other at each accelerator pulse, in order to 
provide $\ks$ decays while cancelling the effect of any left-right asymmetry 
of either the beam line or the detection apparatus. 
$\kl$ and $\ks$ decays were distinguished on the basis of their reconstructed 
transverse coordinates at the regenerator plane.

The detector setup was switched between charged and neutral mode by adding or 
removing a lead photon converter: for $\pi^0 \pi^0$ decays the direction of at 
least one converted photon was required to be measured in the drift-chamber 
magnetic spectrometer, thus allowing the measurement of the kaon transverse 
momentum with respect to the beam axis, so that inelastically regenerated 
events could be identified and removed. 
A lead-glass calorimeter was used for the measurements of the photon energies
and impact point positions, leading to a 6.5 MeV/$c^2$ kaon mass resolution;
the longitudinal decay point position was determined as the weighted average
of the two (consistent) ones obtained by imposing the $\pi^0$ mass constraint 
to $\gamma \gamma$ pairs.

Muon veto counters were used to suppress $K_{\mu3}$ decays.  
The subtracted backgrounds due to inelastically regenerated $\ks$ were at the 
level of 13\% for neutral and 2\% for charged decays, while backgrounds from 
$\kl$ 3-body decays were at the 8\% and 3\% level respectively; the 
uncertainties on these background subtractions were the largest sources of 
systematic errors.

The irreducible difference in the detector acceptance for $\kl$ and $\ks$ 
decays (of order 10\%) arising from the different lifetimes was corrected by 
Monte Carlo simulation. 
The value of $\epsilon'$ was extracted from a fit of the ratios of regenerated 
to vacuum events in kaon momentum bins for each of the four modes, from which 
the magnitude of the coherent regeneration amplitude, assumed to follow a 
power-law behaviour as a function of kaon momentum, was also determined; the
phase of the regeneration amplitude was used as an input to the fit.

With 3152 \mbox{$\kl \To \pi^0 \pi^0$} events the result was $\Rea(\epseps)=$ 
\mbox{$(-4.6 \pm 5.3 \pm 2.4)$} $\cdot 10^{-3}$, the first error being 
statistical and the second systematic.

\subsection{\bf Recent experiments on $\Rea(\epseps)$}
        
\subsubsection*{\emph{E731 at FNAL}}
        
The E731 experiment, performed at FNAL at the end of the '80s, adopted the 
double-beam technique as its predecessor E617, with a different detector. 

After a short test run in 1985 (E731A), from which $\Rea(\epseps)=$ 
\mbox{$(3.2 \pm 2.8 \pm 1.2) \cdot 10^{-3}$} was obtained \cite{E731A}, the 
experiment was extensively upgraded.
The original setup required at least one of the photons from the 
$\pi^0 \pi^0$ decay to convert in a thin lead sheet, which was added in the 
middle of the 40 m long evacuated decay region during neutral mode running, 
in order to obtain the decay vertex position by tracking the $e^+ e^-$ pair.
The upgraded detector allowed running without this requirement, therefore 
collecting a much higher statistics for the neutral mode.
Most of the data-taking period, in 1987 and 1988, alternated collection of 
$\pi^+\pi^-$ and $\pi^0 \pi^0$ decays, with different proton beam intensities;
in the last 20\% of the data-taking period all four decay modes were collected
simultaneously, further reducing the sensitivity to differential biases.

The regenerator and an upstream absorber were placed alternately on one of the 
two neutral beams at every accelerator pulse.
A lead-glass calorimeter was used to detect $\pi^0 \pi^0$ decays, while
rejecting the \mbox{$\kl \To 3\pi^0$} background down to 1.8\% and, together 
with the magnetic spectrometer used for $\pi^+ \pi^-$ measurement, to suppress 
$K_{e3}$ background down to the 0.3\% level in the vacuum beam. 
The $K_{\mu3}$ background was rejected by muon veto counters.

Events were assigned to the vacuum or the regenerator beam according to the 
extrapolated kaon direction (for $\pi^+ \pi^-$ decays) or using the 
energy-weighted average photon impact position (for $\pi^0 \pi^0$ decays).
This procedure is affected by background due to incoherently regenerated 
$\ks$, which are produced at finite angle: veto counters in the regenerator 
helped in suppressing events resulting from inelastic reactions.
Such background was at the 0.2\% level for $\pi^+ \pi^-$ decays from the 
regenerator beam, measured by extrapolating the kaon squared transverse 
momentum ($p_T^2$) distribution. For neutral decays this background was at the 
2-3\% level in the regenerator as well as in the vacuum beam (due to 
large-angle \ks scattering), and was subtracted by modelling the distributions 
in Monte Carlo using the measured $p_T^2$ spectra from charged decays as input.

To maximise the \mbox{$\kl \To \pi^0 \pi^0$} statistics, the fiducial region 
for the neutral decay mode was longer than the one for charged decays.
In order to better control the large acceptance correction induced by the 
different lifetimes, all the limiting geometrical apertures were defined by 
active veto elements; the correction was based on accurate Monte Carlo 
simulation of the apparatus, checked against large data samples of 
\mbox{$\kl \To 3\pi^0$} and $K_{e3}$ decays.

The direct $CP$ violation parameter was extracted by a global fit (performed in
kaon momentum bins in the 40-160 GeV/$c$ range) of the ratio of $\pi\pi$ 
events in the two beams, corrected for acceptance; the longitudinal vertex 
range used was 110-137 m from the target for the charged mode and 110-152 m
from the target for the neutral mode.
In the fit, the regeneration amplitude at a reference kaon momentum, and the 
exponent of its assumed power-law dependence from such momentum were left as 
free parameters, as well as a parameter describing the energy dependence of 
the kaon absorption in the absorber and regenerator. 
$CPT$ symmetry was assumed in the fit, the phase of $\epsilon'$ was set to the 
value given by the $\pi\pi$ phase shift analysis, while $\Delta m$ and 
$\tau_S$ were set at the values measured by the same experiment.
Averaging the results of each bin, the result was much less sensitive to the
energy dependence of the absorption cross section.

The result on direct $CP$ violation \cite{E731_eprime} with the total sample of
410000 \mbox{$\kl \To \pi^0 \pi^0$} and 329000 \mbox{$\kl \To \pi^+ \pi^-$} 
decays was $\Rea(\epseps)=$ \mbox{$(0.74 \pm 0.52 \pm 0.29)$} $\cdot 10^{-3}$  
(the first error being statistical and the second systematic), consistent with 
no direct $CP$ violation within the errors.

The dominant systematic errors were due to the imperfect knowledge of the 
electromagnetic calorimeter response as a function of energy, accidental 
activity effects in the detector and the knowledge of the acceptance 
correction; a thorough discussion of the E731 experiment can be found in 
\cite{E731_long}.

\subsubsection*{\emph{NA31 at CERN}}

The NA31 experiment, performed at CERN at the same time as the E731 experiment,
detected simultaneously $\pi^0 \pi^0$ and $\pi^+ \pi^-$ decays originating 
from two alternating neutral beams produced by 450 GeV/$c$ protons impinging 
onto targets located at different longitudinal positions. 
The \kl beam was produced by an intense proton beam on a target placed at
$\approx$ 240 m from the detector, while the \ks beam was produced by an 
attenuated (by a factor $\sim 3 \cdot 10^{-4}$) proton beam impinging on a 
similar target assembly, mounted on a movable support, which could be 
positioned at a distance $\approx$ 80-130 m from the detector, inside the 
evacuated decay region. 
The two neutral beams were collinear and contained within an evacuated beam
pipe passing through a central hole in the detectors, in order to reduce 
\kl elastic scattering, \ks regeneration and neutron interactions to negligible
levels.

The non-magnetic detector was based on two large four-plane drift chambers 
and a liquid-argon/lead sandwich calorimeter, read-out in sets of transverse 
($x$ and $y$) strips. This was followed by an iron-plastic scintillator
hadronic calorimeter.
Plastic scintillator hodoscopes were used for triggering. Veto and muon 
counters helped reducing background from \kl semi-leptonic decays. 

Data were collected alternating the \kl and \ks beams. At the average kaon 
momentum of 100 GeV/$c$ the \kl decay distribution is essentially flat 
($\beta \gamma c \tau_L \simeq$ 3 km); by moving the \ks target along the 50 m 
long decay region, the effective longitudinal decay vertex distribution for \ks
($\beta \gamma c \tau_S \simeq$ 5 m) was made similar to that of $\kl$, 
therefore making the relative acceptance correction, performed in bins of kaon 
energy and decay position, very small in the analysis.

The subtracted background, only significant for $\kl$, was due to three-body 
$\kz$ decays, mainly in the neutral channel where it was below 3\%.
With no magnetic analysis, a significant fraction (about 40\%) of good 
$\pi^+\pi^-$ events were eliminated by the cuts used to reject the main 
$K_{e3}$ background, based on the energy deposition in the calorimeters; in 
the later data-taking runs a transition radiation detector was added to the 
setup, strongly reducing the $K_{e3}$ background with a large gain in 
efficiency for the $\pi^+ \pi^-$ mode.
Event losses due to activity in the detector uncorrelated with the kaon decay
(``accidentals''), at the level of a few percent, were made similar for the
two decay modes.

The experiment took data in 1987, 1988 and 1989, and its result on direct $CP$ 
violation \cite{NA31_eprime} with the total sample of 428000
\mbox{$\kl \To \pi^0 \pi^0$} decays was $\Rea(\epseps)=$ \mbox{$(2.30 \pm 
0.65) \cdot 10^{-3}$}, indicating the existence of direct $CP$ violation.
The dominant systematic errors were due to the understanding of biases related 
to accidental activity in the detector, $\kl$ background subtraction in the
neutral mode, and the knowledge of the absolute energy scale and its stability 
for neutral decays.

\subsubsection*{\emph{KTeV at FNAL}}

The final results of E731 and NA31 were not in good agreement, their 
probability to be consistent being 7.7\%, and left the fundamental question of
the existence of direct $CP$ violation not confirmed: while NA31 had a 3.5 
standard deviation signal, the world average $\Rea(\epseps)=$ \mbox{$(14.5 \pm
7.8) \cdot 10^{-3}$} was only 1.9 standard deviations from zero, after 
inflating the error to take the disagreement ($\chi^2 = 3.2$) into account
according to the PDG recipe \cite{PDG_2002}.
This situation stimulated both the FNAL and the CERN groups to design new and 
higher precision experiments, in order to clarify the matter, aiming at a 
reduction of the experimental errors by a factor $\approx$ 3. 


The KTeV program at FNAL comprises the E832 experiment for the search of 
direct $CP$ violation in neutral $K$ decays, and the E799 experiment for 
studies of $CP$ violation in rare $\kl$ decays, sharing most of the 
experimental set-up.
The E832 experiment uses the double-beam technique of its predecessors, with 
several important improvements. All four decay modes are collected at the same
time, therefore allowing a better control of several systematic effects. 
The uncertainties related to inelastic scattering and regeneration are reduced
by using an active lead-scintillator regenerator, which allows the vetoing of 
any event with an energy release detected in the device.

\begin{figure}[htb!]
\begin{center}
    \epsfig{file=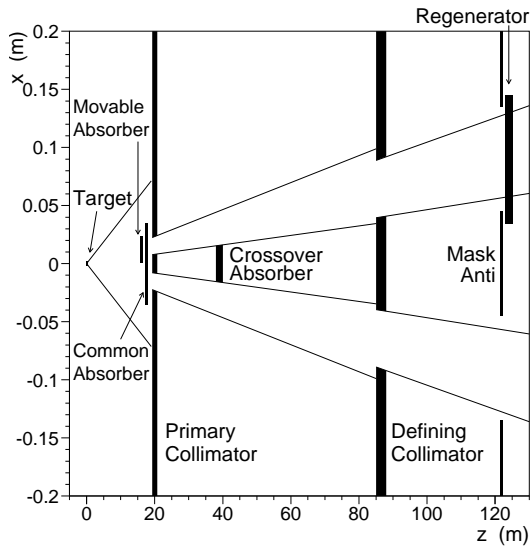,width=0.6\textwidth}
  \caption{Scheme of the KTeV double beam arrangement.}
  \label{fig:ktev_beams}
\end{center}
\end{figure}

As in previous experiments, both the 2 hadronic interaction lengths thick 
regenerator and an upstream absorber are moved alternately on either of the 
twin beams at each accelerator pulse, therefore resulting in an effective 
cancellation of any bias related to spatial detector asymmetries (see fig.
\ref{fig:ktev_beams}). 
The twin neutral beams are produced by $3 \cdot 10^{12}$ 800 GeV/$c$ protons 
in a 20 s spill every minute, providing a 0.9 MHz kaon flux in front of the 
regenerator.
The average magnitude of the regeneration amplitude is 3\%, and in the 
regenerator beam the $\kl$-related component accounts for 20\% of the decay 
rate; in the ``vacuum'' beam $\kl$ decays dominate, less than 1\% being due to
the residual $\ks$ component. 

\begin{figure}[hbt!]
\begin{center}
    \epsfig{file=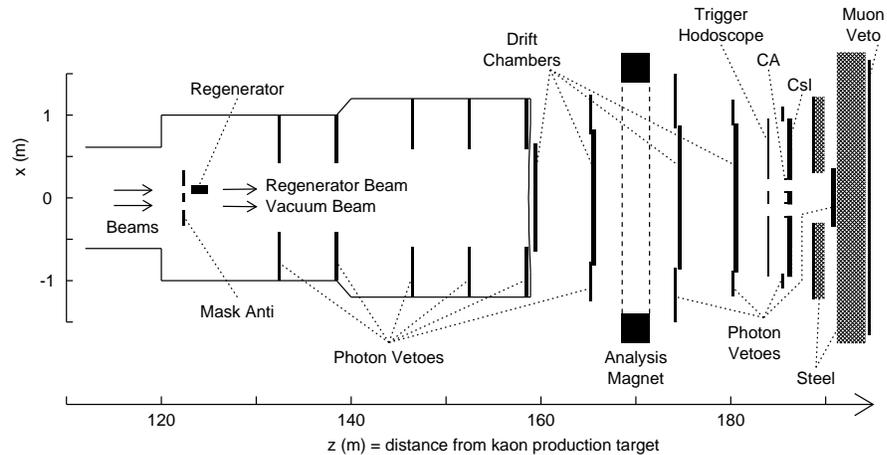,width=1.0\textwidth}
  \caption{Scheme of the KTeV detector.}
  \label{fig:ktev_detector}
\end{center}
\end{figure}

The detector, placed at $\sim$ 160 m from the target (see fig. 
\ref{fig:ktev_detector}), has as main components a magnetic spectrometer with 
a 410 MeV/$c$ kick based on proportional wire chambers and a 
high-performance pure CsI crystal electromagnetic calorimeter with excellent 
energy resolution and small non-Gaussian resolution tails, which allows a 
very accurate measurement of $\pi^0 \pi^0$ decays.

Decays of diffractively regenerated $\ks$ ($\sim$ 1\% of the coherently 
regenerated ones) are reduced in the data by kinematic cuts in the analysis, 
while the dominant inelastically regenerated component (a factor 100 larger 
than the coherent one) is reduced by vetoing on activity in the active 
regenerator. 

Backgrounds in the vacuum beam include misidentified 3-body $K$ decays, at
the level of 0.09\% in the $\pi^+ \pi^-$ mode and 0.11\% in the $\pi^0 \pi^0$ 
mode. Scattering in the regenerator (and in the collimators, one order of
magnitude less) are the major sources of background in the regenerator beam, 
at the level of 0.08\% in the $\pi^+ \pi^-$ mode and 1.2\% in the $\pi^0 
\pi^0$ mode, but are also the largest background source ($\sim 0.4\%$) in the 
vacuum beam for the neutral mode, where kaons from the regenerator beam 
scattered at large angles can be reconstructed in the vacuum beam, leading to
a counting error in the double ratio.

Kaons in the momentum range 40 to 160 GeV/$c$, decaying in an evacuated region 
between 110 and 158 m from the target, are used in the analysis, which is 
performed in 10 GeV/$c$ wide kaon momentum bins, in order to reduce the 
sensitivity to detector acceptance effects and to the knowledge of the 
spectra, while also allowing to account for the momentum dependence of the 
regeneration amplitude. 
No binning is used in the longitudinal decay position.

The result is extracted from a fit of the event yields in the four modes 
entering the double ratio, using acceptance functions determined by 
Monte Carlo simulation; the acceptance correction induces a shift of $\sim 85
\cdot 10^{-4}$ on $\Rea(\epseps)$, mostly (85\%) due to geometry.
Only 5\% of the collected events are actually used for the measurement, most 
of them being instead checked to understand the detector acceptance and 
efficiency, and to model them in the simulation.

Other factors entering the fit are the momentum-dependent part of the 
vacuum-to-regenerator kaon flux ratio (measured from \mbox{$\kl \To \pi^+ 
\pi^- \pi^0$} decays), the kaon lifetimes and mass difference (fixed), the 
phases of the $CP$-violating amplitude ratios $\phi_{+-}$ and $\phi_{00}$ 
(also fixed to the super-weak value $\phi_{SW}$, assuming $CPT$), and the 
regeneration amplitude, for which a power-law dependence on kaon momentum is 
assumed, from which the phase is determined using theoretical input.

E832 collected data both in 1996-97 (3.3 million \mbox{$\kl \To \pi^0 \pi^0$} 
decays) and in 1999 (a sample of similar size).
Results from the analysis of the 1996-97 sample are available 
\cite{KTeV_eprime1} \cite{KTeV_eprime2}:
\Dm{
  \Rea(\epseps) = (2.071 \pm 0.148_{\mathrm{\,stat}} \pm 
    0.239_{\mathrm{\,syst}}) \cdot 10^{-3} = (2.07 \pm 0.28) \cdot 10^{-3}
}
from the global fit with $\chi^2 =  27.6$ with 21 degrees of freedom.

The data was also analyzed with a weighting technique similar to the one used
by the NA48 experiment (see below), which does not depend on an accurate 
Monte Carlo acceptance correction; with this approach the statistics is 
effectively reduced by a factor 1.7, and the obtained result is fully 
consistent with the one quoted above.

This measurement of direct $CP$ violation by KTeV is fully consistent with the 
NA31 measurement, and at 2 standard deviations from the E731 one. The error
appears to be dominated by systematics, although part of it can be reduced by 
including the 1999 data sample; the largest systematic uncertainties
are due to the reconstruction and the knowledge of the background for the 
$\pi^0 \pi^0$ channel. Improvements on these systematics are being pursued, to 
better exploit the $1 \cdot 10^{-4}$ statistical error which can be reached 
with the total sample.

\subsubsection*{\emph{NA48 at CERN}}

\begin{figure}[htb!]
\begin{center}
    \epsfig{file=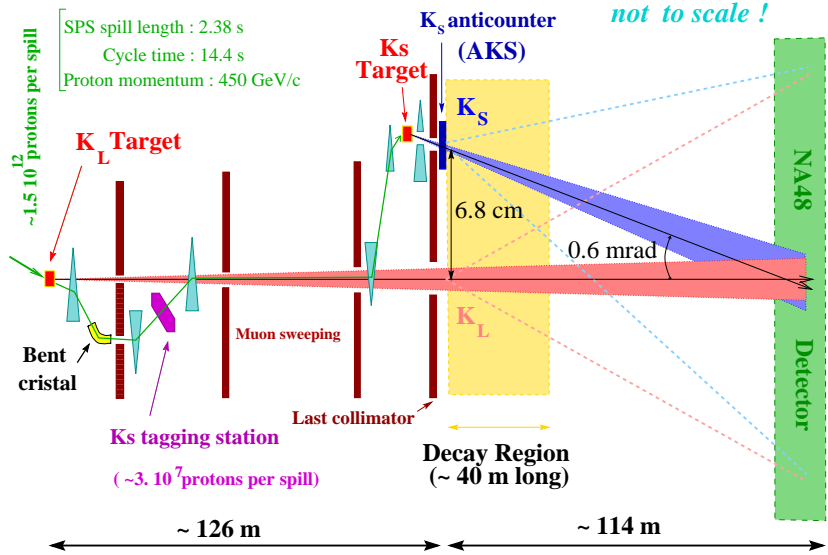,width=0.9\textwidth}
  \caption{Scheme of the NA48 beam arrangement.}
  \label{fig:na48_beams}
\end{center}
\end{figure}

The NA48 experiment at CERN adopted a double-target approach as NA31, but with 
several important differences with respect to its predecessor, the most 
important being that all four decay modes were detected simultaneously.
The neutral beam for $\ks$ decays is obtained by a selecting a small fraction 
($\sim 2 \cdot 10^{-5}$) of the primary 450 GeV/$c$ protons, used to produce
the \kl beam on a target at $\approx$ 240 m from the detector, channelling 
them in a single bent crystal and steering them on a secondary target close to 
the beginning of the fiducial decay region, at $\approx$ 120 m from the
detector. 
Two quasi-collinear neutral beams, converging at a 0.6 mrad angle at the 
centre of the detector, are therefore simultaneously available, and after
emerging from the evacuated decay region they cross the experimental apparatus 
always remaining within a thin vacuum pipe, in order to avoid interactions of 
the large neutron and photon flux (see fig. \ref{fig:na48_beams}).
The proton targeting angles are chosen in such a way that the spectra of
decaying $\ks$ and $\kl$ are similar, and any small residual difference is 
irrelevant for the analysis performed in small kaon momentum bins.

The experiment collects all four decay modes simultaneously, and the 
identification of $\ks$ and $\kl$ decays is performed by using a time 
coincidence technique: an array of scintillator counters, placed on the 
20 MHz secondary proton beam directed to the $\ks$ target, allows the precise 
measurement of the time of passage of each single proton. By comparing such 
times with the event time provided by high-resolution detectors, events for 
which a proton is found in a 4 ns wide coincidence window are flagged as $\ks$.
Such technique does not induce any asymmetry in first order, but only a small
dilution effect (corrected for); small higher-order effects due to 
charged-neutral asymmetries in the time measurement tails or rate-induced 
effects were accurately measured.

The close target is not movable, thus allowing for a much improved collimation 
and shielding than in NA31; nevertheless, the acceptance correction due to the 
different lifetime of $\ks$ and $\kl$ is made small also in NA48 by using only 
$\kl$ decays in the region where also $\ks$ decays are present, and by 
weighting offline the $\kl$ events as a function of the decay proper time, to 
have very similar longitudinal decay distributions at the price of losing 
statistics in the smallest $CP$-violating sample of decays.

\begin{figure}[hbt!]
\begin{center}
    \epsfig{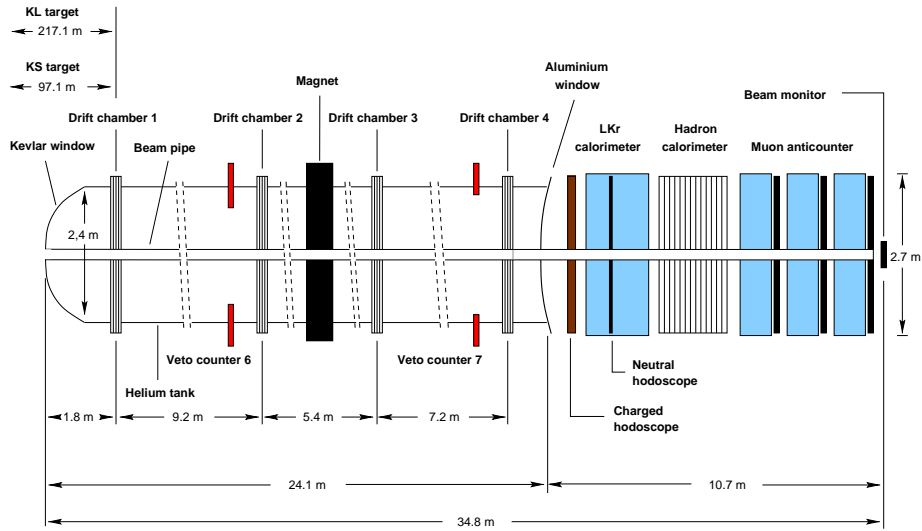}
  \caption{Scheme of the NA48 detector.}
  \label{fig:na48_detector}
\end{center}
\end{figure}

Another important difference with respect to NA31 is the presence of a 
magnetic spectrometer with four 4-view large drift chambers (see fig.
\ref{fig:na48_detector}), which allows a significant reduction in the 
backgrounds due to semi-leptonic $\kl$ decays in the charged mode.
$\pi^0 \pi^0$ decays are measured by a quasi-homogeneous liquid Krypton
electromagnetic calorimeter with projective tower structure, working as a 
ionization chamber, whose excellent energy, space and time resolutions are 
complemented by a very good stability, uniformity and ease of calibration.

Backgrounds are almost exclusively due to misidentified 3-body \kl decays, 
and are at the level of 0.06\% for neutral and 0.2\% for charged \kl decays,
while they are negligible for \ks decays. A 0.1\% contribution due to 
beam scattering is measured and subtracted from the \mbox{$\kl \To \pi^0 
\pi^0$} sample.

The result is extracted by averaging the double ratio measured in 5 GeV/$c$ 
wide kaon momentum bins, in the range 70-170 GeV/$c$, for decays in a 
longitudinal region corresponding to 3.5 $\tau_S$ in proper time.

NA48 collected data for the direct $CP$ violation measurement in 1997 ($0.5
\cdot 10^6$ \mbox{$\kl \To \pi^0 \pi^0$} decays) \cite{NA48_eprime_97}, 1998, 
1999 ($3 \cdot 10^6$) \cite{NA48_eprime_9899}. 
After the implosion of the vacuum pipe and the complete rebuilding of the 
damaged drift chambers, a new data-taking period in 2001 with a better duty
cycle ($\times 1.8$) and reduced instantaneous intensity ($-30\%$) was used to 
perform systematic cross-checks on rate-related effects and to complement the 
statistics ($1.5 \cdot 10^6$ \mbox{$\kl \To \pi^0 \pi^0$} decays) 
\cite{NA48_eprime_01}.
The final combined result from all the collected data is \cite{NA48_eprime_01}:
\Dm{
  \Rea(\epseps) = (1.47 \pm 0.14_{\mathrm{stat}} \pm 
  0.09_{\mathrm{\,stat/syst}} \pm 0.15_{\mathrm{\,syst}}) \cdot 10^{-3} = 
  (1.47 \pm 0.22) \cdot 10^{-3}
}
where the first quoted error is purely statistical, the second is the one 
induced by the finite statistic of the control samples used to study 
systematic effects, and the third is purely systematic.
The largest systematic uncertainties are due to the reconstruction of the
$\pi^0 \pi^0$ decays and the knowledge of the $\pi^+ \pi^-$ trigger efficiency.


The measurement of direct $CP$ violation by NA48 is well consistent with the 
NA31 and the KTeV results.

\subsection{$\phi$ factories}

A different technique for studying decays of $K$ mesons, exploiting their
production in correlated coherent pairs, was proposed a long time ago 
\cite{Coerenti}, and is actually being pursued in the KLOE experiment at the 
DA$\Phi$NE $e^+ e^-$ storage ring in Frascati and, with lower luminosity,
at the VEPP-2M storage ring in Novosibirsk.

The $\phi$ meson, produced practically at rest, decays with $\sim$ 34\% 
probability in a $\ks \kl$ pair, providing an intense source of monochromatic
back-to-back $\ks$ and $\kl$ beams (and $K^+ K^-$ pairs, with 49\% branching
ratio).
By considering decays of neutral kaon pairs to $\pi \pi$, one can extract
information on direct $CP$ violation in a different way. Since the $\phi$ is
a $C$-odd meson, $C$-conservation in its decay to pairs of neutral kaons, 
induced by strong interactions, implies that only $\ks \kl$ or $\kz \kzb$ 
states are allowed in its decay, independently from the validity of $CP$ or 
$CPT$ symmetries:
\Ea{
  & | \phi \rangle \To \frac{1}{\sqrt{2}} 
    \left[ |\kzb(\Vec{p}) \kz(-\Vec{p}) \rangle - 
    | \kz(\Vec{p}) \kzb(-\Vec{p}) \rangle \right] = \nonumber \\
  & \frac{1}{\sqrt{2}} \frac{1+|\epsilon|^2}{1-\epsilon^2}
    \left[ | \ks(\Vec{p}) \kl(-\Vec{p}) \rangle - 
    | \kl(\Vec{p}) \ks(-\Vec{p}) \rangle \right]
}
where $\Vec{p}$ denotes the kaon momentum.
This feature allows to tag either the mass eigenstate or the strangeness (at 
that time) of a decaying kaon, by detecting respectively a $\pi\pi$ decay 
close to the interaction point (at the level of validity of $CP$ symmetry) or 
the lepton charge in a semi-leptonic decay (at the level of validity of the 
$\Delta S = \Delta Q$ rule) for its companion.
Angular momentum conservation in the decay of the spin 1 $\phi$ forces the 
kaon pair to be in a p-wave orbital angular momentum state; Bose symmetry then 
implies that the two kaons cannot decay simultaneously to the same $\pi \pi$ 
final state, since two identical spinless bosonic systems cannot be in an 
antisymmetric state.
Both the above features are independent from the validity of the $CP$ and 
$CPT$ symmetries.

Bose symmetry alone allows in principle to detect direct $CP$ violation: if 
one of the two kaons decays to $\pi^+ \pi^-$, the other one must be, at that 
time, that particular combination of $\kz$ and $\kzb$ which cannot decay to 
$\pi^+ \pi^-$. In absence of direct $CP$ violation this combination is just 
$K_2$, which cannot decay to $\pi^0 \pi^0$ either. Therefore, an event in 
which one detects simultaneous decays to $\pi^+ \pi^-$ and $\pi^0 \pi^0$, 
while allowed by Bose symmetry, would be evidence for direct $CP$ violation: 
the amplitude for simultaneous decay to $\pi^+ \pi^-$ and $\pi^0 \pi^0$ is
actually proportional to $\eta_{+-}-\eta_{00}$.
Experimental resolution will blur the definition of simultaneity, but the 
study of the decay distribution to the $\pi^+\pi^- \pi^0 \pi^0$ state as a 
function of the decay time difference $\Delta t \equiv t_{+-} - t_{00}$,
integrating on their sum $t_{+-}+t_{00}$, can provide information on direct 
$CP$ violation \cite{Coerenti} \cite{Coerenti2} \cite{Coerenti3}.
Such distribution (neglecting terms of order $|\epsilon|^2$)
\Ea{
  & I(\pi^+\pi^-,\pi^0\pi^0;\Delta t) = \nonumber \\
  & \int dt_{+-} \, dt_{00} \, 
    \delta(t_{+-} - t_{00} - \Delta t) \left| 
    A(\phi \To \pi^+\pi^-, \pi^0\pi^0; t_{+-},t_{00}) \right|^2 
    \simeq \nonumber \\
  & \frac{2 \Gamma_S(\pi^+\pi^-) \Gamma_S(\pi^0\pi^0)}{\Gamma_S+\Gamma_L}
    e^{-(\Gamma_S+\Gamma_L)|\Delta t|/2} \left[
    |\eta_{+-}|^2 e^{\Delta \Gamma \Delta t/2} + \right. \nonumber \\
  & \left.  |\eta_{00}|^2 e^{-\Delta \Gamma \Delta t/2} -
    2\Rea \left( \eta_{+-}^*\eta_{00} e^{i\Delta m \Delta t} \right) \right]
}
is roughly constant for $\Delta t$ between about 10 and 100 $\tau_S$, and is 
symmetric for $\Delta t \rightarrow -\Delta t$ if $\epsilon'=0$.
Indeed the intensity asymmetry, for small $|\epseps|$ and neglecting 
$|\omega|$, is given by
\Ea{ \Ds 
  & A_I(\pi^+\pi^-,\pi^0\pi^0;\Delta t) \equiv 
    \frac{I(\pi^+\pi^-,\pi^0\pi^0;\Delta t)-I(\pi^+\pi^-,\pi^0\pi^0;-\Delta t)}
    {I(\pi^+\pi^-,\pi^0\pi^0;\Delta t)+I(\pi^+\pi^-,\pi^0\pi^0;-\Delta t)} 
    \simeq \nonumber \\
  & \frac{3 \Rea(\epseps) \sinh(\Delta \Gamma \Delta t/2) -  
    3 \Imm(\epseps) \sin(\Delta m \Delta t)}
    {[1-\Rea(\epseps)][\cosh(\Delta \Gamma \Delta t/2)-
    \cos(\Delta m \Delta t)]}
}
so that one is sensitive to the imaginary part of $\epseps$ by using a limited 
range of $\Delta t$, while the asymmetry for $\Delta t \rightarrow \infty$ is 
only sensitive to the real part of $\epseps$, and contains the same 
information as the double ratio of decay widths.

Partially integrated rate asymmetries can also be used 
\Dm{
  A_\Gamma(t) \equiv
  \frac{\Gamma(0<\Delta t<t)-\Gamma(-t<\Delta t<0)}
  {\Gamma(0<\Delta t<t)+\Gamma(-t<\Delta t<0)}
}
where $\Gamma(t_1<\Delta t<t_2) \equiv \int_{t_1}^{t_2} d(\Delta t) 
I(\Delta t)$. The asymptotic value of such asymmetry for $\pi^+\pi^-, 
\pi^0\pi^0$ decays reduces to
\Dm{
  A_\Gamma(\pi^+\pi^-,\pi^0\pi^0;\infty) \simeq
  3 [\Rea(\epseps) - 2(\Gamma_L/\Gamma_S) \Imm(\epseps)]
}
and also has only a limited sensitivity to the imaginary part of $\epseps$.

In principle, by fitting the shape of the $I(\Delta t)$ distribution directly, 
one could extract both the real and the imaginary parts of $\epseps$; it is
perhaps only of academic interest to point out that such shape analysis, 
being also sensitive to the phase difference $\phi_{+-}-\phi_{00}$, could 
distinguish the cases in which, even if $\epsilon' \neq 0$, one has 
$|\eta_{00}|=|\eta_{+-}|$ and therefore the double ratio method would give a 
null result.
On the other hand, this approach is very sensitive to experimental resolution 
effects on the measurement of the decay time (\emph{i.e.} flight path) 
difference, 
particularly in the region around $\Delta t=0$ where the distribution is 
rapidly changing, being also the region which is sensitive to $\Imm(\epseps)$.
Estimates of achievable accuracies with this approach at a $\phi$ factory 
\cite{Dambrosio_dafnehb} are $\sim 2 \cdot 10^{-4}$ and $\sim 3 \cdot 10^{-3}$ 
for the real and imaginary parts of $\epseps$ respectively.

In experiments at a $\phi$ factory, in which \kl and \ks are produced at the 
same point, detector acceptance (and its uniformity) is an important issue, 
since the \kl lifetime corresponds to about 340 cm. Interferometric 
measurements require a very accurate and precise knowledge of the decay vertex 
position.

Radiative decays of the $\phi$ with soft photons generate a $C$-even 
background due to \mbox{$\phi \To \gamma (\kz \kzb)_{C=+1}$}, and therefore 
$|\ks\ks \rangle$ or $|\kl\kl \rangle$ final states (and $|\ks(\Vec{p}) 
\kl(-\Vec{p}) \rangle + |\kl(\Vec{p}) \ks(-\Vec{p}) \rangle$ if $CPT$ symmetry 
is violated).
Although such processes are expected to be suppressed by large factors 
$O(10^7)$ with respect to the non-radiative $\phi$ decay to $C$-odd $\kz \kzb$ 
pairs, they can dilute, and in some cases fake, the asymmetries discussed 
above, and their effect has to be taken into account \cite{Isidori}.

Apart from the interferometric technique discussed above, also at $\phi$ 
factories direct $CP$ violation can be measured by the double ratio method, as 
in fixed target experiments.
The statistical error on $\Rea(\epseps)$ from the measurement of $A_\Gamma$ is
$\sim 1/(3\sqrt{N})$ ($N$ being the number of $\phi \rightarrow \pi^+\pi^-,
\pi^0\pi^0$ decays).
In order to reach the $10^{-4}$ level on $\Rea(\epseps)$, integrated 
luminosities of the order of $\sim$ 10 fb$^{-1}$ are required, with both 
techniques, corresponding to about 2 years running at $5 \cdot 10^{32}$ 
cm$^{-2}$ s$^{-1}$ luminosity.


The VEPP-2M collider at Novosibirsk, with $5 \cdot 10^{30}$ cm$^{-2}$ s$^{-1}$
luminosity, collected 33 pb$^{-1}$ at the $\phi$ centre of mass energy in 
1999-2000.
The two installed experiments are CMD-2 \cite{CMD-2}, with cylindrical 
drift chambers and calorimeters in a 1.5 T super-conducting magnet, and the 
non-magnetic SND \cite{SND}, with smaller drift chambers and a spherical 
NaI(Tl) crystal calorimeter ($0.9 \cdot 4\pi$ acceptance); these experiments 
collected $2 \cdot 10^6$ $\ks \kl$ decays and $1 \cdot 10^6$ $K^+K^-$ decays 
in the above period \cite{Blinov}.
An upgraded machine VEPP-2000, which should reach peak luminosity $1 \cdot 
10^{32}$ cm$^{-2}$ s$^{-1}$, is in preparation at Novosibirsk.

\begin{figure}[htb!]
\begin{center}
    \epsfig{file=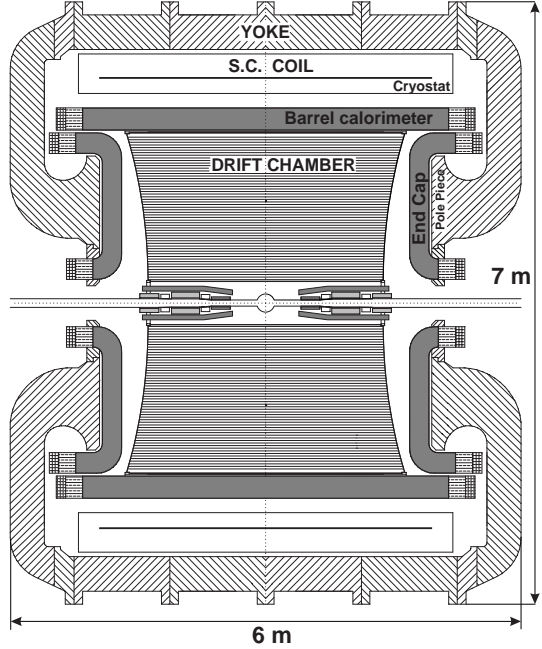,width=0.6\textwidth}
  \caption{Schematic drawing of the main elements of the KLOE detector.}
  \label{fig:kloe_detector}
\end{center}
\end{figure}

The DA$\Phi$NE collider at Frascati reached in 2002 a luminosity $\simeq 8 
\cdot 10^{31}$ cm$^{-2}$ s$^{-1}$, steadily increasing toward the design 
value of $5 \cdot 10^{32}$ cm$^{-2}$ s$^{-1}$, and the KLOE experiment 
\cite{KLOE} collected $\simeq$ 500 pb$^{-1}$ so far. 
The KLOE detector (see fig. \ref{fig:kloe_detector}) consists of a large 
volume drift chamber surrounded by a lead-scintillating fibre electromagnetic 
calorimeter, both enclosed in a super-conducting solenoid providing a 2 T 
$\cdot$ m field integral. The tracking detector, with stereo views, is rather 
light and works with a helium-based gas mixture to minimize regeneration and 
scattering. 
The calorimeter is sensitive to low photon energies and has excellent time 
resolution, which is crucial since particle identification uses velocity 
measurements.
No results on $\Rea(\epseps)$ are yet available; a ten-fold increase in data
is required to reach a statistical error in the $10^{-4}$ range.

\subsection{The CPLEAR approach}

The CPLEAR experiment \cite{CPLEAR} at the CERN antiproton ring performed an 
extensive set of measurements in the neutral kaon system with a different 
technique, analogous to one used today in heavy-meson experiments at hadron 
machines: by exploiting the strangeness-conserving associate production 
reactions (driven by strong interactions)
\Dm{
  \overline{p}p \, (\textrm{at rest}) \To 
    \left\{ \begin{array}{l} 
    K^+ \pi^- K^0 \\ K^- \pi^+ \overline{K}^0
    \end{array} \right.
}
which amount to $0.4\%$ of the total $\overline{p}p$ cross-section at rest,
the initial strangeness of the produced neutral kaons could be known, and 
their time evolution and flavour-dependent symmetries could be studied.
Also, by exploiting an initial state of precisely known energy and momentum,
the neutral kaon direction and momentum could be determined by the
production kinematics.

The partial decay rates to a state $f$ accessible to both $\kz$ and $\kzb$, 
for states being initially $\kz$ or $\kzb$, indicated as $\Gamma_f(t)$ and 
$\overline{\Gamma}_f(t)$ respectively, do evolve in time in such a way that 
their time-dependent asymmetry is 
\Ea{ \Ds
  & A_{CP}^{(f)}(t) \equiv \frac{\Gamma_f(t) - \overline{\Gamma}_f(t)}
    {\Gamma_f(t) + \overline{\Gamma}_f(t)} = \nonumber \\
  & \frac{[-2\Rea(\beps)/(1+|\beps|^2)](e^{-\Gamma_St} + |\eta_f|^2 
    e^{-\Gamma_Lt}) + 2|\eta_f| e^{-(\Gamma_S+\Gamma_L)t/2} 
    \cos(\Delta m t - \phi_f)}
    {(e^{-\Gamma_St} + |\eta_f|^2 e^{-\Gamma_Lt}) - 
    [4\Rea(\beps)/(1+|\beps|^2)]|\eta_f|e^{-(\Gamma_S+\Gamma_L)t/2} 
    \cos(\Delta m t - \phi_f)}
}

Any asymmetry between the instantaneous partial decay rates for states being 
initially $\kz$ or $\kzb$ to decay to a given $\pi \pi$ state (\emph{i.e.} 
either $\pi^+ \pi^-$ or $\pi^0 \pi^0$) could be used to extract a measurement 
of direct $CP$ violation \cite{CPLEAR_prop}: the initial value (at $t=0$) of 
this asymmetry is indeed
\Ea{
  & A_{CP}^{(\pi \pi)}(0) \simeq 
    2 \Rea(\eta_{\pi\pi}) -2 \Rea(\epsilon) \simeq \left\{
    \begin{array}{ll} 
    +2 \Rea(\epsilon') & (\textrm{for $\pi^+ \pi^-$}) \\
    -4 \Rea(\epsilon') & (\textrm{for $\pi^0 \pi^0$}) 
    \end{array} \right.
}
Due to the finite experimental resolution on the proper decay time, in order 
to be able to measure an asymmetry with a statistically significant sample of 
events one has to resort to time-integrated measurements. The sensitivity to 
direct $CP$ violation effects is largest for short integration times: this is 
readily understood since any effect of $CP$ violation due to $\kz--\kzb$ 
mixing is relatively less important for $t \ll \tau_S$, before strangeness 
oscillations can take place, while direct $CP$ violation effects are 
independent of time.

Neglecting terms of order $|\epsilon|^3$, the time-integrated asymmetry
between time $0$ and $T$ is given for short times ($T \ll \tau_S$) by
\cite{Amelino}
\Dm{
  I^{(\pi\pi)}(T) \equiv \int_0^T A_{CP}^{(\pi \pi)}(t) \, dt \simeq 
    -2 \Rea(\epsilon) + 2 \Rea(\eta_{\pi\pi}) 
    \frac{\Gamma_S T}{1-e^{-\Gamma_S T}} \quad (T \ll \tau_S)
}
in which, as discussed above, the term independent from $T$ only depends on 
$\epsilon'$, while mixing only contributes at finite times. 
However, in practice, the information which can be obtained on 
$\Rea(\epsilon')$ is drastically limited by statistics, and the uncertainty 
on the knowledge of $\Rea(\epsilon)$.

Another way of extracting the information on direct $CP$ violation is by
considering the difference of the two integrated asymmetries for $\pi^+\pi^-$ 
and $\pi^0\pi^0$ states which, in the same approximation, is given by 
\Dm{
  I^{(\pi^+\pi^-)}(T) - I^{(\pi^0\pi^0)}(T) \simeq 
  \frac{6 \Gamma_S T}{1-e^{-\Gamma_S T}} \Rea(\epsilon') 
  \quad (T \ll \tau_S)
}
For intermediate integration times $(\tau_S \ll T \ll \tau_L)$, the approximate
expression for the time-integrated asymmetry (in the limit $\Imm(\epseps) 
\simeq 0$ and $\phi(\epsilon) \simeq \pi/4$) is 
\Dm{
  I^{(\pi\pi)}(T) \simeq 
    \left\{ \begin{array}{ll} 
    2 \Rea(\epsilon) [1+2 \Rea(\epseps)] & (\textrm{for $\pi^+ \pi^-$}) \\
    2 \Rea(\epsilon) [1-4 \Rea(\epseps)] & (\textrm{for $\pi^0 \pi^0$}) 
  \end{array} \right.
}
and consequently from the ratio of the two asymmetries one obtains, in the 
case $\Rea(\epseps) \ll 1$
\Dm{
  \Rea(\epseps) \simeq 
  \frac{1}{6} \left( 1- \frac{I^{(00)}(T)}{I^{(+-)}(T)} \right)
}
This allows to extract $\Rea(\epseps)$ without the need to measure 
$\Rea(\epsilon)$ to the (unattainable) level of precision which would be 
necessary using only one of the decay modes, \emph{e.g.} from 
$I^{(\pi^+\pi^-)}$.
Given the fact that $I^{(\pi^+\pi^-)} \simeq 5 \cdot 10^{-3}$ the number of 
events needed to reach the same statistical accuracy as with the double ratio  
method used in the comparison of $\ks$ and $\kl$ decays, is $\sim 4 \cdot 
10^4$ times larger, in practice again an insurmountable handicap.

For $\pi^+ \pi^-$ decays, in which the decay vertex position can be measured
with good accuracy, the modulus $|\eta_{+-}|$ and phase $\phi_{+-}$ can be  
extracted from the fit of the $\kz,\kzb$ proper decay time distributions.
Without information on photon directions, the same cannot be done for $\pi^0 
\pi^0$ decays to any reasonable accuracy, and the integrated decay rates 
between $0$ and 20 $\tau_S$ can be used only for those, to obtain 
\Dm{
  |\epseps| \simeq \frac{1}{10} \left[ 2 - 
  \frac{I^{(\pi^0\pi^0)}(T \approx 20 \tau_S)}{\Rea(\eta_{+-})} \right]
}

Clearly, the determination of $|\eta_{+-}|$ and $|\eta_{00}|$ by the separate 
measurement of $\pi^+ \pi^-$ and $\pi^0 \pi^0$ decays could also allow a 
measurement of direct $CP$ violation through the usual double ratio of decay
rates.
The above amplitude ratios could be measured either ignoring the initial 
strangeness of the kaon, in which case the decay rate reduces to the sum of
two exponentials, from which $|\eta_f|^2$ can be extracted, or by fitting the 
interference term in the rate asymmetry in the proper decay time region around 
$\approx 14 \tau_S$. The latter approach is more sensitive due to the small
value of $|\eta_f|$, but the systematic uncertainties are very different in 
the two cases.

Intense fluxes of flavour-tagged $\kz$ and $\kzb$ were produced by stopping
low-energy anti-protons (200 MeV/$c$, $10^6 \, \overline{p}/s$) from the CERN 
LEAR ring in a low-density gaseous hydrogen target.
The detector consisted in a cylindrical tracking chamber, a threshold Cerenkov
counter to identify charged kaons, and an electromagnetic calorimeter made 
of lead plates inter-spaced with streamer tubes, all enclosed in a 
solenoidal magnet providing a constant 0.44 T field.
The experimental acceptance extended to 20 $\tau_S$ on average (60 cm).

The differences in the interactions of $K^+ \pi^-$ and $K^- \pi^+$ with the 
detector required all the observables to be independently normalised for 
$\kz$ and $\kzb$, to avoid systematic asymmetries; these can only 
be important close to the production target, before strangeness oscillations 
have taken place. Inefficiencies are less dangerous since they can only 
``dilute'' the measurement of an existing asymmetry but cannot induce a fake 
one by themselves.

The final results of CPLEAR on the amplitude ratios are \cite{CPLEAR_eta}
\Ea{
  |\eta_{+-}| = (2.264 \pm 0.035) \cdot 10^{-3} \quad &&
  \quad |\eta_{00}| = (2.47 \pm 0.39) \cdot 10^{-3} 
}
The systematic errors for the $\eta_{+-}$ measurement arise primarily from the 
imperfect knowledge of the regeneration amplitude due to detector material, 
(affecting the $\kz,\kzb$ rates through the interference with the $\ks$ 
amplitude), from the subtraction of the background due to semi-leptonic $\kl$ 
decays, and the imperfect knowledge of the experimental time resolution 
function; the latter, and its perturbation by photons not related to the 
neutral kaon decay process, were the main sources of systematic errors for the
$\eta_{00}$ measurement.

From the above ratios one would get naively (without taking into account 
correlated uncertainties) $\Rea(\epseps) =$ \mbox{$(-31.7 \pm 62.9)$} $\cdot 
10^{-3}$.

Strangeness-tagged kaon decays also allow the measurement of several other 
kinds of asymmetries, which could be useful to disentangle $CP$ violation in 
the decay amplitudes from the contribution due to the strangeness mixing 
\cite{Amelino}.

\subsection{Comments on strategies of experimental approach}

The combination at face value of the world averages of the independent 
measurements of $|\eta_{+-}|$ and $|\eta_{00}|$ \cite{PDG_2002} gives
\Dm{
  \Rea(\epseps) = (8 \pm 16) \cdot 10^{-3} 
}
To go significantly below this level of error, dedicated experiments are
required, and only the ones based on the double ratio method provided precise 
results, so that in the following only these will be discussed.

The experimental techniques to measure $\epsilon'$ evolved steadily in time, 
with improvements being dictated by the systematic limitations found in the 
previous generation of experiments, in a continuous effort to increase the 
level of accuracy; clear trends can be identified, with each collaboration 
attacking the most critical sources of systematic errors and eliminating by 
clever design the elements which lead to larger uncertainties in previous 
results.

\begin{table}
\begin{center}
\vspace{6cm}
\rotatebox{90}{
\begin{tabular}{|l|c|c|c|c|c|c|}
\cline{1-7}
  & BNL  & FNAL & FNAL & CERN & FNAL & CERN \\
  & E749 & E617 & E731$^{\dagger}$ & NA31 & KTeV$^{\dagger \dagger}$ & NA48 \\
\cline{1-7} 
$p_p$ (GeV/$c$) & 
  28 & 400 & 800 & 450 ($\kl$) & 800 & 450 \\
           & & & & 360 ($\ks$) & & \\
$p$ intensity ($s^{-1}$) &
  $3 \cdot 10^{11}$ & $7 \cdot 10^{12}$ & 
  $0.15-1 \cdot 10^{11}$ & $0.4 \cdot 10^{11}$ ($\kl$) &
  $1.5 \cdot 10^{11}$ & $6.3 \cdot 10^{11}$ ($\kl$) \\
  & & & & $1.2 \cdot 10^7$ ($\ks$) & & $1.2 \cdot 10^7$ ($\ks$) \\
\ks flux (s$^{-1}$) & 
  $\sim 10^2$ & $\sim 20$ &
  $0.9-6 \cdot 10^2$ & $0.7 \cdot 10^3$ &
  $0.3 \cdot 10^2$ & $1.1 \cdot 10^2$ \\
\kl flux (s$^{-1}$) & 
  $3 \cdot 10^6$ & $7 \cdot 10^5$ & 
  $1.3-8.3 \cdot 10^6$ & $0.4 \cdot 10^6$ &
  $1.8 \cdot 10^5$ & $8.4 \cdot 10^6$ \\
$n/K$ in beam & 
  $\approx$ 30 & 10 & $\approx$ 1 & 8-9 & 0.8 & $\approx$ 5 \\
Fid. region (m[$\tau_S$]) &
  1.2 [2.1] & 13 [3] &  
  42 [11] ($\pi^0\pi^0$) & 46.8 &
  48 [13] & 13-32 [3.5] \\
& & & 17 [4.5] ($\pi^+\pi^-$) & & & \\ 
$p_K$ (GeV/$c$) & 
  7-14 & 50-140 ($\pi^0\pi^0$)& 40-160 & 60-180 & 40-160 & 70-170 \\
&& 30-120 ($\pi^+\pi^-$) &&&& \\
$\sigma(z_{\mathrm{decay}})$ (m) & 
  0.12 ($\pi^0 \pi^0$) & 1.7 ($\pi^0 \pi^0$) & 
  $\approx$ 1 ($\pi^0 \pi^0$)  & $\approx$ 1.2 ($\pi^0 \pi^0$) & 
  0.25 ($\pi^0 \pi^0$) & 0.5 ($\pi^0 \pi^0$) \\
  & $\approx$ 0.5 ($\pi^+ \pi^-$) & &
  $\approx 0.15$ ($\pi^+ \pi^-$) & $\approx$ 0.8 ($\pi^+ \pi^-$) & 
  $\approx 0.2$ ($\pi^+ \pi^-$) & $\approx$ 0.5 ($\pi^+ \pi^-$) \\
$\sigma(E_\gamma)/E_\gamma$ &
  8\% & 3\% & 1.0\% & 1.4\% & 0.6\% & 0.8\% \\
$\sigma(m_{\pi^+\pi^-})$ (MeV/$c^2$) & 
  15 & 4.5 & 3.5 & $\approx$ 20 & 1.6 & 2.5 \\
$N(\kl \To \pi^0 \pi^0)$ ($10^{3}$) & 
  1.1 & 3.2 & 410.3 & 428 & 3348 & 4837 \\
  \quad bkg(non-$\pi\pi$)/tot & 
    17.5\% & 7.6\% & 1.8\% & 2.67\% & 0.11\% & 0.06\% \\
  \quad bkg($\pi\pi$)/tot & 
    -- & 2.6\% & 3.4\% & -- & 0.38\% & 0.1\% \\
$N(\kl \To \pi^+ \pi^-)$ ($10^{3}$) & 
  8.1 & 10.6 & 329.0 & 1142 & 11126 & 21554 \\
  \quad bkg(non-$\pi\pi$)/tot & 
    $\approx$ 7\% & 3.0\% & 0.34\% & 0.54\% & 0.1\% & 0.16\% \\
  \quad bkg($\pi\pi$)/tot & 
    -- & 0.4 \% & -- & 0.09\% & 0.01\% & -- \\
$N(\ks \To \pi^0 \pi^0)$ ($10^{3}$) & 
  3.2 & 5.7 & 800.0 & 2254 & 5556 & 7370 \\
  \quad bkg(non-$\pi\pi$)/tot & 
    1.2\% & 0.5\% & 0.05\% & 0.07\% & -- & -- \\
  \quad bkg($\pi\pi$)/tot & 
    1.5\% & 12.7\% & 2.6\% & -- & 1.22\% & -- \\
$N(\ks \To \pi^+ \pi^-)$ ($10^{3}$) & 
  19.9 & 25.8 & 1060.7 & 5541 & 19291 & 31830 \\
  \quad bkg(non-$\pi\pi$)/tot & 
     $\approx 0$ & 0.2\% & -- & -- & -- & -- \\
  \quad bkg($\pi\pi$)/tot & 
    0.2\% & 1.7\% & 0.15\% & 0.03\% & 0.08\% & -- \\
Main syst. & 
  background & background & 
  $\gamma$ energy & accidental & 
  $\pi^0 \pi^0$ rec. & $\pi^0 \pi^0$ rec. \\
Data taking &
  && 1985-1988 & 1986-1989 & 1996-1999 & 1997-2001 \\
Pubbl. date & 
  1985 & 1985 & 1990-1993 & 1988-1993 & 1999- & 1999-2002 \\
$\Rea(\epseps)$ ($\times 10^4$) & 
  $17 \pm 72 \pm 43$ & $-46 \pm 53 \pm 24$ & 
  $7.4 \pm 5.2 \pm 2.9$ & $23.0 \pm 6.5$ &
  $20.71 \pm 1.48 \pm 2.39$ & $14.7 \pm 1.4 \pm 1.7$ \\
\cline{1-7}
\end{tabular}
}
\caption{Comparison of double ratio experiments (see text). 
$\dagger$ E731A result not included. $\dagger \dagger$ 1997 only.}
\end{center}
\label{tab:comparison}
\end{table}

Table \ref{tab:comparison} presents a schematic comparison of several 
parameters for some experiments measuring the double ratio: in this table are
reported the (average) momentum and (instantaneous) intensities of the primary
proton beam and of the kaons, the amount of neutrons in the neutral beams, 
some indicative (average) detector resolution parameters, the size and 
background fractions of the collected samples.

In the measurements of direct $CP$ violation in neutral kaon decays through 
the double ratio method, several possible sources of systematic errors arise:
\begin{enumerate}
\item Imperfect knowledge of relative $\kl$ and $\ks$ fluxes.
\item Differences in the triggering, detection or reconstruction efficiencies 
for $\pi^+ \pi^-$ and $\pi^0 \pi^0$ decays.
\item Detector acceptance differences for $\kl$ and $\ks$ decays.
\item Differential resolution effects.
\item Changes in detector properties during the data taking period.
\item Rate and accidental-induced differential effects.
\item Contamination by $\kl$ scattered in a regenerator and $\ks$ regenerated 
inelastically or by neutrons in the beam.
\item Imperfect background subtractions.
\item Imperfect knowledge of the fiducial regions in the kaon phase-space.
\end{enumerate}
We will discuss the above issues and the techniques used to address them (not 
necessarily in the above order, since many of them are tightly interlinked), 
focusing on the recent experiments at hadronic machines, which provided 
results on direct $CP$ violation.

Apart from coherent production of \kl \ks pairs, no hadronic interaction can 
provide a pure \ks beam, but only a coherent mixture of \ks and \kl in equal 
proportions (thanks to $CPT$ symmetry, independently from $CP$ conservation in 
strong interactions, and independently from the $\kz/\kzb$ production cross 
section ratio). While beams with any required $\kl$ purity are easily 
available at a suitable distance from the production target, ``\ks beams'' are 
invariably 50\%-50\% \ks--\kl mixtures; nevertheless, when considering 
$\pi \pi$ decays, the large ratio of partial decay widths (due to the 
approximate validity of $CP$ symmetry) results in a large effective 
suppression of \mbox{$\kl \To \pi \pi$} decays with respect to \ks in the 
region close to the production target (by a factor $|\eta|^2 \simeq 5 
\cdot 10^{-6}$ at the target).
The above considerations make clear that the intrinsic impurity of the beams 
has to be taken into account in the analysis, this being however a trivial 
task.
Intense neutral beams have usually rather wide momentum spectra, and at the
higher end of the spectrum \ks could still account for a sizable fraction of 
$\pi\pi$ decays, even at rather large distances from the production target.
On the other hand a fraction of \kl contributes $\pi \pi$ decays in a ``\ks 
beam'', depending on the distance from their production point at which the 
fiducial region starts and its size.
The term describing \ks--\kl interference in the expression for the yield of 
$\pi \pi$ decays must also to be taken into account: the coefficient of this 
term contains the already mentioned ``dilution factor'', which has to be 
measured experimentally.
 
All but the earliest experiments are independent from the knowledge of the 
absolute flux normalisation, by detecting at the same time at least two of the 
four decay channels which enter the double ratio (\emph{i.e.} having 
simultaneous 
$\kl$ and $\ks$ beams and/or detecting at the same time $\pi^+ \pi^-$ and  
$\pi^0 \pi^0$ decays). 
Note that this is true for the double ratio measurement, while in order to 
measure the ``single ratios'' (\emph{i.e.} $|\eta_{+-}|^2$ and 
$|\eta_{00}|^2$) the knowledge of flux normalisation is required.

From an experimental point of view, the detection of charged and neutral 
decays involves different parts of the apparatus, and is therefore subject to 
rather different instrumental effects. On the other hand, clearly there cannot 
be any instrumental difference in the detection of $\ks$ or $\kl$ decays to 
the same final state, provided they occur at the same place and time, which 
however is not usually the case in real experiments.

When $\kl$ and $\ks$ decays to a given mode are collected at the same time, 
but separately for charged and neutral decays, the $\kl/\ks$ flux ratio is 
irrelevant for the measurement of $R$, provided such ratio is exactly the same 
during the detection periods dedicated to the two modes. 
In this case $R$ is computed by dividing the ``single ratios'', known up to a 
(common) multiplicative constant related to the flux ratio. 

By using a double-beam technique with a regenerator or the production of 
coherent kaon pairs, $\kl/\ks$ flux normalisation is (approximately or 
exactly) not an issue. 
In the case of the FNAL twin beam experiments (E617, E731, E832-KTeV), the 
$\kl/\ks$ beam intensity ratio is constant, provided the geometry of the beam 
line elements and the regenerator properties do not change in time; the same 
is of course true for $\kl \ks$ pairs produced at a $\phi$-factory, so that,  
for what concerns primary flux cancellation, simultaneous charged and neutral 
mode data collection is not an unavoidable requirement: any instrumental 
charged-neutral detection difference has no effect on the double ratio in this 
case.

When measuring simultaneously the charged and neutral decays, with $\kl$
and $\ks$ collected at different times, flux normalisation is irrelevant
as long as the relative detection efficiency for the two decay modes is
stable in time and independent from the (different) environmental conditions 
(\emph{e.g.} rates and accidental activity) when running with the different 
beams.
In this case the double ratio is computed as the ratio of the $\pi \pi$ charge
ratios ($\pi^+\pi^-/\pi^0\pi^0$), known up to a (common) multiplicative 
constant related to the detection efficiencies; this was the case of the NA31
experiment at CERN.

In the CERN NA48 double-beam approach, although the secondary beam used to 
produce $\ks$ is derived from the primary one used to produce $\kl$, some 
limited residual $\kl/\ks$ beam intensity variations (at the level of $\approx 
10\%$) were present, due to different transmission and focusing of the beams; 
these variations could bias the measurement if coupled to time variations in 
the relative charged-neutral detection efficiencies on the same time scale; 
the possible effects had therefore to be measured and controlled, and 
eventually cancelled by intensity-ratio weighting.

The latest generation of experiments (KTeV and NA48, but also E731 in its 
last phase) were simultaneously sensitive to all four decay modes, therefore 
making flux cancellation more effective; the following discussion will focus 
mostly on these experiments, in which most of the corrections to be applied to 
the measured double ratio are zero to first order; some interesting discussion
can be found in \cite{Roads}.

Even for experiments with simultaneous \ks and \kl sources, detector 
acceptance differences for their decays are in general present. 
If the geometrical parameters of the two beams (angular divergence and 
emittance distribution) are equal, and their incidence on the detector is the 
same, either intrinsically (because they coincide spatially at the detector 
plane, as in the CERN experiments) or statistically (because they alternate 
positions, as in the FNAL experiments), the acceptance for a given final state 
only depends on the kaon momentum and on its longitudinal decay position. 

The angular parameters of the beams can be made similar by collimation, and 
are identical in case of coherent regeneration.
The momentum spectra of the two beams are usually rather similar, either due to
the choice of suitable proton targeting angles (as in NA48) or because of the
fact that the squared regeneration amplitude has a $1/p_K^{\alpha}$ dependence 
($p_K$ being the incident $\kl$ momentum, and $\alpha \simeq 1.2$) quite 
matching the $1/p_K$ kinematic factor which appears in the spectrum of 
decaying $\ks$ (as in regenerator-based experiments). 
For this reason the acceptance dependence on the kaon momentum can be safely 
taken into account by performing the analysis in kaon energy bins, with 
resolution effects being kept under control by the choice of adequate bin 
sizes.

On the other hand, longitudinal decay point distributions are intrinsically 
very different for $\ks$ and $\kl$ of the same momentum, whatever the 
production point; binning in such a variable could give rise to biases 
induced by bin to bin migration effects caused by the finite (and different 
for charged and neutral modes) experimental resolution, coupled with the very 
different distributions. 
For the above reason, acceptance cancellation can only be achieved by 
artificially changing one (or both) the distributions to make them similar; 
this was performed in NA31 and some previous experiments by collecting data 
with the $\ks$ target positioned at different longitudinal positions and 
combining the measurements, and in NA48 by weighting the $\kl$ events at the 
analysis stage and trading off statistical power in the procedure, also not 
using $\kl$ decays occurring at longitudinal positions in which no $\ks$ 
decays are available. 
The FNAL experiments chose instead not to have any intrinsic acceptance 
cancellation, gaining in statistics at the price of having to deal with a 
large acceptance correction, and therefore a very accurate Monte Carlo 
simulation of the apparatus, checked against large samples of $K_{e3}$ and 
$3\pi^0$ decays.

The general experimental layout for these high-energy experiments is rather 
similar, consisting of a long, evacuated decay region followed by a 
detector region filled with helium to reduce multiple scattering, kaon 
regeneration and photon conversion and, in the case of KTeV, also the 
interactions of the neutral beam particles. 
The magnetic spectrometer, as light as possible in order to limit the impact on
the electromagnetic calorimeter performance, requires a large-gap magnet 
providing a uniform transverse momentum kick. 
Excellent electromagnetic calorimetry, in terms of resolution, uniformity and
calibration, is an essential requirement in order to reduce backgrounds and to
control the energy and position scales for neutral events.
Veto elements around the active detector region are used to reduce trigger
rates and backgrounds due to incompletely contained events, while others 
behind absorber walls are used for muon background suppression.
Among the advantages of using high-energy ($\sim$ 100 GeV) kaon beams, besides 
the possibility of exploiting better resolutions for photon detection, is the
fact that the small solid angle acceptance allows planar detectors, which can 
be kept accurately under control for what concerns their geometry and 
positions.

Triggering and reconstruction efficiencies are in general different for
charged and neutral decays: the knowledge of their ratio is however not
needed as long as the $\kl$ and $\ks$ decays are collected simultaneously.
In particular, if the short- and long-lived kaon decays originate from beams 
which have different properties, care should be taken to check any possible
effect due to local rate differences: different detector illuminations or 
different time structures, which could lead to a systematically different 
time distribution of charged and neutral decays, can potentially result in a 
bias when coupled to detector or trigger efficiency variations on the same 
scale.

Trigger efficiencies have to be continuously monitored, by collecting large
enough downscaled fractions of ``minimum-bias'' triggers.
Triggering and detection efficiencies have in general a non-null correlation 
(partially different for charged and neutral modes) with instantaneous rates 
in the detectors, so that they should be checked for variations on all the 
time scales on which $\kl/\ks$ time variations can be expected. 

The time variation of detector performances could induce biases when coupled 
with variations of the $\kl/\ks$ flux ratio on the same time scale. 
Local inefficiencies due to malfunctioning wire chamber wires or calorimeter 
cells have to be continuously monitored; they were avoided in KTeV by data 
rejection and immediate repair, and tracked in time by NA48, to be taken into 
account in analysis and detector simulation. 
The periodic reversal of the spectrometer magnetic field helps reducing any 
residual left-right asymmetry which could couple to any spatial asymmetry of 
the two beams.

Differential effects induced by accidental activity, due to kaon decays or 
other particles in the beams, can also cause the same kind of effects, 
resulting in a non-linear dependence of the number of collected events on the
beam intensity. The accidental activity, while largely affecting both beams at
the same level, is clearly higher in a regenerator experiment such as KTeV.
Both experiments study these effects by using the software overlay of events
collected randomly in time at a rate proportional to the instantaneous beam 
intensity.
The production of $\ks$ by using a close target is a more efficient technique 
than the use of a regenerator, requiring a lower proton intensity to have the 
same kaon yield, and therefore resulting in a lower level of accidental 
activity. However, with this technique the region of the close target must be 
appropriately shielded to avoid unwanted accidental particles hitting the 
detector; in particular, possible effects induced by the simultaneous presence 
of other particles, produced by the same proton interaction from which the 
decaying $\ks$ emerged, and therefore ``in-time'' with respect to the decay, 
should be checked as potential sources of bias. 
On the other hand, a regenerator is also a source of large unwanted 
backgrounds.

The experimental requirement of making simultaneous \kl and \ks beams as 
similar as possible to avoid any bias, is of course at odds with the necessity 
of being able to distinguish the \kl or \ks nature of the decay to actually 
form the double ratio\footnote{The ultimate arrangement of measuring kaon 
decays from a \emph{single} neutral beam in the region around 12 $\tau_S$ from 
the production target, where the yields of $\pi \pi$ decays from \kl and \ks 
are comparable, is also the one in which mass eigenstates are intrinsically 
indistinguishable!}.
The origin of a single $\pi \pi$ decay has to be attributed to a $\ks$ or 
$\kl$ on an event by event basis, in order to measure the double ratio: this 
clearly requires two distinct beams (in time or space).

In the KTeV twin beams arrangement, the two quasi-parallel beams (1.6 mrad
divergence) are well separated in the transverse plane along all the decay 
region and also at the detector; the transverse kaon position can therefore be 
used to identify the beam from which the event originated. 
For $\pi^+\pi^-$ decays, the transverse position of the extrapolated kaon 
trajectory at the regenerator plane, measured with good accuracy, is used; for 
$\pi^0 \pi^0$ decays, measured in a destructive detector, the kaon trajectory 
is not known and the energy-weighted impact position of the four photons,
corresponding to the intercept of the kaon trajectory with the calorimeter 
plane\footnote{Actually with a plane at a position corresponding to the 
energy-weighted average of the longitudinal shower development in the 
detector, which depends on the decay vertex position for a non-projective 
calorimeter such as the one of KTeV, which is however rather compact (50 cm
thick), thus reducing the dependence on such an effect.}, must be 
used: even with a good resolution in this variable, much smaller than the beam 
separation thanks to the high-performance calorimeter ($\sim$ 1 mm vs. 
$\approx$ 300 mm separation of the $9 \times 9$ cm$^2$ beams), this approach 
cannot discriminate against kaons which are scattered at large angle in the 
regenerator or in collimators, which give the largest contribution to the 
background, to be measured and subtracted.
As already mentioned, any instrumental asymmetry due to the separation of the 
beams at the detector plane is statistically cancelled by alternating the 
regenerator on the two beams at every accelerator pulse.

In the NA48 approach with very close \kl and \ks beams converging at the 
centre of the detector ($\simeq$ 60 mm separation at the beginning of the 
fiducial region, 0.6 mrad angle), illumination differences are intrinsically 
very small, depending only on the different beam divergences, and the 
assignment of an event to either beam is performed with a time-of-flight 
technique on tagged protons. 
This approach requires precise event time information from the detectors, and 
the only asymmetries which can be induced in the double ratio are the ones due 
to different time measurement tails for charged and neutral decays, leading to 
asymmetric mis-tagging of a \ks decay. The above systematic effect, as well 
as the small ones due to effective rate differences for charged and neutral 
decays, induced by rate-dependent trigger efficiencies, had to be checked and 
corrected for using an alternative tagging method, possible for $\pi^+\pi^-$, 
based on the measured kaon trajectory at the collimator plane, which clearly 
identifies the beam from which the event originates.

Background due to 3-body kaon decays is an intrinsically asymmetric component 
entering the double ratio, being decay mode dependent and different for $\kl$ 
and $\ks$. 
Using redundant information in the reconstruction of the final state, this 
background can be subtracted, but the accuracy of the procedure is 
intrinsically limited by the statistical fluctuations of the subtracted number
of events, and also systematically limited by the fact that the distribution 
of the background in the signal region for the relevant variables must be 
extrapolated using some model (possibly extracted from the data).
High-resolution detectors are therefore used in order to be able to impose 
tighter cuts on the signal, thereby reducing the background fraction before the
subtraction.

All experiments are affected by background due to the dominant 3-body decay 
modes of the intense \kl beams: $\pi^+ \pi^- \pi^0$, $\pi^\pm e^\mp \nu$ 
($K_{e3}$) and $\pi^\pm \mu^\mp \nu$ ($K_{\mu3}$) for the charged mode, 
$3\pi^0$ for the neutral mode.

For the charged mode, while the $\pi^+ \pi^- \pi^0$ background can be 
subtracted with relative ease by magnetic analysis, due to its good signature 
in the detector, semi-leptonic modes with an undetected neutrino usually 
contaminate the signal region for the relevant kinematic variables.
Particle identification is used both to reject such semi-leptonic backgrounds  
and to positively identify the residual contamination, in order to subtract it:
transition radiation detectors or calorimetric information combined with
momentum measurement are used for $e^\pm$ identification, and muon veto 
counters to identify $\mu^\pm$. 

A fraction of the $\pi^+ \pi^- \gamma$ decays is unavoidably included in the 
$\pi^+ \pi^-$ sample: the irreducible inner bremsstrahlung contribution, 
dominant for small photon energies, however appears in the same $\kl/\ks$ ratio
as the non-radiative decay from which it derives, and as such cannot change
the measured double ratio at all. 
The direct emission component, which is important for higher photon energies 
($\sim$ 1.5\% of the $\pi^+ \pi^-$ mode for $E_\gamma >$ 20 MeV 
\cite{KTeV_pipigamma}), being only present for $\kl$ could create a bias on 
the double ratio; however a good $\pi^+ \pi^-$ invariant mass resolution 
allows this component to be rejected from the sample.

In experiments with a close target, one also gets hyperon decays from the 
neutral beam produced there, since hyperon lifetimes are comparable to that of
\ks: the contamination of the $\pi^+ \pi^-$ sample by misidentified 
\mbox{$\Lambda \To p \pi^-$} decays can be made very small by rejecting the 
kinematic region corresponding to large momentum asymmetry of the two charged 
particles in the laboratory system.

The background in the neutral mode due to \mbox{$\kl \To 3\pi^0$} decays is 
partly suppressed by photon veto detectors surrounding the decay volume; the 
remaining background fraction with missing or superimposed photons in the 
detector is reduced and subtracted by using the kinematic constraints of the 
\mbox{$\kz \To 2 \pi^0 \To 4\gamma$} decay chain. 
Assuming zero transverse momentum of the decaying kaon with respect to the 
beam axis (a good approximation for collimated beams in the $\sim$ 100 GeV/$c$ 
momentum range), one of the three mass constraints is used to compute 
the longitudinal decay vertex position, leaving two more for background 
rejection and control: defining $E_i$ as the photon energies and $r_{ij}$ as 
the distance between two photon clusters ($i,j={1,4}$), the longitudinal 
distance $d$ of the decay vertex from the calorimeter is computed in KTeV as 
the weighted average of the two values 
\Dm{
  d_{ij} = \frac{\sqrt{E_i E_j r^2_{ij}}}{m_{\pi^0}}
}
corresponding to the photon pairing which gives the best vertex agreement, 
and in NA48 as
\Dm{
  d = \sum_{i>j} \frac{\sqrt{E_i E_j r^2_{ij}}}{m_K}
}

Since the reconstructed decay vertex position in the neutral mode is obtained
by constraining the invariant masses of sets of photons, background events due 
to $3\pi^0$ decays with missing energy in the detector are reconstructed 
downstream of their true decay position (smaller $d$ values): the neutral 
background fraction then generally increases in the region closer to the 
detector. For this reason, depending on the size and the location of the 
fiducial region, the amount of $3\pi^0$ background in the neutral channel can 
be quite different.

In this context, \mbox{$\kl \To \pi^0 \pi^0 \gamma$} is a negligible 
background ($< 6 \cdot 10^{-3}$ of the $\pi^0 \pi^0$ mode). 

In regenerator-based experiments, a different additional class of backgrounds 
arises from scattering in the regenerator itself, and can possibly affect both 
the \kl and \ks samples.
Apart from the coherently regenerated $\ks$ component, a diffractively 
regenerated $\ks$ component is produced at non-zero angle with respect to the 
incident $\kl$ beam, and inelastic $\kl$ interactions can contribute large 
backgrounds as well. The latter are usually eliminated by veto counters in the 
regenerator detecting recoil charged particles, while the former have to be 
subtracted by analyzing the kaon transverse momentum distribution.
Kaon scattering is induced by the presence of matter along the kaon path:
regenerators and collimators (in a ratio $\approx$ 10 to 1 for KTeV).
Scattered kaons, besides requiring a correction when the extrapolated kaon 
impact position on the detector is used to tag the nature of the kaon, have
to be subtracted because of their different regeneration properties and 
acceptance, compared to unscattered ones.
Any effect induced by scattering, however, is intrinsically the same for both
decay modes of a given kaon type ($\ks$ or $\kl$), and therefore cancels in 
the double ratio to the extent at which the analysis cuts can be made equal 
for $\pi^+\pi^-$ and $\pi^0\pi^0$; the residual correction to be applied is 
due to the unavoidable differences in the selection of the two channels, 
mostly performed using different detectors. Moreover, the scattering 
background can be studied and measured in the charged mode, where more 
information on the event is usually available due to tracking, and with the 
same caveats mentioned above, this knowledge can be applied to the neutral 
mode.

One more potential source of background is the interaction of the intense flux 
of photons and neutrons with material along the beam line.
In the case of KTeV, photons are partially absorbed by a 14 radiation lengths 
thick lead slab after the target; the neutron flux, reduced to below the kaon 
one by more absorbers, is largely symmetric among the different modes.
In NA48, with a higher neutron and photon flux ratio with respect to kaons, 
the use of an evacuated beam pipe crossing all the detectors avoids that the 
neutral beams traverse any significant amount of material.

The presence of holes in the detectors is also a difference among KTeV and 
NA48: while in both experiments the neutral beams themselves do not touch the
active part of the calorimeters, they do cross an active part of the drift
chambers in KTeV (although, since calorimeter information is necessary also 
for charged decays, a fraction of this area is excluded from the acceptance),
and not in NA48. Since most of the particle rate (both due to the neutral 
particles in the beams and to kaon decay products) is sharply peaked close to 
the beam(s) in the transverse plane, one has either to worry about detector 
efficiencies as a function of the distance to the beam axis, or to cope with 
acceptance effects due to slight differences in beam illumination close to 
the hole itself (the bulk of the acceptance correction in NA48); in both cases 
the effects have to be modelled by Monte Carlo simulation which includes the
hole in acceptance due in particular to the requirement of identifying 
electrons by their showering in the calorimeters, in order to reject 
background from $K_{e3}$ decays.

One important systematic issue is that of the knowledge of the relative
accuracy of the measurements for kinematic variables on which cuts are 
performed, and for which $\ks$ and $\kl$ have different distributions. 
This is the case for the kaon momentum and particularly for its longitudinal 
decay vertex position, so that the energy and longitudinal (but also 
transverse) scales of the charged and neutral detectors have to be known 
accurately.
The spectrometer scales are directly determined by its well known geometry, 
and checked with the value of the reconstructed kaon mass for $\pi^+ \pi^-$ 
decays (or the $\Lambda$ mass in $p\pi^-$ decays), the adjustment of which 
fixes the overall momentum scale, linked to the absolute value of the 
spectrometer magnetic field.
For NA48, in which the $\ks$ and $\kl$ kaon spectra differences can be larger,
the kaon momentum is reconstructed using the ratio of pion momenta and their 
opening angle at the decay vertex, in a way which is independent from the 
knowledge of the absolute value of the magnetic field.

For neutral decays, the decay vertex position is determined by imposing 
kinematic constraints for the decay process, and is therefore linked to the 
absolute energy scale of the calorimeters as discussed above. This scale can 
be measured to a limited level of accuracy by calibration with $e^\pm$ beams, 
but ultimately has to be fitted from \mbox{$\ks \To \pi \pi$} data.
This is done (for charged decays as well) by adjusting the reconstructed 
positions of well defined detector edges for decays in which such a detector 
did not fire. The accuracy and the stability of this adjustment procedure 
(at the $10^{-4}$ level) as a function of time and of kinematic 
variables determines the systematic error, and was one of the main reasons for 
building high performance calorimeters.
When a Monte Carlo is not used for the largest part of the acceptance 
correction as in NA48, the sensitivity of the result to the knowledge of the 
above scales, which directly affects both the position and the size of the 
fiducial region, with a potentially large effect on the steep $\ks$ vertex 
distributions, is greatly reduced by fixing the most sensitive upstream edge 
of such region with a hardware cut (defined by a veto counter); event weights 
in NA48 are defined in terms of the measured kaon proper lifetime, instead of 
position, which is seen to be independent of the absolute energy scale of the 
calorimeter.
In both experiments the neutral energy scales are also checked, at different
points, by exploiting the prompt $3\pi^0$ decays of $\eta$ mesons produced on 
the vacuum window (KTeV) or on thin polyethylene targets during special 
$\pi^-$ beam runs (NA48): the reconstruction of the known decay vertex 
position for such events allows further checks on the energy scale and
non-linearities.

\begin{figure}[hbt!]
\begin{center}
    \epsfig{file=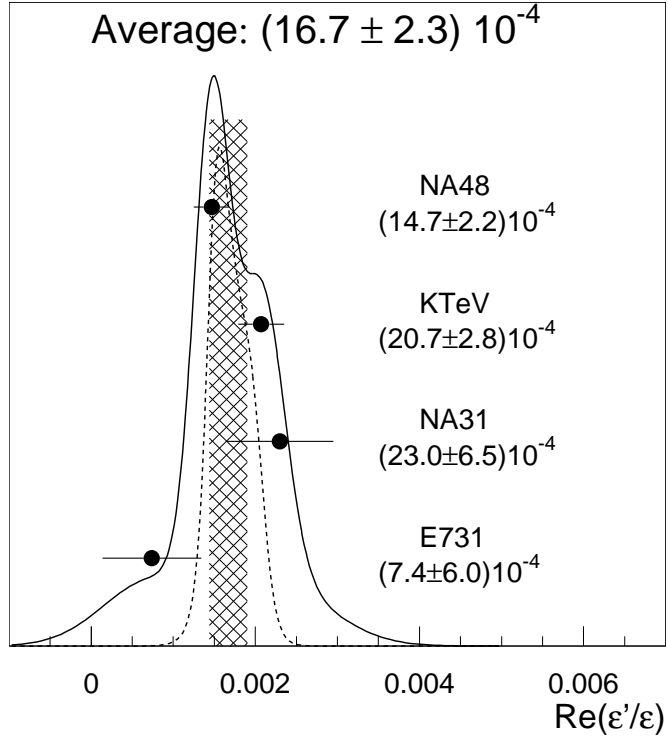,width=0.9\textwidth}
    \label{fig:ideo}
    \caption{Ideogram of recent published $\Rea(\epseps)$ measurements. 
The curves show (unnormalised) probability distributions according to the PDG 
procedure \cite{PDG_2002} (solid line) or a Bayesian ``skeptical'' approach 
\cite{D'Agostini} (dashed line).}
\end{center}
\end{figure}

The world average of $\epseps$ measurement is 
\Dm{
  \Rea(\epseps) = (16.7 \pm 2.3) \cdot 10^{-4}
}
where the error has been inflated by a factor 1.44 according to the procedure 
adopted by the PDG \cite{PDG_2002}, due to the poor $\chi^2$ value of 6.2 
(with 3 degrees of freedom). The probability of the four most precise 
measurements to be consistent is 10\%, and it varies between 7\% and 20\% when 
a single measurement is ignored. A graphical depiction of the present data is 
shown in figure \ref{fig:ideo}. 

\begin{figure}[hbt!]
\begin{center}
    \epsfig{file=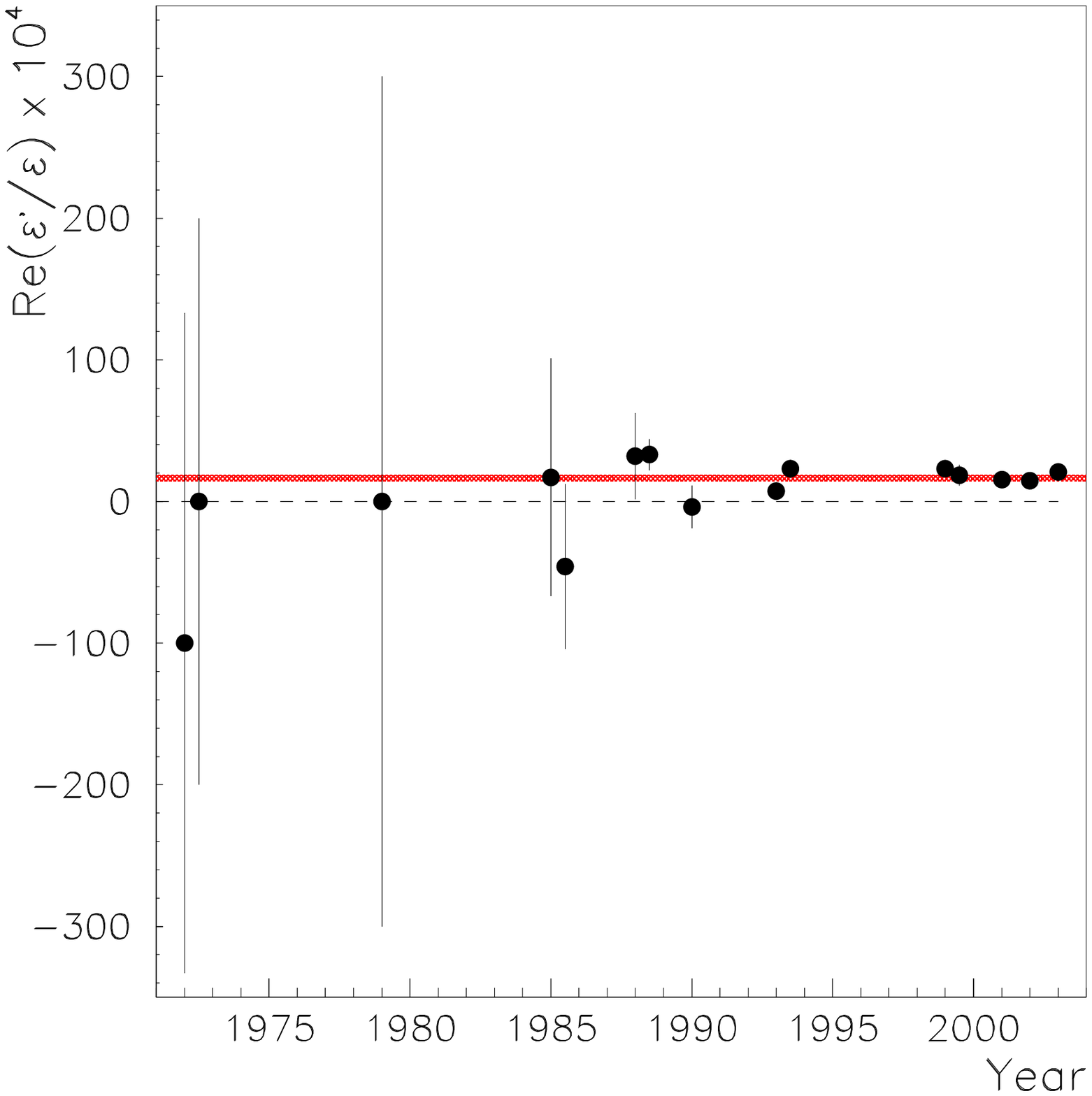,width=0.8\textwidth}
    \label{fig:history}
    \caption{Time evolution of $\Rea(\epseps)$ measurements. The horizontal
band represents the current world average.}
\end{center}
\end{figure}

Despite the somewhat unsatisfactory consistence, such averaged result is still 
more than 7 standard deviations from zero, therefore proving at last the 
existence of direct $CP$ violation in neutral kaon decays, after a long 
history of measurements, graphically summarised in figure \ref{fig:history}.

The value of $\Rea(\epseps)$ usually quoted by experiments is derived from the
double ratio of partial decay widths according to the simple relation 
(\ref{eq:R}), which neglects terms of order $|\epsilon'|^2$ and also 
$|\omega|$; the proper comparison with theoretical predictions should be done 
retaining the small correction due to the latter, so that the actual formula 
becomes
\Dm{
  \Rea(\epseps) = \frac{1}{6} 
  \left(1-\frac{|\omega|}{\sqrt{2}} \cos(\delta_2-\delta_0) \right) 
  \left(1- \left| \frac{\eta_{00}}{\eta_{+-}} \right|^2 \right)
}
This represents a $\sim 2\%$ correction, completely negligible compared to the 
present uncertainties in the theoretical computation, and also to the present 
experimental accuracy, but not to the size of the corrections which the 
experiments do consider to quote the central value.
The experimental value of $\Rea(\epseps)$ to be compared with theoretical 
predictions is therefore
\Dm{
  \Rea(\epseps) = (16.3 \pm 2.3) \cdot 10^{-4}
}
and it should be noted that the uncertainty on the value of $|\omega|$ (see 
\cite{Gardner} and references therein) hardly affects any comparison with 
theory in itself, since the empirical value of such parameter is used both in 
the computation and in extracting the value of $\Rea(\epseps)$ from the 
experiments.

With no constraint on the relative phase of $\epsilon'$ and $\epsilon$ is
imposed, the relation at the same level of approximation is
\Ea{
  & 1- \left| \frac{\eta_{00}}{\eta_{+-}} \right|^2 = 
    6 \, \Rea(\epseps) \left(1+\frac{|\omega|}{\sqrt{2}} 
    \cos(\delta_2-\delta_0) \right) - \nonumber \\
  & \Imm(\epseps) \left(3\sqrt{2} |\omega| \sin(\delta_2-\delta_0) \right)
}
but the above phase difference is $\sim \delta_2-\delta_0+\pi/2-\phi_{SW} =
(-1.2 \pm 1.5)^\circ$ and therefore the last term can be safely neglected.

While the order of magnitude of the measured value of $\Rea(\epseps)$ is not 
at odds with what can be expected in the Standard Model, it does not pose any 
strong constraint on its underlying picture of $CP$ violation, since the 
computation of the hadronic part of the decay process is not yet completely 
under theoretical control. 
The common expectation is that improvements in the accuracy of lattice QCD 
computations will ultimately allow precise quantitative comparisons to be 
performed\footnote{It is interesting to note, however, that the two most 
recent lattice QCD computations of $\epseps$ in the SM are in gross 
disagreement with the experimental measurement \cite{eps_lattice}.}.

Being evidence of direct $CP$ violation, the measured non-zero value of 
$\Rea(\epseps)$ translates to an asymmetry between $CP$-conjugate processes
\Ea{ \Ds
  & \frac{\Gamma(\kz \To \pi^+ \pi^-)-\Gamma(\kzb \To \pi^+ \pi^-)}
    {\Gamma(\kz \To \pi^+ \pi^-)+\Gamma(\kzb \To \pi^+ \pi^-)} = 
    (5.56 \pm 0.84) \cdot 10^{-6} \\
  & \frac{\Gamma(\kz \To \pi^0 \pi^0)-\Gamma(\kzb \To \pi^0 \pi^0)}
    {\Gamma(\kz \To \pi^0 \pi^0)+\Gamma(\kzb \To \pi^0 \pi^0)} = 
    (-11.1 \pm 1.7) \cdot 10^{-6} 
}
which makes the effect of direct $CP$ violation more self-evident.

\subsection{Other neutral $K$ decays}

\subsubsection*{$\mathit{K_{S,L} \To 3\pi}$}

Experimentally, \mbox{$K \To 3\pi$} decays \cite{Zemach} are analyzed in terms 
of two adimensional, Lorentz-invariant, independent variables $u,v$ (sometimes 
called $Y,X$):
\Ea{
  u \equiv \frac{s_3-s_0}{m_\pi^2} \quad && 
  \quad v \equiv \frac{s_1-s_2}{m_\pi^2} 
}
having defined 
\Ea{
  & s_i \equiv (p_K - p_i)^2 \\
  & s_0 \equiv (s_1+s_2+s_3)/3 = m_K^2/3 + \sum_i m_{\pi_i}^2/3
}
where $p_K$ and $p_i$ are the kaon and pion four-momenta, with the convention
that the index $i=3$ refers to the ``odd'' pion (the neutral one in $\pi^+ 
\pi^- \pi^0$).
Decay distributions are expressed as a power series in $u,v$, usually stopping
at second order due to the limited phase space (Q-values being in the 75-93 MeV
range):
\Dm{
  |A(K \To 3\pi)|^2 \propto 1 + gu + hu^2 + jv + kv^2 + fuv
}
where the parameters $g,h,j,k,f$ are usually referred to as ``Dalitz plot
slopes''. Note that in case of identical pions in the final state the 
odd terms in $u,v$ are intrinsically undefined and other terms may coincide 
(\emph{e.g.} for $3\pi^0$ decays there are no $g,j,f$ terms and a single 
quadratic slope term).
Any non-zero linear slope parameter for $v$ ($j \ne 0$ or $f \ne 0$) would be
evidence of $CP$ violation.

\begin{table}
\begin{center}
\begin{tabular}{|l|c|c|c|c|}
\hline
& $\kl \To \pi^+ \pi^- \pi^0$ & $\ks \To \pi^+ \pi^- \pi^0$ &
  $\kl \To \pi^0 \pi^0 \pi^0$ & $\ks \To \pi^0 \pi^0 \pi^0$ \\
\hline
BR  & $(12.58 \pm 0.19)\%$ & $(3.2^{+1.2}_{-1.0}) \cdot 10^{-7}$ &
      $(21.08 \pm 0.27)\%$ & $< 3 \cdot 10^{-7}$  \\
&& $(2 \pm 10) \cdot 10^{-9}$$(\dagger)$ && (90\% CL)\cite{NA48_3pi} \\
$g$ & $(0.678 \pm 0.008)$ & & $-$ & $-$ \\
$h$ & $(0.076 \pm 0.006)$ & & $(-5.0 \pm 1.4) \cdot 10^{-3}$ & \\
$j$ & $(1.1 \pm 0.8)\cdot 10^{-3}$ & & $-$ & $-$ \\
$k$ & $(9.9 \pm 1.5)\cdot 10^{-3}$ & & see $h$ & see $h$ \\
$f$ & $(4.5 \pm 6.4)\cdot 10^{-3}$ & & $-$ & $-$ \\
\hline
\end{tabular}
\caption{Experimental values of \mbox{$\kz \To 3\pi$} branching ratios and 
decay parameters, from \cite{PDG_2002} except where otherwise noted. 
$(\dagger)$: $CP$-violating component.}
\label{tab:3pi_neutral}
\end{center}
\end{table}

The possible $3\pi$ final states for neutral kaon decays are 
$\pi^+ \pi^- \pi^0$ and $\pi^0 \pi^0 \pi^0$; the measured branching ratios 
and slopes are reported in Table \ref{tab:3pi_neutral}.

Bose symmetry only allows isospin states I=1,3 for $3\pi^0$, so that this is
a pure $CP=-1$ eigenstate. Since the $\kl$ readily decays to $3\pi^0$, the (so 
far unobserved) transition \mbox{$K_S \To 3\pi^0$} requires $CP$ violation, 
and the decay \mbox{$K_1 \To 3\pi^0$} would imply direct $CP$ violation.
If direct $CP$ violation is not dominant in such decay, the expected branching 
ratio is
\Dm{
  BR(\ks \To 3\pi^0) = BR(\kl \To 3\pi^0) \frac{\tau_S}{\tau_L} |\epsilon|^2 
  \simeq 1.9 \cdot 10^{-9}
}

The $CP$ eigenvalue of the $\pi^+ \pi^- \pi^0$ state is $(-1)^{l+1}$, where 
$l$ is the eigenvalue of the orbital angular momentum between the two charged 
pions; such a state can have therefore both $CP=+1$ (for I=0,2) and $CP=-1$ 
(for I=1,3), so that both kaon $CP$ eigenstates can decay into it without 
violating $CP$; an angular momentum analysis is required to measure $CP$ 
violation in such decays.
The I=0,2 states must have a non-zero orbital angular momentum between any 
pair of pions, so that kaon decays into them are strongly suppressed by the 
centrifugal barrier, due to the limited phase space available. This is the 
case for the $CP$-conserving \mbox{$K_S \To (\pi^+ \pi^- \pi^0)_{I=0,2}$} 
transition, which being odd under the exchange of the $\pi^+ \pi^-$ momenta 
($v \To -v$), gives a vanishing contribution to the decay amplitude when 
integrated over the whole Dalitz plot (or at its centre): a measurement of 
\mbox{$\ks \To \pi^+ \pi^- \pi^0$} in such a region would be an indication of 
$CP$ violation.

The following $CP$-violating quantities are usually defined:
\Ea{
  && \eta_{+-0} \equiv \left.
    \frac{A(\ks \To \pi^+ \pi^- \pi^0)}
         {A(\kl \To \pi^+ \pi^- \pi^0)} \right|_{u=v=0} \\
  && \eta_{000} \equiv \left.
    \frac{A(\ks \To 3\pi^0)}
         {A(\kl \To 3\pi^0)} \right|_{u=v=0} \\
  && \eta^v_{+-0} \equiv \left.
    \frac{\partial A(\kl \To \pi^+ \pi^- \pi^0)/\partial v}
         {\partial A(\ks \To \pi^+ \pi^- \pi^0)/\partial v} \right|_{u=v=0} 
}
Since the $CP$-conserving \mbox{$\ks \To \pi^+\pi^-\pi^0$} amplitude vanishes 
at the centre of the Dalitz plot due to Bose symmetry, $\eta_{+-0}$ can also 
be defined as the ratio of the \kl to \ks transition amplitudes to the $CP=-1$ 
eigenstate.

Since $\pi^+\pi^-\pi^0$ is not a $CP$ eigenstate, $\ks--\kl$ interference in
the partial decay rate to this final state is an indication for $CP$ violation 
only when it is present after integration over the internal angle variables.
The $CP$-conserving decay amplitude for \mbox{$\kl \To \pi^+ \pi^- \pi^0$} 
interferes with both the $CP$-conserving and the $CP$-violating $\ks$ decay
amplitudes to the same final state, but the interference terms are 
respectively odd and even in $v$, and this feature can be used to separately 
measure them: as mentioned, the $CP$-conserving term vanishes upon integration 
over the whole phase space, allowing the extraction of the $CP$-violating 
interference term
\Dm{
  \eta_{+-0} = 
  \frac{\int du dv \, A^*(\kl \To \pi^+\pi^-\pi^0)
    A(\ks \To \pi^+\pi^-\pi^0;CP=-1)}
  {\int du dv \, |A(\kl \To \pi^+\pi^-\pi^0)|^2}
}

In analogy to the case of $\pi \pi$ decays, one often writes 
\Ea{
  \eta_{+-0} = \epsilon + \epsilon'_{+-0} \quad && \quad 
  \eta_{000} = \epsilon + \epsilon'_{000}
}
in order to have parameters independent from the phase convention.

When the final state is an isospin eigenstate with definite permutation 
symmetry (which would be the case for $3\pi^0$ in absence of $\Delta I>1/2$ 
transitions), or in absence of final-state interactions, one has (assuming 
$CPT$ symmetry)
\Dm{
  \epsilon'_f = i \left[ \frac{\Imm[A(\kz \To f)]}{\Rea[A(\kz \To f)]} -
    \frac{\Imm(A_0)}{\Rea(A_0)} \right]
}
where $A_0 \equiv A(\kz \To (\pi\pi)_{I=0})$, for any $CP=+1$ eigenstate $f$.
One can see that in this case a non-null imaginary part of $\eta_f$ 
requires either $CP$ violation in a decay amplitude or in the interference of 
mixing and decay.
In the general case, the presence of either of these kinds of $CP$ violation 
can originate a difference of $\eta_f$ for different states $f$. The  
measurement of such a difference with high precision presents an experimental 
challenge, since cancellations of systematic effects cannot be usually
achieved; while such a measurement cannot disentangle the two above types of 
$CP$ violation, it would be an indication of direct $CP$ violation.

The experimental study of $CP$ violation in $3\pi$ decays of the neutral kaons
requires intense sources of $\ks$ mesons: such studies can be performed by
analyzing decays from intense $\kz$ beams in the first few lifetimes after 
production (E621 at FNAL \cite{E621_3pi}, NA48 at CERN \cite{NA48_3pi}), by 
comparing $\kz$ and $\kzb$ decays of strangeness-tagged neutral kaons (CPLEAR 
at CERN \cite{CPLEAR}), or with tagged $\ks$ at a $\phi$ factory (KLOE at 
DA$\Phi$NE \cite{Dambrosio_dafnehb}).

The very small expected branching ratios $O(10^{-9})$ make the detection of 
$CP$ violation in these channels very challenging; present experiments were 
only able to put upper limits on the $\eta_{3\pi}$ parameters for $\ks$ 
($\eta_{+-0},\eta_{000}$), without detecting $CP$ violation even at the level 
of $\epsilon$.

By using strangeness-tagged neutral kaon decays, the event yield asymmetry 
for opposite strangeness of the kaon at production time can be studied as a 
function of proper time $t$:
\Ea{ \Ds 
  & A_{CP}^{(3\pi)}(t) \equiv
    \frac{\Gamma(\kz \To 3\pi)-\Gamma(\kzb \To 3\pi)}
    {\Gamma(\kz \To 3\pi)+\Gamma(\kzb \To 3\pi)} = \nonumber \\
  & -2 \Rea(\epsilon_S) + 2 |\eta_{3\pi}| \cos(\Delta m t + \phi_{3\pi}) 
    e^{-\Delta \Gamma t/2}
}
where $\phi_{3\pi}$ is the phases of $\eta_{3\pi}$).
The value of $\eta_{3\pi}$ can be extracted by fitting the $\ks--\kl$ 
interference term, while a non-zero asymmetry at $t=0$ is an indication of
direct $CP$ violation.

While direct $CP$ violation in $\pi \pi$ decays of neutral kaons is suppressed 
by the $\Delta I=1/2$ rule, this need not be the case for other decay modes, 
but it turns out to be true also for the $3\pi$ mode, in first approximation. 
In the Standard Model, the parameters $\epsilon'_{+-0}$ and $\epsilon'_{000}$, 
can be substantially enhanced over their lowest-order chiral perturbation 
theory estimate (the Li-Wolfenstein relation $|\epsilon'_{+-0}| \simeq 
|\epsilon'_{000}| \simeq 2|\epsilon'|$, valid at the centre of the Dalitz plot 
for $\pi^+\pi^-\pi^0$) due to the presence of higher-order contributions not 
suppressed by the $\Delta I=1/2$ rule, but they still remain small with 
respect to $\epsilon$ (see \cite{Cheng} and references therein).

The measurements of $\eta_{+-0}$ are dominated by the CPLEAR results obtained
with this technique \cite{CPLEAR_3pi}:
\Ea{
  & \Rea(\eta_{+-0}) = (-2 \pm 7^{+4}_{-1}) \cdot 10^{-3} \\
  & \Imm(\eta_{+-0}) = (-2 \pm 9^{+2}_{-1}) \cdot 10^{-3} 
}
corresponding to a limit $|\eta_{+-0}|<17 \cdot 10^{-3}$ at 90\% CL, which 
translates to $BR_{CPV}(\ks \To \pi^+\pi^-\pi^0) < 6.3 \cdot 10^{-8}$ for the
$CP$-violating part.
For $\eta_{000}$ a preliminary NA48 result \cite{NA48_3pi} gives
\Ea{
  & \Rea(\eta_{000}) = (-26 \pm 10 \pm 5) \cdot 10^{-3} \\
  & \Imm(\eta_{000}) = (-34 \pm 10 \pm 11) \cdot 10^{-3}
}
and when $\Rea(\eta_{000})$ is fixed to $\Rea(\epsilon)$ one has a limit
$|\eta_{000}|<29 \cdot 10^{-3}$ at 90\% CL, corresponding to $BR(\ks \To 3
\pi^0) < 3.1 \cdot 10^{-7}$.

$CP$ violation (even of the indirect type) is not yet seen in these channels, 
and higher statistic experiments would be required for a search of direct 
$CP$ violation.

In experiments at $\phi$ factories one could study the interference terms of 
the relative time intensity distributions when one kaon decays to $3\pi$ and 
the other semi-leptonically \cite{Dambrosio_dafnehb}; this approach is 
statistically more powerful since the interference term can be proportional to 
$\eta$ instead of $|\eta|^2$. 
Using the same notation as for the discussion of $\pi \pi$ decays, the 
time-dependent asymmetry
\Ea{ \Ds 
  & A_{3\pi}(\Delta t) = \frac{I(3\pi, l^+ \pi^- \nu;\Delta t)-
    I(3\pi, l^- \pi^+ \overline{\nu};\Delta t)}
    {I(3\pi, l^+ \pi^- \nu;\Delta t)+
    I(3\pi, l^- \pi^+ \overline{\nu};\Delta t)} = \nonumber \\
  & \frac{2 \Rea(\epsilon) e^{\Delta \Gamma \Delta t/2} - 
    2\Rea(\eta_{3\pi} e^{i\Delta m \Delta t})}
    {e^{\Delta \Gamma \Delta t/2} + 
    |\Gamma_S(3\pi)/\Gamma_L(3\pi)|^2 e^{-\Delta \Gamma \Delta t/2}}
}
reduces to $2\Rea(\epsilon)$ for large positive $\Delta t$ but has a larger 
sensitivity to $\epsilon'_{3\pi}$ for $\Delta t <0$.

\subsubsection*{$\mathit{K_{S,L} \To \pi \pi \gamma}$}

Among the final states accessible to the neutral kaons, the radiative 
$\pi^+ \pi^- \gamma$ mode, with 
\Ea{ 
  & BR(\ks \To \pi^+ \pi^- \gamma) \simeq 1.8 \cdot 10^{-3} \\
  & BR(\kl \To \pi^+ \pi^- \gamma) \simeq 4.4 \cdot 10^{-5}
} 
has also been considered for searches of direct $CP$ violation. 
Since $\pi\pi\gamma$ is not a pure $CP$ eigenstate, $CP$ violation cannot be 
detected as a simple violation of a selection rule, although as for all 
non-leptonic decay modes, the measurement of an interference effect between 
$\ks$ and $\kl$ is evidence \cite{pipigamma_sehgal} for $CP$ 
violation\footnote{This is true when the final state is summed over internal 
angle variables and polarizations.}.

As for all radiative decays, a (dominant) fraction of its rate, corresponding 
to low-energy photons, is unavoidably included in the measurements of the 
corresponding non-radiative mode, due to the finite energy threshold for 
photon detection \cite{KLOE_R}.

The decay of neutral kaons to the $\pi^+ \pi^- \gamma$ final state can occur 
through a so called \emph{inner-bremsstrahlung} (IB) process, \emph{i.e.} an 
electric odd-multipole photon emission from one of the pions in the $\pi^+ 
\pi^-$ state. This electromagnetic process is $CP$-conserving, and therefore 
the (approximate) $CP$ symmetry strongly hinders the 
\mbox{$\kl \To \pi^+ \pi^- \gamma$} decay to proceed in this way (the $CP$ 
eigenvalue of $\pi\pi\gamma$ states is $(-1)^{l+1}$ for electric multipoles 
$El$ and $(-1)^l$ for magnetic ones $Ml$, \emph{i.e.} $CP=+1$ for E1, M2, 
$\ldots$, and $CP=-1$ for M1, E2, $\ldots$).
Other contributions to $\pi^+ \pi^- \gamma$ decays of neutral kaons due to 
radiation from the decay vertex can be present, and are usually called 
\emph{direct emission} (DE) terms; these are generally smaller than IB, and 
lowest multipolarity (dipole, $l=1$) terms strongly dominate. 
$CP$ symmetry strongly suppresses any M1(E1) DE term in $\ks$($\kl$) decays to 
$\pi^+\pi^-\gamma$.

The photon energy spectra for the two types of emission (IB and DE) are 
different, therefore allowing the measurement of their relative importance in 
$\kl$ decays. Experimental results \cite{PDG_2002} show that the DE term in 
$\kl$ decays is mainly of M1 nature.

Any possible $CP$-violating asymmetry between $\pi^+$ and $\pi^-$ in this 
decay would require the interference of $\pi\pi$ states of opposite parity, 
and therefore both odd and even multipoles; a DE E2 term is therefore the 
necessary ingredient to have such kind of effects, which are therefore 
expected to be strongly suppressed by the smallness of the multipole expansion 
parameter.
Charge asymmetries of this kind larger than 2.4\% are excluded at 90\% CL by
KTeV \cite{KTeV_pipigamma}. 
In the case in which the photon polarization is observed, a sign asymmetry in 
the angle between the direction of such polarization and the $\pi^+\pi^-$ 
plane also requires higher order multipoles.
Ignoring the above asymmetries, only the lowest-order E1 and M1 DE terms can 
be considered, as will be done in the following.

While the \mbox{$\ks \To \pi^+ \pi^- \gamma$} decay is dominated by the 
unsuppressed IB process, \mbox{$\kl \To \pi^+ \pi^- \gamma$} receives two 
competing contributions: the IB, induced through the $K_1$ component 
(suppressed by the approximate $CP$ symmetry), and the (intrinsically smaller, 
mainly M1) DE from the dominant $K_2$ component, which does not interfere with 
the previous one in the total rate.

A $CP$-violating amplitude ratio can be defined for a $CP$ eigenstate such as
\Dm{
  \eta_{+-\gamma} \equiv 
  \frac{A(\kl \To \pi^+ \pi^- \gamma;E1)}{A(\ks \To \pi^+ \pi^- \gamma;E1)}
}
where only the lowest order multipole has been considered.

The proper decay time distribution of $\pi^+ \pi^- \gamma$ decays from a 
mono-energetic neutral kaon beam is described by 
\Ea{
  & I(\pi^+\pi^-\gamma;t) \propto  C_S \, e^{-\Gamma_S t} + 
   [C_L^{(IB)} |\eta_{+-\gamma}|^2 + C_L^{(DE)}] e^{-\Gamma_L t} + \nonumber \\
  & C_{\mathrm{int}} D(p_K) |\eta_{+-\gamma}| 
    \cos (\Delta m t - \phi_{+-\gamma}) \, 
    e^{-(\Gamma_S+\Gamma_L)t/2} 
}
where $\phi_{+-\gamma}$ is the phase of $\eta_{+-\gamma}$.
The $C_S$ term describes $\ks$ (IB-dominated, E1) decays, and the $C_L^{(IB)}$ 
and $C_L^{(DE)}$ terms $\kl$ decays due to IB (E1, indirect $CP$-violating) 
and DE (mostly $CP$-conserving M1, but possibly also direct $CP$-violating E1) 
respectively. 
The coefficient $C_{\mathrm{int}}$ of the interference term is multiplied by 
the ``dilution factor'' $D(p_K)$ describing the incoherent mixture of $\kz$ 
and $\kzb$ present in the beam at the production point, defined earlier.

In absence of direct $CP$ violation, the dominant DE contribution to 
\mbox{$\kl \To \pi^+ \pi^- \gamma$} decay is of M1 multipolarity ($CP$ 
conserving \mbox{$K_2 \To \pi^+ \pi^- \gamma$} decay) and does not interfere 
with the IB one ($CP$ conserving E1 \mbox{$K_1 \To \pi^+ \pi^- \gamma$} 
decay), so that the $\ks--\kl$ interference term $C_{\mathrm{int}}$ (due to E1 
amplitudes) is unaffected by $CP$ violation, which only modifies the 
coefficient of the $e^{-\Gamma_L t}$ term. In this case the $CP$-violating 
amplitude ratio $\eta_{+-\gamma}$ is equal to $\eta_{+-} \simeq \epsilon$, and 
the ratio of the E1 $\pi^+ \pi^- \gamma$ branching ratio to the $\pi^+ \pi^-$ 
branching ratio is clearly the same for $\ks$ and $\kl$.

An E1 direct emission contribution from the decay of the dominant $K_2$ 
component of the $\kl$ would be a direct $CP$-violating term, which would 
affect the interference term and shift the measured value of $\eta_{+-\gamma}$ 
away from $\epsilon$.

Since the inner bremsstrahlung contribution is determined by the amplitude for
the corresponding non-radiative decay \mbox{$\kl \To \pi^+ \pi^-$}, one has, 
at first order in the ratio of DE/IB $\ks$ decay amplitudes \cite{Isidori}:
\Ea{ \Ds 
  & \eta_{+-\gamma} = \eta_{+-} + \epsilon'_{+-\gamma} \simeq \nonumber \\
  & \eta_{+-} + \frac{(\beps-\eta_{+-})A(K_1 \To \pi^+\pi^-\gamma;E1)+
    A(K_2 \To \pi^+\pi^-\gamma;E1)}{A(\ks \to \pi^+\pi^-\gamma;E1(IB))}
}
Even if the direct $CP$ violating term $\epsilon'_{+-\gamma}$ is not suppressed
by the $\Delta I=1/2$ rule as it happens for $\epsilon'$, it contains the 
small factor given by the ratio of the direct emission and the inner 
bremsstrahlung amplitudes; since also in this case the two interfering 
amplitudes have widely different magnitude, theoretical predictions for 
$\epsilon'_{+-\gamma}$ are in the $10^{-5}$ range \cite{Isidori}.

Evidence for direct $CP$ violation in $\pi^+ \pi^- \gamma$ decays of neutral 
kaons could therefore be obtained by a non-zero difference in the measured 
fractions of radiative (IB) to non-radiative $\pi^+ \pi^-$ decays for $\ks$ 
and $\kl$: such a measurement would require a precise subtraction of the DE 
component in $\kl$ decays and either a pure $\ks$ beam or a precise 
subtraction of the $\kl$ component in a mixed beam, both difficult at the 
level of accuracy required.
A better approach is that of analyzing $\pi^+ \pi^- \gamma$ decays at short 
proper times (close to production target or regenerator), where the 
interference term dominates on the $\kl$ decay terms, and fit their proper 
decay time distributions to extract the $\eta_{+-\gamma}$ parameter, searching 
for differences from $\epsilon$. 
In the limit in which $\ks$ decays are given by the IB term only, the 
$\eta_{+-\gamma}$ parameter which is measured is actually the ratio of the 
$CP$-violating part of the $\kl$ decay amplitude over the ($CP$-conserving) 
$\ks$ decay amplitude.
This kind of measurement was actually performed by the FNAL experiments E731 
\cite{E731_pipigamma} and E773 \cite{E773_pipigamma} for decays downstream of 
a regenerator. The amount of $CP$ violation measured is consistent with what 
is expected from indirect $CP$ violation only \cite{PDG_2002}:
\Ea{
  & |\eta_{+-\gamma}| = (2.35 \pm 0.07) \cdot 10^{-3} \\
  & \phi_{+-\gamma} = (44 \pm 4)^\circ
}
and assuming any difference between $\eta_{+-\gamma}$ and $\eta_{+-}$ to be due
to direct $CP$ violation, E731 \cite{E731_pipigamma} quoted 
\Dm{
  |\epsilon'_{+-\gamma}|/|\epsilon| < 0.3 \quad (90\% \, CL)
}

By measuring the polarization of the photon in the final state, more 
information could be obtained: in presence of M1 DE terms, the decay 
amplitudes depend on the photon polarization; however, assuming $CPT$ symmetry 
and neglecting higher order multipoles, $CP$ violation could induce a net 
photon polarization in pure $\pi^+ \pi^- \gamma$ decays only if there are 
differences in the final-state interactions of the interfering two-pion states 
(with different isospin): approximate $CP$ symmetry and IB dominance strongly 
suppress any such effect in $\ks$ decays, while they could be present in $\kl$ 
decays.

In the time evolution of a generic $\ks--\kl$ mixture, an oscillating 
net photon polarization is expected, independently from $CP$ violation effects.
Downstream of a regenerator it would be possible to reach, for some given 
value of proper time, complete photon polarization; in presence of direct $CP$ 
violation this value of proper time could be different for the two 
polarization states.
The experimental difficulties of photon polarization measurements however 
make the above measurements less appealing (but see the following section on
$\pi^+ \pi^- e^+ e^-$).

The $\pi^+ \pi^- \gamma$ decay could provide new information on $CP$ 
violation, although any direct $CP$-violating contribution is predicted to be 
rather small in the SM (\emph{e.g.} the authors of \cite{pipigamma_chiral} 
quote $|\epsilon'_{+-\gamma}/\epsilon|<0.02$ at best).

In experiments with correlated kaon pairs at a $\phi$ factory one could 
exploit the time difference distribution for radiative $\pi^+\pi^-$ and 
semi-leptonic decays to extract information on $\eta_{+-\gamma}$ 
\cite{Dambrosio_dafnehb}:
\Ea{
  & I(\pi^\pm l^\mp \nu,\pi^+\pi^-\gamma;\Delta t<0) \simeq 
    \frac{\Gamma_L(\pi^\pm l^\mp \nu) \Gamma_S(\pi^+\pi^-\gamma)}
    {\Gamma_S+\Gamma_L} = \nonumber \\ 
  & \left[ 
    \frac{\Gamma(\kl \To \pi^+\pi^-\gamma)}{\Gamma(\ks \To \pi^+\pi^-\gamma)}
    e^{-\Gamma_L |\Delta t|} + e^{-\Gamma_S |\Delta t|} \pm 
    \right. \nonumber \\
  & \left. 2 e^{-(\Gamma_S+\Gamma_L) |\Delta t|/2} 
    |\Rea(\eta_{+-\gamma})| \cos(\Delta m |\Delta t| - \phi_{+-\gamma}) 
    \right]
}

For \mbox{$K_{S,L} \To \pi^0 \pi^0 \gamma$} decays no IB contribution is 
present, and Bose symmetry forbids odd multipole contributions. The rates 
being highly suppressed (predicted BR $< 10^{-8}$ \cite{Isidori}), these 
decays (never observed so far) are not very promising for $CP$ violation 
studies.

\subsubsection*{$\mathit{K_{S,L} \To \pi^+ \pi^- e^+ e^-}$}

The decays \mbox{$K_{S,L} \To \pi^+ \pi^- e^+ e^-$} are expected to proceed 
through an intermediate state $\pi^+ \pi^- \gamma^*$, followed by internal 
conversion, and are therefore related to the $\pi^+ \pi^- \gamma$ decays 
discussed above. Although suppressed by two orders of magnitude with respect 
to those, the decays discussed here have the advantage of giving easier 
experimental access to the polarization of the (virtual) photon, which can 
induce asymmetries in the orientation of the $e^+ e^-$ decay plane with 
respect to the $\pi^+ \pi^-$ one.

The angular distribution in the angle $\phi$ between the normals to the 
$\pi^+ \pi^-$ and $e^+e^-$ planes can be parameterized as 
\Dm{
  \frac{d\Gamma}{d\phi} = I_1 \cos^2 \phi + I_2 \sin^2 \phi + 
  I_3 \sin \phi \cos \phi
}
where the $\sin \phi \cos \phi$ term, which changes sign for $\phi \To -\phi$, 
contains the interference between the two dominant contributions to the decay, 
the indirect $CP$-violating inner bremsstrahlung (E1) and the $CP$-conserving 
direct emission (M1).
Other small contributions could come from a (direct $CP$-violating) E1 DE term
and through a $CP$-conserving ``charge-radius'' \mbox{$\kl \To \ks \gamma$} 
transition.

\begin{table}
\begin{center}
\begin{tabular}{|l|c|c|}
\hline
& $\ks \To \pi^+ \pi^- e^+ e^-$ & $\kl \To \pi^+ \pi^- e^+ e^-$ \\
\hline
BR        & $(4.7 \pm 0.3) \cdot 10^{-5}$ & $(3.38 \pm 0.13) \cdot 10^{-7}$ \\
Asymmetry & $(0.5 \pm 4.3)$\% & $(13.5 \pm 1.5)$\% \\
\hline
\end{tabular}
\caption{Experimental data on \mbox{$\kz \To \pi^+ \pi^- e^+ e^-$} decays, 
from \cite{PDG_2002} \cite{KTeV_pipiee_new} \cite{NA48_pipiee}.}
\label{tab:pipiee}
\end{center}
\end{table}

Table \ref{tab:pipiee} summarizes the available experimental information: a 
large asymmetry ($\approx$ 14\%) in the above-mentioned angle $\phi$ was 
predicted \cite{pipiee_theory1} and actually observed \cite{KTeV_pipiee} in 
$\kl$ decays, and not in $\ks$ decays \cite{NA48_pipiee} as expected. 

Introducing the angle $\theta_e$ between the three-momentum of the $e^+$ and 
that of the di-pion system, as measured in the $e^+ e^-$ rest frame, the 
differential decay rate can be written (dropping terms proportional to 
$m_e^2$) as \cite{pipiee_theory2} 
\Ea{ \Ds
  & \frac{d\Gamma}{d\cos \theta_e d\phi} = A_1 + A_2 \cos 2\theta_e + 
    A_3 \sin^2 \theta_e \cos 2\phi + A_4 \sin 2\theta_e \cos \phi 
    \quad + \nonumber \\
  & \quad A_5 \sin \theta_e \cos \phi + A_6 \cos \theta_e + \nonumber \\
  & \quad A_7 \sin \theta_e \sin \phi + A_8 \sin 2\theta_e \sin \phi +
    A_9 \sin^2 \theta \sin 2\phi
}
where the terms with coefficients $A_4,A_7$ and $A_9$ are $CP$-violating.
The only numerically relevant coefficient in the SM is $A_9$, which generates 
the decay plane asymmetry discussed above. 
While $A_4$ and $A_9$ are mainly induced by indirect $CP$ violation, $A_7$ only
contains direct $CP$ violation contributions, which would induce an asymmetry
in the distribution of decays in $\sin \theta_e \sin \phi$.
Unfortunately the ratio of direct to indirect $CP$ violation for this decay is 
predicted to be at most of order $10^{-3}$ \cite{pipiee_theory2} in the 
Standard Model.
A measurement of direct $CP$ violation from the angular distribution in the 
\mbox{$\kl \To \pi^+ \pi^- e^+ e^-$} decay would require a much larger 
statistics than the current world data sample ($\sim 6 \cdot 10^3$ events), 
besides a very accurate knowledge of the detection angular acceptance, and 
looks currently beyond reach.

$CP$ violation effects can also show up in the rare decays of neutral kaons to 
four leptons, with \cite{PDG_2002} \cite{KTeV_Moriond03} \cite{KTeV_eemumu}
\Ea{
  & BR(\kl \To e^+ e^- e^+ e^-) = (4.06 \pm 0.22) \cdot 10^{-8} \\
  & BR(\kl \To \mu^+ \mu^- e^+ e^-) = (2.7 \pm 0.3) \cdot 10^{-9}
}
in the distribution of the angle between the planes of the two $l^+ l^-$ pairs 
in the kaon rest frame.
Such distributions have been actually measured with limited precision by KTeV 
\cite{KTeV_eeee} \cite{KTeV_Moriond03} \cite{KTeV_eemumu}, but the statistics 
does not allow to identify the presence of $CP$ violation in these decays.

\subsubsection*{$\mathit{\kl \To \pi^0 l\overline{l}}$}

An intense theoretical and experimental activity has been (and is being) 
devoted to the study of the decays \mbox{$\kl \To \pi^0 l \overline{l}$} 
(where $l$ is a lepton, $e$, $\mu$ or $\nu$), since such loop-dominated decays 
(so far unobserved) are expected to be mostly $CP$-violating, with a 
direct-indirect $CP$ violation hierarchy rather different from the one of 
$\pi\pi$ decays. Moreover, in some cases their properties can be related with 
reasonable confidence to the fundamental parameters of the theory, thus 
allowing sensitive searches for new physics.
Contrary to the case of $\pi\pi$ decays, the presence of a single hadron in 
the final state reduces the theoretical difficulties in linking the elementary 
amplitudes, expressed in terms of quarks, to the measured ones for physical 
mesons: the hadronic part of the amplitude can be extracted from the 
experimental knowledge of the \mbox{$K^+ \To \pi^0 e^+ \nu$} decay process
via isospin symmetry.

Within the Standard Model, the amplitudes for these decays receive three
different kinds of contributions (see \emph{e.g.} \cite{Isidori}). 
The first contribution, only present for final states with charged leptons, 
is the one induced by \mbox{$K \To \pi \gamma^* \gamma^*$} transitions, 
the only one which can be $CP$-conserving. Theoretical predictions of such 
contribution are unreliable at present, but the experimental measurements of 
the \mbox{$\kl \To \pi^0 \gamma \gamma$} decay allow to bound it: two recent 
measurements of such decay mode \cite{KTeV_pigg} \cite{NA48_pigg} are poorly 
consistent among them in this respect, and can lead to predictions for the 
contribution to $\kl \To \pi^0 l^+ l^-$ which can differ by an order of 
magnitude (see \cite{Gabbiani} and references therein).
The $CP$ conserving contribution is predicted to be similar for the electron 
and the muon modes.

The second and most interesting contribution is a short-distance ``direct'' 
$CP$ violating one\footnote{Since there is a single hadron in the final state, 
and negligible phases from final-state interactions, this is actually $CP$ 
violation in the interference of mixing and decay, containing both direct and 
indirect $CP$ violation \cite{BurasFleischer}.}; in the SM this component, 
dominated by diagrams with top-quark loops, can be predicted with good 
precision as a function of the CKM mixing matrix parameters, and is smaller by 
a factor $\simeq 5$ for the muon mode than for the electron one.

The third contribution is a $CP$-violating one, mostly induced by \mbox{$K \To 
\pi \gamma^* (Z^*)$} transitions ($l^+ l^-$ in the $J^{PC} = 1^{--}$ state), 
which is largely a manifestation of indirect $CP$ violation due to the $K_1$
component of $\kl$. This component is negligible for the final state with two 
neutrinos \cite{Littenberg}, since the diagram with a virtual $Z$ is heavily 
suppressed ($\sim 10^{-7}$) with respect to that with a virtual photon. The
latter cannot be predicted in a reliable way, but can be determined by a 
measurement of \mbox{$\ks \To \pi^0 l \overline{l}$}:
\Dm{
  BR_{\mathrm{ind}}(\kl \To \pi^0 l^+ l^-) = 
    |\epsilon|^2 \frac{\tau_L}{\tau_S} BR(\ks \To \pi^0 l^+ l^-)
}
For the electron mode, the $\ks$ decay was measured by the NA48/1 experiment, 
which presented a preliminary result \cite{NA48_pi0ee}
\Dm{
  BR(\ks \To \pi^0 e^+ e^-) = 
    (5.8^{+2.8}_{-2.3} \pm 0.3 \pm 0.8_{\mathrm{theo}}) \cdot 10^{-9}
}
where the first error is statistical, the second systematic, and the third
one is due to the theoretical uncertainty in the extrapolation to the full 
phase space\footnote{A form factor derived from \cite{Portoles} was used.}
This corresponds to 
\Dm{
  BR_{\mathrm{ind}}(\kl \To \pi^0 l^+ l^-) = 
  17.7^{+8.6}_{-6.9} \cdot 10^{-12}
}
for the indirect $CP$-violating contribution to the corresponding $\kl$ decay.
The indirect $CP$-violating amplitude can interfere with the direct one, so 
that a two-fold ambiguity remains in the relation between the branching ratio 
and the ``direct'' $CP$-violating contribution, depending on the relative sign 
of the two $CP$-violating components; the expectation for the total $\kl \To 
\pi^0 e^+ e^-$ branching ratio is now $1 \div 4 \cdot 10^{-11}$, but in case 
of destructive interference with the large indirect $CP$-violating component 
the branching ratio could have almost no sensitivity to the ``direct'' 
$CP$-violating part.

\begin{table}
\begin{center}
\begin{tabular}{|l|c|c|c|}
\hline
& $\kl \To \pi^0 e^+ e^-$ & $\kl \To \pi^0 \mu^+ \mu^-$ &
$\kl \To \pi^0 \nu \overline{\nu}$ \\
\hline
$BR_{\textrm{CPC}}$ & 
$0.3 \div 7 \cdot 10^{-12}$ & $(0.3 \div 7) \cdot 10^{-12}$ & $0$ \\
$BR_{\textrm{CPV dir}}$ & 
$4 \cdot 10^{-12}$ & $8 \cdot 10^{-13}$ & $3 \cdot 10^{-11}$ \\
$BR_{\textrm{CPV ind}}$ & 
$(0.15 \div 15) \cdot 10^{-11}$ & $(0.3 \div 30) \cdot 10^{-12}$ & 
$6 \cdot 10^{-15}$ \\
$BR_{\mathrm{tot}}$ &
$(3 \div 10) \cdot 10^{-12}$ & $(4 \div 10) \cdot 10^{-12}$ & 
$3 \cdot 10^{-11}$ \\
\hline
$BR_{\mathrm{exp}}$ (90\% CL) &
$< 2.8 \cdot 10^{-10}$ &
$< 3.8 \cdot 10^{-10} \, (\dagger)$ &
$< 5.9 \cdot 10^{-7} \, (\dagger)$ \\
\hline
$BR(l^+ l^- \gamma \gamma)$ & $6 \cdot 10^{-7}$ & $1 \cdot 10^{-8}$ & $-$ \\
\hline
\end{tabular}
\caption{Indicative ranges of the Standard Model theoretical expectations 
(\cite{Littenberg_lectures} and references therein), experimental limits 
from KTeV \cite{PDG_2002} \cite{KTeV_Moriond03} \cite{KTeV_pi0ee} 
\cite{KTeV_pi0mumu} \cite{KTeV_pi0nunu} for \mbox{$\kl \To \pi^0 l 
\overline{l}$} branching ratios, and branching ratios of the $l \overline{l} 
\gamma \gamma$ background. $(\dagger)$: 1997 sample only.}
\label{tab:pill_neutral}
\end{center}
\end{table}

Standard Model predictions for the branching ratios of \mbox{$\kl \To \pi^0 l
\overline{l}$} decays are in the $10^{-11}-10^{-12}$ range; those for the 
related \mbox{$\ks \To \pi^0 l^+ l^-$} decay are in the $10^{-8}-10^{-10}$ 
range; such expectations and the current experimental limits are summarized in 
table \ref{tab:pill_neutral}.

From the experimental point of view, the search for \mbox{$\kl \To \pi^0 
e^+e^-$} requires high rate experiments with very good electromagnetic 
calorimetry and photon detection. The largest backgrounds are due to the 
accidental superposition of $2\pi^0$ and $3\pi^0$ decays with single or double 
Dalitz decay of a $\pi^0$, and to the radiative Dalitz decay 
\mbox{$\kl \To e^+e^- \gamma \gamma$} (``Greenlee'' background 
\cite{Greenlee}, with $BR =$ \mbox{$(6.0 \pm 0.3) \cdot 10^{-7}$}) which 
can only be reduced by the $\pi^0$ mass constraint on the $\gamma \gamma$ 
pair; the measurement of \mbox{$\kl \To \pi^0 e^+e^-$} will ultimately require 
a reliable background subtraction and therefore an even higher sensitivity
than dictated by the branching ratio.

For the \mbox{$\kl \To \pi^0 \mu^+ \mu^-$} decays the background due to the 
muonic radiative Dalitz decay is a factor 60 smaller than for the $e^+e^-$ 
mode, \mbox{$BR(\kl \To \mu^+ \mu^- \gamma \gamma)$} = \mbox{$(1.0 \pm 0.7) 
\cdot 10^{-8}$}), but the kinematic cuts used to reduce it are also less 
effective, and the prediction for the ``direct'' $CP$-violating components is 
a factor $\approx 5$ smaller.

As far as direct $CP$ violation is concerned, the detection of \mbox{$\kl \To 
\pi^0 l^+ l^-$} decays needs to be complemented by accurate measurements 
of other kaon decays, since only the short-distance term can be predicted 
reliably. Additional experimental input, such as measurements of $e^+e^-$ 
energy distribution asymmetries or the $\ks--\kl$ interference term in the 
time dependence of the rate, would disentangle the different contributions.

The decay \mbox{$\kl \To \pi^0 \nu \overline{\nu}$} is the most interesting 
one: neglecting neutrino masses and assuming lepton flavour conservation, the 
final state is a pure $CP=+1$ eigenstate ($\nu \overline{\nu}$ pair produced 
in a state with angular momentum 1), so that there is no $CP$-conserving 
contribution, while the indirect $CP$-violating one is expected to be strongly 
suppressed in the Standard Model \cite{Littenberg} and can be bound by the  
measurement of \mbox{$K^+ \To \pi^+ \nu \overline{\nu}$}. The decay is 
therefore dominantly due to mixing-induced $CP$ violation, and as such its 
decay rate can be predicted reliably. 
The process occurs in the SM through second-order weak interactions only: weak 
penguin diagrams and W-exchange box diagrams in which only the top-quark loops 
are relevant; the branching ratio can be predicted with very good ($\sim 
10^{-2}$) accuracy, depending only on the top-quark mass, the strong coupling 
constant $\alpha_S$ and CKM matrix elements, so that its measurement could put 
strong constraints on the flavour mixing structure. 
With the present knowledge of the mixing matrix the prediction is 
\mbox{$BR(\kl \To \pi^0 \nu \overline{\nu})$}=\mbox{$(2.6 \pm 1.2) \cdot 
10^{-11}$}.
An indirect limit can be obtained in a model-independent way from the 
measurement of the related \mbox{$K^+ \To \pi^+ \nu \overline{\nu}$} 
\cite{Grossman}: $BR(\kl \To \pi^0 \nu \overline{\nu}) < $\mbox{$1.7 \cdot 
10^{-9}$}.

Clearly, this channel is also the most challenging from an experimental point 
of view: its signature is a single $\pi^0$ in the detector, and to suppress 
the \mbox{$\kl \To \pi^0 \pi^0$} background with two missing photons, which 
has a branching ratio $10^8$ times larger\footnote{``Yesterday's signal is 
today's calibration and tomorrow's background''.}, a hermetic and highly 
efficient ($\sim 1 \cdot 10^{-4}$) photon detector is required, but 
photo-nuclear interactions pose an intrinsic limit to the efficiency for low 
energy photons. 
The best branching ratio limit to date was obtained by the KTeV experiment 
\cite{KTeV_pi0nunu}, requiring the Dalitz decay of the $\pi^0$ in order to be 
able to reconstruct the decay vertex from the $e^+e^-$ pair and cut on the 
$\gamma \gamma $ invariant mass. While the above requirement reduces the 
sensitivity by two orders of magnitude, without imposing it the same 
experiment obtained \cite{KTeV_pi0nunu_gg}, in a 1-day special run with a 
``pencil'' beam, a worse limit (by a factor 2.7), background-dominated by beam 
neutron interactions with the material in front of the detector.
Extrapolations show that the use of the $\pi^0 \To \gamma \gamma$ decay mode 
and improvements in background suppression are required for future searches
(see \emph{e.g.} \cite{Kettell} for a recent review).

Two major projects are underway to search for this decay (\cite{E391a} and 
\cite{KOPIO}, see also \cite{KAMI}) with an expected sensitivity of a few tens
of events. 

\begin{figure}[htb!]
\begin{center}
    \epsfig{file=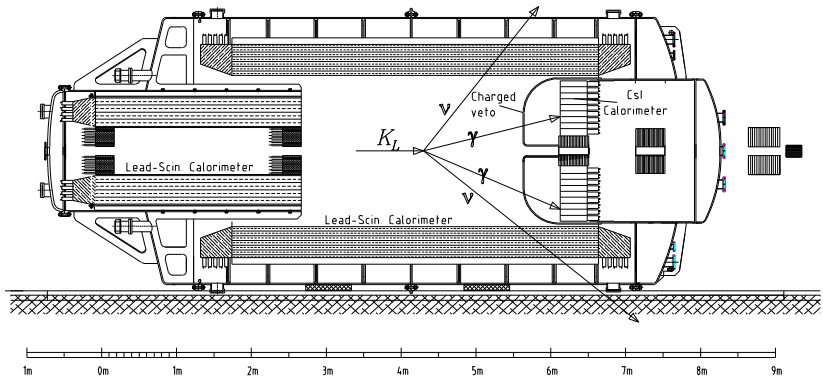,width=0.9\textwidth}
  \caption{Schematic drawing of the E391a experimental apparatus.}
  \label{fig:e391a}
\end{center}
\end{figure}

One experiment is planned to run at the new J-PARC 50 GeV proton synchrotron 
at Tokai (beam expected in 2008); its approach is going to be tested by the 
E391a pilot project \cite{E391a} running at the 12 GeV KEK PS in 2004. 
The experimental technique is based on a high transverse momentum ($p_T>120$ 
MeV/$c$) $\pi^0$ selection, making use of a well collimated, low energy 
(2 GeV/$c$) ``pencil'' beam, free from hyperon background, entering a 
hermetic, highly evacuated double decay region, used to suppress the beam halo 
and to reject decays occurring upstream of the fiducial volume (see fig. 
\ref{fig:e391a}). 
The advantage of this approach is the relatively high acceptance for the 
signal ($\sim 8\%$).
Photon detection is based on a CsI calorimeter efficient down to energies of 
1 MeV, and the efficiency requirements on the lead-scintillator veto counters
for low energy photons are reduced to the $10^{-4}$ level by suppressing the 
fully neutral $\pi^0\pi^0$ and $\gamma \gamma$ backgrounds with the $p_T$ cut.
The single event sensitivity of E391a is estimated to be above the level of the
Standard Model predictions ($\sim 2 \cdot 10^{-10}$ with less than 1 
background event), while 1000 events would be expected (for $BR = 3 \cdot 
10^{-11}$) in the full-scale experiment (with $\sim 16\%$ acceptance).

\begin{figure}[hbt!]
\begin{center}
    \epsfig{file=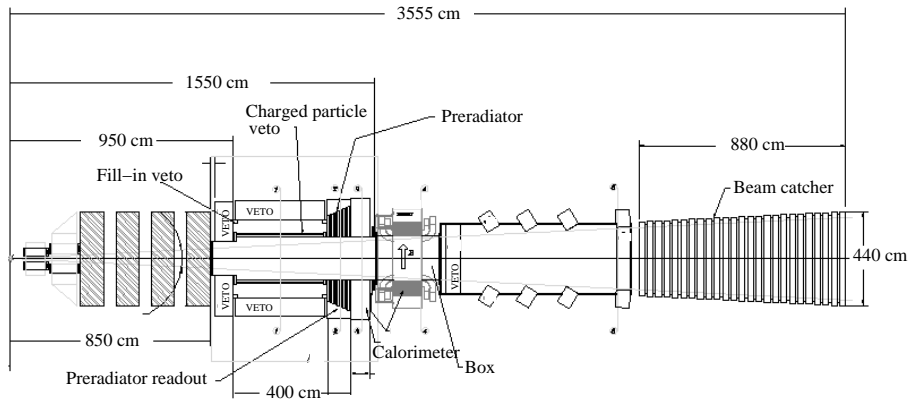,width=1.0\textwidth}
  \caption{Schematic drawing of the KOPIO experimental apparatus.}
  \label{fig:kopio}
\end{center}
\end{figure}

The KOPIO (E926) experiment \cite{KOPIO} at the 24 GeV BNL AGS, planned for 
2006, aims at a full kinematic reconstruction of the event by using a 2 
radiation length thick pre-radiator to measure photon directions, and an 
intense 800 MeV/$c$ RF-microbunched neutral beam (200 ps wide bunches every 
40 ns) to obtain the $\kl$ momentum by time-of-flight measurement from the 
electromagnetic ``shashlyk'' calorimeter (see fig. \ref{fig:kopio}). The soft 
spectrum for the neutral beam is such that \kl and neutrons are well below the 
$\pi^0$ hadro-production threshold, although the beam region will be in high 
vacuum. The beam will be very small in one of the two transverse dimensions, 
thus providing an additional constraint for the decay vertex reconstruction 
without limiting its acceptance.
While also this approach reduces the requirements on the vetoing efficiency to 
manageable levels, a complete charged particle and photon veto system, 
including a forward ``beam catcher'' with aerogel Cerenkov counters, will 
complement the detector.
With a $\sim 1.6\%$ acceptance, 50 signal events are expected in 3 years of 
running, with a signal to background ratio of 2.

Both collaborations have already shown encouraging results concerning the 
required beam and detector performances: E391a had a successful engineering 
run in 2002 on a neutral beam, during which the calorimeter was calibrated;
KOPIO had results close and sometimes better than required on the beam 
micro-bunching and the pre-shower angular resolution.

For the $\kl \To \pi^0 \mu^+ \mu^-$ decay, the use of additional information 
obtained from the measurement of muon polarizations through their decay 
asymmetries can be conceived; this has been proposed \cite{Diwan} as a 
superior way to disentangle the different contributions and extract the $CP$ 
violation in the decay amplitudes. 
Indeed, the $P$-odd longitudinal polarization of the $\mu^+$ in the $\mu^+ 
\mu^-$ rest frame is found for this decay to be proportional to the 
$CP$-violating phase of the $K_2 \To \pi^0 \gamma \gamma$ 
amplitudes\footnote{By comparison, the muon polarization transverse to the 
decay plane is (barring FSI effects) $T$-violating by itself, but does not 
allow to disentangle the indirect $CP$-violating contribution.}.
Numerical estimates of the expected polarizations confirm that they can be 
large, as expected from the prediction of comparable sizes for the terms in 
the decay amplitude.

\subsubsection*{$\mathit{K_{S,L} \To \gamma \gamma}$ and 
$\mathit{K_{S,L} \To l^+ l^-}$}

The $\gamma \gamma$ final state from $\kz$ decay is not a $CP$-eigenstate,
but a superposition of two $CP$-eigenstates which are distinguished by photon
polarizations being parallel or orthogonal:
\Ea{
  & | (\gamma \gamma)_{CP=+1} \rangle \equiv 
    | (\gamma \gamma)_{\parallel} \rangle = 
    \frac{1}{\sqrt{2}} [ |LL\rangle + |RR\rangle ] \\
  & | (\gamma \gamma)_{CP=-1} \rangle \equiv 
    | (\gamma \gamma)_{\perp} \rangle = 
    \frac{1}{\sqrt{2}} [ |LL\rangle - |RR\rangle ] 
}
where $R,L$ indicate right or left photon helicity. 
One can define two $CP$-violating amplitude ratios:
\Ea{ \Ds
  & \eta_{+} = \frac{A(\kl \To (\gamma\gamma)_{CP=+1})}
    {A(\ks \To (\gamma\gamma)_{CP=+1})} = \epsilon + \epsilon'_{+} \\
  & \eta_{-} = \frac{A(\ks \To (\gamma\gamma)_{CP=-1})}
    {A(\kl \To (\gamma\gamma)_{CP=-1})} = \epsilon + \epsilon'_{-} 
}
Direct $CP$ violation is expected to be larger by more than an order of 
magnitude than in $\pi \pi$ decays (see \cite{Isidori} and references therein),
particularly for the $CP=-1$ final state.

The decay rates for $\ks$ and $\kl$ decays to this states are small and 
similar, since \cite{PDG_2002} \cite{NA48_gg} :
\Ea{
  & BR(\ks \To \gamma \gamma) = (2.77 \pm 0.07) \cdot 10^{-6} \\
  & BR(\kl \To \gamma \gamma) = (5.93 \pm 0.08) \cdot 10^{-4}
}
and therefore $CP$ violation experiments are rather difficult.

One possibility for measuring $CP$ violation in these decays would be the
study of the time dependence of the decay rate in a $CP$ eigenstate for 
strangeness-tagged beams: the $\ks--\kl$ interference term changes sign for 
initially pure $\kz$ or $\kzb$. 
Unfortunately, the measurement of photon polarizations to study 
$CP$-eigenstates induces large suppression factors, making such experiments 
very difficult, given the small branching ratios. 
If photon polarizations are not measured, the total rates can be studied, in 
which $CP$-violating asymmetries between $\kz$ and $\kzb$ are also expected 
due to $\ks--\kl$ interference: large amounts of tagged decays ($>10^6$) would 
be required to measure any direct $CP$-violating effect, making also these 
measurements very challenging. 
Very intense kaon beams would allow such studies, by analyzing the decays
in the first few $\ks$ lifetimes from the production target.

\begin{table}
\begin{center}
\begin{tabular}{|l|c|c|}
\hline
Decay & Branching ratio & Notes \\
\hline
$\kl \To e^+ e^-$ & $9^{+6}_{-4} \cdot 10^{-12}$ & 
4 evts. (BNL E871 1998) \\
$\kl \To \mu^+ \mu^-$ & $(7.25 \pm 0.16) \cdot 10^{-9}$ &
$6.2 \cdot 10^3$ evts. (BNL E871 2000) \\
\hline
$\ks \To e^+ e^-$ & $< 1.4 \cdot 10^{-7}$ (90\% CL) &
(CPLEAR 1997) \\
$\ks \To \mu^+ \mu^-$ & $< 3.2 \cdot 10^{-7}$ (90\% CL) &
(CERN 1973) \\
\hline
\end{tabular}
\caption{Experimental data \cite{PDG_2002} for the \mbox{$\kz \To l 
\overline{l}$} branching ratios.}
\label{tab:kll}
\end{center}
\end{table}

The \mbox{$\kz \To \gamma \gamma$} transitions are also the main contribution 
to the \mbox{$\kz \To l^+ l^-$} decay amplitudes, in proportion to the lepton 
mass.
Only $\kl$ decays to these channels have been observed so far (see table 
\ref{tab:kll}). 
$CP$-eigenstates can be defined as for the $\gamma \gamma$ final state.
In presence of final-state interactions, $CP$-conserving and $CP$-violating 
amplitudes can interfere and lead to $CP$ violation which can be observed by 
measuring a non-zero asymmetry corresponding to a net longitudinal 
polarization of either lepton: 
\Dm{
  \langle P_l \rangle = \frac{N(l^-;R)-N(l^-;L)}{N(l^-;R)+N(l^-;L)}
}
where $N(l^-;R,L)$ are the numbers of $l^-$ emitted with positive or negative
helicity. In practice only $\mu^+$ polarization can be measured; 
predictions of this asymmetry for \mbox{$\kl \To \mu^+\mu^-$} decays are in 
the $10^{-3}$ range within the SM \cite{Isidori}; the direct $CP$-violating 
contribution is however estimated to be negligible with respect to the 
indirect one, expected at the level of $\sim 2 \cdot 10^{-3}$ (see 
\cite{Diwan_pol} and references therein).

\section{Charged $K$ decays}

For charged kaons, electric charge conservation forbids mixing: any difference 
between $K^+$ and $K^-$ decay parameters would be a signal of (direct) $CP$ 
violation. 
As usual, at least two interfering decay amplitudes with different strong and
weak phases are required in order to generate an observable asymmetry.
This rules out the dominant decay channels accounting for 93\% of the decay 
width: leptonic and semi-leptonic decays, as well as $\pi \pi$ decays for 
which (neglecting small electromagnetic isospin-breaking effects 
\cite{Gerard} or possible violations of Bose statistics \cite{Greenberg}) the 
final state can only be in the single $I=2$ isospin state.
$CP$ violation effects can only be detected in other decay channels, 
\emph{e.g.} in the $3\pi$ channels, which account for most of the remaining 
decay width; in this case however, the limited phase space available reduces 
the possible size of the strong phase shifts, and therefore of the expected 
asymmetries.

\subsection{Partial rate asymmetries}

The $3\pi$ states are the ones with largest branching ratios in which $CP$ 
violation effects can be expected to be present.

\begin{table}
\begin{center}
\begin{tabular}{|l|c|c|}
\hline
& $K^+ \To \pi^+ \pi^+ \pi^-$ & $K^- \To \pi^- \pi^- \pi^+$ \\
\hline
BR & \multicolumn{2}{c|}{$(5.576 \pm 0.031)\%$} \\
$g$ & $(-0.2154 \pm 0.0035)$ & $(-0.217 \pm 0.007)$ \\
$h$ & $(0.012 \pm 0.08)$ & $(0.010 \pm 0.06)$ \\
$k$ & $(-0.0101 \pm 0.0034)$ & $(-0.0084 \pm 0.0019)$ \\
\hline
& $K^+ \To \pi^0 \pi^0 \pi^+$ & $K^- \To \pi^0 \pi^0 \pi^-$ \\
\hline
BR & \multicolumn{2}{c|}{$(1.73 \pm 0.04)\%$} \\
$g$ & $(0.685 \pm 0.033)$ & $(0.642 \pm 0.057)$ \\
$h$ & $(0.066 \pm 0.024)$ & $(0.064 \pm 0.040)$ \\
$k$ & $(0.0197 \pm 0.0054)$ & $(0.006 \pm 0.004)$ \\
\hline
\end{tabular}
\caption{Experimental values of \mbox{$K^\pm \To 3\pi$} branching ratios and 
decay parameters \cite{PDG_2002} \cite{Protvino}; see also \cite{HyperCP_3pi}.}
\label{tab:3pi_charged}
\end{center}
\end{table}

The phenomenological description of these decays was discussed for the case of 
neutral kaons; the experimental values for branching ratios and Dalitz plot 
slopes are summarised in table \ref{tab:3pi_charged}.

\begin{table}
\begin{center}
\begin{tabular}{|c|c|l|}
\hline
\multicolumn{3}{|c|}{$K^\pm \To \pi^\pm \pi^0 \gamma$} \\
\hline
$A_\Gamma$ & $(0.9 \pm 1.3)\%$ & Direct, 
$\sim 4 \cdot 10^4$ evts. (Argonne 1969) \\
\hline
\multicolumn{3}{|c|}{$K^\pm \To \pi^\pm \pi^\pm \pi^\mp$} \\
\hline
$A_\Gamma$ & $(0.07 \pm 0.12)\%$ & Direct, 
$3.2 \cdot 10^6$ evts. (BNL 1970) \\
$A_g$ & $(-7.0 \pm 5.3)\cdot 10^{-3}$ & Direct,
$3.2 \cdot 10^6$ evts. (1970) (see also \cite{HyperCP_3pi}) \\
$A_h$ & $(0.1 \pm 4.4)$ & \\
$A_k$ & $(0.09 \pm 0.20)$ & \\
\hline
\multicolumn{3}{|c|}{$K^\pm \To \pi^0 \pi^0 \pi^\pm$} \\
\hline
$A_\Gamma$ & $(0.0 \pm 0.6)\%$ & Direct, 
$\sim 1.6 \cdot 10^4$ evts. (Argonne 1969) \\
$A_g$ & $(0.032 \pm 0.050)$ & (see also \cite{Denisov}) \\
$A_h$ & $(0.02 \pm 0.36)$ & \\
$A_k$ & $(0.53 \pm 0.26)$ & \\
\hline
\multicolumn{3}{|c|}{$K^\pm \To \mu^\pm \nu(\overline{\nu})$} \\
\hline
$A_\Gamma$ & $(-0.54 \pm 0.41)\%$ & Direct measurement, 
(BNL 1967) \\
\hline
\multicolumn{3}{|c|}{$K^\pm \To \pi^\pm \pi^0$} \\
\hline
$A_\Gamma$ & $(0.8 \pm 1.2)\%$ & Direct measurement,
$4 \cdot 10^3$ evts. (BNL 1973) \\
\hline
\multicolumn{3}{|c|}{$K^\pm \To \pi^\pm \mu^+ \mu^-$} \\
\hline
$A_\Gamma$ & $(-0.02 \pm 0.12)$ & Direct measurement,
$\sim 10^2$ evts. \cite{HyperCP_pimumu} \\
\hline
\end{tabular}
\caption{Experimental data on direct $CP$-violating asymmetries in $K^\pm$ 
decays, from \cite{PDG_2002} unless otherwise noted.}
\label{tab:3pi_CP}
\end{center}
\end{table}

The rate and slope asymmetries are defined as
\Ea{
  & A_\Gamma^{(f)} \equiv \frac{\Gamma(K^+ \To f)-\Gamma(K^- \To \overline{f})}
    {\Gamma(K^+ \To f)+\Gamma(K^- \To \overline{f})} \\
  & \quad A_g^{(f)} \equiv \frac{g(K^+ \To f)-g(K^- \To \overline{f})}
    {g(K^+ \To f)+g(K^- \To \overline{f})} 
}
and similarly for the other slope parameters.
Due to the limited phase space available for the decays, quadratic slopes are 
generally smaller than the linear ones; as a consequence partial rate 
asymmetries, which do not get any contribution from the integral of the linear 
terms over the whole Dalitz plot, tend to be suppressed with respect to slope 
asymmetries. Standard Model predictions for such asymmetries are generally at
the $10^{-4}$ level or below, while experimental limits are at least an order
of magnitude higher as shown in table \ref{tab:3pi_CP}.

Several experiments in the '70s measured partial decay rate asymmetries for
charged kaons, using either absolute $K^\pm$ flux normalisation obtained with 
differential Cerenkov counters, or exploiting other sets of inclusive decays 
as normalisation.

More recently, the HyperCP (E871) experiment at FNAL, dedicated to the study 
of $CP$ violation asymmetries in hyperon decays, collected in 1997 and 1999 
a large sample of $\pi^\pm \pi^+ \pi^-$ decays of charged kaons ($\approx 3.9 
\cdot 10^8$ $K^+$ and $\approx 1.6 \cdot 10^8$ $K^-$).
A small fraction ($\sim$ 10\%) of these decays has been analyzed to measure 
the $A_g$ slope asymmetry, resulting in a preliminary value \cite{HyperCP_3pi} 
\Dm{
  A_g^{(\pi^\pm \pi^+ \pi^-)} = (2.2 \pm 1.5 \pm 3.7) \cdot 10^{-3}
}
the first error being statistical and the second systematic (in the 
denominator the value $2g$ from \cite{PDG_2002} was used).
For this preliminary study the dominant systematic effects were induced by the 
knowledge of the magnetic fields, the efficiency differences of parts of the 
detector for charge-conjugate states, and secondary beam effects. 
The difference of $K^+$ and $K^-$ momentum spectra, as well as the different 
interactions of $\pi^+$ and $\pi^-$ in the spectrometer, were also found to be 
sources of spurious asymmetries.

The ISTRA+ experiment at Protvino collected both \mbox{$K^+ \To \pi^+ \pi^0 
\pi^0$} and \mbox{$K^- \To \pi^- \pi^0 \pi^0$} decays, which are being 
analyzed.
From a sample of $\sim 5 \cdot 10^5$ $K^\pm$ decays (50\% of the available 
data sample) a preliminary result was reported \cite{Denisov}:
\Dm{
  A_g^{(\pi^\pm \pi^0 \pi^0)} = (-0.3 \pm 2.5) \cdot 10^{-3}
}
where the quoted error is only the statistical one.

Two other experiments have as a main goal the measurement of slope asymmetries 
in charged kaon decays.
The NA48/2 experiment \cite{NA48/2} uses essentially the NA48 detector on a 
new beam line, with two simultaneous and collinear unseparated charged meson 
beams with narrow momentum spectrum (60 GeV/$c$ $\pm$ 5\%); a 400 GeV/$c$ 
primary intensity of $1 \cdot 10^{12}$  protons per pulse provides $\sim 5 
\cdot 10^6$ simultaneous $K^\pm$ entering the fiducial decay volume every 
16.8 s.
Even as the fluxes and compositions of the two beams are different, the 
concurrent detection of $\pi^\pm \pi^+ \pi^-$ decays in the same detector, 
coupled with the periodic reversal of the spectrometer magnetic field, will 
allow a good cancellations of systematic effects in the measurement. 
A sample in excess of $10^9$ $\pi^\pm \pi^+ \pi^-$ decays is expected in a 120 
day run, leading to a statistical sensitivity of the order $2 \cdot 10^{-4}$ 
on $A_g$, with systematic effects kept under control at the same level. The 
experiment starts taking data in 2003, and will also collect significant 
samples of $\pi^\pm \pi^0 \pi^0$ decays, for which the slope asymmetry can 
also be measured with similar sensitivity and rather different systematics.

The OKA experiment \cite{OKA} will exploit a new Protvino U-70 PS 
RF-separated beam (based on CERN-Karlsruhe separators used at CERN in the 
70's). A 70 GeV/$c$ primary protons intensity of $1 \cdot 10^{13}$ per pulse 
will provide either $4 \cdot 10^6$ $K^+$ or $1.3 \cdot 10^6$ $K^-$ of 12.5 or 
18 GeV/$c$ momentum entering the fiducial region every 9 s.
An electromagnetic calorimeter based on lead glass and PWO crystals, and a
beam spectrometer made of proportional chambers and drift tubes in a large 
aperture magnet, providing a 3 T$\cdot$m field integral, will be the main 
elements of the detector, derived from the SPHINX, ISTRA+ and GAMS setup.
Although no simultaneous $K^\pm$ beams will be available, the periodic change 
of polarities of all the beam-line elements will help in controlling many 
systematics.
The beam line is under construction and the first beam is foreseen for 2004; 
$\sim 4 \cdot 10^{11}$ charged $K$ decays are expected in 3 months of data 
taking. The estimated statistical error on the measurement of the 
$CP$-violating slope asymmetry in $K^\pm \To \pi^+ \pi^+ \pi^-$ decays is 
$\sim 1 \cdot 10^{-4}$ in a 3 months run.

In the radiative decay \mbox{$K^\pm \To \pi^\pm \pi^0 \gamma$} ($BR \simeq$  
\mbox{$2.8 \cdot 10^{-4}$}), the inner bremsstrahlung contribution 
is suppressed by the $\Delta I=1/2$ rule, and the interference with an 
electric dipole direct emission term could give rise to a rate asymmetry, or 
to asymmetries in the Dalitz plot or in the photon spectrum, indications of 
direct $CP$ violation. 
Predictions for such asymmetries are however at most $10^{-4}$ in the SM 
\cite{Isidori}, and the current experimental limit on the rate asymmetry
is two orders of magnitude higher.

Similar estimates hold for the rate asymmetry of \mbox{$K^\pm \To \pi^\pm 
\gamma \gamma$}, which has an even smaller branching ratio ($BR \simeq 1.1   
\cdot 10^{-6}$) and is experimentally obscured by a $\pi^+ \pi^0$ background 
with a rate $2 \cdot 10^4$ times larger.

Rate asymmetries in the \mbox{$K^\pm \To \pi^\pm e^+e^-$} decays (with $BR  
\simeq 3 \cdot 10^{-7}$) are also predicted at the $O(10^{-5})$ level in the 
Standard Model \cite{Ecker}, and $K^\pm$ asymmetries in the $e^+ e^-$ 
invariant mass distributions in these decays, which could reach $\sim 
10^{-4}$, are also out of reach of forthcoming experiments.
For the \mbox{$K^\pm \To \pi^\pm \mu^+ \mu^-$} decay the measurement of 
asymmetries involving muon polarization could also be conceived (see following 
section), but the tiny branching ratio makes this approach unattractive.
None of the above measurements looks therefore experimentally accessible at 
present.

\subsection{$T$-odd correlation experiments}

In semi-leptonic \mbox{$K \To \pi l \nu$} ($K_{l3}$) decays, with large 
branching ratios, the V-A structure of the weak current only allows two form 
factors\footnote{No experimental evidence of scalar or tensor form factors was 
detected so far.}, which are relatively real if $T$ symmetry holds. $K_{e3}$ 
experiments are not sensitive to one of the form factors, due to the smallness 
of the electron mass with respect to the kaon mass, so that measurable $CP$ 
violation effects can only be expected in $K_{\mu3}$ decays.

The lepton polarization in the direction transverse to the $(\pi,l)$ decay 
plane is 
\Dm{
  P_T(l) \propto \langle \Vec{S}_l \cdot \Vec{p}_l \times \Vec{p}_\pi \rangle
}
where $\Vec{S}_l$ is the lepton polarization vector and $\Vec{p}_l, 
\Vec{p}_\pi$ the lepton and pion three-momenta respectively.
The above quantity is odd under the so-called ``naive'' time reversal 
(inversion of spin and momenta), and violates time reversal invariance if 
final-state interactions are neglected, being proportional to the imaginary 
part of the form factors ratio $\Imm(\xi)$. 
In \mbox{$K^\pm \To \pi^0 l^\pm \nu$} decays, such final-state interactions 
are expected to be very small in the Standard Model ($P_T < 10^{-5}$), and 
a significant non-zero transverse polarization measurement in this decay would 
be both a true signal of time reversal violation (and therefore of direct $CP$ 
violation if $CPT$ symmetry is valid), and indication for new physics.
The transverse polarization induced by Standard Model final-state interactions 
in \mbox{$\kl \To \pi^\mp l^\pm \nu(\overline{\nu})$} decays is larger 
$O(10^{-3})$, at the level of accuracy reached by experiments in the '80s.

In practice only $\mu^+$ polarization can be measured reliably, so that 
experiments have been performed both in \mbox{$\kl \To \pi^- \mu^+ \nu$} and 
\mbox{$K^+ \To \pi^0 \mu^+ \nu$} decays.

The more accurate experiments use detectors with cylindrical symmetry, in 
which kaons decay in flight \cite{BNL_kmu3} or (for $K^+$) after stopping in a 
target \cite{KEK_kmu3}. In both cases events in which the detector (and beam) 
axis lies in the ($\Vec{p}_\mu, \Vec{p}_\pi$) decay plane are selected.
Toroidal magnetic fields guide decay muons to stop into polarimeters, without 
affecting the transverse component of their polarization; the muon 
polarization is correlated to the positron emission direction in its decay, 
and any component transverse to the decay plane induces an asymmetry in the 
measurement of the left and right (counterclockwise and clockwise with 
respect to the beam direction) positron counters. In decay in-flight 
experiments, muons were made to precess in a magnetic field parallel to the 
beam line before their decay was detected, thus eliminating systematic 
differences in detector efficiencies at the price of a lower analyzing power; 
left-right asymmetries in the polarimeters could be controlled by the accurate 
reversal of the axial precession magnetic field, which determined one of the 
dominant systematic uncertainties.
Also, events with different orientation of the decay plane are expected to 
exhibit opposite asymmetries, thus providing a powerful systematic check.

The most recent experiment, E246 at KEK \cite{KEK_kmu3}, analyzed $\sim 8.3
\cdot 10^6$ $\pi^0 \mu^+ \nu$ decays of 660 MeV/$c$ $K^+$ stopped in a 
scintillating fibre target, extracting the asymmetry from a double ratio of 
clockwise and anticlockwise event rates for events with opposite decay plane 
orientations, in order to reduce the sensitivity to systematic errors 
which can potentially induce spurious asymmetries.
Only experimental effects introducing a net screw asymmetry around the beam
direction can mimic $T$ violation: the largest systematic errors in this 
experiment arise from the asymmetry in the magnetic field shapes and the
misalignment of the polarimeter.

Experiments have given null polarization results so far \cite{Morse} 
\cite{E246_new}
\Ea{
  P_T(\mu) = (1.7 \pm 5.6) \cdot 10^{-3} \quad && (\kl \To \pi^- \mu^+ \nu) \\
  P_T(\mu) = (-1.12 \pm 2.34) \cdot 10^{-3} \quad && (K^+ \To \pi^0 \mu^+ \nu) 
}
corresponding to \cite{PDG_2002} \cite{E246_new}
\Ea{
  & \Imm(\xi)_{\kl} = -0.007 \pm 0.026 \\
  & \Imm(\xi)_{K^+} = -0.0028 \pm 0.0075 
}
Improved experiments have been proposed, which could reach $ \sim 1 \cdot 
10^{-4}$ error on the transverse polarization, a factor $\sim 20$ better than
achieved by present experiments.

Measurements of transverse muon polarization in other decays have been 
considered in the literature (see \cite{Diwan_pol} for a thorough discussion).
$K^+ \To \mu^+ \nu \gamma$ decays ($BR \simeq 5.5 \cdot 10^{-3}$) can be 
studied in the experiments designed to collect $K^+ \To \pi^0 \mu^+ \nu$, with 
minor changes to the setup; transverse polarizations at the $10^{-4}$ level 
are expected in the Standard Model due to final-state interactions, and the 
KEK E246 experiment recently reported the first measurement of such 
quantity \cite{E246_munug}:
\Dm{
  P_T(\mu) = (-0.64 \pm 1.85) \cdot 10^{-2} \quad (K^+ \To \mu^+ \nu \gamma)
}

For the $K^+ \To \pi^+ \mu^+ \mu^-$ decay ($BR \sim 8 \cdot 10^{-8}$) the 
contribution to the transverse muon polarization due to Standard Model final 
state interactions can be $\sim 10^{-3}$, while spin-correlations involving 
both muons' polarizations are cleaner signals of $T$ violation but prohibitive 
from an experimental point of view, because of the tiny branching ratio and 
the large background from $K^+ \To \pi^+ \pi^+ \pi^-$.

$T$-odd correlations (in the sense explained above) involving only momenta can 
be studied in 4-body decays with distinguishable particles in the final state: 
an example are the radiative semi-leptonic decays \mbox{$K \To \pi l \nu 
\gamma$}, which have branching ratios in the $10^{-3} \div 10^{-4}$ range for 
$\kl$ decays and in the $10^{-4} \div 10^{-5}$ range for $K^\pm$ decays. 
For $\kl$, electromagnetic final-state interactions could be expected to 
induce a fake signal, which should be largely absent for $K^\pm$ decays. 
Unfortunately any $CP$ violating effect in such radiative decays is suppressed 
by the dominance of the inner bremsstrahlung process; the necessary direct 
emission component, not yet observed in these decays, is expected at the level 
of a few percent of the former, and generally larger for the muonic decays.
A measurement of the $T$-odd asymmetry in the $\Vec{p}_\pi \cdot 
\Vec{p}_e \times \Vec{p}_\gamma$ distribution for \mbox{$K^- \To \pi^0 e^- \nu 
\gamma$} decays by the ISTRA experiment \cite{Bolotov} gave a null result: 
$0.03 \pm 0.08$ with 192 events. 
The OKA experiment at Protvino \cite{OKA} plans to measure the $T$-odd
asymmetry in the $\Vec{p}_\pi \cdot \Vec{p}_\mu \times \Vec{p}_\gamma$ 
distribution for the (so far unobserved) \mbox{$K^+ \To \pi^0 \mu^+ \nu 
\gamma$} decays (the predicted branching ratio is $\sim 2 \cdot 10^{-5}$ in 
the SM \cite{Bijnens}), on an expected sample of $\sim 10^5$ events.

\begin{table}
\begin{center}
\begin{tabular}{|c|c|l|}
\hline
Decay & BR & Notes \\
\hline 
$\kl \To \pi^\pm e^\mp \nu \gamma$ & $(3.53 \pm 0.06) \cdot 10^{-3}$ & 
$15 \cdot 10^3$ evts. (KTeV 2001) \\
$\kl \To \pi^\pm \mu^\mp \nu \gamma$ & $(5.7 \pm 0.7) \cdot 10^{-4}$ & 
$2.5 \cdot 10^2$ evts. (NA48 1998) \\
\hline
$K^\pm \To \pi^0 e^\pm \nu \gamma$ & $(2.65 \pm 0.20) \cdot 10^{-4}$ & 
192 evts. (ISTRA 1986) \\
& $< 5.3 \cdot 10^{-5}$ (DE) & \\
$K^\pm \To \pi^0 \mu^\pm \nu \gamma$ & $< 6.1 \cdot 10^{-5}$ (90\% CL) & 
(Argonne 1973) \\
\hline
\end{tabular}
\caption{Experimental data \cite{PDG_2002} for the radiative semi-leptonic
decays of kaons.}
\label{tab:krad}
\end{center}
\end{table}

Table \ref{tab:krad} summarizes the experimental information available on
these decays.

$CP$ violation measurements in other decays of charged kaons have been 
considered in the literature, such as the rare process \mbox{$K^\pm \To 
\mu^\pm \nu e^+ e^-$} ($BR \simeq 1.3 \cdot 10^{-7}$), in which $T$-odd 
correlations related to muon polarization could be probed.
Such measurements are however very difficult in practice, due either to large 
backgrounds or tiny branching ratios, which make polarization experiments 
prohibitive\footnote{The BNL E865 experiment \cite{E865_4l} recently collected 
$2.2 \cdot 10^3$ \mbox{$K^\pm \To \mu^\pm \nu e^+ e^-$} decays, a 150-fold 
increase on the previous world sample, which could allow some $CP$ violation 
studies to be performed.}

As already mentioned, independently from final-state interaction effects, an 
unambiguous direct $CP$ violation signal could be obtained by measuring a 
difference in the magnitude of any $T$-odd correlation for $K^+$ and $K^-$.

\section{Other meson systems}

\subsection{Phenomenology}

Heavy flavoured meson systems, \emph{i.e.} not self-conjugate mesons 
containing quarks heavier than the strange one, are the subject of very active 
investigations.
In particular $B$ meson decays, in which the effects of the three 
quark generations can be present at tree level, are a promising arena for
the search of $CP$ violation effects, since the asymmetries are less 
suppressed by the smallness of quark family mixing. 
The focus of the studies are the mixing-induced $CP$-violating decay rate 
asymmetries for neutral mesons, which in some cases can be expressed in a 
reliable way in terms of the elementary phases in the theory.
Unfortunately, the measurements of such asymmetries involve specific final 
states, whose branching ratios for such high mass systems are usually very 
small, requiring large statistics to be experimentally detected, this being 
the reason for the building of the so-called ``$B$ factories''.
Also in heavy meson systems direct $CP$ violation effects are difficult to 
predict theoretically from first principles, due to the non-perturbative 
physics of hadronization and final-state interactions.

The quantity $x \equiv \Delta m/\overline{\Gamma}$ (where $\overline{\Gamma}$
is the average decay width of the two mass eigenstates) determines whether the
flavour oscillations are observable; it can have rather different values for 
the various neutral meson systems \cite{PDG_2002}: 
\Ea{
  & x_K = 0.948 \\
  & x_D < 2.9 \cdot 10^{-2} \quad (95\% \, CL) \\
  & x_{B_d} = 0.755 \\
  & x_{B_s} > 19 \quad (95\% \, CL) 
}

When compared to kaons, neutral $B$ mesons have a much larger set of decay 
modes available, due to their larger mass, but they allow a more limited 
variety of experimental approaches for their study: the reason is that for 
them 
\Dm{
  \left| \frac{\Delta \Gamma}{\Delta m} \right| \sim 
  \left| \frac{\Gamma_{12}}{M_{12}} \right| \ll 1
\label{eq:smallmix}
}
This can be understood in a simple way as due to the fact that the width 
difference originates from the physically accessible decay modes common to 
$M$ and $\overline{M}$, which are only a very small fraction of the total 
available to the heavy meson.

While for lighter mesons $\Delta F=2$ transitions cannot be computed in a 
reliable way, for heavy ones such as the neutral $B$, with masses distant 
from the region of hadronic resonances, the mass and lifetime differences, 
linked to the off-diagonal elements of the effective Hamiltonian, are more 
tractable from a theoretical point of view, and the relation 
(\ref{eq:smallmix}) can be shown to be valid in a rather model-independent way.

Such relation implies that $|y| \equiv |\Delta \Gamma/\overline{\Gamma}| 
\sim |x \Gamma_{12}/M_{12}|$ is also very small for $B_d$ mesons, and could 
reach 1-10\% for $B_s$ mesons depending on the value of $x_s$.
Experimentally \cite{PDG_2002}, for heavy mesons:
\Ea{
  & (|\Delta \Gamma|/\overline{\Gamma})_D = (0.003 \pm 0.022) \\
  & (|\Delta \Gamma|/\overline{\Gamma})_{B_d} = (0.008 \pm 0.041) \\
  & (|\Delta \Gamma|/\overline{\Gamma})_{B_s} < 0.3
}

The two physical states have therefore very similar decay widths and it is
not possible to perform experiments with just one of them.
For this reason the formalism used for heavy mesons does not use the 
amplitudes for physical states and their ratios $\eta_f$, but only the 
amplitudes for flavour eigenstates.

The relation among $|\Gamma_{12}|$ and $|M_{12}|$ also implies that the
measure of $CP$ violation in the mixing
\Dm{
  1- \left| \frac{1-\beps}{1+\beps} \right| = 
  1- \left| \frac{q}{p} \right| \simeq 
  \frac{1}{2} \left| \frac{\Gamma_{12}}{M_{12}} \right|
  \sin \left[ \arg \left( \frac{\Gamma_{12}}{M_{12}} \right) \right]
}
has a sensitivity to the $CP$-violating phase difference of the off-diagonal 
terms $M_{12}$ and $\Gamma_{12}$ which is suppressed by the smallness of 
$|\Gamma_{12}/M_{12}| \sim |\Delta \Gamma/\Delta m| \ll 1$.

$CP$-violating decay rate asymmetries are defined in general as\footnote{Note 
the conventional sign difference with respect to previous definitions, 
originating from the fact that the flavour of the decaying meson is usually 
determined by the \emph{opposite} flavour of an associated particle produced 
in the same reaction.}
\Dm{
  A_{CP}^{(f)}(t) \equiv 
    \frac{\Gamma(\overline{M} \To \overline{f};t) - \Gamma(M \To f;t)}
         {\Gamma(\overline{M} \To \overline{f};t) + \Gamma(M \To f;t)}
}

To measure $CP$ violation in $M--\overline{M}$ mixing, one can measure the 
meson flavour at a given time and exploit a decay mode (or set of modes) which 
cannot support $CP$ violation in the decay amplitudes.

The time-dependent asymmetry of total (inclusive) decay rates for 
flavour-tagged mesons gives such a measure of $CP$ violation in the mixing:
\Ea{ \Ds 
  & A_{CP}^{\mathrm{(incl)}}(t) = 
    \frac{\Gamma(\overline{M}(t) \To \mathrm{all})-
    \Gamma(M(t) \To \mathrm{all})}
    {\Gamma(\overline{M}(t) \To \mathrm{all})+
    \Gamma(M(t) \To \mathrm{all})} = \nonumber \\
  & \frac{2\Rea(\beps)}{1+|\beps|^2} \left[
    \frac{-\cosh(\Delta \Gamma t/2)+ y\sinh(\Delta \Gamma t/2)+ 
    \cos(\Delta m t)+ x\sin(\Delta m t)}
    {\cosh(\Delta \Gamma t/2)- y\sinh(\Delta \Gamma t/2)} \right]
\label{eq:Ainclusive}
}
where $M(t),\overline{M}(t)$ indicate states which were flavour-tagged as $M,
\overline{M}$ at $t=0$; expression (\ref{eq:Ainclusive}) is valid at first 
order in $2\Rea(\beps)/(1+|\beps|^2)$.
The time-integrated version vanishes by $CPT$ symmetry, and in the limit $y 
\simeq 0$ (which is expected to be a good approximation for $B$ mesons): 
\Dm{
  A_{CP}^{\mathrm{(incl)}}(t) \simeq \frac{2\Rea(\beps)}{1+|\beps|^2} 
  [ x \sin(\Delta m t) - 2 \sin^2(\Delta m t/2)]
}

The partial decay rate asymmetry for ``wrong'' (mixed) flavour-specific decays 
of flavour-tagged mesons, \emph{i.e.}
\Dm{
  A_{CP}^{(M)}(t) = 
    \frac{\Gamma(\overline{M}(t) \To f)-\Gamma(M(t) \To \overline{f})} 
    {\Gamma(\overline{M}(t) \To f)+\Gamma(M(t) \To \overline{f})} = 
    \frac{|p/q|^2|A_f|^2-|q/p|^2|\overline{A}_{\overline{f}}|^2}
    {|p/q|^2|A_f|^2+|q/p|^2|\overline{A}_{\overline{f}}|^2}
\label{eq:Amix}
}
where $M,\overline{M}$ cannot decay to $f,\overline{f}$ respectively, is seen 
to be independent of time. 
While this asymmetry cannot separate $CP$ violation in the mixing and in 
the decay, it can happen that, due to the absence of final-state interactions, 
$CPT$ symmetry itself imposes $|A_f|=|\overline{A}_{\overline{f}}|$ so that 
the latter type of $CP$ violation is not possible: this is the case for 
semi-leptonic decays in the Standard Model.
The above asymmetry (\ref{eq:Amix}) is then a pure measurement of $CP$ 
violation in the mixing:
\Dm{
  A_{CP}^{(M)} = 
  \frac{1-|q/p|^4}{1+|q/p|^4} = \frac{4\Rea(\beps)}{1+|\beps|^2}
  \quad \quad (|A_f| = |\overline{A}_{\overline{f}}|)
}

The asymmetry for ``right'' (unmixed) flavour-specific decays gives instead
\Dm{
  A_{CP}^{(U)}(t) = 
    \frac{\Gamma(\overline{M}(t) \To \overline{f})-\Gamma(M(t) \To f)} 
    {\Gamma(\overline{M}(t) \To \overline{f})+\Gamma(M(t) \To f)} =
    \frac{|\overline{A}_{\overline{f}}|^2 - |A_f|^2}
    {|\overline{A}_{\overline{f}}|^2 + |A_f|^2}
}
which however in many cases is forced to be null by $CPT$ symmetry.

In the case of neutral mesons, apart from time-integrated asymmetries, the 
analysis of mixing-induced $CP$ violation can also give information on direct 
$CP$ violation.
The time-dependent asymmetry for decays to a common $CP$ eigenstate $f$, for 
mesons tagged as $\overline{M}$ or $M$ (at time $t=0$), can have a non-zero 
value when any kind of $CP$ violation is present.
Neglecting terms of second order in the $CP$-violating parameters, the full
expression for such asymmetry is
\Dm{
  A_{CP}^{(f)}(t) \simeq
    \frac{\mathcal{A}_{CP}^{\mathrm{(mix)}} \cosh(\Delta \Gamma t) + 
    \mathcal{A}_{CP}^{\mathrm{(m/d)}} \cos(\Delta m t) + 
    \mathcal{A}_{CP}^{\mathrm{(int)}} \sin(\Delta m t)}
    {\cosh(\Delta \Gamma t/2) -
    \mathcal{A}_{\Delta \Gamma} \sinh(\Delta \Gamma t/2)}
}
where $\mathcal{A}_{CP}^{\mathrm{(mix)}}$ expresses $CP$ violation in the 
mixing, $\mathcal{A}_{CP}^{\mathrm{(m/d)}}$ $CP$ violation in either the 
mixing or in the decay amplitudes, $\mathcal{A}_{CP}^{\mathrm{(int)}}$ $CP$ 
violation in the interference of mixing and decay, and 
$\mathcal{A}_{\Delta \Gamma}$ is induced by the decay width difference. 

The expression for the pure mixing term is 
\Dm{
  \mathcal{A}_{CP}^{\mathrm{(mix)}} = 
  \frac{2\Rea(\beps)}{1+|\beps|^2} = 
  \frac{1-|q/p|^2}{1+|q/p|^2}
}
If $CP$ violation in the mixing is small ($|q/p| \simeq 1$), at least when 
compared to the mixing-induced one, this term can be neglected; this is proved 
to be a good approximation for $B_d$ mesons \cite{BABAR_CPT}, so that when
considering this system one often puts $\mathcal{A}_{CP}^{\mathrm{(mix)}} = 0$.

The expressions for the other terms appearing in the asymmetry are
\Ea{
  & \mathcal{A}_{CP}^{\mathrm{(m/d)}} = 
    \frac{|\lambda_f|^2-1}{|\lambda_f|^2+1} \\
  & \mathcal{A}_{CP}^{\mathrm{(int)}} = 
    \frac{2\Imm(\lambda_f)}{|\lambda_f|^2+1} \\
  & \mathcal{A}_{\Delta \Gamma} = \frac{2\Rea(\lambda_f)}{|\lambda_f|^2+1} 
}

For a final state $f$ it is convenient to introduce the complex parameter 
\Dm{
  \lambda_f \equiv \frac{q}{p} \frac{\overline{A}_f}{A_f} = 
  \left( \frac{1-\beps}{1+\beps} \right) \frac{\overline{A}_f}{A_f} = 
  \eta_{CP}(f) \frac{q}{p} \frac{\overline{A}_{\overline{f}}}{A_f}
}
(independent from the choice of phase convention for the $M,\overline{M}$ 
states) where $\beps$ is the $CP$ impurity parameter appearing in the 
expression of physical states in terms of flavour eigenstates, and $A_f \equiv 
A(M \To f)$, $\overline{A}_f \equiv A(\overline{M} \To f)$. The last equality 
refers to a $CP$ eigenstate with eigenvalue $\eta_{CP}(f)$, and in this case 
$\lambda_f$ is a measure of $CP$ violation; one can have a non-zero 
$CP$-violating decay asymmetry if the phase of $\lambda_f$ is different from 
$0$ or $\pi$, or if its modulus is different from 1.
The first case can arise if the ``mixing'' phase 
\Dm{
  \phi_M \equiv \arg \left( \frac{1-\beps}{1+\beps} \right) 
}
is different from the ``decay'' phase 
\Dm{
  \phi_D^{(f)} \equiv \arg \left( \frac{\overline{A}_f}{A_f} \right)
}
so that $\mathcal{A}_{CP}^{\mathrm{(int)}} \neq 0$.
This is mixing-induced $CP$ violation (which can be both direct and indirect), 
as can be seen from the fact that the corresponding term in the decay 
asymmetry only builds up in time starting from zero at $t=0$, as 
$M--\overline{M}$ flavour mixing gets in effect. 

The second case can arise either because $|(1-\beps)/(1+\beps)| \neq 1$, 
\emph{i.e.} (indirect) $CP$ violation in the mixing due to the $CP$ impurity 
of the mass eigenstates (independent of the decay channel $f$), or because 
$|\overline{A}_f/A_f| \neq 1$, \emph{i.e.} (direct) $CP$ violation in the 
decay due to the interference of competing amplitudes with different phases 
(possibly specific to the decay mode $f$); in both cases
$\mathcal{A}_{CP}^{\mathrm{(m/d)}} \neq 0$. 
Clearly, both effects can also be present at the same time (while the two 
ratios entering the definition of $\lambda_f$ have unphysical phases which 
depend on the convention chosen for the $M,\overline{M}$ states, their moduli 
have a physical meaning), and they both give rise to a decay asymmetry which 
is maximal at $t=0$, and gets washed out with time, due to flavour mixing.

If the width difference is neglected ($y \simeq 0$, see also \cite{BABAR_CPT}) 
one can set $\mathcal{A}_{\Delta \Gamma}=0$, and this assumption will be made 
in what follows.
The expression for the asymmetry in absence of $CP$ violation in the mixing 
then reduces to a sum of two terms, one proportional to the cosine and one to 
the sine of $\Delta m \, t$, whose coefficients are bounded to lie within the 
$\mathcal{A}_{CP}^{\mathrm{(m/d)}2} + \mathcal{A}_{CP}^{\mathrm{(int)}2} 
\le 1$ region.
When integrating over a time which is long on the time scale of flavour mixing 
(\mbox{$t \gg 1/\Delta m$}), both oscillating terms get averaged to zero; this 
is experimentally the case when the flavour mixing period is not much larger 
than the lifetime of the meson.

If $CP$ violation in the mixing is small (as is the case for $\bz,\dz$ mesons 
in the Standard Model) one can write 
\Dm{
  \frac{q}{p} = \frac{1-\beps}{1+\beps} = e^{i\phi_M}
}
as a pure phase: $\beps$ is then purely imaginary, and as such it can be 
redefined to zero by a flavour rotation.

In presence of a single elementary decay amplitude (or of several amplitudes 
with the same phase), $CP$ violation in the decay cannot be present and the 
ratio of $\overline{M},M$ amplitudes is also a pure phase
\Dm{
  \frac{\overline{A}_f}{A_f} = \eta_{CP}(f) e^{i\phi_D^{(f)}}
}
When these two conditions are satisfied the cosine term in $A_{CP}^{(f)}(t)$ 
is absent, and the decay asymmetry directly measures the (sine of the) sum of 
the phases of the $M \leftrightarrow \overline{M}$ mixing process and the 
\mbox{$M \To f$} decay process:
\Dm{
  A_{CP}^{(f)}(t) = \eta_{CP}^{(f)} \sin[\phi_M + \phi_D^{(f)}] 
  \sin(\Delta m \, t)
}

When considering decays to a $CP$ self-conjugate set of quarks, the final state
is in general a superposition of states with opposite $CP$ eigenvalues, 
contributing opposite mixing-induced asymmetries which partially cancel; this
is only possible when there is more than one particle with non-zero spin in 
the final state, or more than two particles, otherwise angular momentum 
conservation forces the $CP$-parity of the final state to a single value.

It is seen that, in the limit in which $CP$ violation in the mixing is 
negligible, the appearance of a non-zero cosine term in the decay asymmetry is 
an indication of direct $CP$ violation, due to interfering amplitudes with 
different phases.
In this case the quantity $\lambda_f$ contains the strong phases of the two 
elementary amplitudes, and therefore it is much harder to relate to the 
dynamical parameters of the underlying decay; at the same time the 
coefficient of the sine term is no longer simply related to the phases of the 
weak amplitudes (a product of elements of the quark mixing matrix in the SM).
The largest direct $CP$ violation effects are expected for decays to which two
amplitudes of comparable magnitude contribute, \emph{i.e.} when the lowest 
order contribution is suppressed. 

Time-integrated decay asymmetries are measures of direct $CP$ violation for
neutral or charged mesons:
\Dm{
  A_{CP}^{(f)} = \int_0^\infty dt \, A_{CP}^{(f)}(t) = 
  \frac{1-y^2}{1+x^2} \frac{|\lambda_f|^2 -1 +2x \, \Imm(\lambda_f)}
  {|\lambda_f|^2 +1 -2y \, \Rea(\lambda_f)}
}
which are however suppressed if $x$ is much larger than 1 or $y^2$ is close 
to 1.

$CP$-violating decay asymmetries can also be formed with untagged decays
(see \emph{e.g.} \cite{Branco}), but their measurement cannot disentangle $CP$ 
violation in the mixing and in the decay amplitudes.
The untagged time-integrated asymmetry is
\Dm{
  A_{CP}^{(M)} = - \frac{\mathcal{A}_{CP}^{\mathrm{(mix)}} (x^2+y^2) -
    \mathcal{A}_{CP}^{\mathrm{(dir)}} [1 + x^2 -
    \mathcal{A}_{CP}^{\mathrm{(mix)}2} (1-y^2)]}
    {1+x^2 -\mathcal{A}_{CP}^{\mathrm{(mix)}2} (1-y^2) + 
    \mathcal{A}_{CP}^{\mathrm{(dir)}} \mathcal{A}_{CP}^{\mathrm{(mix)}} 
    (x^2+y^2)}
}
where 
\Dm{
  \mathcal{A}_{CP}^{\mathrm{(dir)}} = \frac{|A_f|^2 - |\overline{A}_f|^2}
  {|A_f|^2 + |\overline{A}_f|^2}
}
Direct $CP$ violation in heavy meson decays could also be studied without  
using any flavour tag, and therefore with a large statistical advantage, by  
analysing their decays to self-conjugate final states in which distinct, 
$CP$-conjugate resonances can be identified \cite{Gardner_untagged}: an 
example is \mbox{$B^0 \To \pi^+ \pi^- \pi^0$} in which the presence of both 
$\rho^+ \pi^-$ and $\rho^- \pi^+$ intermediate states can allow the 
manifestation of direct $CP$ violation, as Dalitz plot asymmetries, even in 
absence of any strong phases.

The study of decay asymmetries for inclusive modes has been proposed 
\cite{Inclusive}, not only because of the larger event yields, but also 
because by summing over all the states corresponding to a set of quantum 
numbers which are conserved by the strong interactions, the knowledge of the 
final-state interaction phases is not required to compute the asymmetries. 
For the same reason, though, such asymmetries do not probe direct $CP$ 
violation.

Considering decays of coherent $M \overline{M}$ meson pairs, the asymmetries 
are expressed as a function of $\Delta t$, the time difference between the 
decay of the meson and that of its companion used to determine its flavour 
(at that time); the formul\ae are the same as above, with $t$ replaced by 
$\Delta t$, which can also be negative. 
Mixing-induced $CP$ violation can be observed as an asymmetry of the $A_{CP}$ 
distributions with respect to $\Delta t=0$ for decays of mesons of either 
flavour. When integrating such distributions symmetrically over $\Delta t$ to 
get the total rate, mixing-induced $CP$ violation, proportional to 
$\sin(\Delta m \Delta t$, cannot give an observable effect, while the other 
two kinds of $CP$ violation can.

In experiments using $M \overline{M}$ pairs, $CP$ violation in the mixing is 
measured by comparing the yields of positive and negative same-sign di-lepton 
events originating from (flavour-specific) semi-leptonic decays (for which no 
direct $CP$ violation is present):
\Dm{
  A_{CP}^{(l)} = \frac{N(l^+l^+)-N(l^-l^-)}{N(l^+l^+)+N(l^-l^-)}
}

For antisymmetric $M \overline{M}$ pairs, such as the ones produced by the 
decay of a vector quarkonium resonance ($J^{PC}=1^{--}$), in the limit in 
which $CP$ violation in the mixing is negligible, the measurement of a decay 
of the pair to two $CP$ eigenstates with the same eigenvalue would be an 
indication of direct $CP$ violation, since in such case the physical meson 
states can be chosen as $CP$ eigenstates. Unfortunately, for heavy mesons the 
branching ratios to $CP$ eigenstates are usually very small, so that the 
requirement of having both mesons decaying to such modes represents a heavy
penalty which makes this approach unappealing.

The general ($CPT$-symmetric) expression \cite{Coerenti2} for the decay rate 
to two final states $f_1,f_2$ as a function of their time difference 
$\Delta t = t_1-t_2$, reduces, in the limit $\Delta \Gamma \simeq 0$ to
\Ea{ \Ds 
  & I(f_1,f_2;\Delta t) \propto
    \left( 1-e^{-|\Delta t|\overline{\Gamma}} \right) \quad \\
  & \quad \left[ \mathcal{A}_C \cos^2(\Delta m \Delta t/2) + 
    \mathcal{A}_S \sin^2(\Delta m \Delta t/2) +
    \mathcal{A}_I \sin(\Delta m \Delta t) \right]
}
where
\Ea{
  & \mathcal{A}_C = 4 \left| \frac{\lambda_1 -\lambda_2}
    {(1+\lambda_1)(1+\lambda_2)} \right|^2 \\
  & \mathcal{A}_S = 4 \left| \frac{1-\lambda_1 \lambda_2}
    {(1+\lambda_1)(1+\lambda_2)} \right|^2 \\
  & \mathcal{A}_I = 2 \Imm \left( \frac{(1-\lambda_1)(1-\lambda_2^*)}
    {(1+\lambda_1)(1+\lambda_2^*)} \right) = 
    4 \frac{(1-|\lambda_1|^2) \Imm(\lambda_2) - 
    (1-|\lambda_2|^2) \Imm(\lambda_1)}{|1+\lambda_1|^2 |1+\lambda_2|^2}
}
If $f_1=f_2$, or if $CP$ violation is only present in the mixing ($\lambda_1 = 
\lambda_2$), one has $\mathcal{A}_C = 0$ and $\mathcal{A}_I = 0$; if there is 
no $CP$ violation in the mixing nor in the decay ($|\lambda_{1,2}|=1$) then 
$\mathcal{A}_I=0$. In both cases the expression for $I(f_1,f_2;\Delta t)$ is
symmetric with respect to $\Delta t=0$, so that any $\Delta t$ asymmetry in 
such distribution 
\Dm{
  A_I(f_1,f_2;\Delta t) \equiv 
    \frac{I(f_1,f_2;\Delta t)-I(f_1,f_2;-\Delta t)}
    {I(f_1,f_2;\Delta t)+I(f_1,f_2;-\Delta t)} \simeq
    \frac{\mathcal{A}_I \sin(\Delta m \Delta t)}
    {\mathcal{A}_+ + \mathcal{A}_- \cos(\Delta m \Delta t)}
}
where 
\Ea{
  & \mathcal{A}_+ = 2 \left| \frac{1+\lambda_2}{1-\lambda_2} \right|^2 
    \left[ 1 + |\lambda_1|^2 + |\lambda_2|^2 + 
    |\lambda_1|^2 |\lambda_2|^2 - 4\Rea(\lambda_1) \Rea(\lambda_2) \right] \\
  & \mathcal{A}_- = 2 \left| \frac{1+\lambda_2}{1-\lambda_2} \right|^2 
    \left[ -1 + |\lambda_1|^2 + |\lambda_2|^2 - 
    |\lambda_1|^2 |\lambda_2|^2 - 4\Imm(\lambda_1) \Imm(\lambda_2) \right]
}
is non-zero only if both mixing-induced and one other kind of $CP$ violation 
are simultaneously present: for heavy meson systems, in which $CP$ violation 
in the mixing is negligible, this requires the presence of (direct) $CP$ 
violation in the decay.

Comparisons between the values of the parameters extracted from the study of 
time-dependent asymmetries in neutral meson decays to different $CP$ 
eigenstates can also provide evidence for direct $CP$ violation. 
Mixing-induced $CP$ asymmetries arise because of a difference between the 
decay and the mixing phases and, as mentioned, the choice of phase convention 
always allows the former to be shifted to zero, at the expense of the latter.
Since in general decay phases depends on the particular mode, when they are
non-zero one cannot find any phase convention in which \emph{all} the 
$CP$-violating phases only appear in the mixing amplitudes, thus giving 
evidence for direct $CP$ violation. 
By comparing the decay of flavour-tagged mesons to two different $CP$ 
eigenstates $f_1$ and $f_2$, if $CP$ violation would only be due to state 
mixing one would get, for the coefficients of the sine term in the 
time-dependent decay asymmetries:
\Dm{
  \mathcal{A}_{CP}^{\mathrm{(int)}}(f_1) = 
  \eta_{CP}(f_1) \eta_{CP}(f_2) \mathcal{A}_{CP}^{\mathrm{(int)}}(f_2) 
}
where $\eta_{CP}(f_{1,2})$ are the $CP$ eigenvalues of the two states.
Any deviation from the above relation is a signal of direct $CP$ violation,
which can be expressed as
\Dm{
  \eta_{CP}(f_1) \lambda_{f_1} \neq \eta_{CP}(f_2) \lambda_{f_2}
}
This is (as for $\epsilon'$ in the neutral $K$ system) a measure of direct 
$CP$ violation which can be non-zero also in presence of a single elementary 
decay amplitude for each process, and in absence of final-state interactions 
(\emph{i.e.} when $|\overline{A}_{f_1}/A_{f_1}|=1=
|\overline{A}_{f_2}/A_{f_2}|$), since one is comparing two different decays: 
the difference $\Imm(\lambda_{f_1}) \neq \Imm(\lambda_{f_2})$ can only be 
induced by a difference in the weak phases of the decay amplitudes 
$\phi_D^{(f)}$.
A super-weak scenario is defined as one in which a choice of phase convention 
exists in which all such phases can be made simultaneously real.

When the final state is an incoherent mixture of $CP$-even and $CP$-odd 
states, asymmetries are diluted by a factor $|1-2r|$, $r$ being the fraction of
events with a given $CP$ eigenvalue; this is the case for decays to two-body 
final states in which at least one of the particles has spin, so that states 
with different orbital angular momentum can have different $CP$ eigenvalues. 
The measurements of the asymmetries for such mixed states require an angular
momentum analysis in order to either estimate the fractions of the two $CP$
eigenstates on a statistical basis (\emph{i.e.} measure an effective 
$\eta_{CP}$ parameter), or to weight the events in the asymmetry fit according 
to the decay configuration.

When final states are considered which are not $CP$ eigenstates, but still can
be reached by both $M$ and $\overline{M}$ mesons, the expressions for the 
asymmetries and the extraction of weak phases from them become more 
complicated \cite{Aleksan}; $CP$ violation can still be probed by any 
difference between the time-dependent decay distributions for \mbox{$M \To f$} 
and \mbox{$\overline{M} \To \overline{f}$}, or by considering the inclusive 
state $f + \overline{f}$, but the need to consider two different final states 
makes the experimental investigation more demanding.
By comparing the $M$ and $\overline{M}$ decay distributions to a final state 
which is not a $CP$ eigenstate, as a function of $|\Delta t|$ for mesons 
produced in correlated pairs, direct $CP$ violation can be probed either by 
the shape of the distributions or by their normalization difference.

$T$-odd correlations in heavy meson decays have also been studied 
\cite{GolowichB}, such as \mbox{$\langle \Vec{S}_\tau \cdot \Vec{p}_\tau 
\times \Vec{p}_c \rangle$} in the inclusive semi-leptonic decays 
\mbox{$b \To c \tau \overline{\nu}$}, which requires measuring both momentum 
and polarization of the emitted $\tau$ lepton in events with a quark jet. 
Other $T$-odd correlations could be measured, such as the ones involving $c$ 
quark polarization in the $b$ decays, which require the measurement of 
exclusive channels in which the charmed quark hadronizes to a $D^*$, decaying 
to $D\pi$ and leading to a 4-body final state, in which triple correlations 
can be built just from the momenta. 
Exclusive decays of heavy mesons to vector particles, such as 
\mbox{$\bz \To K^* l^+ l^-$}, also allow $T$-odd correlations to be studied, 
although the tiny branching ratios make experimental measurements very 
challenging.

In general, in the SM these correlations are estimated to give larger signals 
for $B$ meson decays than for $K$, but no $CP$ violation effect has been 
measured in this way so far.

\subsection{Experimental considerations}

As mentioned, searches for direct $CP$ violation can be performed by measuring 
the time-integrated partial rate asymmetries $A_{CP}^{(f)}$ in any 
decay\footnote{We omit here and after the superscript $(f)$.} of charged 
mesons or (if $CP$ violation in the mixing is negligible) in decays of neutral 
mesons to \emph{flavour-specific} final states. When using meson pairs 
(correlated or not), the relative normalization is not an issue, and if the 
decay is \emph{flavour-specific}, or \emph{self-tagging} ($M \nrightarrow 
\overline{f}$, $\overline{M} \nrightarrow f$, which is always the case for 
charged meson decays) such asymmetries are simply given by the asymmetry of 
measured (untagged) $\overline{f}$ and $f$ events:
\Dm{
  A_{CP}^{(U)} = \frac{N(\overline{f})-N(f)}{N(\overline{f})+N(f)}
}
We remark again that for neutral mesons this asymmetry would be non-zero also 
in case of $CP$ violation in the mixing.

When considering instead states $f$ which can be reached by both $M$ and 
$\overline{M}$, flavour tagging information on the decaying meson is 
necessary; in this case the time-integrated asymmetry still contains diluted 
information on $CP$ violation:
\Dm{
  A_{CP} = \frac{N(\overline{M} \To f)-N(M \To f)}
    {N(\overline{M} \To f)+N(M \To f)}
}
where $M$, $\overline{M}$ refer to the meson flavour at the tagging time 
(different from the decay time) so that this asymmetry becomes 
\Dm{
  A_{CP} = \frac{\mathcal{A}_{CP}^{\mathrm{(m/d)}} + 
    x \mathcal{A}_{CP}^{\mathrm{(int)}}}{1+x^2}
}
(where $x \equiv \Delta m/\Gamma$), ignoring again any $CP$-violating effect 
in the mixing.

Since the statistical error of the measurement is $\sigma(A_{CP}) \sim 
1/\sqrt{N}$, ($N$ being the measured number of events), the relevant quantity 
defining the observability of an asymmetry for a given decay mode is the 
product $BR \cdot A_{CP}^2$ ($BR$ being the branching ratio).

The experimental measurement of asymmetries is also diluted in presence of 
background (assuming no $CP$ asymmetry in the background), so that the 
measured asymmetry $A_{CP}^{\mathrm{meas}}$ is related to the one of the 
signal by
\Dm{
  A_{CP}^{\mathrm{meas}} = A_{CP} \frac{N_S}{N_S+N_B}
}
where $N_S$ and $N_B$ are the numbers of signal and background events 
respectively. 
Taking also into account the detection efficiency $\varepsilon(f)$, the number 
$N$ of mesons required to measure an asymmetry at $n$ standard deviations from 
zero in the decay to the final state $f$ is
\Dm{
  N > \frac{n^2}{\varepsilon(f) A_{CP}(f)^2 BR(f)}(1+N_B/N_S)
}
from which one can argue that the study of rare decay modes for which however 
relatively large asymmetries are expected is favoured.

If the meson decay time is measured, the study of the shape of time-dependent 
asymmetries allows the search of $CP$ violation as discussed in detail in the 
previous section; this kind of measurement only became possible in recent 
times thanks to the development of precise vertex detectors with $\mu$m 
resolutions.

\subsection{Light unflavoured mesons}

No mixing effects are present for the ($CP$ self-conjugate) unflavoured light 
mesons, which are discussed here for completeness.
The possibility of $CP$ violation in the \mbox{$\pi \To \mu \To e$} weak decay 
chain has been considered; final-state interactions due to weak interactions 
induce very small phases, leading to very tiny effects in the Standard Model.
Direct comparison of $\pi^+$ and $\pi^-$ require pions to decay in vacuum to
overcome the large asymmetric effects due to pion interactions with matter 
(as compared to antimatter).

By comparing the $\mu^+$ and the $\mu^-$ polarizations, as measured by the 
oscillation amplitudes $A_\pm$ of the $e^\pm$ counting rate in a magnetic 
field, in the muon $g-2$ experiments \cite{Picasso}, a (direct) $CP$ asymmetry 
limit was obtained \cite{Kaplan}:
\Dm{
   -0.01 < \frac{A_+ - A_-}{A_+ + A_-} < 0.02
}

\begin{table}
\begin{center}
\begin{tabular}{|l|c|c|}
\hline
Channel & BR limit (90\% CL) & Notes \\
\hline
$\eta \To \pi^+ \pi^-$  & $3.3 \cdot 10^{-4}$ & CMD-2 1999 \\
$\eta \To \pi^0 \pi^0$  & $4.3 \cdot 10^{-4}$ & CMD-2 1999 \\
$\eta \To 4\pi^0$       & $6.9 \cdot 10^{-4}$ & Crystal Ball 2000 \\
\hline
$\eta' \To \pi^+ \pi^-$ & $2 \cdot 10^{-2}$ & Bubble ch. 1969 \\
$\eta' \To \pi^0 \pi^0$ & $9.4 \cdot 10^{-4}$ & GAMS-2000 1987 \\
\hline
\end{tabular}
\caption{Limits on $CP$-violating decays of $\eta,\eta'$ mesons 
\cite{PDG_2002}.}
\label{tab:etas}
\end{center}
\end{table}

$CP$-forbidden decays of the $\eta(548)$ and $\eta'(958)$ self-conjugate 
mesons have also been searched for: since such mesons decay (strongly) to an 
odd number of pions, the observation of their decay to an even number of pions 
would be evidence of direct $CP$ violation.
Table \ref{tab:etas} summarizes the current experimental limits for these
decays.

\subsection{$D$ mesons}

In the system of neutral $D$ mesons, all $CP$ violation effects are expected 
to be very small in the Standard Model, $O(0.01)$: $\dz--\dzb$ mixing is known 
to be very small and can be described rather well by physics of the first two 
quark generations only; the top-quark loops, which in the Standard Model 
induce the largest $CP$ violating effects in the strange and beauty mesons, 
are here absent. 
The $c$ quark can decay without suppression due to the small inter-generation 
mixing, and the lifetime of charged mesons is relatively short when compared 
to that of $B$ mesons (scaled with their masses).
In Cabibbo-allowed (such as $D^+,D^0 \To K^-X$) and doubly Cabibbo-suppressed 
(such as $D^0 \To K^+ \pi^-$) decays of charmed mesons,
there are usually no competing amplitudes in the Standard Model capable 
of inducing sizable direct $CP$ violating asymmetries in the decay rates.
The known indirect $CP$ violation contribution in $\kz$ decays can induce a 
rate difference in the decays \mbox{$D^\pm \To \ks \pi^\pm$} equal to $-2 \Rea
\epsilon_K$, and any difference with respect to this value would be a signal 
of (direct) $CP$ violation in the $D$ decay process.

The search for direct $CP$ violation in singly Cabibbo-suppressed decays of
charmed mesons (such as $D^0 \To K^+ K^-, \pi^+ \pi^-$, or $D^+ \To \ks K^+$) 
is also very challenging, as far as Standard Model predictions are concerned 
(at the $\sim 10^{-3}$ level); searches for $CP$ violation in $D$ meson decays 
are therefore considered sensitive probes for new physics.
 
Partial rate asymmetries are defined as
\Dm{
  A_{CP}^{(f)} = \frac{N(D^+,\dz \To f)-N(D^-,\dzb \To \overline{f})}
    {N(D^+,\dz \To f)+N(D^-,\dzb \To \overline{f})} 
}

In photo-production experiments, such partial rate asymmetries are measured by 
normalizing to Cabibbo-allowed decays, since the production rates of $\dz$ 
and $\dzb$ are different (although in this way $CP$ asymmetries from new 
physics in the normalization mode could mask a signal in the 
Cabibbo-suppressed mode). 
Searches for direct $CP$ violation in Cabibbo-allowed decay modes (not expected
in the Standard Model) actually look for differences in asymmetries
among different decay channels.

By considering both ``right sign'' $\dz \To \overline{f}$, $\dzb \To f$ decays 
and ``wrong sign'' ones $\dz \To f$, $\dzb \To \overline{f}$, where $f$ and
$\overline{f}$ are in general not $CP$ eigenstates ($|\overline{f}\rangle =
CP|f\rangle$), one can study the ``wrong sign'' time-dependent decay rates for 
flavour-tagged neutral $D$ mesons:
\Ea{ \Ds
  & r(\tau) \equiv \frac{\Gamma(D^0(\tau) \To f)}{|\overline{A}_f|^2} = 
    \frac{1}{2} \left| \frac{q}{p} \right|^2 \left[ e^{-\tau} 
    (1/|\lambda_f|^2+1) \cosh(y \tau) - \right. \nonumber \\
  & \left. 2\Rea(1/\lambda_f) \sinh(y \tau) +
    (1/|\lambda_f|^2-1) \cos(x \tau) -
    2\Imm(1/\lambda_f) \sin(x \tau) \right] \\
  & \overline{r}(\tau) \equiv 
    \frac{\Gamma(\overline{D}^0(\tau) \To \overline{f})}
    {|A_{\overline{f}}|^2} = 
    \frac{1}{2} \left| \frac{p}{q} \right|^2 \left[ e^{-\tau} 
    (|\lambda_{\overline{f}}|^2+1) \cosh(y \tau) - \right. \nonumber \\
  & \left. 2\Rea(\lambda_{\overline{f}}) \sinh(y \tau) + 
    (|\lambda_{\overline{f}}|^2-1) \cos(x \tau) -
    2\Imm(\lambda_{\overline{f}}) \sin(x \tau) \right]
}
where the decay time is measured in units of the $D^0$ mean life: $\tau \equiv 
\overline{\Gamma} t$ (here $\Delta m = m_+-m_-$, $\Delta \Gamma = \Gamma_+ - 
\Gamma_-$, where the subscripts indicate the $CP$ eigenvalue).
For flavour-specific (\emph{e.g.} semi-leptonic) decays, and for small $CP$
violation one has $1/|\lambda_f| \ll 1$ and $|\lambda_{\overline{f}}| \ll 1$; 
in the limit of small mixing ($x \ll 1$, $y \ll 1$) the above expressions 
reduce to
\Ea{ \Ds 
  & r(\tau) \simeq \frac{e^{-\tau}}{4} \left| \frac{q}{p} \right|^2
    (x^2+y^2) \tau^2 \\
  & \overline{r}(\tau) \simeq \frac{e^{-\tau}}{4} \left| \frac{p}{q} \right|^2
    (x^2+y^2) \tau^2
}
so that if there is no $CP$ violation in the mixing one has in this case 
$r(\tau) = \overline{r}(\tau)$.

The phenomenology for hadronic decays of neutral $D$ mesons is complicated by 
the possibility of doubly Cabibbo-suppressed decays (``wrong sign'') leading 
to the same final states. 
To leading order in the $CP$-violating parameters:
\Ea{
  & r(\tau) = e^{-\tau} \left[ \frac{R_D}{(1+A_D)^2} + \right. \nonumber \\
  & \left. \sqrt{R_D}\frac{1+A_M}{1+A_D} 
    \Rea \left[(y+ix)e^{i(\delta+\phi_D+\phi_M)}\right] \tau + 
    \frac{1}{2} R_M (1+A_M)^2 \tau^2 \right] \\
  & \overline{r}(\tau) = e^{-\tau} \left[ R_D(1+A_D)^2 + \right. \nonumber \\
  & \left. \sqrt{R_D}\frac{1+A_D}{1+A_M} 
    \Rea \left[(y+ix)e^{i(\delta-\phi_D-\phi_M)}\right] \tau +
    \frac{1}{2} \frac{R_M}{(1+A_M)^2} \tau^2 \right]
}
where $R_D$ is a measure of the double Cabibbo suppression in the amplitude 
\Dm{
  \frac{A_f}{A_{\overline{f}}} = -\sqrt{R_D} e^{-i\delta}
}
(with $\delta$ a strong phase difference between the two amplitudes), $R_M$ is
a measure of the double Cabibbo suppressed mixing rate
\Dm{
  \int_0^\infty dt \, \Gamma(\dz(t) \To f) = R_M |\overline{A}_f|^2
}
and the real parameters $A_M, A_D$ and $\phi_M+\phi_D$ characterize $CP$ 
violation in the mixing, in the decay and in the interference of the two 
processes respectively:
\Ea{ \Ds 
  & \frac{q}{p} = (1+A_M) e^{i\phi_M} \\
  & \frac{\overline{A}_f}{A_{\overline{f}}} = (1+A_D) e^{i\phi_D}
}
In the limit of $CP$ conservation $A_M$, $A_D$ and $\phi_M+\phi_D$ are all zero
and 
\Dm{
  r(\tau) = \overline{r}(\tau) = e^{-\tau} \left[
    R_D + \sqrt{R_D} \Rea[(y+ix)e^{i\delta}] \tau + 
    \frac{1}{2} R_M \tau^2 \right]
}
For $CP$ eigenstates, $f$ and $\overline{f}$ coincide, and one has (for 
$CP|f\rangle=+|f\rangle$):
\Ea{
  r(\tau) \propto \left| e^{-i(m_+-i\Gamma_+/2)\tau/\overline{\Gamma}} + 
    \eta_f e^{-i(m_--i\Gamma_-/2)\tau/\overline{\Gamma}} \right|^2 \\
  \overline{r}(\tau) \propto \left| 
    e^{-i(m_+-i\Gamma_+/2)\tau/\overline{\Gamma}} -  
    \eta_f e^{-i(m_--i\Gamma_-/2)\tau/\overline{\Gamma}} \right|^2 
}
where $\eta_f = (1-\lambda_f)/(1+\lambda_f)$ describes $CP$ violation of all
three types; for the opposite $CP$ eigenvalue the above formula holds with $1 
\leftrightarrow 2$ and $\eta_f \rightarrow 1/\eta_f$.

By analyzing decays of $D^0$ mesons produced from the (strong) decay chain   
\mbox{$D^{*+} \To D^0 \pi^+$} and \mbox{$D^{*-} \To \overline{D}^0 \pi^-$}, 
the charge of the (slow) pion accompanying the neutral $D$ meson can be used 
to identify the initial flavour; this flavour-tagging technique has no 
significant bias for experiments performed at colliders in which quarks and 
anti-quarks are produced in pairs \cite{CLEO_D}, and mis-tagging rates are 
usually $O(10^{-3})$. The decay of charmonium states to open charm, such as 
$J/\Psi(3770) \To \dz \dzb$, can also be exploited to tag the flavour of 
neutral $D$ mesons.

Systematic effects in the measurements can arise from asymmetries in the 
fitting of the signal and background components, and from flavour-tagging 
asymmetries due to charge-dependent biases in the detector acceptance and
efficiency.

\begin{table}
\begin{center}
\begin{tabular}{|l|c|c|l|}
\hline
Channel & BR & $A_{CP}$ & Notes \\
\hline
$D^+ \To \ks \pi^+$              & $1.4\%$ &
  $-0.016 \pm 0.017$             & FOCUS \\
$D^+ \To \ks K^+$                & $2.9 \cdot 10^{-3}$ &
  $0.07 \pm 0.06$                & FOCUS \\
$D^+ \To K^+ K^- \pi^+$          & $8.8 \cdot 10^{-3}$ &
  $0.002 \pm 0.011$              & E687, E791, FOCUS \\
$D^+ \To K^+ \overline{K}^{0*}$  & $4.2 \cdot 10^{-3}$ &
  $-0.02 \pm 0.05$               & E687, E791 \\
$D^+ \To \phi \pi^+$             & $6.1 \cdot 10^{-3}$ & 
  $-0.014 \pm 0.033$             & E687, E791 \\
$D^+ \To \pi^+ \pi^+ \pi^-$      & $3.1 \cdot 10^{-3}$ &
  $-0.02 \pm 0.04$               & E791 \\
\hline
$D^0 \To \pi^+ \pi^-$            & $1.4 \cdot 10^{-3}$ & 
  $0.021 \pm 0.026$              & CLEO, E791, FOCUS \\ 
$D^0 \To \pi^0 \pi^0$            & $8.4 \cdot 10^{-4}$ &
  $0.001 \pm 0.048$              & CLEO \\
$D^0 \To K^\pm \pi^\mp$          & $1.5 \cdot 10^{-4}$ &
  $0.02^{+0.19}_{-0.20}$         & CLEO \\
$D^0 \To \ks \pi^0$              & $1.14\%$ &
  $0.001 \pm 0.013$              & CLEO \\
$D^0 \To K^+ K^-$                & $4.1 \cdot 10^{-3}$ & 
  $0.005 \pm 0.016$              & CLEO, E687, E791, FOCUS \\ 
$D^0 \To \phi \ks$               & $4.7 \cdot 10^{-3}$ &
  $-0.03 \pm 0.09$               & CLEO \cite{CLEO_DKpi} \\
$D^0 \To \ks \ks$                & $3.6 \cdot 10^{-4}$ &
  $-0.23 \pm 0.19$               & CLEO \cite{CLEO_DKpi} \\
$D^0 \To K^\mp \pi^\pm \pi^0$    & $13\%$ & 
  $-0.03 \pm 0.09$               & CLEO \\ 
$D^0 \To K^\pm \pi^\mp \pi^0$    & $5.6 \cdot 10^{-4}$ & 
  $0.09^{+0.25}_{-0.22}$         & CLEO \\ 
\hline
\end{tabular}
\caption{Measurements of direct-$CP$ violating asymmetries in $D$ meson 
decays, from \cite{PDG_2002} except where indicated otherwise.}
\label{tab:directD}
\end{center}
\end{table}

\begin{figure}[hbt!]
\begin{center}
    \epsfig{file=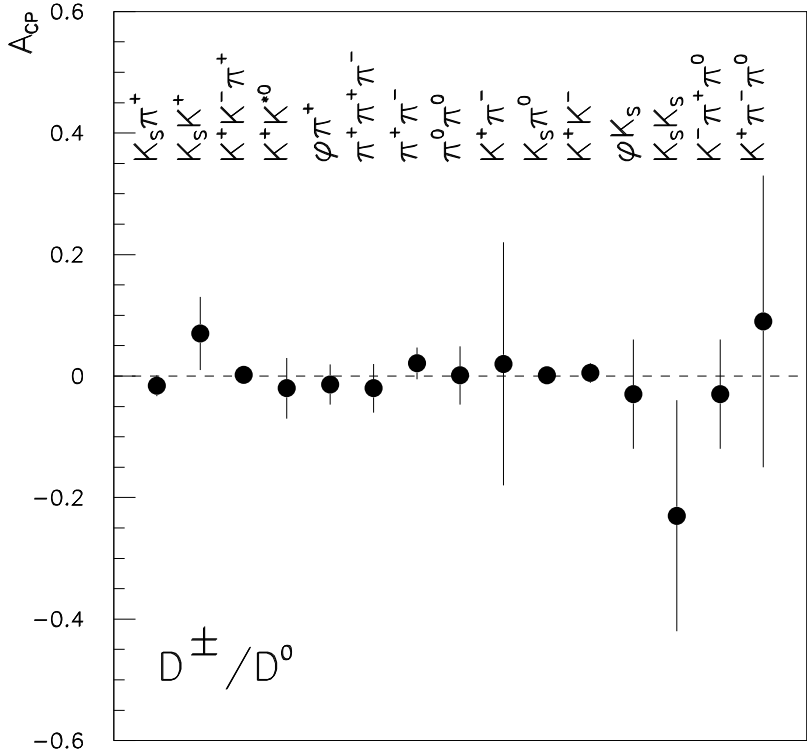,width=1.0\textwidth}
  \caption{Graphical representation of time-integrated $CP$ asymmetries for 
$D^\pm$ and $\dz(\dzb)$ decays.}
  \label{fig:acp_d}
\end{center}
\end{figure}

While no $CP$ violation effects have been detected in the $D$ meson system so
far, table \ref{tab:directD} and fig. \ref{fig:acp_d} summarize the 
experimental results of searches for direct $CP$ violation in their decays.

\subsection{$B$ mesons}

In the system of neutral $B$ mesons (see \emph{e.g.} \cite{Fleischer} for a 
recent review), $CP$ violation effects in the mixing are generally expected to 
be small, $\lesssim O(10^{-2})$.
The reason is that the total decay width difference $\Delta \Gamma$ of the two 
physical states is very small, being induced by the final states common to $B$ 
and $\overline{B}$ decays, which have small $O(10^{-3})$ branching ratios for 
such a system with many open decay channels: $\Delta \Gamma/\Gamma \ll 1$, 
combined with the empirical fact that $|\Delta m| \sim \Gamma$ results in 
$|\Gamma_{12}| \ll |M_{12}|$ and therefore for the asymmetries due to mixing
\Dm{
  A_{CP}^{\mathrm{(mix)}} \propto 1- \left| \frac{q}{p} \right|^2 \propto 
    \Imm(\Gamma_{12}/M_{12}) \lesssim 
    \Delta \Gamma/\Delta m \lesssim 10^{-3}
}
In other terms, in the Standard Model, when final-state interaction effects 
are small, $M_{12}$ and $\Gamma_{12}$ acquire their phases from the same 
combination of CKM matrix elements, leading to a small relative phase and 
small asymmetries. 
Precise theoretical predictions are however rather difficult.

$CP$ violation in mixing has not been experimentally detected yet for neutral 
$B$ mesons. The $CP$-violating impurity in the physical $B_d$ meson states is 
measured to be \cite{PDG_2002}
\Dm{
  \frac{\Rea(\beps_{B_d})}{1+|\beps_{B_d}|^2} = (0 \pm 4) \cdot 10^{-3} 
}
from the (time-integrated or time-dependent) charge asymmetry in like-charge 
di-lepton events from semi-leptonic decays, and from the analysis of the 
time-dependent asymmetry of inclusive decays, using samples where the initial
flavour state is tagged (see also \cite{BABAR_CPT}).
For $B_s$ mesons flavour oscillations have not been observed yet.

Mixing-induced $CP$ violation is instead expected to give large $O(1)$ effects 
in several neutral $B$ meson decay asymmetries. For channels in which the decay
is dominated by a single elementary amplitude (and therefore $CP$ violation 
effects in the decay are negligible), such asymmetries can be related in a
direct way to the basic parameters of the underlying theory, with small 
theoretical uncertainties. 
This is among the main reasons for the great interest in the measurement of 
$CP$ violation in $B$ decays.

In the Standard Model, direct $CP$ violation in $B$ meson decays is expected 
to occur in charmless hadronic decays, in which \emph{tree} and \emph{penguin} 
\mbox{$b \To u$} decay amplitudes of comparable magnitudes can interfere.
Studied channels include \mbox{$B_d^0 \To K^+ \pi^-$}, \mbox{$B^+ \To K^+ 
\pi^0$}, \mbox{$B^+ \To \ks \pi^+$} (together with their $CP$-conjugates, of 
course).
Direct $CP$ violation is instead expected to be negligible for modes to which 
only either \emph{tree} or \emph{penguin} graphs contribute: some examples are 
respectively \mbox{$\bz \To J/\Psi \kz$} and \mbox{$\bz \To \phi \kz$}; 
searches for direct $CP$ violation in these channels are therefore sensitive 
to physics beyond the Standard Model.

$CP$ violation in the $B$ meson system has been mostly studied in $e^+ e^-$ 
collider experiments (\emph{e.g.} the LEP experiments, CLEO at CESR, BABAR at 
SLAC, BELLE at KEK), but significant results have been obtained in experiments 
at hadronic machines (\emph{e.g.} CDF at the TeVatron).
The $e^+ e^-$ environment, when compared to the hadronic one, provides a high 
signal to background ratio, cleaner events and a low interaction rate; the 
absolute production cross sections being however much smaller, very high 
luminosities are required.

$B$ mesons can be studied at colliders by exploiting their non-resonant 
inclusive associate production ($b\overline{b}X$ final states): this is the 
case at hadron colliders and high-energy $e^+e^-$ colliders.

At $e^+ e^-$ ``B-factories'', pairs of $B_d-\overline{B}_d$ mesons (both 
neutral and charged) are produced by decays of the $\Upsilon(4S)$ bottomonium 
state (just above the open beauty threshold) at its formation energy of 10.6 
GeV: the branching ratio of this resonance to $B \overline{B}$ pairs is above 
96\% and the high instantaneous luminosities ($5-8 \cdot 10^{33}$ cm$^{-2}$ 
s$^{-1}$) provide copious sources of $B$ mesons.
The time-dependent $CP$-violating asymmetries depend only on the time 
difference $\Delta t$ of the two meson decays: the short lifetime of the $B$ 
mesons, $c \tau_B \sim$ 500 $\mu$m, corresponding to only 23 $\mu$m when 
produced from the decay of a $\Upsilon(4S)$ at rest, requires very precise 
vertex detectors for its measurement. 
The measurement of the distance between the two decay vertexes at a symmetric 
collider (usually the production point cannot be determined with sufficient 
accuracy) only gives the sum of the decay times.
By using asymmetric beam energies, the Lorentz boost of the final state 
($\langle \beta \gamma \rangle$ = 0.56 at SLAC with 3.1+9 GeV $e^+e^-$ beam
energies, 0.43 at KEK with 3.5+8 GeV) allows the experimental measurement of 
time-dependent decay asymmetries, since the decay time difference is then 
given by $\Delta z \simeq \beta \gamma c \Delta t$, with an average 
separation between the two vertexes of 260 or 200 $\mu$m respectively, while 
the typical resolution achieved by the vertex detectors is about 180 $\mu$m. 

While the study of time-dependent $\bz--\bzb$ meson asymmetries was performed 
at LEP, when running at the $\Upsilon(4S)$ resonance one can profit from the 
tight kinematic constraints which help in reducing the background; moreover,  
the combinatorial component is less important due to the absence of 
fragmentation products.
The first evidence for (mixing-induced) $CP$ violation outside the neutral 
kaon system was actually provided by the measurements of such asymmetries in 
\mbox{$B_d^0,\overline{B}_d^0 \To J/\Psi \ks$} decays \cite{BABAR_ind} 
\cite{BELLE_ind}. 

While the flavour eigenvalue of $B^\pm$ mesons is determined by their 
charge, as measured from the decay products, the flavour eigenvalue for 
neutral $B$ mesons, apart from flavour-specific decay modes, has to be 
determined by exploiting their associate production.
In experiments using correlated pairs of $B$ mesons, the flavour tagging of a 
decaying $B$ meson provides flavour information for the opposite $B$ at the 
same time.
In experiments where $B$ mesons are produced incoherently, their flavour (at a
given time for neutral mesons) can be determined by that of the companion $b$ 
hadron; the flavour tagging information obtained in such a way by a companion 
neutral meson is degraded by the flavour mixing, in contrast with what happens 
with correlated pairs where mixing does not affect the tagging.
The $B$ meson flavour can also be determined by exploiting its production from 
decays of excited $B^{**}$ states through the decay chain \mbox{$B^{**} \To 
B \pi$}: in reconstructed $B^{**}$ meson decays one obtains the flavour of 
the $\bz$ from the charge of the accompanying charged pion (\mbox{$B^{**+} \To 
B_d \pi^+$}, \mbox{$B^{**-} \To \overline{B}_d \pi^-$}).

Experiments at B-factories, discussed in the following, use large acceptance 
asymmetric detectors, which rely on excellent vertex tracking to identify 
secondary vertexes, discriminate against light quark vertexes, and measure 
decay time differences.
Good particle identification over a wide kinematic range is also an important 
requirement: flavour tagging relies on the identification of electrons and
muons from semi-leptonic decays, and kaons as well; moreover, the selection of 
exclusive decay channels needs good $K/\pi$ discrimination.
Both the BABAR \cite{BABAR} and the Belle \cite{BELLE} detectors have silicon
vertex detectors, central drift chambers with helium-based gas to minimize 
multiple scattering, CsI(Tl) crystal electromagnetic calorimeters, and muon 
detection systems outside super-conducting solenoids providing 1.5 T magnetic 
field.
Both experiments use Cerenkov detectors for hadron identification, based on 
silica quartz bars for BABAR, and on silica aerogel counters for Belle.
The analysis is based on the full reconstruction of one $B$ meson to a given 
final state, while the opposite one is partially reconstructed in order to 
determine its flavour through the identification of decay leptons, kaons or 
soft charged pions from $D^*$ decays. The interesting decays are identified 
kinematically by constraining their centre of mass energies and momenta, and
exploiting particle identification. Maximum likelihood fits which include the
tagging efficiency parameters, background fractions and shapes in several
discriminating variables, are used to extract event yields and asymmetries.

Direct $CP$ violation has been searched for, both by measuring time-integrated 
decay asymmetries and by fitting the cosine term in time-dependent 
asymmetries.

The first approach is simpler: after event reconstruction, background
rejection and signal fitting, the measured event asymmetry is formed.
For non-self-tagging modes, such as \mbox{$\bz(\bzb) \To K^0 \pi^0$}, $B$ 
flavour tagging information is needed, and flavour mixing dilutes the 
measurement by a factor $1/(1+x^2)$ as mentioned.

The flavour of the $B$ meson can be distinguished either by its charge or by 
the electric charge of one of its decay products (secondaries for states 
containing only neutral particles such as \mbox{$\bz \To \phi K^{0*} \To \phi 
K^\pm \pi^\mp$}), by exploiting some selection rule. 
For neutral $B$ mesons, semi-leptonic decays are mostly used, based on the 
$\Delta B = \Delta Q$ rule (valid to high accuracy in the Standard Model): 
the charge of the energetic lepton identifies the flavour of the meson at the 
time of its decay (\mbox{$\bz \To l^+ X$}, \mbox{$\bzb \To l^- X$}), although 
the \mbox{$b \To c \To l$} decay results in opposite lepton charges; 
alternatively, the charge of reconstructed charm mesons (\mbox{$B_d^0 \To 
D^{*-}$, $B_s^0 \To D_s^-$}), detected from the charge of the soft pions 
from \mbox{$D^{*\pm} \To \dz(\dzb)\pi^\pm$}, or that of kaons from \mbox{$b 
\To c \To s$} decays, can be used, being all correlated with the flavour of 
the decaying $B$, to different extent.

Particle identification, achieved with multiple redundant measurements, is 
crucial for these measurements, as well as its charge symmetry together with
that of track reconstruction. 
The measured $CP$-violating asymmetries are diluted by the errors due to the 
mis-tagging of the $B$ meson flavour, \emph{i.e.} by the (single or double) 
mis-identification of the final state particles, and have to be corrected 
accordingly: 
\Dm{
  A_{CP}^{\mathrm{meas}} = (1-2w) A_{CP}
}
where the wrong-tag fraction $w$ (ranging from a few percent for the best
signatures to several tens of percent for poor tagging information).
The mis-tag probabilities for different samples can be extracted from the fit 
of the time-dependent asymmetries, or extracted from the time-integrated 
analysis of flavour-specific final states: assuming that the flavour of the
fully reconstructed meson is correctly identified, the measured 
time-integrated fraction of mixed events $\chi^{\mathrm{meas}}$ is given by
\Dm{
  \chi^{\mathrm{meas}} = \chi + (1-2\chi) w
}
where $\chi = x^2/[2(1+x^2)]$.
Fake asymmetries can be induced in case the mis-tagging probabilities are  
different for $\bz$ and $\bzb$.

Large backgrounds affect most of the channels which have very small branching 
ratios, and they are usually the dominant source of systematic error; in 
order to have small statistical errors on the signal samples, precise 
measurements of the backgrounds are also required, so that a fraction of data
taking has to be spent off the resonance.
Important systematic cross-checks are given by the measurements of the 
asymmetries for the background events, which in most cases have to be 
consistent with zero.

\begin{table}
\begin{center}
\begin{tabular}{|l|c|c|l|}
\hline
Channel & BR & $A_{CP}$ & Notes \\
\hline
$B^+ \To \pi^+ \pi^0$  & $5.4 \cdot 10^{-6}$ & $-0.07 \pm 0.15$ & 
BABAR \cite{BABAR_K+pi0}, BELLE \cite{BELLE_Tomura} \\ 
$B^+ \To K^+ \pi^0$    & $1.3 \cdot 10^{-5}$ & $0.01 \pm 0.12$ & 
BABAR \cite{BABAR_K+pi0}, BELLE \cite{BELLE_Tomura}, CLEO \\ 
$B^+ \To K_S \pi^+$    & $1.1 \cdot 10^{-5}$ & $-0.02 \pm 0.09$ & 
BABAR \cite{BABAR_KSpi+}, BELLE \cite{BELLE_K0pi+}, CLEO \\ 
$B^+ \To K^+ \eta'$    & $7.7 \cdot 10^{-5}$ & $0.02 \pm 0.04$ & 
BABAR \cite{BABAR_etapK+}, BELLE \cite{BELLE_etapK+}, CLEO \\ 
$B^+ \To K^{*+} \eta$  & $2.2 \cdot 10^{-5}$ & $-0.05^{+0.25}_{-0.30}$ & 
BELLE \cite{BELLE_Tomura} \\
$B^+ \To \omega \pi^+$ & $5.4 \cdot 10^{-6}$ & $-0.20 \pm 0.19$ & 
BABAR, CLEO \\ 
$B^+ \To \omega K^+$   & $5.3 \cdot 10^{-6}$ & $-0.21 \pm 0.28$ & 
BELLE \cite{BELLE_omegaK} \\
$B^+ \To J/\Psi \pi^+$ & $4.2 \cdot 10^{-5}$ & $-0.01 \pm 0.13$ &
BABAR, BELLE \cite{BELLE_Jpsi} \\ 
$B^+ \To J/\Psi K^+$   & $1.0 \cdot 10^{-3}$ & $-0.007 \pm 0.018$ &
BABAR, BELLE \cite{BELLE_Jpsi}, CLEO \\ 
$B^+ \To \Psi' K^+$    & $6.6 \cdot 10^{-4}$ & $-0.10 \pm 0.07$ &
BELLE \cite{BELLE_Jpsi}, CLEO \\
$B^+ \To \phi K^+$     & $1.2 \cdot 10^{-5}$ & $0.04 \pm 0.09$ & 
BABAR \cite{BABAR_phiK+} \\
$B^+ \To \phi K^{*+}$  & $1.2 \cdot 10^{-5}$ & $0.16 \pm 0.17$ & 
BABAR \cite{BABAR_phiK*+} \\
$B^+ \To K^{*+} \gamma$ & $3.8 \cdot 10^{-5}$ & $0.05 \pm 0.09$ &
BELLE \cite{BELLE_Suzuki} \\
$B^+ \To D_1 K^+$      & & $0.10 \pm 0.15$ & 
BABAR \cite{BABAR_D1K+}, BELLE \cite{BELLE_DK+} \\
$B^+ \To D_2 K^+$      & & $-0.19 \pm 0.18$ &
BELLE \cite{BELLE_DK+} \\
$B^+ \To D_{CP} K^+$   & & $0.06 \pm 0.18$ & 
BABAR \cite{BABAR_Sciolla} \\
$B^+ \To \pi^+ \pi^+ \pi^-$ & $1.1 \cdot 10^{-5}$ & $-0.39 \pm 0.35$ &
BABAR \cite{BABAR_3pi} \\
$B^+ \To K^+ \pi^+ \pi^-$ & $5.7 \cdot 10^{-5}$ & $0.01 \pm 0.08$ &
BABAR \cite{BABAR_3pi} \\
$B^+ \To K^+ K^+ K^-$ & $3.4 \cdot 10^{-5}$ & $0.02 \pm 0.08$ &
BABAR \cite{BABAR_3pi} \\
\hline
$B_d^0 \To K^+ \pi^-$    & $1.8 \cdot 10^{-5}$ & $-0.09 \pm 0.04$ & 
BABAR \cite{BABAR_pipi}, BELLE \cite{BELLE_Tomura}, CLEO \\ 
$B_d^0 \To \ks \pi^0$    & $5.4 \cdot 10^{-6}$ & $0.04 \pm 0.12$ &
BABAR \cite{BABAR_phiK*+} \\
$B_d^0 \To \eta K^{*0}$  & $2.1 \cdot 10^{-5}$ & $0.17^{+0.28}_{-0.25}$ & 
BELLE \cite{BELLE_Tomura} \\
$B_d^0 \To \phi K^{*0}$  & $1.1 \cdot 10^{-5}$ & $0.04 \pm 0.12$ & 
BABAR \cite{BABAR_phiK*+} \\
$B_d^0 \To \rho^+ \pi^-$ & $2.5 \cdot 10^{-5}$ & $-0.18 \pm 0.09$ & 
BABAR \cite{BABAR_Sciolla} \\
$B_d^0 \To \rho^+ K^-$   & $7.3 \cdot 10^{-6}$ & $0.28 \pm 0.19$ &
BABAR \cite{BABAR_Sciolla} \\
$B_d^0 \To D^{*+} D^-$   & $8.8 \cdot 10^{-4}$ & $-0.03 \pm 0.12$ &
BABAR \cite{BABAR_D*+D-} \\
$B_d^0 \To K^{*0} \gamma$ & & $-0.06 \pm 0.07$ & 
BELLE \cite{BELLE_Suzuki}, CLEO \\
\hline
$B \To K^{*} \gamma$ & $4.2 \cdot 10^{-5}$ & $-0.01 \pm 0.07$ &
BABAR, CLEO \\
$b \To s \gamma$         & $3.3 \cdot 10^{-4}$ & $-0.08 \pm 0.11$ & 
CLEO \\
\hline
\end{tabular}
\caption{Measurements of time-integrated direct $CP$-violating asymmetries in 
$B$ meson decays (from \cite{PDG_2002} unless otherwise indicated).
Here $CP(D_1)=+1$, $CP(D_2)=-1$, $D_{CP}= D_1,D_2$.}
\label{tab:Bdirect}
\end{center}
\end{table}

\begin{figure}[hbt!]
\begin{center}
    \epsfig{file=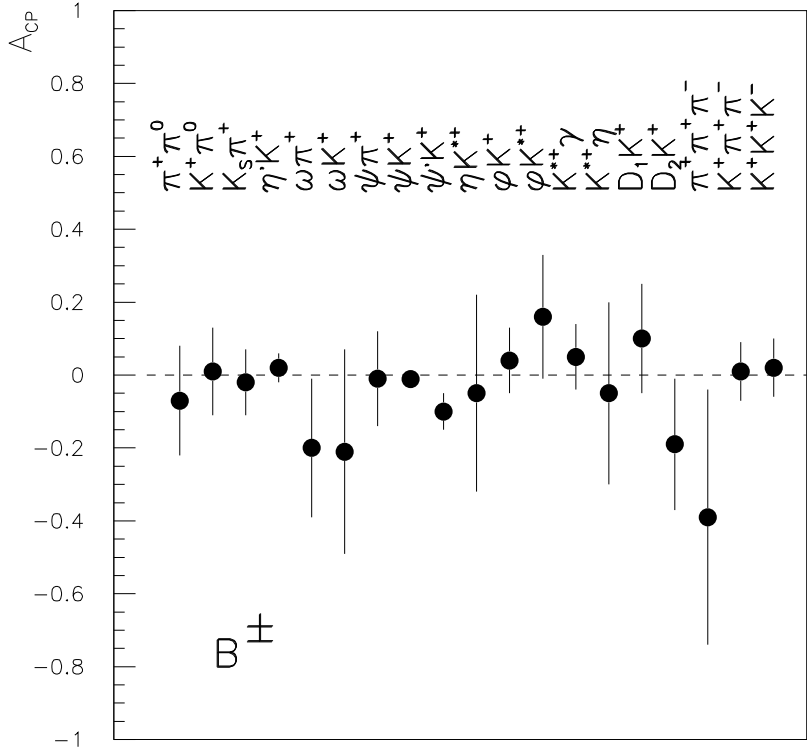,width=1.0\textwidth}
  \caption{Graphical representation of time-integrated $CP$ asymmetry 
measurements for $B^\pm$ decays.}
  \label{fig:acp_bc}
\end{center}
\end{figure}

\begin{figure}[hbt!]
\begin{center}
    \epsfig{file=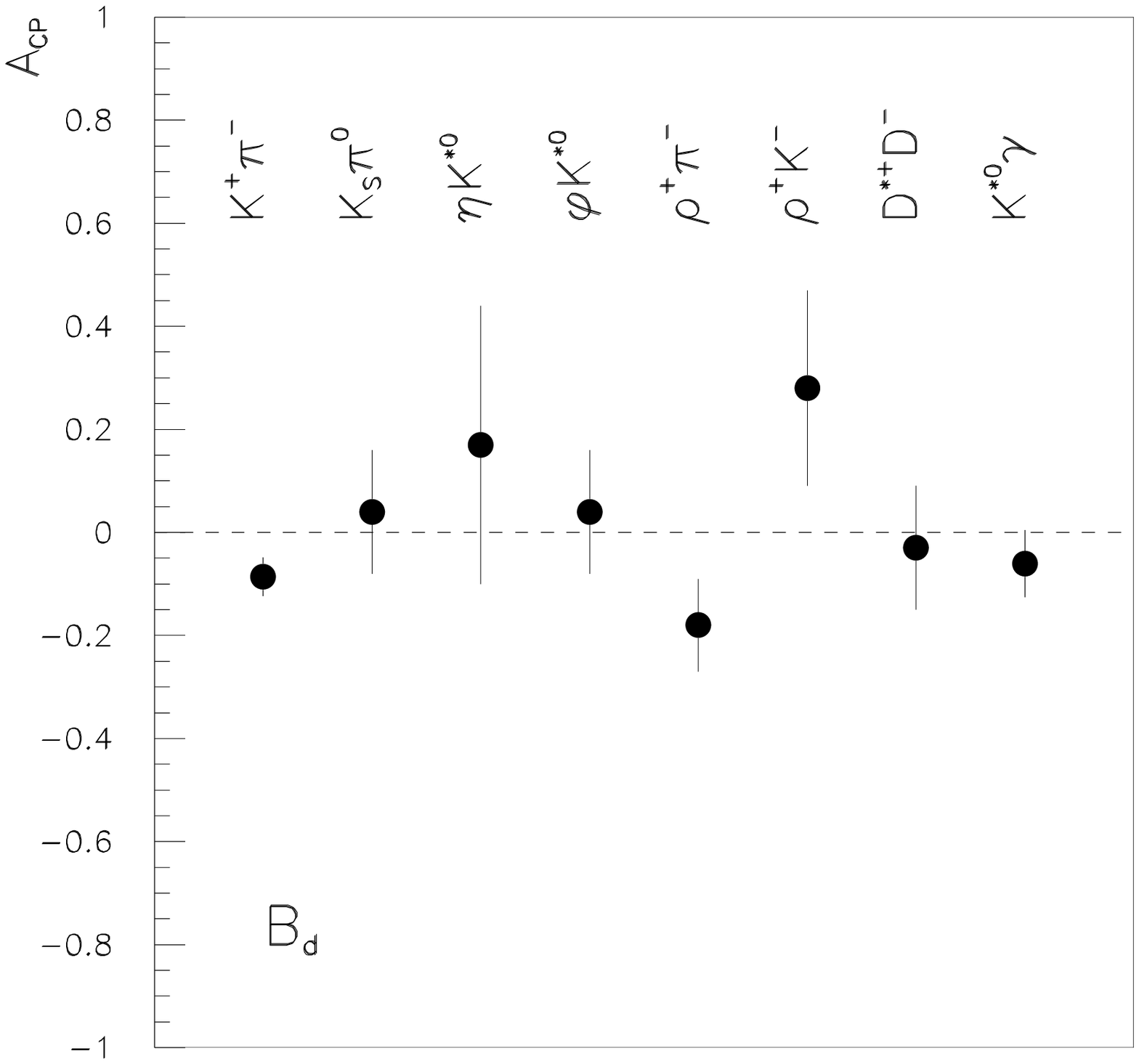,width=1.0\textwidth}
  \caption{Graphical representation of time-integrated $CP$ asymmetry 
measurements for $B^0_d(\overline{B}^0_d)$ decays.}
  \label{fig:acp_bn}
\end{center}
\end{figure}

\begin{table}
\begin{center}
\begin{tabular}{|l|c|c|l|}
\hline
Channel & BR & Value & Notes \\
\hline
$B_d^0 \To \pi^+ \pi^-$  & $4.5 \cdot 10^{-6}$ & $0.51 \pm 0.23$ & 
BABAR \cite{BABAR_pipi}, BELLE \cite{BELLE_pipi} \\
$B_d^0 \To \eta' \ks$    & $3.1 \cdot 10^{-5}$ & $0.08 \pm 0.18$ &
BELLE \cite{BELLE_phiks} \\
$B_d^0 \To \rho^+ \pi^-$ & $2.5 \cdot 10^{-5}$ & $-0.36 \pm 0.18$ & 
BABAR \cite{BABAR_Sciolla} \\
$B_d^0 \To \phi \ks$     & $5.4 \cdot 10^{-6}$ & $0.17 \pm 0.68$ &
BABAR \cite{BABAR_Monchenault}, BELLE \cite{BELLE_phiks} \\
$B_d^0 \To K^+ K^- \ks$  & $1.5 \cdot 10^{-5}$ & $-0.40 \pm 0.43$ & 
BELLE \cite{BELLE_phiks} \\
$B_d^0 \To J/\Psi \pi^0$ & $2.1 \cdot 10^{-5}$ & $-0.31 \pm 0.28$ & 
BABAR \cite{BABAR_psipi0}, BELLE \cite{BELLE_psipi0} \\
$B_d^0 \To D^{*+} D^{*-}$ & & $-0.02 \pm 0.28$ & 
BABAR \cite{BABAR_Sciolla} \\
$B_d^0 \To c\overline{c} \ks$    & $---$ & $-0.052 \pm 0.042$ & 
BABAR \cite{BABAR_ccs}, BELLE \cite{BELLE_ccs}\\
\hline
\end{tabular}
\caption{Measurements of the direct $CP$ violation coefficient 
$\mathcal{A}_{CP}^{\mathrm{(m/d)}}$ in time-dependent asymmetries of neutral 
$B$ meson decays (from \cite{PDG_2002} unless otherwise indicated, see also
\cite{HFAG}). Errors are scaled according to the PDG recipe.}
\label{tab:Bdirect_td}
\end{center}
\end{table}

\begin{figure}[hbt!]
\begin{center}
    \epsfig{file=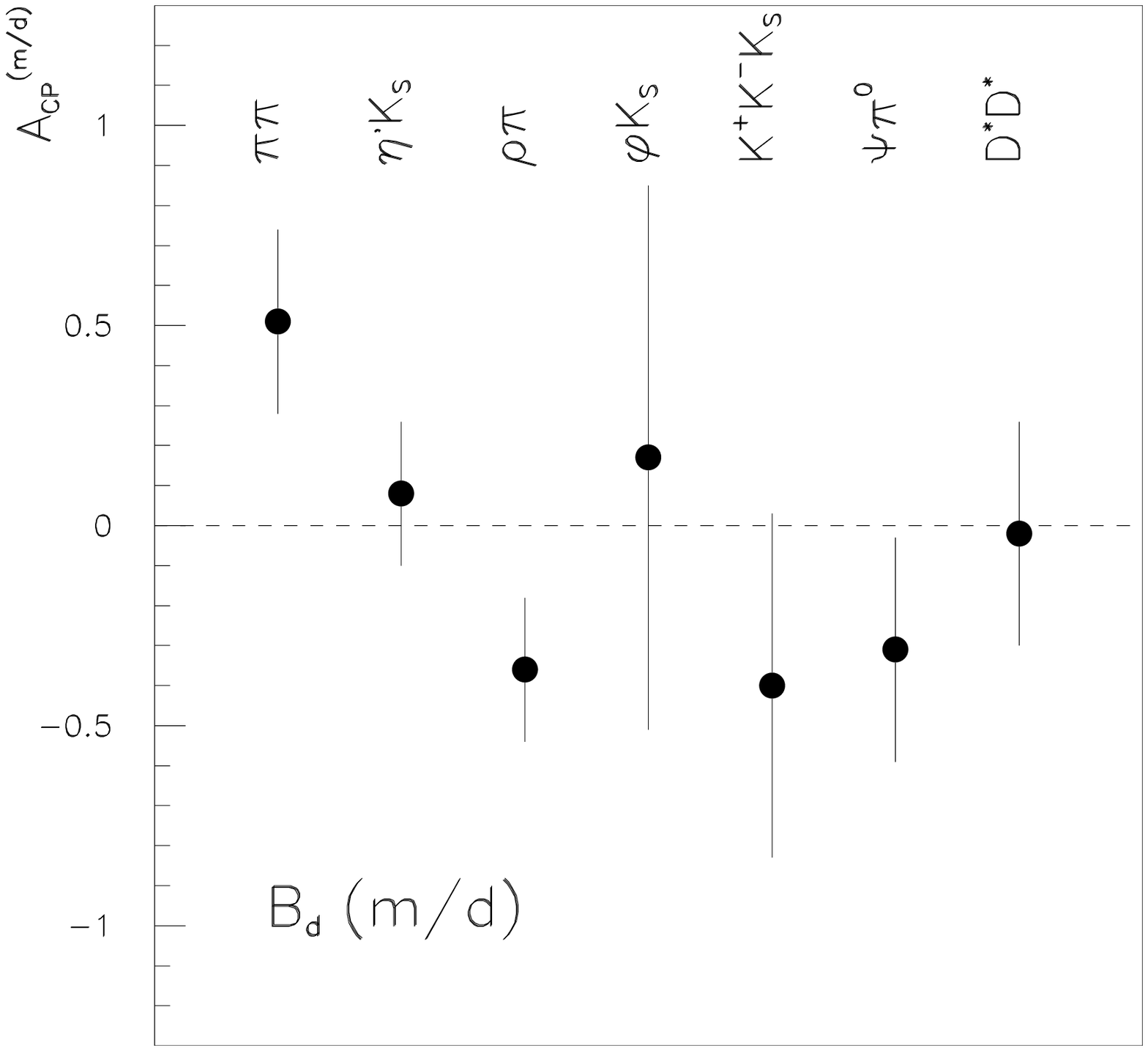,width=1.0\textwidth}
  \caption{Graphical representation of the coefficient of the cosine term in 
time-dependent $CP$ asymmetries for $B^0_d(\overline{B}^0_d)$ decays.}
  \label{fig:cos}
\end{center}
\end{figure}

Results on $CP$-violating asymmetries for several \emph{flavour-specific} final
states have been reported by CLEO (with $\approx 1 \cdot 10^7$ $B 
\overline{B}$ events), BABAR and BELLE (more than $8 \cdot 10^7$ $B 
\overline{B}$ events each, so far); no clear evidence of direct $CP$ violation 
has been found so far; table \ref{tab:Bdirect} and figures \ref{fig:acp_bc},
\ref{fig:acp_bn} summarise the available experimental information on the 
time-independent $CP$ asymmetries, while table \ref{tab:Bdirect_td} and figure 
\ref{fig:cos} refer to the direct $CP$-violating coefficient of the 
time-dependent ones.

Due to the limited size of the statistical samples, in some cases in which no
significant direct $CP$ violation is expected in the Standard Model, the 
time-dependent asymmetry is analysed by assuming $|\lambda_f|=1$ to get the 
most precise result on the underlying parameters of the theory, related to 
the mixing-induced $CP$-violating term. 

A hint of direct $CP$ violation in the $B$ system comes from the BELLE fit of 
the \mbox{$B_d^0 \To \pi^+ \pi^-$} time-dependent decay distribution 
\cite{BELLE_pipi}, in which a non-zero cosine term is found:
\Ea{
  & \mathcal{A}_{CP}^{\mathrm{(m/d)}} = +0.77 \pm 0.27 \pm 0.08 \\
  & \mathcal{A}_{CP}^{\mathrm{(int)}} = -1.23 \pm 0.41^{+0.08}_{-0.07}
}
(the first error being statistical and the second systematic). Such result 
is not confirmed by the BABAR analysis \cite{BABAR_pipi} of the same channel 
with similar statistics ($\sim 160$ signal events):
\Ea{
  & \mathcal{A}_{CP}^{\mathrm{(m/d)}} = +0.30 \pm 0.25 \pm 0.04 \\
  & \mathcal{A}_{CP}^{\mathrm{(int)}} = +0.02 \pm 0.34 \pm 0.05
}
The BELLE result for $\mathcal{A}_{CP}^{\mathrm{(m/d)}}$ is 2.7 standard 
deviations from zero, and corresponds to a 2.2 standard deviation indication 
for direct $CP$ violation, irrespective of the value of mixing-induced $CP$ 
violation. This result however lies outside the physically allowed region 
$\mathcal{A}_{CP}^{\mathrm{(m/d)}2} + \mathcal{A}_{CP}^{\mathrm{(int)}2} 
\le 1$; imposing the above constraint the result is 
$\mathcal{A}_{CP}^{\mathrm{(m/d)}} = +0.57$. 
The BABAR experiment quotes a 90\% confidence interval on the coefficient of 
the cosine term as $[-0.12, +0.72]$, and the naive average of the two results 
is $\mathcal{A}_{CP}^{\mathrm{(m/d)}} = 0.51 \pm 0.23$ after scaling the 
error to account for their poor consistency.
Clearly one has to wait for more data in order for the situation to be 
clarified.

\begin{table}
\begin{center}
\begin{tabular}{|l|c|c|l|}
\hline
Channel & $\eta_{CP}$ & Value & Notes \\
\hline
$B_d^0 \To \pi^+ \pi^-$  & +1 & $-0.47 \pm 0.61$ & 
BABAR \cite{BABAR_pipi}, BELLE \cite{BELLE_pipi} \\
$B_d^0 \To \eta' \ks$    & --1 & $0.34 \pm 0.34$ & 
BELLE \cite{BELLE_phiks} \\
$B_d^0 \To \phi \ks$     & --1 & $-0.38 \pm 0.41$ &
BABAR \cite{BABAR_Monchenault}, BELLE \cite{BELLE_phiks} \\
$B_d^0 \To K^+ K^- \ks$  & +1$(\dagger)$ & $0.49 \pm 0.55$ & 
BELLE \cite{BELLE_phiks} \\
$B_d^0 \To J/\Psi \pi^0$ & +1 & $-0.43 \pm 0.49$ & 
BABAR \cite{BABAR_psipi0}, BELLE \cite{BELLE_psipi0} \\
$B_d^0 \To J/\Psi \ks$ & --1 & $0.76 \pm 0.06$ & 
BABAR \cite{BABAR_ccs}, BELLE \cite{BELLE_ccs} \\
$B_d^0 \To J/\Psi \kl$ & +1 & $-0.75 \pm 0.12$ &
BABAR \cite{BABAR_ccs}, BELLE \cite{BELLE_ccs} \\
$B_d^0 \To \Psi' \ks$ & --1 & $0.69 \pm 0.24$ & 
BABAR \cite{BABAR_ccs} \\
$B_d^0 \To \Psi K^{*0}$ & +1$(\dagger)$ & $-0.15 \pm 0.40$ & 
BABAR \cite{BABAR_ccs} \\
  & & $-0.19 \pm 0.56$ & (Effective for CP=+1) \\
$B_d^0 \To D{^*+} D^{*-}$ & +1$(\dagger)$ & $0.32 \pm 0.47$ & 
BABAR \cite{BABAR_Sciolla} \\
  & & $0.37 \pm 0.55$ & (Effective for CP=+1) \\
$B_d^0 \To c\overline{c} \ks$    & --1 & $0.75 \pm 0.06$ & 
BABAR \cite{BABAR_ccs}, BELLE \cite{BELLE_ccs}\\
\hline
\end{tabular}
\caption{Measurements of the mixing-induced $CP$ violation coefficient 
$\mathcal{A}_{CP}^{\mathrm{(int)}}$ in time-dependent asymmetries of neutral 
$B$ meson decays to $CP$ eigenstates (from \cite{PDG_2002} unless otherwise 
indicated, see also \cite{HFAG}). Errors are scaled according to the PDG 
recipe. $(\dagger)$: dominant $CP$ eigenvalue.}
\label{tab:Binduced}
\end{center}
\end{table}

\begin{figure}[hbt!]
\begin{center}
    \epsfig{file=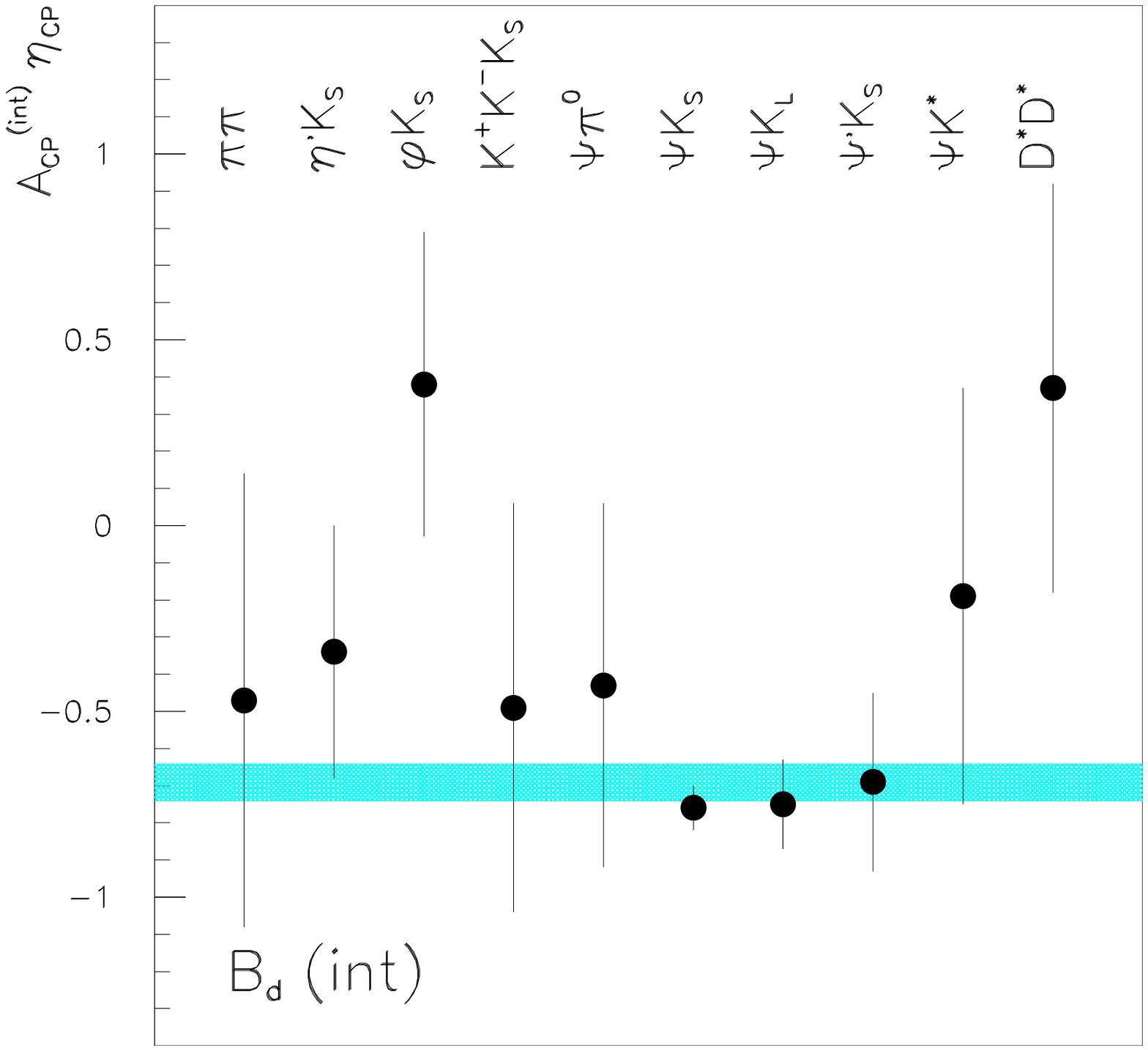,width=1.0\textwidth}
  \caption{Graphical representation of the measured coefficient of the sine 
term (multiplied by the $CP$-parity $\eta_{CP}$ of the final state) in 
time-dependent $CP$ asymmetries for $B^0_d(\overline{B}^0_d)$ decays to $CP$ 
eigenstates. The horizontal band represent the experimental average for all 
modes.}
  \label{fig:sin}
\end{center}
\end{figure}


As previously mentioned, precise measurements of mixing-induced asymmetries in 
decays to different $CP$ eigenstates, even if some of such asymmetries turn 
out to be zero (\emph{i.e.} no $CP$ violation), could give an indication of 
direct $CP$ violation, by the comparison with the precisely measured asymmetry 
in the charmonium-$\ks$ final state \cite{Bigi}. 
Table \ref{tab:Binduced} and figure \ref{fig:sin} summarise the available 
measurements: the probability of consistence is about 6\%, no significant 
evidence of direct $CP$ violation is seen yet. While the future increase in 
the statistical accuracy of the measurements is likely to reveal at some time 
differences among the above asymmetries, it is difficult to predict their 
pattern, since direct $CP$ violation at present eludes the theoretical efforts 
of a precise estimation.

The CLEO experiment also studied the inclusive process \mbox{$B \To X_s 
\gamma$}, where $X_s$ is any final state containing a strange quark 
\cite{CLEO_bsg}, by considering events with high energy photons ($2.2 \div 
2.7$ GeV): in this region of phase space there is little background from other 
$B$ decay processes, although a significant continuum contribution has to be 
subtracted, by using several event shape variables. The measured (null) 
asymmetry also contains a small contribution (estimated to be $\sim$ 2\%) from 
a possible \mbox{$b \To d \gamma$} asymmetry.

The experimental study of the heavier $B^0_s$ and $B^\pm_c$ mesons requires 
centre of mass energies above the $\Upsilon(4S)$. $B^0_s$ mesons have been
studied both at $e^+ e^-$ (LEP, SLD) and $p \overline{p}$ (CDF) collider 
experiments; no signal of flavour oscillations has been identified yet, and 
no $CP$ violation effects have been reported.
These systems and their $CP$ violation effects will be studied extensively at 
future hadron collider experiments, LHC-b \cite{LHCb} (but also ATLAS, CMS) at 
the CERN LHC and B-TeV \cite{BTeV} at the Fermilab TeVatron.

High-energy hadron colliders have the advantage of large $b\overline{b}$
production cross sections (100-500 $\mu$b), but the number of unrelated 
particles produced in each collision will be very large, making the analysis 
rather challenging.
The measured decay-rate asymmetries have to be corrected for the detection 
efficiency asymmetries, and at pp colliders also for the different production 
rates of $\bz$ and $\bzb$ mesons.

The trigger systems, based on transverse momentum and secondary vertexes, will 
be key elements for the physics performance; B-TeV plans to use the vertex 
detector already at the first trigger level.
Both the LHC-b and the B-TeV detectors are single-arm spectrometers, with 
vertex detectors based on silicon strips for the former and silicon pixels for 
the latter; both experiments have aerogel RICH's for particle ID and are 
expected to run at $\sim 2 \cdot 10^{32}$ cm$^{-2}$ s$^{-1}$ luminosity, 
starting in 2007.


\section{Other systems}

\subsection{Hyperon decays}

Soon after the discovery of parity violation, it was realized 
\cite{HyperonsCP} that weak hyperon decays could be an interesting laboratory 
for studying discrete symmetry violations.
Since baryon number conservation effectively forbids mixing, any $CP$ 
violation effect in baryon decays would be a signal of direct $CP$ violation.

As for kaons, transitions described by gluonic penguin diagrams are thought to 
give rise to ineliminable phase differences between the decay amplitudes of 
hyperons and anti-hyperons.
In the decays of $\Xi$ and $\Lambda$, such differences can be measured through 
the interference between amplitudes of different final-state angular 
momentum ($S$ and $P$ wave), with different final-state interaction phases.
Parity violation induces observable asymmetries in the angular decay 
distribution of final states from polarized hyperons, and in the polarization
of the final state particles themselves.
A single amplitude strongly dominates hyperon semi-leptonic decays in the SM,
so that no significant $CP$ violation effects are expected there.

In the non-leptonic decay \mbox{$Y \To B \pi$} (where $Y$ is a hyperon and 
$B$ a baryon), if the parent $Y$ is polarized, the angular distribution of $B$ 
in the rest frame of the parent (averaging over the daughter polarization 
states) is non-isotropic:
\Dm{
  \frac{dN}{d\Omega} = \frac{1}{4\pi} (1+ \alpha_Y \Vec{P}_Y \cdot 
    \hat{\Vec{p}}_B)
}
and its polarization vector is
\Dm{
  \Vec{P}_B = 
    \frac{(\alpha_Y + \hat{\Vec{p}}_B \cdot \Vec{P}_Y) \hat{\Vec{p}}_B - 
    \beta_Y (\hat{\Vec{p}}_B \times \Vec{P}_Y) - 
    \gamma_Y \hat{\Vec{p}}_B \times (\hat{\Vec{p}}_B \times \Vec{P}_Y)}
    {1+\alpha_Y + \hat{\Vec{p}}_B \cdot \Vec{P}_Y}
}
where $\Vec{P}_{Y,B}$ are the polarization vectors of the hyperon and 
secondary baryon, and $\hat{\Vec{p}}_B$ is the unit vector of the secondary 
baryon momentum in the parent hyperon rest frame.
If the parent hyperon $Y$ is unpolarized, the polarization of $B$ reduces to
$\Vec{P}_B = \alpha_Y \hat{\Vec{p}}_B$, and its angular distribution is 
isotropic in the rest frame of the parent.
$\alpha_Y, \beta_Y, \gamma_Y$ are the decay asymmetry parameters ($\alpha_Y^2 
+ \beta_Y^2 + \gamma_Y^2 = 1$), related to the 
interference of the $S$ and $P$ wave decay amplitudes $A_S$ and $A_P$:
\Ea{ 
  & \alpha_Y = \frac{2 \Rea(A_S^* A_P)}{|A_S|^2+|A_P|^2} \\
  & \beta_Y  = \frac{2 \Imm(A_S^* A_P)}{|A_S|^2+|A_P|^2} \\
  & \gamma_Y = \frac{|A_S|^2-|A_P|^2}{|A_S|^2+|A_P|^2} 
}
Since $\alpha_Y$ and $\beta_Y$ change sign under a $CP$ transformation, when
comparing the decays of a hyperon and its antiparticle one can form three 
$CP$-violating observables:
\Ea{ 
  \Ds \Delta \equiv \frac{\Gamma_Y-\Gamma_{\overline{Y}}}
    {\Gamma_Y+\Gamma_{\overline{Y}}} \quad &
  \Ds A \equiv \frac{\alpha_Y+\alpha_{\overline{Y}}}
    {\alpha_Y-\alpha_{\overline{Y}}} & \quad 
  \Ds B \equiv \frac{\beta_Y+\beta_{\overline{Y}}}
    {\beta_Y-\beta_{\overline{Y}}}
}
where $\Gamma_Y \propto |A_S|^2 + |A_P|^2$ is the partial decay rate, and 
$\overline{Y}$ refers to the anti-hyperon decay.
The $A,B$ observables measure direct $CP$ violation as the difference in the
amount of parity violation in the decays of particles and anti-particles.

Also for hyperon decays, theoretical predictions are difficult because of 
hadronic uncertainties. The $A_{\Xi}$ and $A_{\Lambda}$ asymmetry parameters 
range from $10^{-4}$ to $10^{-5}$ in the Standard Model, and they can be 
larger in some super-symmetric models, even when direct $CP$ violation in kaon 
decays is still predicted to be small.

Experimentally, $A$ is the more accessible observable, requiring the 
measurement of the daughter polarization in the decay of unpolarized hyperons,
or of the angular asymmetry parameters in the decay of polarized hyperons. 
$B$ is generally predicted to be larger than $A$, and $B' \equiv (\beta_Y-
\beta_{\overline{Y}})/(\alpha_Y-\alpha_{\overline{Y}})$ is also independent of 
the final-state interaction phases, while $\Delta$ is predicted to be much 
smaller.

Early approaches to the measurement of $CP$ violation in hyperon decays focused
on the measurement of the $A$ asymmetry in $\Lambda,\overline{\Lambda}$ decays.
Three experiments were performed, and the statistically-limited null results 
have a precision of 1-2\% (see table \ref{tab:hyperons}).

If the secondary baryon is itself a hyperon, one can exploit the analyzing 
power of its parity-violating weak decay to get information on its 
polarization without performing any spin measurement.
Considering the decay of unpolarized \mbox{$\Xi^- \To \Lambda \pi^-$} 
followed by \mbox{$\Lambda \To p \pi^-$}, the angular decay distribution of 
the proton (averaging over its unmeasured polarization) actually measures the 
$\Lambda$ polarization, induced by the $\alpha_{\Xi}$ parameter:
\Dm{
  \frac{dN}{d \cos \theta_{\Lambda}} \propto 
  1 + \alpha_{\Xi} \alpha{\Lambda} \cos \theta_{\Lambda}
}
where $\theta_{\Lambda}$ is the angle between the proton direction in the
$\Lambda$ rest frame and the $\Lambda$ direction in the $\Xi$ rest frame.
In this way the slope of the angular anisotropy of the protons in the 
$\Lambda$ rest frame actually allows a measurement of the asymmetry
\Dm{
  A_{\Xi \Lambda} = 
    \frac{\alpha_{\Xi} \alpha_{\Lambda} -
    \alpha_{\overline{\Xi}} \alpha_{\overline{\Lambda}}}
         {\alpha_{\Xi} \alpha_{\Lambda} +
    \alpha_{\overline{\Xi}} \alpha_{\overline{\Lambda}}} \simeq
    A_{\Xi} + A_{\Lambda}
}

Each of the $A$ asymmetries requires strong phase differences for the 
interfering decay amplitudes: while the $p\pi^-$ phase shift difference is 
measured to be $\delta_P-\delta_S = (7.1 \pm 1.5)^\circ$, the $\Lambda \pi$
one is not measured and is predicted to be small, so that $A_{\Xi \Lambda}$ 
should be dominated by $A_{\Lambda}$.

The decays of $\Xi^-$ and $\overline{\Xi}^+$ discussed above have been 
studied with the CLEO II detector at the CESR $e^+ e^-$ storage ring, running 
at a centre of mass energy just below the $\Upsilon(4S)$ resonance. 
Inclusively-produced $\Xi^-$ (and $\overline{\Xi}^+$) were reconstructed in 
the $\Lambda \pi^-$ decay mode\footnote{Here and after the two $CP$-conjugate 
modes are implied.}, with \mbox{$\Lambda \To p \pi^-$}; particle 
identification was provided by a combination of ionization energy loss and 
time-of-flight information.
The number of reconstructed $\Xi$ events after subtracting the $\approx$ 7\%
combinatorial background was $8.4 \cdot 10^3$.
An unbinned maximum likelihood fit to the $\cos \theta_{\Lambda}$ distribution 
was then performed using a fit function determined by a large Monte Carlo 
sample, generated with fixed values of the asymmetry parameters and suitably 
weighted.
Several systematic effects in the measurement do cancel in forming the 
asymmetry, which is also insensitive to any $\Xi$ production polarization.
The result \cite{CLEO_hyperons} is 
\Dm{
  A_{\Xi \Lambda} = (-5.7 \pm 6.4 \pm 3.9) \cdot 10^{-2}
}
the first error being statistical and the second systematic.

The HyperCP (E781) experiment at Fermilab is dedicated to the measurement of 
the combined $CP$ asymmetry parameter $\alpha_{\Xi\Lambda}$ using the same 
decay chain.
$\Xi^-$ (and $\overline{\Xi}^+$) are produced at 0$^\circ$ by $7.5 \cdot 10^9$
protons/s (of 800 GeV/$c$ momentum) on a fixed target, and charged secondaries 
of 150 GeV/$c$ $\pm$ 25\% are magnetically selected. 
Parity conservation in the strong interaction induced production process 
enforces the $\Xi$ polarization to be zero when they are produced at zero 
angle.
After a 13 m evacuated decay volume, the decay products are detected in a 
MWPC-based magnetic asymmetric spectrometer. 
The polarities of the collimator and spectrometer magnet were periodically
reversed to switch from $\Xi^-$ to $\overline{\Xi}^+$ running.
The data collected in 1997 and 1999 contains about $2 \cdot 10^9$ $\Xi^-$ and 
$0.5 10^{-9} \overline{\Xi}^+$ reconstructed decays, corresponding to a 
statistical sensitivity of $\sim 2 \cdot 10^{-4}$ on $A_{\Xi \Lambda}$.

The fact that the $\Lambda$ helicity frame used to define the asymmetry changes
from event to event reduces any sensitivity to local acceptance asymmetries
of the detector. Two different analysis techniques are being pursued: one which
uses Monte Carlo samples to correct for the detector acceptance, and one in 
which no acceptance correction is applied and the events are weighted to reduce
any difference between the $\Xi^-$ and $\overline{\Xi}^+$ samples.

The preliminary analysis of a sample corresponding to $\approx$ 2\% of the
total has been presented \cite{HyperCP}, with the result
\Dm{
  A_{\Xi \Lambda} = (-0.7 \pm 1.2 \pm 0.6) \cdot 10^{-3}
}
where the first error is statistical and the second systematic, being presently
limited by statistics. 

The largest systematic effects affecting this preliminary result are the 
imperfect reversal of the magnetic fields in the apparatus and the effect of 
the earth's one, the imperfect symmetry of the detector efficiencies, and 
effects related to rates or the different interaction properties of 
$\pi^+$ vs. $\pi^-$ and $p$ vs. $\overline{p}$. Backgrounds, at the level of 
0.3\% are also different for the two $CP$ conjugate states. 
The understanding of several of the above systematic effects is limited by the 
size of the analyzed sample and is therefore expected to improve.

When its full data sample will be completely analyzed, the HyperCP 
experiment will provide by far the best test of (direct) $CP$ asymmetries in 
hyperon decays.
However most probably it will not reach the sensitivity corresponding to the 
Standard Model predictions; hence (although theoretical difficulties linked to 
hadronic physics makes precise predictions difficult) any positive signal 
would be a signal of new physics.

\begin{table}
\begin{center}
\begin{tabular}{|l|c|c|}
\hline
Reaction & Asymmetry & Notes \\
\hline
$p\overline{p} \To \Lambda X$ &
$A_{\Lambda} = -0.02 \pm 0.14$ & CERN R608 (1985) \\
$e^+e^- \To J/\Psi \To \Lambda \overline{\Lambda}$ &
$A_{\Lambda} = 0.01 \pm 0.10$ & Orsay DM2 (1988) \\
$p\overline{p} \To \Lambda\overline{\Lambda}$ &
$A_{\Lambda} = 0.013 \pm 0.022$ & CERN PS185 (1996) \\
$p \mathrm{Be} \To \Xi^- X \To \Lambda \pi^- X$ &
$A_{\Xi \Lambda} = 0.012 \pm 0.014$ & FNAL E756 (2000) \\
$e^+ e^- \To \Xi^- X \To \Lambda \pi^- X$ &
$A_{\Xi \Lambda} = -0.057 \pm 0.075$ & 
CLEO prel. (2000) \cite{CLEO_hyperons} \\
$p \mathrm{Be} \To \Xi^- X \To \Lambda \pi^- X$ &
$A_{\Xi \Lambda} = (-0.7 \pm 1.4) \cdot 10^{-3}$ &
HyperCP prel. (2002) \cite{HyperCP} \\
\hline
\end{tabular}
\caption{Experimental results on $CP$ violating asymmetries in hyperon decays 
(from \cite{PDG_2002} unless otherwise indicated).}
\label{tab:hyperons}
\end{center}
\end{table}

Table \ref{tab:hyperons} summaries the experimental information available on
(direct) $CP$ violation asymmetries in hyperon decays; no signals of (direct) 
$CP$ violation have been found so far.

As mentioned above, similar observables would be accessible by studying the 
decays of heavy baryons ($\Lambda_b$, $\Xi_b$), produced at hadron machines or 
future $e^+ e^-$ colliders running at the $Z$ centre of mass energy.

$T$-odd triple product correlations in two-body decays, involving the hyperon 
polarization (in general non-zero in production) could have large values in 
the SM \cite{Bensalem}: one example is $\langle \Vec{p}_K \cdot 
\Vec{S}_{\Lambda_b} \times \Vec{S}_p \rangle$ for the decay \mbox{$\Lambda_b 
\To p K^-$}, which however requires the measurement of proton polarization; 
for similar decays such as \mbox{$\Xi_b \To \Sigma^+ K^-$} one could exploit 
the self-analyzing decay of the secondary hyperon to build $T$-odd 
correlations using the final state momenta.
Other kinds of $T$-odd correlations, not involving the polarization of the 
parent particle, are also possible for heavy baryon two-body decays, since 
their high mass allows vector mesons in the final states: asymmetries 
involving the polarization of both final state particles in \mbox{$\Lambda_b 
\To p K^{*-}$} decays are an example, although the SM predicts tiny 
asymmetries in such channels.
No significant $CP$ violation is expected in the SM for $T$-odd correlations 
in (three-body) semi-leptonic decays, where a single amplitude strongly 
dominates.

A possible dependence of the angular decay distribution on the $T$-odd 
correlation $\langle \Vec{S}_\Sigma \cdot (\Vec{p}_e \times \Vec{p}_\nu) 
\rangle$ in the semi-leptonic \mbox{$\Sigma^- \To n e^- \overline{\nu}$} decay 
was also probed at the FNAL hyperon beam, with a null result for the relevant 
coefficient $C = (0.11 \pm 0.10)$.

\subsection{Lepton decays}

The recent increasing evidence for neutrino oscillations led to a 
re-examination of the possibility of $CP$ violation in the leptonic sector. 
Such a phenomenon is not present in the Standard Model for charged leptons, 
but appears in several of its extensions; models predicting lepton flavour 
violation often imply also $CP$ violation in lepton decays, usually as a 
consequence of the interference between transitions mediated by (some 
additional) scalar boson and the ones induced by the weak currents.

The analysis of muon decay parameters shows no signs of $CP$ violation in that 
process, such as the $e^+$ polarization transverse to the plane defined by
the $\mu^+$ polarization and the $e^+$ momentum: $P_T(e^+) = 0.007 \pm 0.023$
\cite{PDG_2002}; scalar or tensor interactions would be required for such a 
polarization to appear.

$CP$ violation in semi-hadronic $\tau$ decays has also been investigated 
recently (see \emph{e.g.} \cite{CLEO_tau}).
By postulating a term in the decay amplitude mediated by scalar exchange, the 
relative phase of its coupling constant with respect to the Standard Model 
($W^\pm$) term induces $CP$-odd terms in the decay distributions (which as 
usual require a difference in the strong phases for the vector and scalar 
exchange in order to be non-zero).
Both spin-dependent and spin-averaged $CP$-odd terms can be present; the former
can be measured by exploiting the spin correlation of $\tau$ leptons produced 
in \mbox{$e^+e^- \To \tau^+ \tau^-$}. 
Several $CP$-odd observables can be considered, and their average values have 
been found to be consistent with zero, with errors which reach at best the 2\% 
level in \mbox{$\tau^\mp \To \pi^\mp \pi^0 \nu_\tau (\overline{\nu}_\tau)$} 
and \mbox{$\tau^\mp \To \kz \pi^\mp \nu_\tau (\overline{\nu}_\tau)$} decays.

\subsection{Neutrino oscillations}

The properties of neutrinos are experimentally difficult to determine, since 
they are only affected by the weak interaction. On the other hand, this same
property makes them a very interesting system to extract information on the 
fundamental parameters of such interaction, without complications due to 
strong or electromagnetic effects; moreover the relationship with the theory 
is much more direct, since neutrinos are closer to being measurable free 
states than quarks, which are permanently confined within hadrons in a 
complicated and poorly known way. 

Evidence for neutrino flavour oscillations (and therefore, indirectly, for 
non-degenerate neutrino masses) is now compelling, and therefore the issue of 
$CP$ violation in the lepton sector is being actively considered.
If neutrinos are non-degenerate in mass, a lepton mixing matrix which 
describes the relation of the physical (mass) eigenstates to the flavour 
eigenstates can be defined; as in the case of quarks, complex matrix elements 
could induce $CP$ violation effects, and with three lepton flavours one complex
phase remains which cannot be eliminated by field redefinitions. 
$CP$ violation could be expected to give large effects in such a system, the 
intrinsic properties of which are completely determined by the $CP$-violating 
weak interactions. Moreover, results on neutrino oscillations indicate that, 
contrary to the case of quarks, one or more of the neutrino mixing angles 
could be large: it should be kept in mind that $CP$-violating effects require 
all three angles to be non-zero, while confirmed experimental indications of 
such a property only exist for two of them, as determined by solar and 
atmospheric neutrino experiments.

If neutrinos are Majorana particles (self-conjugate), two additional 
$CP$-violating phases appear in the mixing matrix (for 3 flavours); such 
phases do not produce any observable effect in oscillations, and only affect 
the rates for neutrinoless double-beta decay: the observation of such process, 
together with independent information on individual neutrino masses, could in 
principle allow to extract information on these phases, although this might 
turn out to be difficult in practice \cite{NoMajorana}.


$CP$ violation effects \cite{CPneutrinos} can be present both in the time 
evolution of neutrinos in flavour space (oscillations) and in their 
interactions with matter (detection); the latter is clearly problematic to 
study, due to the macroscopic $CP$ asymmetry of the detection system.
$CP$ violation in neutrino oscillations can be experimentally measured by
comparing the probability of a neutrino, prepared in a given flavour state, 
to be detected after propagation in vacuum as a given different flavour, with 
the same probability for an antineutrino:
\Dm{
  P(\overline{\nu}_\alpha \To \overline{\nu}_\beta) \neq 
  P(\nu_\alpha \To \nu_\beta)
}
indicates $CP$ violation, if $CPT$ symmetry is valid.

As usual, $CPT$ symmetry imposes some constraints on the relations among the 
above probabilities: among these is the fact that $CP$ violation effects in 
neutrino oscillations, implying differences in transition probabilities between
different flavour states, can be studied only in appearance experiments, since
the probability of a neutrino to remain in the same flavour eigenstate,  
measured in disappearance (reactor, solar, atmospheric) experiments, is equal 
to that for an antineutrino, if $CPT$ symmetry is valid. 

In order to have observable effects, it is also necessary that all three 
families do actively participate to the oscillations: one of the mixing angles 
is however known not to be large ($\sin(\vartheta_{13})<0.16$).
Moreover, if the hierarchy of neutrino masses is such that, for a given 
experimental arrangement, one has a negligible effect from one of the mass 
differences ($\Delta m_{ij}^2L/2E \ll 1$, where $\Delta m_{ij}$ ($i,j=1,2,3$) 
is the mass difference between the two neutrino species, $L$ the distance 
between the production and detection point, and $E$ the average neutrino 
energy), the $CP$-violating asymmetry becomes negligible.
It is usually assumed therefore that only long-baseline experiments, sensitive 
to all three neutrino masses, can aim at directly measuring $CP$ violation 
effects with neutrinos \cite{Bernabeu}.

The measurable $CP$ violation effects in neutrino oscillations arise from the
interference of two neutrino mixing amplitudes with different weak phases and
different oscillation phases (due to the mass differences).
The $CP$-asymmetric matter effects in long-baseline experiments add more
complications, so that the simple $CP$-odd asymmetries between oscillation
probabilities are no longer a sufficient signal for $CP$ violation.

The best and less model-dependent approach to search for $CP$ violation in 
neutrino oscillations seems to be the direct search for a $CP$ asymmetry 
between $\nu_e \To \nu_\mu$ transitions and their $CP$-conjugates. This could 
be done exploiting future intense muon storage facilities (``neutrino 
factories''), presently under study (see \emph{e.g.} \cite{Mezzetto}), 
providing very large fluxes of neutrinos, with only two different flavours and 
opposite helicity ($\nu_\mu, \overline{\nu}_e$ or $\overline{\nu}_\mu, 
\nu_e$), searching for the appearance of wrong sign muons and comparing the 
results for runs with opposite muon charges.

The sensitivity to $CP$-violating asymmetries when using intense conventional
$\nu_\mu$ neutrino beams would be limited by the intrinsic contamination of 
$\overline{\nu}_\mu$ ($\sim$ 1\%) and $\nu_e$ ($\sim$ 0.5\%), although this 
can be accurately predicted for a primary proton energy below the kaon 
production threshold.

The search of $CP$ violation in neutrinos is clearly a very interesting field 
of research, since its detection would establish that such phenomenon is not a 
peculiarity of quarks. 
However, precise estimates of the size of expected effects require the 
knowledge of all the neutrino mixing angles, and it is clear that $CP$ 
violation studies will require new facilities for sensitive studies.

\subsection{Electric dipole moments}

Studies of $CP$ violation are not restricted to unstable elementary particles, 
but have been pursued, both in experiments and in theory, also for the
constituents of ordinary matter. 
The most relevant example is the search for static, permanent electric dipole 
moments (EDM) of elementary, non-degenerate systems. Such a quantity is odd 
under both parity and time reversal transformations, and it indicates a 
violation of those symmetries if it has a non-zero value: in such systems the 
EDM vector can only be parallel to the only intrinsic vector, \emph{i.e.} 
the particle spin, which however has opposite transformation properties with 
respect to $P$ and $T$. 
The search for electric dipole moments of elementary particles has a long 
history dating back to the '50s; an exhaustive discussion of the subject 
can be found in \cite{Khriplovich}.

Experimental approaches are based on detecting the electric ``Zeeman effect''
when the system is placed in an external electric field, and the key issue is 
the reduction of spurious effects in the measurement; a typical measurement is 
the change in the Larmor precession frequency of the system for parallel or
anti-parallel magnetic and electric fields.

To avoid the experimental complications of working with a particle with 
non-zero electric monopole moment (charge), neutral particles were mostly 
studied: several experiments are devoted to the measurement of the neutron 
EDM with ultra-cold neutrons ($T \sim$ 2 mK). 
In an atomic system the rearrangement of charges in an external electric 
field could shield its effects, but in heavy paramagnetic atoms relativistic 
effects lead instead to significant enhancements of the EDM induced by 
valence electrons, and very sensitive searches are performed to search for EDM 
in Cs and $^{205}$Tl atoms. EDM in diamagnetic atoms would be driven instead 
by the nuclear spin direction, and searches in $^{129}$Xe, $^{199}$Hg and 
other atoms, and also heavy polar molecules (such as TlF), are an active 
target of study.

\begin{table}
\begin{center}
\begin{tabular}{|c|c|}
\hline
$e^-$      & $d_e = (6.9 \pm 7.4) \cdot 10^{-28} \, e$ cm \\
$\mu^\pm$  & $d_\mu = (3.7 \pm 3.4) \cdot 10^{-19} \, e$ cm \\
$\tau^\pm$ & $-2.2 \cdot 10^{-17} < \Rea(d_\tau) < 4.5 \cdot 10^{-17} \, e$ 
cm (95\% CL) \cite{Belle_dtau} \\
           & $-2.5 \cdot 10^{-17} < \Imm(d_\tau) < 0.8 \cdot 10^{-17} \, e$ 
cm (95\% CL) \cite{Belle_dtau} \\
$p$        & $d_p = (-3.7 \pm 6.3) \cdot 10^{-23} \, e$ cm \\
$n$        & $d_n < 6.3 \cdot 10^{-26} \, e$ cm (90\% CL) \\
\hline
\end{tabular}
\caption{Experimental limits on electric dipole moments of elementary
particles, from \cite{PDG_2002} unless otherwise indicated.}
\label{tab:edm}
\end{center}
\end{table}

No positive signal of permanent EDM has been found so far; table \ref{tab:edm} 
summaries the current experimental limits.

If $CPT$ symmetry is valid, $T$ violation is equivalent to $CP$ violation, but 
in this case its classification as direct or indirect discussed above is not 
really appropriate: non-zero EDMs would be a signal of $T$ violation in the  
static properties of a system and its interaction with the environment.

\subsection{Production asymmetries}

Electric dipole moments of unstable leptons have also been searched for by
studying their coupling to photons in $e^+ e^-$ formation: the most general
Lorentz invariant electromagnetic coupling is described by 
\Dm{
  F_1(q^2) \gamma^\mu + F_2(q^2) \frac{i}{2m_l} \sigma^{\mu \nu} q_\nu - 
  F_3(q^2) \sigma^{\mu \nu} \gamma_5 q_\nu
}
where $\gamma_\mu, \gamma_5$ e $\sigma^{\mu \nu}$ are Dirac matrices and their
combinations, $m_l$ is the lepton mass and $q_\nu$ the photon four-momentum. 
$F_1(0)$ is the electric charge, $F_2(0)$ the anomalous magnetic moment 
$(g-2)/2$ and $F_3(0)$ the ratio of electric dipole moment and electric 
charge, which is zero if $P$ or $T$ symmetry holds.

The study of the differential cross section for the reaction 
\mbox{$e^+ e^- \To \tau^+ \tau^-$} (when dominated by virtual photon 
production) allows to put limits on the $\tau$ electric dipole moment. 
$T$ violation would allow $\tau^+ \tau^-$ production in a $CP$-odd state 
which, thanks to the relatively long lifetime of the $\tau$ lepton and the 
fact that its decay does not depend from the EDM value, would manifest as 
different momentum correlations, with respect to the dominant production 
mechanism in a $CP$-even state, for the final states from the decay of the 
$\tau$ lepton pair. 
Photon angular and energy distributions in the $e^+ e^- \To \tau^+ \tau^- 
\gamma$ state are also sensitive to the electric dipole moment.

When the production mechanism is through a virtual $Z$, several $T$-odd 
observables sensitive to anomalous couplings of the $Z$ boson (weak dipole 
moment) can be built \cite{Nachtmann}, by using the $\tau$ polarization 
vectors which can be determined by their self-analyzing decays; polarized 
beams would allow more $T$-odd correlations to be probed.

Anomalous couplings of the electroweak gauge bosons, $W^+ W^- Z \gamma$,
$Z \gamma \gamma^*$, $ZZ^* \gamma$, $ZZZ^*$, have been studied in $e^+ e^-$ 
and $p \overline{p}$ interactions with several final states, by detailed 
analysis of production cross sections and angular or energy distributions.
Such anomalous couplings, which are zero at tree level in the Standard Model,
could in principle introduce $CP$ violation effects, for which limits have been
set, mostly by LEP \cite{PDG_2002}.
No $CP$-violating effects have been detected so far.

$CP$ violation measurements in the $t \overline{t}$ production processes at 
present $p\overline{p}$ colliders ($10^4 \div 10^5$ $t\overline{t}$/year at the
TeVatron) or future $pp, e^+ e^-, \mu^+ \mu^-, \gamma \gamma$ colliders 
($10^7 \div 10^8$ $t\overline{t}$/year at LHC, $10^5 \div 10^6$ 
$t\overline{t}$/year at a future $e^+ e^-$ linear collider), have been 
discussed in the literature (see \cite{toptop} for a comprehensive review).
The study of top quark production presents some advantages: due to its large 
mass, this quark decays before hadronizing, so that its production and decay 
processes are not masked by non-perturbative physics; this also means that any 
$CP$ violation in these processes is of the direct type. For the same reason, 
the spin of the top quark is an important observable (just as for leptons) 
which can be analyzed in its decay, particularly the semi-leptonic ones 
(\mbox{$t \To b l^+ \nu$}).
Moreover, while $CP$-violating effects in top-quark production are expected to
be small in the Standard Model, the large mass of the $t$ makes such 
processes highly sensitive to several other possible sources of $CP$ violation
from new physics.

If the \mbox{$t \To b W^+$} decay dominates the other decay modes of the top 
quark, partial rate asymmetries are expected to be small, as they would vanish 
due to $CPT$ symmetry in the limit of a single decay mode; other decay modes 
would have larger asymmetries, but their rates are expected to be so small 
to make measurements difficult.
Decay asymmetries which are not suppressed by the above argument are the ones 
obtained by integrating only over a part of the phase space; one example is 
the lepton average energy asymmetry in \mbox{$t \To b l \nu$} decays
\Dm{
  A_{CP}^{(E)} \equiv \frac{\langle E(l^+) \rangle - \langle E(l^-) \rangle}
  {\langle E(l^+) \rangle + \langle E(l^-) \rangle}
}
Such partially-integrated rate asymmetries in semi-leptonic decays turn out to 
be proportional to the lepton mass, so that the $l=\tau$ channel is usually 
considered. Asymmetries involving the lepton polarization in semi-leptonic 
decays are expected to be larger, in a model-independent way; examples are
\Ea{ \Ds
  & A_y \equiv \frac{N(\tau^+;S_y>0)-N(\tau^+;S_y<0)+
    N(\tau^-;S_y>0)-N(\tau^-;S_y<0)}
  {N(\tau^+;S_y>0)+N(\tau^+;S_y<0)+N(\tau^+;S_y>0)-N(\tau^-;S_y<0)} \\
  & A_z \equiv \frac{N(\tau^+;S_z>0)-N(\tau^+;S_z<0)-
    N(\tau^-;S_z>0)+N(\tau^-;S_z<0)}
  {N(\tau^+;S_z>0)+N(\tau^+;S_z<0)+N(\tau^-;S_z>0)+N(\tau^-;S_z<0)}
}
where the $\tau$ spin components $(S_x,S_y,S_z)$ refer to a $\tau$ rest frame 
in which the $\tau$ momentum is in the $-x$ direction and the $y$ axis is in 
the decay plane along the $b$ direction.

Other production asymmetries which could be studied at an $e^+ e^-$ collider 
are \cite{toptop} 
\Ea{
  & \Ds A_{CP}^{(C)} \equiv \frac{ 
    \int_{\theta_0}^{\pi-\theta_0} d\theta_l 
    \left( \frac{d\sigma^+}{d\theta_l} - 
    \frac{d\sigma^-}{d\theta_l} \right)}
    { \int_{\theta_0}^{\pi-\theta_0} d\theta_l 
    \left( \frac{d\sigma^+}{d\theta_l} + 
    \frac{d\sigma^-}{d\theta_l} \right)} \\
  & \Ds A_{CP}^{(FB)} \equiv \frac{ 
    \int_{\theta_0}^{\pi/2} d\theta_l
    \left( \frac{d\sigma^+}{d\theta_l} + 
    \frac{d\sigma^-}{d\theta_l} \right) - 
    \int_{\pi/2}^{\pi-\theta_0} d\theta_l
    \left( \frac{d\sigma^+}{d\theta_l} + 
    \frac{d\sigma^-}{d\theta_l} \right)}
    { \int_{\theta_0}^{\pi-\theta_0} d\theta_l
    \left( \frac{d\sigma^+}{d\theta_l} + 
    \frac{d\sigma^-}{d\theta_l} \right)} \\
  & \Ds A_{CP}^{(UD)} \equiv \frac{1}{2\sigma(\theta_0)}
    \int_{\theta_0}^{\pi-\theta_0} d\theta_l \left(
    \frac{d\sigma(\Vec{p}_{ly}>0)}{d\theta_l} -
    \frac{d\sigma(\Vec{p}_{ly}<0)}{d\theta_l} \right) \\
  & \Ds A_{CP}^{(LR)} \equiv \frac{1}{2\sigma(\theta_0)}
    \int_{\theta_0}^{\pi-\theta_0} d\theta_l \left(
    \frac{d\sigma(\Vec{p}_{lx}>0)}{d\theta_l} -
    \frac{d\sigma(\Vec{p}_{lx}<0)}{d\theta_l} \right) 
}
where $\sigma^\pm$ are the inclusive cross section for producing positive or
negative leptons ($\sigma = \sigma^+ + \sigma^-$), $\theta_l$ is the 
lepton polar angle with respect to the incident beams' axis, $\theta_0$ a 
suitable cutoff angle, and $\Vec{p}_{lx}, \Vec{p}_{ly}$ are the components of 
the lepton momentum in the centre of mass system in which the $z$ axis is 
defined by the incident particles, the $x$ axis is within the $t\overline{t}$ 
production plane and the $y$ axis is orthogonal to it.

Asymmetries involving top quark polarization have also been considered, such 
as 
\Dm{
  A_S \equiv \frac{\sigma_{LL}-\sigma_{RR}}{\sigma}
}
where the subscripts refer to the $t$ and $\overline{t}$ helicities, which 
could be measured from the angular distributions of the decay products: 
considering \emph{e.g.} semi-leptonic decays \mbox{$t \To b l \nu$}, the 
correlation of top-quark polarization and lepton direction is maximal.

The Standard Model does not predict measurable values for the above 
asymmetries, which could be therefore used as probes of new physics. 

\section{$T$, $CPT$ violation and related issues}

The $CPT$ theorem \cite{LudersPauli} is based on very general principles of 
quantum field theory, and its validity makes $CP$ violation equivalent to a
violation of microscopic time reversibility in elementary physical processes.
Despite its solid foundations, the validity of $CPT$ symmetry has been 
questioned: the possible breakdown of locality at short distances, \emph{e.g.} 
in string theories, of Lorentz invariance, as implemented in explicit models 
\cite{Kostelecky_CPT}, or of quantum mechanics, would undermine the proof of 
its validity. 

$CPT$ symmetry has been tested experimentally by comparing masses, lifetimes,
electric charges and other static properties of particles and anti-particles
in several systems \cite{PDG_2002}.
A complete discussion of $CPT$ tests would be clearly beyond the scope of this 
work (see \emph{e.g.} \cite{BigiSanda} and references therein), and we will
just briefly comment on some of the methods used to check the validity of 
this symmetry, and on how its violation could affect previous considerations.

As for $CP$ violation, in neutral meson systems one also distinguishes between
indirect $CPT$ violation, in the $\Delta F=2$ transitions, and direct $CPT$ 
violation in the $\Delta F=1$ decay amplitudes. The former would lead to 
flavour components of different magnitudes in the physical states $\kl$ and 
$\ks$, described by two mixing parameters 
\Ea{ 
  \beps_S = \beps + \Delta \quad && \quad \beps_L = \beps - \Delta
}
so that $\Delta \neq 0$ indicates $CPT$ violation in the effective Hamiltonian.

An accurate comparison of charge asymmetries in semi-leptonic decays of \kl 
and \ks could give information on the difference of such mixing parameters:
\Dm{
  \delta_{L,S}^{(l)} = \frac{2\Rea(\beps_{L,S})}{1+|\beps_{L,S}|^2} -
  2\Rea(y) \pm 2 \Rea(x_-)
}
where $y$ and $x_-$ parameterise the violation of $CPT$ in the decay 
amplitudes when the $\Delta S=\Delta Q$ rule is valid or not, respectively 
(the above expression is valid at first order in the small quantities 
$|\epsilon_{S,L}|$, $|y|$, $|x_-|$). Any difference among the two charge 
asymmetries above would indicate $CPT$ violation, either ``indirect'' 
($\Rea(\Delta) \neq 0$) or ``direct'' ($y,x_- \neq 0$).
With the inclusion of the statistics collected until 2002, the statistical 
error on the KLOE measurement of the charge asymmetry in semi-leptonic decays 
of $\ks$ will be below 0.01, but still far from the level required for a
significant quantitative test of $CPT$.

The comparison of the measured value of $\Delta$ with the value of 
$\Rea(\epsilon)$ obtained from $\pi \pi$ decays \cite{PDG_2002} also allows to 
put bounds on $CPT$ violation \cite{KTeV_deltal}, \cite{NA48_deltal}:
\Dm{
  \Rea(y + x_- - a_{CPT}) = \Rea \left( \frac{2}{3} \eta_{+-} + 
    \frac{1}{3} \eta_{00} \right) - \frac{\delta_L}{2} = 
    (5 \pm 32) \cdot 10^{-6}
}
where $a_{CPT}$ parameterises direct $CPT$ violation in $\pi\pi$ decays.

Information on $\beps_S$ could also be extracted from the analysis of the 
flavour-tagged kaon decay asymmetry to the final state $\pi^+ \pi^- \pi^0$ in 
a phase space region symmetric for $\pi^+$ and $\pi^-$, for large times (see
eq. \ref{eq:dilution})
\Dm{
  A_{CP}^{(\pi^+\pi^-\pi^0)}(t \To \infty) = 
  1 - \left| \frac{1-\beps_S}{1+\beps_S} \right|^2
}

By using semi-leptonic decays of flavour-tagged kaons, CPLEAR searched for 
$CPT$-violating asymmetries \cite{CPLEAR_CPT}:
\Ea{ \Ds 
  & A_\Delta(t) = \frac{\overline{N}_+(t)-N_-(t)(1+4\Rea(\beps_L))}
    {\overline{N}_+(t)-N_-(t)(1+4\Rea(\beps_L))} +
    \frac{\overline{N}_-(t)-N_+(t)(1+4\Rea(\beps_L))}
    {\overline{N}_-(t)-N_+(t)(1+4\Rea(\beps_L))} 
\label{eq:Adelta}
}
where $N_\pm$ ($\overline{N}_\pm$) is the number of $\kz$ ($\kzb$) decaying to
$\pi^\pm e^\pm \nu(\overline{\nu})$.
The value of $A_\Delta(t)$ for $t \gg \tau_S$ measures $\Rea(\Delta)$,  
independently from the validity of the $\Delta S = \Delta Q$ rule and from
direct $CPT$ violation, while the study of its time dependence allows to 
extract information on $\Imm(\Delta)$.

The phases of the $CP$ violation parameters in the $\pi \pi$ decay modes 
allows checks of $CPT$ violation, due to the fact that direct $CP$ violation 
is very small.
In the limit in which the $(\pi\pi)_{I=0}$ final state saturates the decay
widths of neutral kaons\footnote{Note that in this limit $\epsilon'=0$, and
$CP$ violation in other decay modes is zero.}, the phase of the $\epsilon$ 
parameter is $\phi_{SW}$ if $CPT$ symmetry is valid, so that the phase 
difference $\phi_{+-}-\phi_{SW}$ checks this symmetry.

A recent test of $CPT$ violation by KTeV \cite{KTeV_eprime2} is the 
measurement of 
\Dm{
  \phi_{+-} - \phi_{SW} = (0.61 \pm 1.19)^\circ
}
obtained with a different fit to the longitudinal decay vertex distribution of 
the 1997 $K_{S,L} \To \pi^+ \pi^-$ data used in the $\epsilon'/\epsilon$ 
analysis. 
The error is dominated by the systematics due to the dependence on the 
geometrical acceptance cuts and on the model dependence of the nuclear 
screening corrections which modify the regeneration amplitude.

\begin{figure}[hbt!]
\begin{center}
    \epsfig{file=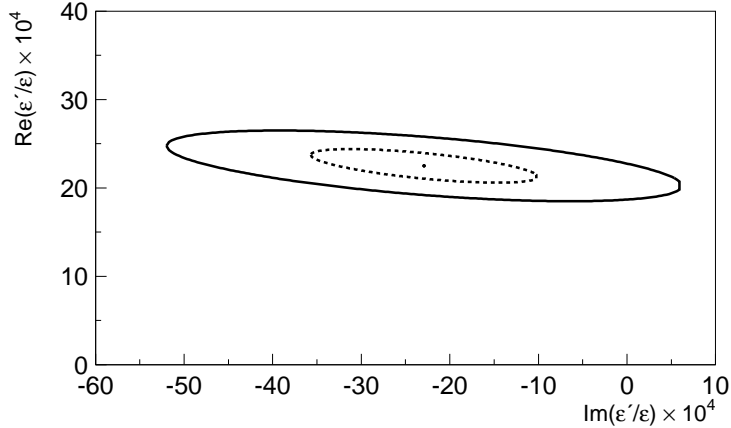,width=0.8\textwidth}
  \caption{KTeV results for $\epseps$ with no $CPT$ assumptions.}
  \label{fig:ktev_epecorr}
\end{center}
\end{figure}

The same experiment also reported a test of direct $CPT$ violation from the 
difference among the phases of $\eta_{00}$ and $\eta_{+-}$:
\Dm{
  \phi_{00}-\phi_{+-} = (0.39 \pm 0.50)^\circ
} 
corresponding to 
\Dm{
  \Imm(\epseps) = (-22.9 \pm 29.1) \cdot 10^{-4}
}
The real part of $\epseps$ extracted from this KTeV analysis, when $CPT$ 
symmetry is not imposed, is
\Dm{
  \Rea(\epseps) = (+22.5 \pm 1.9_{\mathrm{stat}}) \cdot 10^{-4} 
}
where the error is only statistical and larger than in the standard 
analysis, due to the correlation with the imaginary part as shown in fig. 
\ref{fig:ktev_epecorr}.

In the same limit (and at first order in the deviation of the $\phi_{+-},
\phi_{00}$ phases from $\phi_{SW}$), the component of $\Delta$ orthogonal to 
the direction of $\phi_{SW}$ is given by
\Dm{
  \Delta_\perp \simeq |\eta_{+-}| \left( \phi_{SW} - \frac{2}{3} \phi_{+-} -
    \frac{1}{3} \phi_{00} \right) - 
    \frac{\Rea(b_0)}{\Rea(a_0)} \sin(\phi_{SW}) 
}
where the $b_I$ amplitudes measure direct $CPT$ violation in the $K \To 
(\pi\pi)_I$ decay amplitudes:
\Ea{
  & A_I = (a_I + b_I) e^{i\delta_I} \\
  & \overline{A}_I = (a^*_I - b^*_I) e^{i\delta_I}
}
Neglecting $CP$ violation, the $\Delta_\parallel$ component, along the 
direction of $\phi_{SW}$, vanishes at the same level of approximation.
$\Delta_\perp$ and $\Delta_\parallel$ are proportional to the mass and total
decay width difference of $\kz$ and $\kzb$ respectively, so that the limit on
$\Delta_\perp$, \emph{assuming} no direct $CPT$ violation, translates to the 
best limit on $CPT$ symmetry, quoted as\cite{NA48_3pi}
\Ea{ \Ds
  & m(\kz)-m(\kzb) \simeq \frac{2 \Delta m}{\sin(\phi_{SW})} |\eta_{+-}| 
    \left(\phi_{SW} -\frac{2}{3} \phi_{+-} -\frac{1}{3} \phi_{00} \right) 
    = \nonumber \\
  & (-1.7 \pm 4.2) \cdot 10^{-19} \textrm{\, GeV/$c^2$}
}

Analysis based on the constraint obtained from the Bell-Steinberger relation 
\cite{Bell-Steinberger}, which is a consequence of unitarity relating the 
amplitudes for all neutral kaon decay modes to the parameters describing $CP$ 
violation in the effective Hamiltonian for the $\kz--\kzb$ system, allow the 
extraction of $CP$ and $CPT$ violation parameters in the effective Hamiltonian:
\Dm{
  [1+i\tan(\phi_{SW})] [\Rea(\beps)-i\Imm(\Delta)] = \sum_f \alpha_f
}
in which $\alpha_f = (1/\Gamma_S) A^*(\ks \To f) A(\kl \To f) \propto \eta_f 
BR(\ks \To f)$, and the sum runs over all decay channels $f$.
The sum is dominated by the $\pi \pi$ channels, since the hadronic decay modes 
allowed for the \kl in absence of $CP$ violation are suppressed by the small 
$\tau_S/\tau_L$ ratio. 
With this approach $\Imm(\Delta)$ can be obtained with a higher precision than 
that obtained by the $A_\Delta(t)$ measurement, without any dependence on the
amount of direct $CPT$ violation in the dominant decay to $(\pi\pi)_{I=0}$.

The real and imaginary parts of $\Delta$ in the neutral kaon system are 
measured to be zero with an accuracy of $\sim 10^{-4}$ for the real part and 
$\sim 10^{-3}$ for the imaginary part.

It should be noted that, if $CPT$ violation arises through a violation of 
Lorentz invariance, such as can be the case in quantum field theory, the 
parameter $\Delta$ cannot be a constant but must depend on the four-momentum 
\cite{Kostelecky_delta}; moreover, relations between the real and imaginary 
part of $\Delta$ are obtained in explicit models.
A search for a dependence of the phase $\phi_{+-}$ in $K \To \pi^+ \pi^-$ 
decays on the absolute direction of the beam in space, which would appear in a 
fit of such asymmetry as a function of sidereal time due to the Earth's 
rotation, was performed \cite{KTeV_sidereal} with null results.

A thorough (although quite dated) analysis of $CPT$ tests in the neutral kaon
system can be found in \cite{Barmin}. It is important to note that in presence 
of $CPT$ violation, new terms can arise in the expressions for the 
experimental observables which are used to study $CP$ violation. 
As an example, the expressions for the $\epsilon, \epsilon'$ parameters 
parameterising $CP$ violation in $K \To \pi\pi$ decays become
\Ea{ \Ds
  & \epsilon = \beps + i\frac{\Imm(a_0)}{\Rea(a_0)} - \Delta + 
    \frac{\Rea(b_0)}{\Rea(a_0)} \\ 
  & \epsilon' = i \frac{\omega}{\sqrt{2}} 
    \left[ \frac{\Imm(a_2)}{\Rea(a_2)} - \frac{\Imm(a_0)}{\Rea(a_0)} - 
    i \left( \frac{\Rea(b_2)}{\Rea(a_2)} - 
    \frac{\Rea(b_0)}{\Rea(a_0)} \right) \right] 
}
so that in $\epsilon$ both direct and indirect $CPT$ violation terms appear, 
and in $\epsilon'$ a term of direct $CPT$ violation is present which is 
orthogonal to the one describing direct $CP$ violation; after having factored 
out the strong phase shifts in $\omega$, this term is real since it violates 
$CPT$ and $CP$ but not T.

The high-precision study of time-dependent or time-integrated asymmetries for 
pairs of correlated or uncorrelated neutral mesons can provide measurements of 
$CPT$ violating parameters (see \emph{e.g.} \cite{Coerenti3}, 
\cite{Dambrosio_dafnehb}). 
The charge asymmetries of same-sign and opposite-sign di-lepton pairs from 
the decays of ($C$-odd) correlated mesons at $\phi$ or $B$ factories are 
sensitive to $CPT$ violation:
\Dm{
  A_{CPT} = \frac{N_{-+}-N_{+-}}{N_{-+}+N_{+-}} \propto
    \frac{|A(M \To M)|^2-|A(\overline{M} \To \overline{M})|^2}
    {|A(M \To M)|^2+|A(\overline{M} \To \overline{M})|^2} 
\label{eq:ACPT}
}
where $N_{+-}$ ($N_{-+}$) are  the number of di-lepton events in which the 
positive (negative) lepton is emitted before the oppositely charged one
(the allowed decays being $M \To l^+ X$ and $\overline{M} \To l^- X$).
$A_{CPT}$ is non-zero in presence of either $CPT$ violation in the effective 
Hamiltonian ($\Delta \neq 0$) or violation of the $\Delta F = \Delta Q$ rule; 
when the latter is absent the asymmetry is 
\Dm{
  A_{CPT} = -4 \Rea(\Delta) -8 \Imm(\Delta) 
    \frac{\Delta m \Gamma_L}{\overline{\Gamma}^2 + (\Delta m)^2}
}

Limits on the validity of $CPT$ symmetry in $B$ mesons have been obtained from 
inclusive lepton decays of neutral $B$ mesons produced at the $Z$ centre of 
mass energy \cite{OPAL_CPTB}, and with the analysis of the time dependence of 
di-lepton yields for the decays of $\bz \bzb$ pairs produced by $\Upsilon(4S)$ 
(\cite{Belle_dilepton}, \cite{BABAR_CPT} and references therein). 
These tests have not yet reached the level of precision available in the $K$
system: the $\Delta$ parameter is known to be zero with a precision of $\sim
10^{-1}$ for the real part and $\sim 5 \cdot 10^{-2}$ for the imaginary part.

Many tests of the $CPT$-enforced equalities of particle and antiparticle 
properties have been performed, with relative precisions which range from a
few per mille to $\sim 10^{-10}$ for the charge/mass ratio of proton and 
antiprotons; however, in absence of specific $CPT$ violation models, the 
accuracy of the measurements cannot be translated to a level to which the 
symmetry itself is tested to be valid.

As far as $CPT$ symmetry is valid, $CP$ violation is equivalent to $T$ 
violation.
Some of the measurements discussed in the previous sections, however, are 
actual tests of $T$ symmetry, such as those related to transverse polarization 
measurements.
The only positive measurement of $T$ violation to date is that obtained by the
CPLEAR experiment \cite{CPLEAR_CPT} by measuring the asymmetry 
\Dm{
  A_T(t) = \frac{\overline{N}_+(t)-N_-(t)}{\overline{N}_+(t)+N_-(t)}
} 
for flavour tagged neutral kaon decays in the semi-leptonic mode (with the 
same notation used in the definition of $A_\Delta$, eq. (\ref{eq:Adelta})).
The value of this asymmetry for $t \gg \tau_S$ measures time-reversal 
violation if $CPT$ symmetry holds, independently from the validity of the 
$\Delta S = \Delta Q$ rule, but can also be non-zero if $CPT$ (and $CP$) is 
violated and $T$ holds.

The charge asymmetries of same-sign di-lepton pairs from correlated meson 
pairs can also test time reversal symmetry: 
\Dm{
  A_T = \frac{N_{++}-N_{--}}{N_{++}+N_{--}} \propto 
    \frac{|A(\overline{M} \To M)|^2-|A(M \To \overline{M})|^2}
    {|A(\overline{M} \To M)|^2+|A(M \To \overline{M})|^2} \\
}
with the same notation used above in (\ref{eq:ACPT}). This asymmetry is 
non-zero in presence of either $T$ violation or $CPT$ violation in the decay 
amplitudes. 

A long-standing unresolved question, whose discussion is out of the scope 
of this article, is that of the so-called ``strong $CP$ problem'' (see 
\emph{e.g.} \cite{Peccei}): a free $CP$-violating parameter of the Standard 
Model is experimentally constrained to have an ``unnatural'' (and unstable) 
tiny value. Such parameter could induce large $CP$- (and $P$-) violating 
flavour-conserving effects, but the stringent limits on the EDMs of particles 
show that this is not the case. 

Another of the outstanding problems of physics, which is directly related to 
the issue of direct $CP$ violation, is that of explaining the baryon asymmetry 
of the universe.
It is well known that the three necessary ingredients to generate such 
asymmetry are baryon number violation, $C$ and $CP$ violation, and departure 
from thermal equilibrium \cite{Sakharov}. 
The Standard Model contains all such ingredients, but by now it has been made 
clear that it cannot account for the observed value of the baryon asymmetry: 
the mechanism of $CP$ violation linked to the flavour Yukawa couplings is far 
too small to explain the presently observed ratio of matter to radiation in 
the framework of electroweak baryogenesis (see \emph{e.g.} \cite{Huet}).
It is therefore clear that new, and so far unknown, features are essential to 
explain the universe as we observe it, so that also new mechanisms of $CP$ 
violation are expected to exist in a fundamental theory, of which the Standard 
Model is thought to be the low-energy limit. 
Indeed many extensions of the known theory include several new ineliminable 
phases, often even too many, so that some fine tuning of the parameters is 
required for the theoretical predictions to be consistent with the observed 
$CP$ violation.

The above considerations show that the phenomenon of $CP$ violation is very 
likely to be of fundamental importance even at higher energy scales, in 
relation to the search for a more comprehensive theory of microscopic 
phenomena.

\section{Conclusions and Perspectives}

For 37 years the phenomenon of $CP$ violation remained experimentally confined 
to the neutral kaon system, and for a similar period of time it could be 
described by a single parameter, related to a property of the decaying state,
\emph{i.e.} $\kz--\kzb$  mixing, or indirect $CP$ violation. 
The lack of other measurements of $CP$ violation could raise the suspicion of 
it being a peculiar feature of the neutral kaon system, possibly through a 
new, unmeasurably small, fundamental interaction.

In this review, we tried to give an overview of the long path of experimental 
investigations which tried to shed some light on this fundamental issue, and 
which still continue today, with great dedication and intensity.

Today, with the experimental proof of the existence of direct $CP$ violation in
the decays of neutral kaons, and the first measurements of $CP$ violation in
the neutral $B$ meson system, our knowledge has been remarkably improved, 
although this fundamental property of Nature still remains much of a mystery, 
only being detected via neutral meson decays and nowhere else.

The deep mystery shrouding the nature of $CP$ violation can only push 
physicists to increased efforts, both in theory and experiment, aimed at 
reaching a better understanding, with the additional expectation that Nature 
will disclose to us more of her beautifully concealed secrets along the way.

\end{document}